%% file: arxiv_version_with_supplementary_material.tex
\documentclass[twoside,leqno,twocolumn]{article}

\usepackage[letterpaper]{geometry}

\usepackage{ltexpprt}
\usepackage{hyperref}

\usepackage{xurl}
\usepackage{xcolor}

\usepackage{amssymb, amsmath, amsfonts}
\usepackage{mathtools}
\usepackage{caption, subcaption}
\usepackage{tikz}
\usepackage{multirow}
\usetikzlibrary{decorations}
\usetikzlibrary{decorations.pathreplacing}
\usetikzlibrary{positioning}

\DeclareMathOperator*{\argmax}{arg\,max}
\DeclareMathOperator{\len}{len}
\newtheorem{defn}{Definition}[section]

\definecolor{grey0}{gray}{.95}
\definecolor{grey1}{gray}{0.75}
\definecolor{grey2}{gray}{0.6}
\definecolor{nearlywhite}{gray}{0.99}
\definecolor{agreen}{rgb}{.55,.71,0}
\definecolor{ForestGreen}{rgb}{0.13, 0.55, 0.13}
\definecolor{lookslikewhite}{gray}{1.0}
\definecolor{bluegreen}{rgb}{0.0,0.5,0.5}
\definecolor{sowaswieweiss}{gray}{1.0}
\definecolor{adarkergrey}{gray}{.6}
\definecolor{alightergrey}{gray}{.8}

\newcommand{\Histogram}{0.9\linewidth}

\newcommand{\MethodologyI}{.52}
\newcommand{\MethodologyII}{.42}
\newcommand{\radone}{2}
\newcommand{\radtwo}{.7}

\newcommand{\reb}[1]{\textcolor{black}{#1}}

\newcommand{\Ellipse}{\draw[black,very thick,fill=black] (1.59,0) ellipse (.3cm and .1cm)}
\usepackage{algorithm}
\usepackage[noend]{algpseudocode}
\usepackage{colortbl}
\usepackage{paralist}

\usepackage[normalem]{ulem}
\usepackage{soul}

\usepackage{pifont}
\newcommand{\yes}{\ding{51}}%
\newcommand{\no}{\ding{55}}%

\begin{document}

\newcommand\relatedversion{}
\renewcommand\relatedversion{\thanks{The full version of the paper can be accessed at \protect\url{https://arxiv.org/abs/2305.07570}}} 

\title{\Large Feature-aware Manifold Meshing and Remeshing of Point Clouds and Polyhedral Surfaces with Guaranteed Smallest Edge Length}
\author{
	Henriette Lipsch\"utz\thanks{Freie Universit\"at Berlin, Germany}
	\and 
	Ulrich Reitebuch\thanks{Freie Universit\"at Berlin, Germany}
	\and
	Konrad Polthier\thanks{Freie Universit\"at Berlin, Germany}
	\and
	Martin Skrodzki\thanks{TU Delft, The Netherlands; corresponding author: \href{mailto:mail@ms-math-computer.science}{mail@ms-math-computer.science}; supported by the Deutsche Forschungsgemeinschaft (DFG, German Research Foundation) -- 455095046.}
}

\date{}

\maketitle


\fancyfoot[R]{\scriptsize{Copyright \textcopyright\ 2025 by SIAM\\
		Unauthorized reproduction of this article is prohibited}}





\begin{abstract} \small\baselineskip=9pt 		
	Point clouds and polygonal meshes are widely used when modeling real-world scenarios.
	Here, point clouds arise, for instance, from acquisition processes applied in various surroundings, such as reverse engineering, rapid prototyping, or cultural preservation.
	Based on these raw data, polygonal meshes are created to, for example, run various simulations.
	For such applications, the utilized meshes must be of high quality.
	This paper presents an algorithm to derive triangle meshes from unstructured point clouds.
	The occurring edges have a close to uniform length and their lengths are bounded from below.
	Theoretical results guarantee the output to be manifold, provided suitable input and parameter choices.
	Further, the paper presents several experiments establishing that the algorithms can compete with widely used competitors in terms of quality of the output and timing and the output is stable under moderate levels of noise.
	Additionally, we expand the algorithm to detect and respect features on point clouds as well as to remesh polyhedral surfaces, possibly with features.
	
	Supplementary material, an extended preprint, a link to a previously published version of the article, utilized models, and implementation details are made 	\href{https://graphics.tudelft.nl/guaranteed-smallest-edge-length-manifold-meshing}{available online}.
\end{abstract}

\section{Introduction}
\label{sec:IntroductionGeneral}

\noindent Point cloud meshing is an important topic present in different fields of research and in various applications.
Examples include reverse engineering~\cite{helle2021case}, rapid prototyping~\cite{chaudhari2022medical}, or architecture~\cite{3dard_2019}.
A common approach to enable this raw data for further processing is to create a triangle mesh from the point cloud.
The quality of this mesh is, however, affected by outliers, noise, or non-uniform distribution of the input data.
Thus, badly formed mesh elements can become apparent in the resulting geometric model.
They can be long-stretched, thin triangles, so-called slivers, or topological issues.
These faulty representations have to be repaired before the meshes are further processed.

While this issue is rather general and inherent to the workflow, recent research still struggles to circumvent it.
Even when reducing to only a local mesh representation of a given geometry, established methods, such as Delaunay triangulations, do not guarantee to create a manifold mesh of well-shaped triangles~\cite[Sec.~4.4]{wiersma2023fast}.
The present paper aims to close this gap.

We aim to reconstruct a surface from a given point cloud via a sphere-packing approach~\cite{lipschutz2021single}.
The goal is to create a manifold output with guaranteed smallest edge length and with strong consideration of triangle quality provided by a distribution close to uniformity of edge lengths.
Furthermore, as opposed to other meshing approaches, our algorithm works directly on the surface geometry, that is, does not need any parametrization.
Finally, the algorithm performs a greedy disk-growing approach, which enables the processing of the geometry in one pass, making further iterations unnecessary.

A first version of this algorithm has been presented at the 2024 International Meshing Roundtable~\cite{lipschutz2024manifold}.
The contributions of the original article included:
\begin{compactitem}
	\item introduction of a geometric approach suitable to mesh point clouds,
	\item which creates high-quality triangles with edge lengths close to uniformity and of a guaranteed minimum length,
	\item as well as manifold output, provided a suitable input geometry,
	\item in a single sweep over said input,
\end{compactitem}
discussed in Sections~\ref{sec:TheoryAndMethodology} to~\ref{sec:Experiments}.
In this extended version of the article, we build upon the previous contributions and extend the algorithm to handle:
\begin{compactitem}
	\item detection of sharp feature ridges in point clouds,
	\item remeshing of polyhedral surfaces obtaining high-quality meshes with edge lengths close to uniformity,
	\item detection of sharp feature ridges on polyhedral meshes,
\end{compactitem}
presented in Sections~\ref{sec:ExtensionToSurfaceRemeshing} to~\ref{sec:ExperimentsOnPolyhedralSurfaces}.

\section{Related Work}
\label{sec:RelatedWork}

\noindent In the last decades, several attempts were made to reconstruct the ground truth from a given point cloud~$\mathcal{P}$.
The resulting reconstruction depends on the quality of~$\mathcal{P}$, which can include noisy points or normals, outliers, or be sampled non-uniformly.
On top of a reconstruction, the user may ask for guarantees such as correct topology~\cite{Amenta2001power}, or convergence to the ground truth with increasing sampling density~\cite{levin1998approximation}.
Some algorithms guarantee local connectedness of their output~\cite{bernardini1999ball}, while others guarantee their output to stay within the convex hull of the given input~\cite{edelsbrunner1983alpha}.
Other requirements might be, for instance, a result mesh of high quality, that is, consisting of triangles with edge length close to uniformity and vertices of degree close to~$6$.
Finally, the reconstruction should be computed fast.
For an overview of surface reconstruction algorithms, we refer to a recent survey~\cite{huang2022surface}.
The algorithms discussed in the following were chosen for their wide use in the field and will serve as a comparison in Section~\ref{sec:Experiments}.

First, we consider surface reconstruction based on a Poisson equation~\cite{kazhdan2006poisson}, implemented in CGAL~\cite{CGAL2023}.
An implicit function framework is built, where the reconstructed surface appears by extracting an appropriate isosurface.
The output is smooth and robustly approximates noisy data.
Additionally, densely sampled regions allow the reconstruction of sharp features while sparsely sampled regions are smoothly reconstructed.
In later work, these ideas are further developed to create watertight meshes fitting an oriented point cloud by using adaptive, finite elements multi-grid solvers capable of solving a linear system discretized over a spatial domain~\cite{kazhdan2019multi}, implemented in MeshLab~\cite{cignoni2008meshlab}.

Second, the scale-space approach~\cite{CohenSteiner2004AGD}, implemented in CGAL~\cite{CGAL2023}, aims at topological correctness by choosing triangles based on a confidence-based criterion. 
This avoids the accumulation of errors, which is often detected in greedy approaches.
The algorithm is interpolating, and can handle sharp features to a certain extent, but does not come with proven topological correctness.

The advancing front algorithm~\cite{digne2011scale}, implemented in CGAL~\cite{CGAL2023}, handles sets of unorganized points without normal information.
It computes a normal field and meshes the complete point cloud directly, which leads to a high-level reconstruction of details as well as to an accurate delineation of holes in the ground truth.
Therefore, a smoothing operator consistent with the intrinsic heat equation is introduced.
By construction, this approach is almost interpolating and features are preserved given very low levels of noise.

The robust implicit moving least squares (RIMLS) algorithm~\cite{oztireli2009feature}, implemented in MeshLab~\cite{cignoni2008meshlab}, combines implicit MLS with robust statistics.
The MLS approach~\cite{levin1998approximation} is a widely used tool for functional approximation of irregular data.
The development of RIMLS is based on a surface definition formulated in terms of linear kernel regression minimization using a robust objective function which gives a simple and technically sound implicit formulation of the surface.
Thus, RIMLS can handle noisy data, outliers, and sparse sampling, and can reconstruct sharp features.
The number of iterations needed to achieve a reliable result increases near sharp features while smooth regions only need a single iteration.
Furthermore, RIMLS belongs to the set of algorithms producing approximating meshes.

Another approach is based on placing triangles with regard to the restricted Voronoi diagram of a filtered input point set~\cite{boltcheva2017surface}, available via MeshLab~\cite{cignoni2008meshlab}.
This approach has the largest similarity to our algorithm, as we will also employ Voronoi diagrams, however, only to filter points on the tangent plane.
Another shared aspect is that both this and our algorithm work on a set of disks centered at the input points, oriented orthogonal to a guessed or provided normal direction.

All these algorithms come with different guarantees regarding the output.
However, none of these algorithms comes with a guarantee on the edge length, and only some algorithms are guaranteed to provide a manifold mesh.
As we base our surface reconstruction on a set of touching spheres placed on the underlying surface~\cite{lipschutz2021single}, we are able to provide certain theoretical guarantees on the output: given suitable input and parameter choices, our output is always manifold.
Furthermore, the output of our algorithm has a guaranteed minimum edge length, while striving towards uniformity of occurring edge lengths.
For better comparison to other algorithms, we also employ the ``Isotropic Explicit Remeshing'' filter of MeshLab~\cite{cignoni2008meshlab}.
This filter repeatedly applies edge flip, collapse, relax, and refine operations~\cite{hoppe1993Mesh}.
For remeshing polyhedral surfaces, we also make use of the ``Remeshing'' routine provided by the Polygon Mesh Processing (PMP) library~\cite{pmp23}.
Here, the triangle quality is improved by local modifications such as splitting and collapsing of edges and tangential smoothing~\cite{botsch2004remeshing}.
Similar algorithms exist that are based on local operations like edge flips and vertex relocations to achieve anisotropic triangular meshes~\cite{surazhsky2003explicit}.
As opposed to our approach, in their work, the resulting edge lengths depend on the curvature of the geometry remeshed.
Additionally, all of the works listed above require several iterations to obtain an isotropic triangulation while our approach works in a single sweep over the input geometry.
In Sections~\ref{sec:ExperimentalResultsForPointCloudMeshing},~\ref{sec:ExperimentsAndEvaluation_PointClouds}, and~\ref{sec:ExperimentsOnPolyhedralSurfaces}, we will provide a detailed comparison of our algorithm with the works listed here.

When it comes to feature-aware (re)meshing, as a comparison metric between the obtained meshes, we will employ the Hausdorff distance.
This allows us to compare the input, be it a point cloud or a surface mesh, with the (re)meshed version.
An estimate of the one-sided Hausdorff distance is computed by sampling the input and projecting the samples onto the output~\cite{cignoni1998metro}.
A variety of methods has been proposed to identify features on point clouds and meshes.
We follow a basic approach using the variation of normals of two elements~\cite{skrodzki2021large}, allowing us to specify a critical dihedral angle between two normals from which a feature between the two is respected during remeshing.
This follows similar approaches in point cloud and surface smoothing~\cite{yadav2018constraint,yadav2022surface}.

\section{Theory and Methodology}
\label{sec:TheoryAndMethodology}

\noindent Our algorithm aims to reconstruct a manifold~$\mathcal{M}$ from a given point cloud~$\mathcal{P}$.
In order to obtain a manifold mesh with a guaranteed minimum edge length, we first present assumptions and theoretical results on both~$\mathcal{M}$ and~$\mathcal{P}$ in Section~\ref{sec:AssumptionsAndTheory}.
Based on these theoretical results, we present a geometric approach in Section~\ref{sec:Methodology}, which consists of creating a sphere packing from which the output is constructed.
Sections~\ref{sec:Initialization} to~\ref{sec:TriangulatingTheResultingMesh} are devoted to explaining the different steps in detail.

\subsection{Assumptions and Theory}
\label{sec:AssumptionsAndTheory}

\begin{figure}
	\includegraphics[width=\columnwidth]{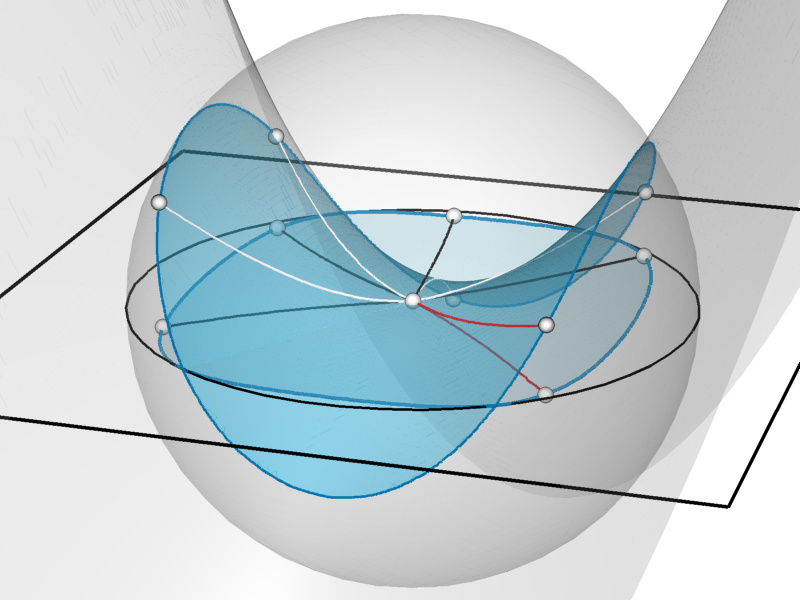}
	\caption{Illustration of Lemma~\ref{lem:projectionConvex}. Intersection of a saddle-shaped surface with a sphere (blue). Six points on the surface are marked as well as their projections to the tangent plane belonging to the center of the sphere.}
	\label{fig:manifoldExplanation}
\end{figure}

\noindent Here, we will derive the assumptions to be made on~$\mathcal{M}$ and~$\mathcal{P}$ to ensure that the constructed surface mesh is manifold.
Let~$\mathcal{M}$ be an orientable, compact $\mathcal{C}^2$-manifold embedded into~$\mathbb{R}^3$, which is assumed to be closed and of finite reach
\begin{align*}
	\rho \coloneqq \inf \left\lbrace \Vert a - m \Vert \mid a \in \mathcal{A}_{\mathcal{M}}\,\land\,m \in \mathcal{M} \right\rbrace \in \mathbb{R}_{>0},
\end{align*}
where $\mathcal{A}_{\mathcal{M}}$ is the \emph{medial axis} of~$\mathcal{M}$ consisting of the points~$q \in \mathbb{R}^3$ satisfying
\begin{align*}
	\min_{p \in \mathcal{M}} \vert q - p \vert = \vert q - \hat{p} \vert = \vert q - \tilde{p} \vert
\end{align*}
for $\hat{p} \neq \tilde{p} \in \mathcal{M}$.
On the manifold~$\mathcal{M}$, we define the \emph{geodesic distance}~$d_{\mathcal{M}}$ as follows:
\begin{defn}
	Let~${\len(f)=\int_0^1\left|f'(t)\right|\:dt}$ denote the length of a curve $f \in \mathcal{C}^1(\lbrack 0,1 \rbrack, \mathcal{M})$ in $\mathcal{M}$. 
	Then the \emph{geodesic} \emph{distance} $d_{\mathcal{M}}$ of $m, m' \in \mathcal{M}$ is defined as	$\inf \left\lbrace \len(f) \mid f \in \mathcal{C}^{1}(\lbrack 0,1 \rbrack, \mathcal{M}): f(0) = m \land f(1) = m' \right\rbrace$.
\end{defn}
Now, for any $p,q \in \mathcal{M}$ such that $\Vert p - q \Vert < 2\rho$, the following estimation holds~\cite[Lemma~3]{boissonnat2019reach}:
\begin{align}
	\label{eq:upperBoundGeodesicDist}
	\Vert p - q \Vert \leq d_{\mathcal{M}}(p,q) \leq 2\rho \arcsin\left( \frac{\Vert p - q \Vert}{2\rho} \right).
\end{align}
Let $T_p\mathcal{M}$ and $T_q\mathcal{M}$ denote the tangent planes at~${p,q \in \mathcal{M}}$.
Lemma 6 in~\cite{boissonnat2019reach} gives an upper bound for the angle~$\sphericalangle\left( T_p\mathcal{M}, T_q\mathcal{M} \right)$ between them,
\begin{align}\label{eq:angleTangentPlanes}
	\sphericalangle\left( T_p\mathcal{M}, T_q\mathcal{M} \right) \leq \frac{d_{\mathcal{M}} (p,q)}{\rho}.
\end{align}
Hence, Inequalities~(\ref{eq:upperBoundGeodesicDist}) and~(\ref{eq:angleTangentPlanes}) imply for the normal vectors $n_p$ at $p$ and $n_q$ at $q$ that
\begin{align}
	\label{equ:AngleInequality}
	\begin{split}
		&\varphi \coloneqq \sphericalangle \left(n_p,n_q \right) \leq 2 \arcsin\left( \frac{\Vert p - q \Vert}{2 \rho} \right)\\
		\Rightarrow &\Vert p - q \Vert \geq r(\varphi) \coloneqq 2 \rho \sin \left( \frac{\varphi}{2} \right).
	\end{split}
\end{align}
For a given angle $\varphi_{\max} \in \left\lbrack 0, \tfrac{\pi}{2} \right\lbrack$, the second part of Equation~(\ref{equ:AngleInequality}) implies that there is a constant $r_{\max} \in \mathbb{R}_{\geq 0}$ such that $\varphi \leq \varphi_{\max}$ if $\Vert p - q \Vert < r_{\max}$. 
Denote by $\mathcal{M}' \coloneqq B_{r_{\max}}(p) \cap \mathcal{M}$ the part of $\mathcal{M}$ that is contained in $\mathcal{M}$ and the ball $B_{r_{\max}}(p)$ centered at~$p$.
Then, the normals~$\reb{n}_q$ of all points $q\in B_{r_{\max}}$ have positive Euclidean scalar product with $\reb{n}_p$.
The assumption that $\mathcal{M}$ is of positive finite reach guarantees that $\mathcal{M}'$ is a single connected component. 
Hence, $\mathcal{M}'$ has a parallel projection to the tangent plane $T_p\mathcal{M}$ without over-folds (Figure~\ref{fig:manifoldExplanation}).
Furthermore, we have:
\begin{lemma}\label{lem:projectionConvex}
	Let $p \in \mathcal{M}$ be a point with normal $\reb{n}_p$. Then, for $r < \rho$, the image of $B_r(p) \cap \mathcal{M}$ under the projection $\pi$ in direction of $\reb{n}_p$ to the tangent plane $T_p \mathcal{M}$ is a convex set.
\end{lemma}

\begin{figure*}
	\centering
	\begin{subfigure}[t]{.24\linewidth}
		\centering
		\scalebox{\MethodologyI}{\input{SplatsOfSizeD}}
		\caption{Splats of radius $s$.}
		\label{fig:SplatsOfSizeD}
	\end{subfigure}
	\begin{subfigure}[t]{0.24\linewidth}
		\centering
		\scalebox{\MethodologyI}{\input{ProjectionToTangentPlane}}
		\caption{Projection to $T_p\mathcal{M}$.}
		\label{fig:ProjectionToTangentPlane}
	\end{subfigure}
	\begin{subfigure}[t]{0.24\linewidth}
		\centering
		\scalebox{\MethodologyI}{\input{VoronoiCells}}
		\caption{Voronoi cells in $T_p\mathcal{M}$.}
		\label{fig:VoronoiCells}
	\end{subfigure}
	\begin{subfigure}[t]{.24\linewidth}
		\centering
		\scalebox{\MethodologyI}{\input{IndividualSplatSizes}}
		\caption{Individual splat radii.}
		\label{fig:IndividualSplatSizes}
	\end{subfigure}
	\caption{
		Illustration of uniform splat size (\ref{fig:SplatsOfSizeD}), the projection of a point $p$ and its vicinity to the tangent plane~$T_p\mathcal{M}$ (\ref{fig:ProjectionToTangentPlane}), and the Voronoi cells with the farthest point circled (\ref{fig:VoronoiCells}), leading to individual splat sizes (\ref{fig:IndividualSplatSizes}).
	}
	\label{fig:MethodologyI}
\end{figure*}
\begin{proof}
	The intersection $\mathcal{I}$ of a closed set $\mathcal{S} \subset \mathbb{R}^d$ having reach $\rho_{\mathcal{S}} > 0$ with a closed ball $B_r(x)$, $r < \rho_{\mathcal{S}}$ and $x \in \mathbb{R}^d$, is geodesically convex in $\mathcal{S}$~\cite[Corollary~1]{boissonnat2019reach}.
	That is, the shortest path between any two points in $\mathcal{I}$ lies itself in the intersection.
	Furthermore, the intersection $\mathcal{M}'$ of $B_{r_{\max}}$ and $\mathcal{M}$ is a topological disk as established above~\cite[Proposition~1]{boissonnat2019reach}.
	
	Here, $\mathcal{I}$ is not empty and consists of a surface patch since $p$ lies on $\mathcal{M}$. 
	Hence, the boundary $\partial \mathcal{M}'$ can be parameterized by a closed curve $\gamma$.
	As $\mathcal{I}$ is geodesically convex, $\gamma$ has positive geodesic curvature.
	The inner product of the normals $\reb{n}_p$ and $\reb{n}_q$ at an arbitrarily chosen point $q \in \partial \mathcal{I}$ is positive: $\langle \reb{n}_p, \reb{n}_q \rangle > 0$, by choice of $r$.
	Therefore, under projection along $N_p$ to the tangent plane $T_p\mathcal{M}$, the sign of curvature is preserved. 
	Hence, the projection $\pi(\gamma)$ is a convex curve.
\end{proof}

Finally, let~$\mathcal{G}$ be a simple graph that can be embedded on~$\mathcal{M}$ such that the vertices in~$\mathcal{G}$ connected by an edge have euclidean distance~$d$.
The connected components remaining after removing~$\mathcal{G}$ from~$\mathcal{M}$ are called \emph{regions}, denoted by~$\mathcal{R}$.
The set of vertices and edges incident to a region~$R\in\mathcal{R}$ is called its \emph{border}, denoted by~$\partial R$.
Note that because~$\mathcal{M}$ is closed, each edge of~$\mathcal{G}$ belongs to the border of exactly two regions.
Also, each vertex of~$\mathcal{G}$ can belong to the borders of several regions at once.
Fix one such region~$R\in\mathcal{R}$.
Lemma~\ref{lem:projectionConvex} implies a choice of points~$q_1, \dotsc, q_k \in \partial R$ is mapped to points~$\pi(q_1), \dotsc, \pi(q_k) \in T_p\mathcal{M}$ in cyclic order, for~$p \in R$ arbitrarily chosen.
Hence, the regions can be extracted correctly with respect to their topology from the cyclic order of the edges at each vertex from the local projection.
Given that the reach criterion is satisfied and given a suitable normal field, we can thus reconstruct a manifold from the input.

\subsection{Methodology}
\label{sec:Methodology}

\noindent Our algorithm extracts a mesh from the input~$\mathcal{P}$ by placing touching spheres of a predefined diameter~$d$ across an approximation of the surface~\cite{lipschutz2021single}.
As the spheres touch, this will guarantee a minimum edge length of the mesh and by the results from Section~\ref{sec:TheoryAndMethodology}, the resulting mesh will be manifold.

We assume to be given unstructured input in form of a point cloud~${\mathcal{P}=\{p_i\mid i=1,\ldots,n\} \subset \mathbb{R}^3}$ with corresponding normals~${\mathcal{N}=\{\reb{n}_{p_i}\mid i=1,\ldots,n\} \subset \mathbb{S}^2}$.
Furthermore, we assume that~$\mathcal{P}$ is sampling an underlying, possibly itself unknown, manifold $\mathcal{M}$ with the properties as listed above.
To approximate the surface, we associate to each point~$p \in \mathcal{P}$ a \emph{splat}~$S_p$ in the shape of a circular disk with radius~$s_p \in \mathbb{R}_{>0}$.
Each~$S_p$ is centered at the respective point~$p$ and placed such that the corresponding normal~$\reb{n}_p$ is orthogonal to~$S_p$.
We assume the radii~$s_p$ to be chosen sufficiently large such that the manifold to be reconstructed is covered.
Here, a manifold is said to be \emph{covered}, if the union of projections of splats to the ground truth covers it.
This is illustrated in Figure~\ref{fig:SplatsOfSizeD}.

\begin{figure*}
	\centering
	\begin{subfigure}[t]{.58\linewidth}
		\centering
		\scalebox{0.72}{\input{SplitBorder_AddInnerVertex}}
		\hspace{0.2cm}
		\scalebox{0.72}{\input{SplitBorder_AddOuterVertex}}
		\caption{Connecting~$v$ by two edges to the same border.}
		\label{fig:borderConnectionsCase1}
	\end{subfigure}
	\begin{subfigure}[t]{.41\linewidth}
		\centering
		\scalebox{0.65}{\input{countryJoin}}
		\caption{Connecting~$v$ to two different borders.}
		\label{fig:borderConectionsCase2}
	\end{subfigure}
	\caption{
		Possibilities when creating new vertices and edge connections in the graph~$\mathcal{G}$: 
		In~\ref{fig:borderConnectionsCase1}, the new vertex~$v$ and its two edges connect elements of the same border.
		Here,~$v$ is created either in the in- or the outside region of the border. 
		In~\ref{fig:borderConectionsCase2}, the new vertex~$v$ is connecting two borders. 
		After introducing~$v$ and its edges, the respective outside regions are still connected, then~$v$ and its edges join  the borders.
		However, if the outside regions are split by~$v$ and its edges, new borders are created which induce the corresponding regions.
	}
	\label{fig:borderConnections}
\end{figure*}
Note that following Lemma~\ref{lem:projectionConvex}, the user has to choose the parameter~$d$ with respect to the reach~\reb{$\rho$} of the input, which can be estimated for point clouds~\cite{huang2013l1}, to ensure a manifold output.
In this sense, choosing~$d$ is always a model-dependent choice of the user. 
If the model has, for instance, been scanned and the user has physical access to the model, a suitable value of~$d$ can be estimated based on the narrowest parts of the model.
Also, the overall point distance in the input point cloud can serve as a means to approach a suitable value of~$d$ from below, for instance by taking the smallest point distance (or a small multiple of it) as value~$d$.
This results in an extremely fine-grained output, which might not be supported by the user's machine memory.
Ultimately, it depends on the use-case scenario of the user how fine-grained they want the output.
In Section~\ref{sec:AuxiliaryBoxDataStructure}, we will discuss a heuristic to iteratively estimate a suitable value for~$d$ from above, aiming for a coarse output that still satisfies the requirements formulated above.

Furthermore, the user also chooses a uniform initial splat size~$s$.
The individual splat sizes will be derived later as described in Section~{\ref{sec:DiscussionOfSplatSize}}.
In the following discussion, we will refer to elements of the input geometry~$\mathcal{P}$ as \emph{points} and to entities created by the algorithm as~\emph{vertices}.

\subsection{Initialization}
\label{sec:Initialization}
\reb{Aside from~$d$ and~$s$, the user provides two \emph{starting vertices} to initialize the algorithm.}
\reb{These vertices are chosen from~$\mathbb{R}^3$ such that the projections of these vertices onto their closest splats are sufficiently close, that is, the distance between the projected vertices is in~$\lbrack d, 2d \rbrack$, so a third vertex having distance~$d$ to both of the starting vertices can be placed by the algorithm.
They do not have to be points from the input~$\mathcal{P}$.
The projected vertices form an initial vertex set of a graph~$\mathcal{G}$ that will ultimately provide the manifold mesh discussed in Section~\ref{sec:AssumptionsAndTheory}.
They can be manually provided or automatically created, for instance, around the maximum $z$-coordinate of the input.
At this stage, $\mathcal{G}$ does not contain any edges.
In the following, positions exactly~$d$ away from at least two already existing vertices are called \emph{vertex candidates}.
The two vertices that are~$d$ away from a candidate are its \emph{parents}.}

\subsection{Disk Growing to Add Vertices of$~\mathcal{G}$}
\label{sec:DiskGrowingToCreateGraphAndRegions}
After initialization, disk growing is performed to create further vertices and edges of $\mathcal{G}$.
A vertex is added from the list of vertex candidates, connected by edges to the two vertices distance~$d$ away \reb{(Figure~\ref{fig:AddingNewVertex})}, and new vertex candidates are added based on the newly placed vertex \reb{(Figure~\ref{fig:NewCandidate})}.
Adding edges to~$\mathcal{G}$ also changes the regions introduced in Section~\ref{sec:AssumptionsAndTheory}: By inserting a new vertex and its two edges, either one border is \emph{split} into two borders or two borders \emph{join} into one (Figures~\ref{fig:borderConnectionsCase1} and~\ref{fig:borderConectionsCase2}).

\begin{figure*}
	\centering
	\begin{subfigure}[t]{0.32\linewidth}
		\centering
		\includegraphics[width=.75\linewidth]{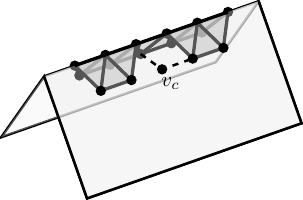}
		\caption{3D view of a tent-like feature, a geometry with reach~$\rho=0$.}
		\label{fig:Roof3D}
	\end{subfigure}
	\hfill
	\begin{subfigure}[t]{0.32\linewidth}
		\centering
		\includegraphics[width=.75\linewidth]{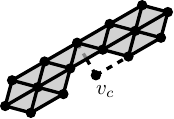}
		\caption{Projection of the graph along an approximated surface normal.}
		\label{fig:RoofProjection}
	\end{subfigure}
	\hfill
	\begin{subfigure}[t]{.33\linewidth}
		\centering
		\input{IntersectingSpheres}
		\caption{Intersection of circle $c$ with splat $S$ gives next vertex candidate.}
		\label{fig:IntersectingSpheres}
	\end{subfigure}
	\caption{
		An input geometry with a vertex candidate~$v_c$~(\ref{fig:Roof3D}), the region projection shows an illegal edge crossing~(\ref{fig:RoofProjection}). Edges to parent vertices are shown dotted.
		Computation of vertex candidate positions~(\ref{fig:IntersectingSpheres}).
	}
	\label{fig:crossingProjection}
\end{figure*}

As shown in Figure~\ref{fig:VisibleSeams}, there might be regions with comparably long borders.
These lead to visible seams in both~$\mathcal{G}$ and its triangulation.
To avoid such seams, we prioritize joining borders over splitting borders.
Thus, we aim to prioritize splits with a larger combinatorial distance between the parent vertices along the border over those with smaller distances.
We do so with a priority assigned to the vertex candidates.

\subsection{Prioritizing of Vertex Candidates}
\label{sec:PoppingAVertexCandidate}
A vertex candidate~$v_c$ is chosen to become a vertex of~$\mathcal{G}$ according to the following priorities, given in decreasing order:
\begin{compactenum}
	\item At least one of the parent vertices has no edges incident to it. 
	(Note: This parent vertex has to be a starting vertex from the initialization.)
	\item At least one of the parent vertices is a vertex with only one edge incident to it.
	\item Inserting $v_c$ and its two edges joins two borders.
	\item Inserting $v_c$ splits a border---prioritize larger distance between parents along the common border.
\end{compactenum}
In all cases, ties are broken by the breadth-first strategy.
To determine the priority of the vertex candidate to be added, it is necessary to know to which of the parent vertices' borders the edges to be introduced will connect.
To find the corresponding border, the candidate edge is projected to the plane defined by the parent vertex and its normal.
Note that because of the results from Section~\ref{sec:AssumptionsAndTheory}, such a projection is possible without over-folds.
Given the prioritization of vertices, these can now be added to the graph~$\mathcal{G}$.

\subsection{Creating a New Vertex}
\label{sec:CreatingANewVertex}
Once a vertex candidate~$v_c$ has been chosen, it is first determined whether there is a vertex~$v$ in~$\mathcal{G}$ such that~$\left\|v_c-v\right\|_2 < d$.
If so, the vertex candidate is discarded.

Next, the priority of~$v_c$ is checked.
In case the vertex does not satisfy the given priority anymore---for instance, because it was created with a parent vertex without any edges incident to it, but the parent vertex gained an edge by now---the vertex candidate's priority is reduced and another vertex candidate is chosen.

Adding~$v_c$ and the corresponding edges to~$\mathcal{G}$ bears one additional problem.
In practice, we do not always know whether the point cloud~$\mathcal{P}$ fulfills the criteria listed in Section~\ref{sec:Methodology}. 
If they are satisfied, the output is guaranteed to be manifold. 
However, the user might have chosen~$d$ too large or the input point cloud might not sample a manifold in the first place.
In either of these cases, all edges created from new vertices still have an edge length of~$d$, but the edges might create non-manifold connections.
Consider Figures~\ref{fig:Roof3D} and~\ref{fig:RoofProjection} for an example of a surface with reach~$\rho=0$.

In these cases, we still want to prevent such faulty connections.
Therefore, we find an approximated surface normal, which will be discussed in Section~\ref{sec:AuxiliaryBoxDataStructure}.
We project~$v_c$ and its prospective edges as well as all edges already existing in the vicinity of~$v_c$ along this normal.
For this projection, the vicinity of~$v_c$ is bounded by~$d$ in normal direction.
We discard~$v_c$ if either of its edges crosses an already existing edge (Figure~\ref{fig:RoofProjection}).
While the algorithm creates manifold output for suitable input point clouds and choices of~$d$, this mechanism improves the output even outside of this regime.
If~$v_c$ has passed these checks,~$v_c$ and the two edges connecting it to its parent vertices are added to~$\mathcal{G}$.

\subsection{Triangulating the Resulting Regions}
\label{sec:TriangulatingTheResultingMesh}
After the disk growing process has finished, the graph~$\mathcal{G}$ provides a set of regions~$\mathcal{R}$. 
On average, each vertex of~$\mathcal{G}$ is connected to approximately four other vertices~\cite[Section~4.2]{lipschutz2021single}.
Therefore, the average border length is approximately four. 
Hence, we are left with the task of triangulating these regions.
In case of surfaces with boundary such as partial scans, we do not want to close the surface by triangulating the interior of the boundary.
Therefore, we give the user the choice to specify a maximal border length~$\partial_{\max}$ that will leave the region as a hole rather than triangulating it.

A region can be irregular in the sense that the inner angle of two consecutive edges can be larger than $180^\circ$.
In such cases, a projection of a single region to a plane is not necessarily a convex polygon.
These inner angles of the faces are found by projecting the edges onto a plane given by the vertex normal.
Then, we triangulate each region by iteratively cutting away the smallest angle as this leads to triangles close to equilateral ones.
Not only have we thereby created a triangulation of the input surface that has guaranteed minimum edge length~$d$, but by the results provided in Section~\ref{sec:AssumptionsAndTheory}, provided that the input and the user-chosen parameters satisfy the restrictions made, the triangulation is also manifold.

\section{Implementation}
\label{sec:Implementation}

\noindent In this section, we will discuss implementation aspects of the algorithm presented above.
In particular, this contains the introduction of data structures for efficient access.
As stated in Section~\ref{sec:Methodology}, we assume to be given a point cloud~$\mathcal{P}$, its normal field~$\mathcal{N}$, user-chosen parameters $d$, $s$, and in case of a surface with boundary, $\partial_{\max}$.
If~$\mathcal{P}$ does not come with a normal field, the user has to estimate one, for instance, via~\cite{mitra2003estimating}.
Furthermore, the user has to choose the implementation-related parameter~$w$ (Section~\ref{sec:WindowSize}).

\subsection{Box Grid Data Structure}
\label{sec:AuxiliaryBoxDataStructure}

\begin{figure}[h]
	\centering
	\begin{subfigure}[t]{0.47\linewidth}
		\centering
		\scalebox{\MethodologyI}{\input{AddingNewVertex}}
		\caption{Adding a new vertex.}
		\label{fig:AddingNewVertex}
	\end{subfigure}
	\begin{subfigure}[t]{.51\linewidth}
		\centering
		\scalebox{\MethodologyII}{\hspace{.3cm}\input{CollectCandidatesTopview}\hspace{-.3cm}}
		\caption{Collecting splats, top view.}
		\label{fig:CollectingSplatsTopView}
	\end{subfigure}
	
	\begin{subfigure}[t]{.47\linewidth}
		\centering
		\scalebox{\MethodologyI}{\input{CollectingSplats}}
		\caption{Collecting normals.}
		\label{fig:CollectingSplats}
	\end{subfigure}
	\hspace{.025\linewidth}
	\begin{subfigure}[t]{.47\linewidth}
		\centering
		\scalebox{\MethodologyI}{\input{NewCandidate}}
		\caption{Finding candidates.}
		\label{fig:NewCandidate}
	\end{subfigure}
	\caption{
		Update steps of the algorithm.
	}
	\label{fig:MethodologyII}
\end{figure}
\begin{figure*}
	\centering
	\begin{subfigure}[t]{0.3\linewidth}
		\centering
		\scalebox{.95}{\input{BorderZero}}
		\caption{Border of length~0.}
		\label{fig:BorderLength0}
	\end{subfigure}
	\hfill
	\begin{subfigure}[t]{0.3\linewidth}
		\centering
		\scalebox{.95}{\input{BorderTwo}}
		\caption{Border of length~4.}
		\label{fig:BorderLength2}
	\end{subfigure}
	\hfill
	\begin{subfigure}[t]{.3\linewidth}
		\centering
		\scalebox{.95}{\input{BorderK}}
		\caption{Border of length~$k+1$.}
		\label{fig:BorderLengthk+1}
	\end{subfigure}
	\caption{
		Different borders and their respective regions:
		Border of length~$0$ consisting of a single vertex and associated to a single, white, surrounding region (\ref{fig:BorderLength0}); 
		border of length~$4$, going back and forth between $v$ and $v'$, associated to a single, white, surrounding region (\ref{fig:BorderLength2});
		cycle of~$k+1$ edges, separating the surface into an inner, light gray and outer, white region (\ref{fig:BorderLengthk+1}).
	}
	\label{fig:bordersAndRegions}
\end{figure*}

\noindent When introducing new vertex candidates (Section~\ref{sec:DiskGrowingToCreateGraphAndRegions}), we need to know all splats close to a given, newly introduced vertex.
In order to have access to these, we build a \emph{box grid data structure} consisting of equal-sized, cubical boxes of side-length~$d$ partitioning the three-dimensional embedding space.
Each box holds a pointer to those input points and their splats that are at most~$d$ away (Figure~\ref{fig:CollectingSplats}). 
This collection of points associated with box~$b_j$ is denoted by~$\mathcal{B}_j$.

As preliminary filter step, we compute an average normal~$\overline{\reb{n}}_{b_j}$ for each box~$b_j$ by summing up the normals of all those points that~$b_j$ has a pointer to, without normalizing the sum.
If~$\left\|\overline{\reb{n}}_{b_j}\right\|_2$ is smaller than~$0.1$, we keep all points in~$b_j$.
If the length is at least~$0.1$, we can assume that enough points agree on a normal direction in this box.
Then, we remove those points~$p$ from the box for which~$\langle \reb{n}_p,\overline{\reb{n}}_{b_j}\rangle < 0$.
We choose a value of~$0.1$ for the length check to filter a small number of points while maintaining coherent normal information.
This will ensure that the following step can succeed.

For each box $b_j$ with at least one associated splat, we compute a \emph{box normal}~$\reb{n}_{b_j}$.
It will be used for projection steps that will ensure manifold properties of the resulting mesh.
This will provide approximated surface normals that allow us to work on data which do not fulfill the requirements listed in Section~\ref{sec:AssumptionsAndTheory} such as the example shown in Figure~\ref{fig:Roof3D}.
To compute an approximation efficiently, we take a finite sampling $\mathcal{S}_F \subset \mathbb{S}^2$ and derive the box normal~$\reb{n}_{b_j}$ as
\begin{align*}
	\label{eq:OptimizationProblem}
	\reb{n}_{b_j} = \argmax_{N\in\mathcal{S}_F}\min_{p_i\in \mathcal{B}_j}\langle \reb{n}_{p_i}, \reb{n}\rangle \approx \argmax_{\reb{n}\in\mathbb{S}^2}\min_{p_i\in \mathcal{B}_j}\langle \reb{n}_{p_i}, \reb{n}\rangle.
\end{align*}
That is, we search for the unit normal that maximizes the smallest scalar product with all point normals associated to the box~$b_j$.
For each box~$b_j$, the scalar product~$\langle \reb{n}_{p_i}, \reb{n}_{b_j}\rangle$ is ideally strictly positive for all points~$p_i \in \mathcal{B}_j$, even in those cases where we did not filter the normals.
Therefore, it allows for a projection onto a plane spanned by~$\reb{n}_{b_j}$ as normal vector such that all points remain positively oriented by their normals.
The newly computed box normal is also used as vertex normal for all vertices lying in $b_j$ from now on.

To achieve a fast lookup, we can either build a uniform grid on the complete bounding box of the input or create a hash structure to only store those boxes that are including input points from~$\mathcal{P}$.
The uniform grid has faster access, but results in many empty boxes and thus large memory consumption.
The hash structure does not use as much memory, but the access is slower.
In our experiments, we utilize the uniform grid structure to be faster as memory consumption can be handled by our test machine\reb{, although the time difference will become significant only for larger models than used here}.

Note that in Section~\ref{sec:TheoryAndMethodology}, we stated that for creating new vertex candidates, we need to traverse all splats that are distance~$d$ away from a given point.
However, here we are collecting all splats that are at distance~$\leq d$ from the box, thus possibly resulting in a higher number of splats to be considered.
That insures that the spheres from Section~\ref{sec:Methodology} are inscribed into the volume within $d$-distances around the boxes (Figure~\ref{fig:CollectingSplatsTopView}).

This leads to the question how to choose a good side-length of the boxes.
As stated above, we use side-length~$d$, that is, their size coincides with the target edge length for the triangulation.
For smaller values, each splat would be associated to more boxes, hence the memory demand would grow.
For larger boxes, there will be many splats associated to each box that do not actually lead to vertex candidates with the currently considered vertex.
Thus, the runtime would grow when checking all splats being far away from the currently considered vertex.
Finally, for larger boxes, it will also become more difficult to compute a suitable box normal for projections.
Hence, we advocate for the middle ground and choose~$d$ as the box size.

\begin{figure*}[t!]
	\centering
	\newlength{\Shampoo}
	\setlength{\Shampoo}{0.15\linewidth}
	\begin{subfigure}[t]{\Shampoo}
		\centering
		\includegraphics[width=0.75\textwidth]{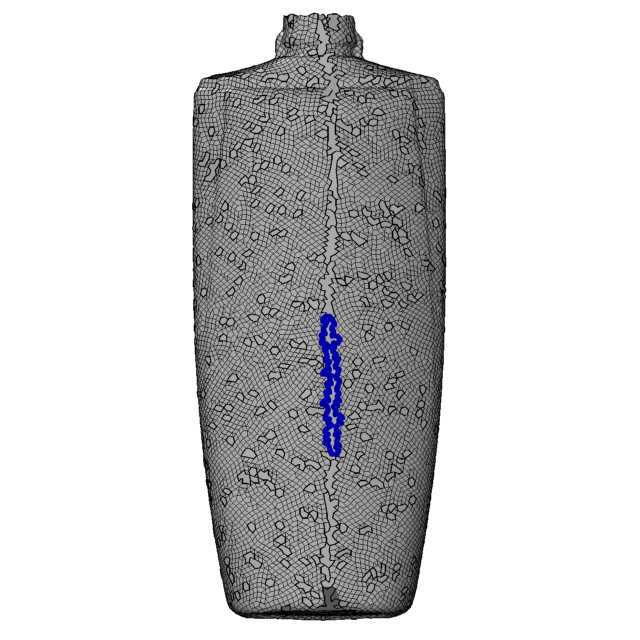}
		\caption[format=hang,justification=centerlast]{Longest border (length~93).}
		\label{fig:VisibleSeamExample}
	\end{subfigure}
	\hfill
	\begin{subfigure}[t]{\Shampoo}
		\centering
		\includegraphics[width=0.75\textwidth]{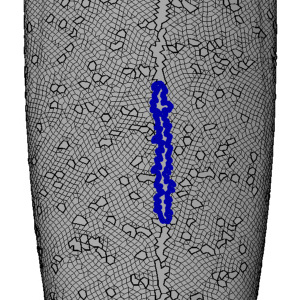}
		\caption{Longest border (close-up).}
		\label{fig:VisibleSeamsAvoided}
	\end{subfigure}
	\hfill
	\begin{subfigure}[t]{\Shampoo}
		\centering
		\includegraphics[width=0.75\textwidth]{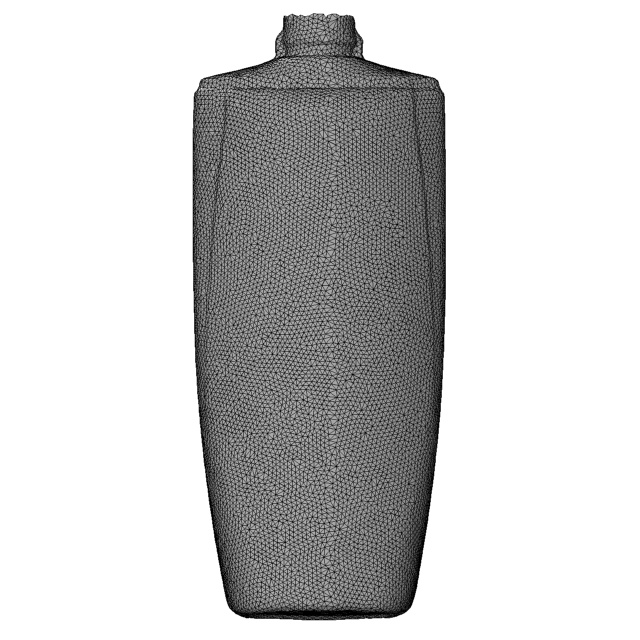}
		\caption{Triangulated regions.}
		\label{fig:VisibleSeamsCloseUp}
	\end{subfigure}
	\hfill
	\begin{subfigure}[t]{\Shampoo}
		\centering
		\includegraphics[width=0.75\textwidth]{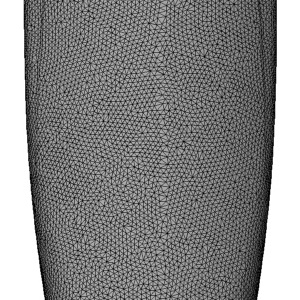}
		\caption{Triangulated region (close-up).}
		\label{fig:VisibleSeamsAvoidedCloseup}
	\end{subfigure}
	\hfill
	\begin{subfigure}[t]{\Shampoo}
		\centering
		\includegraphics[width=0.75\textwidth]{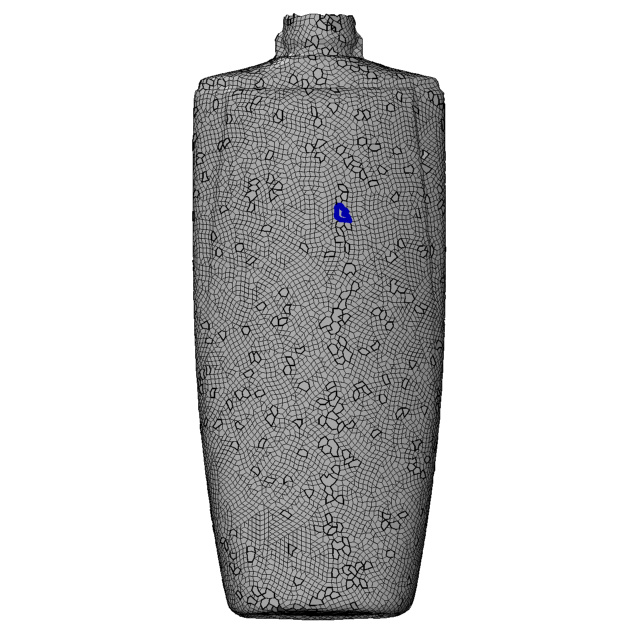}
		\caption{Longest border (length $11$).}
		\label{fig:VisibleSeams_8}
	\end{subfigure}
	\hfill
	\begin{subfigure}[t]{\Shampoo}
		\centering
		\includegraphics[width=0.75\textwidth]{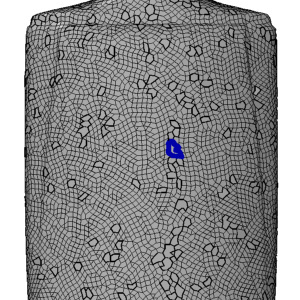}
		\caption{Longest border (close-up).}
		\label{fig:VisibleSeamsAvoidedCloseup_8}
	\end{subfigure}
	\caption{
		Visible seams on the \emph{Bottle Shampoo}. Figures~\ref{fig:VisibleSeamExample} to~\ref{fig:VisibleSeamsAvoidedCloseup} show experimental results obtained by inserting new vertices via pure breadth-first-growing. 
		The seams appear as regions having a high number of border edges compared to all other regions on the surface. Figures~\ref{fig:VisibleSeams_8} and~\ref{fig:VisibleSeamsAvoidedCloseup_8} show results obtained by prioritizing new vertex candidates.
		For better visibility, the target edge length~$d$ was chosen as~$1$.}
	\label{fig:VisibleSeams}
\end{figure*}

The last observation regarding the box normals leads to a heuristic how to check the user's parameter choice of~$d$.
Namely, a choice of~$d$ is considered too large if there is a box whose box normal has negative scalar product with any point normal of a point registered in the box.
This provides a mechanism to alert the user that they have chosen the parameter~$d$ outside of the specifications as provided in Section~\ref{sec:AssumptionsAndTheory} and that the output is thus not guaranteed to be manifold anymore.
This observation allows for the following binary search heuristic to find a suitable, yet not too small value of~$d$ (compare the discussion in Section~\ref{sec:Methodology}): Start with a large initial value~$d_{\text{init}}$, for instance, the bounding box diagonal of the geometry.
Perform a binary search on the interval~$[0,d_{\text{init}}]$ by picking the midpoint of the current interval, building the box grid data structure and evaluating the normal scalar products.
If any is negative, continue the binary search in the lower half of the interval.
If all are positive, the user could either stop, since a suitable value for~$d$ has been found, or continue in the upper half of the interval, if they seek for a coarser output.

\subsection{Window Size}
\label{sec:WindowSize}

\noindent During the disk growing, we maintain a data structure representing the regions' borders.
They consist of oriented half-edge cycles.
Note that this includes degenerate cases, such as a single vertex, interpreted as a border of length~$0$ (Figure~\ref{fig:BorderLength0}).
Each time a new vertex and the two edges to its parent vertices are added to~$\mathcal{G}$, this creates four new half-edges which have to be linked to the existing borders.
To avoid \reb{traversing} very long distances along the borders \reb{when computing priorities of vertex candidates}, we introduce a \emph{window size} $w$ after which the \reb{traversal} is stopped.
A \emph{window} then consists of $2w + 1$ vertices on a common border, running in both directions centered at the vertex currently considered.
This provides a considerable speedup compared to the previous solution~\cite{lipschutz2021single}.

Recall \emph{split} and \emph{join} from Section~\ref{sec:DiskGrowingToCreateGraphAndRegions}.
Note that by cutting the \reb{traversal} at a finite window size, it is not longer possible to distinguish between a split and join operation in all cases.
Preliminary experiments showed that window sizes of~${w\geq8}$ all produced the same quality output, despite not distinguishing splits or joins\reb{, as shown in the supplementary material}.

Furthermore, we experienced that setting the window size to~${w=0}$ immediately creates noticeable negative effects on the result of the algorithm.
In this case, our algorithm defaults to the \emph{breadth-first} strategy of~\cite{lipschutz2021single} and thus creates visible seams on the geometry (Figures~\ref{fig:VisibleSeamExample} to~\ref{fig:VisibleSeamsAvoidedCloseup}).
Starting from window sizes of~${w=2}$ or~${w=3}$, benefits in the quality of the output are apparent as larger visible seams are prevented.
Theoretically, a larger window size will increase the lookup time.
Therefore, in our implementation of the algorithm, we go for~${w=8}$ as a large enough window size to reap its benefits, but a small enough one to not impact the algorithm's run time.

\subsection{Discussion of Splat Size}
\label{sec:DiscussionOfSplatSize}

\begin{figure}[h]
	\centering
	\begin{subfigure}[t]{1.\linewidth}
		\centering
		\includegraphics[width=0.86\textwidth]{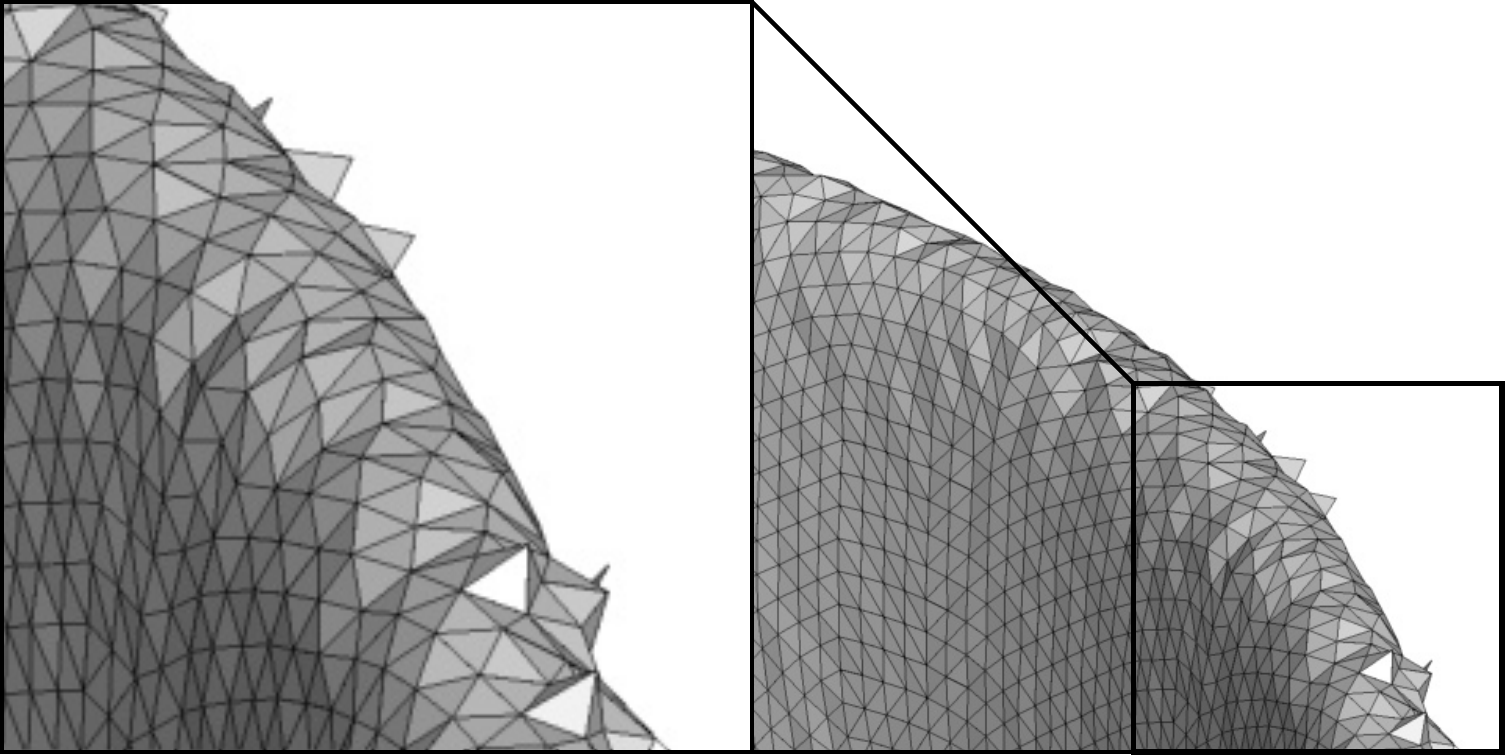}
		\caption{Result with uniform splat sizes.}
		\label{fig:ChinesBowlUniformSplat}
	\end{subfigure}
	\\
	\vspace{0.2cm}
	\begin{subfigure}[t]{1.\linewidth}
		\centering
		\includegraphics[width=0.86\textwidth]{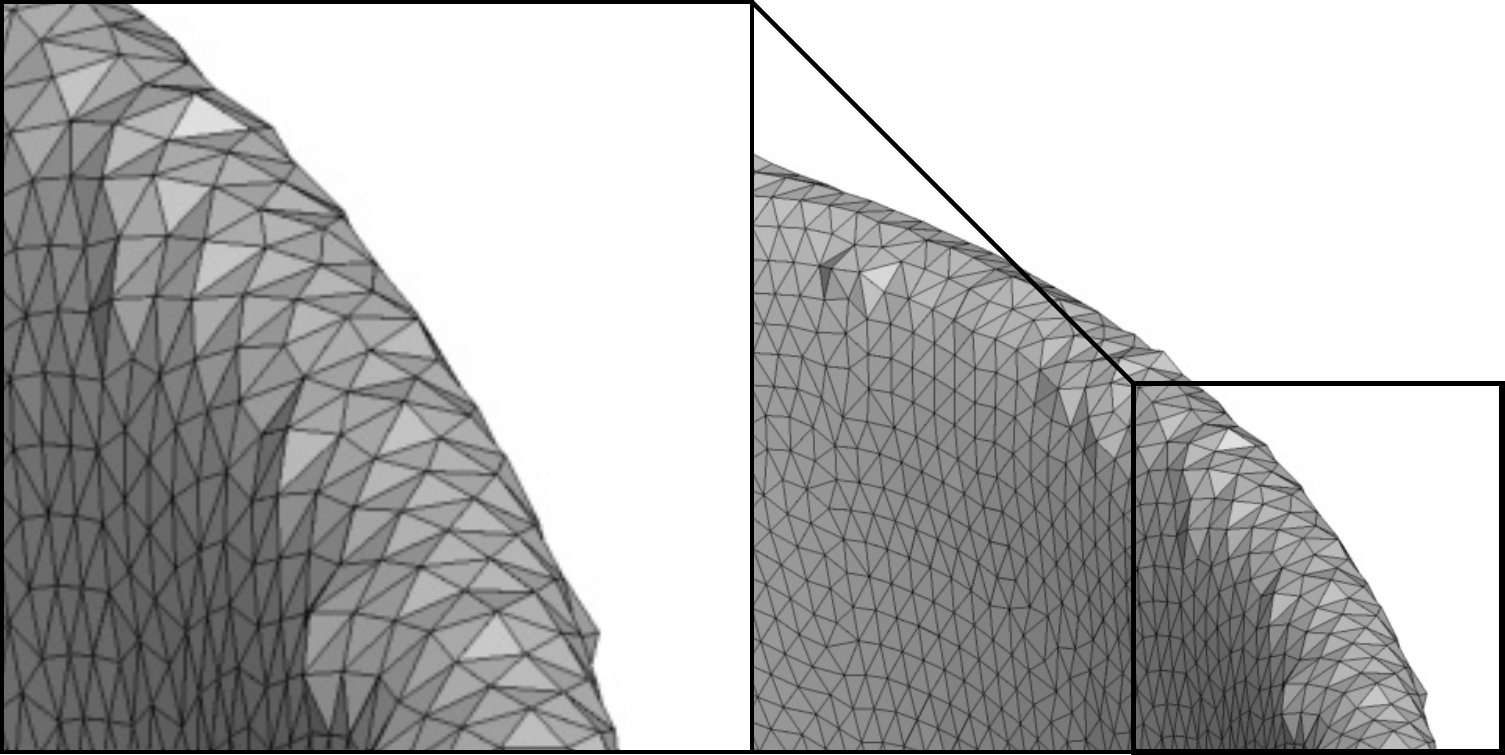}
		\caption{Result with individual splat sizes.}
		\label{fig:ChinesBowlIndividualSplat}
	\end{subfigure}	
	\caption{
		Running our algorithm on the \emph{Bowl Chinese} from~\cite{huang2022surface} with~\ref{fig:ChinesBowlUniformSplat} using a global splat size and~\ref{fig:ChinesBowlIndividualSplat} using a local spat size.
		The letter reduces artifacts in high-curvature regions such as the rim of the bowl. 
	}
	\label{fig:VaryingSplatSizes}
\end{figure}

\noindent In case of non-uniform sampling density, using a global splat size~$s$ might lead to areas covered multiple times. 
For more densely sampled areas, a smaller splat size guarantees the creation of vertices closer to the sampling points.
In our experiments, we saw that in high-curvature regions, smaller splats have small deviation from the surface, while larger splats deviate from the surface significantly.
Hence, when looking for vertex candidates on large splats, the algorithm can place vertices that are somewhat distant to the input points (Figure~\ref{fig:ChinesBowlUniformSplat}).
Therefore, we turn to individual, smaller splat sizes to reduce the deviation of the vertices with respect to an underlying surface represented by the input point cloud.

An additional benefit is that for smaller splat sizes, there are less splats registered per box, which speeds up the algorithm.
However, the individual splats have to have sizes sufficient to cover the underlying geometry.
To find the specific splat size~$s_{p}$ for each point~${p \in \mathcal{P}}$, we use the box data structure (Figure~\ref{fig:IndividualSplatSizes}).
We consider all points~$p_i$ associated to the box containing~$p$.
To map the points~$\lbrace p_i \rbrace$ to $T_p\mathcal{M}$, consider the plane $N_{\perp}$ containing $p$ and $p_i$ and being orthogonal to $T_p\mathcal{M}$.
For each $p_i$, an auxiliary point~$\pi(p_i)$ is determined by rotating $p_i$ around $p$ around the smaller angle in~$N_{\perp}$ until it lies in $T_p\mathcal{M}$.
Hence, $p$ and $\pi(p_i)$ have the same distance~$d_i$ as $p$ and $p_i$ have.
Based on a cyclic sorting around~$p$, we compute a central triangulation, connecting all projections to~$p$ and connecting them pairwise according to their angular sorting (Figure~\ref{fig:ProjectionToTangentPlane}).
For the resulting triangulation, we test whether or not we can flip a central edge to make the incident triangles Delaunay. 
Points~$p_i$, whose edges are flipped, are removed from the following consideration.
For those neighboring points that remain, consider the Voronoi diagram of their triangulation.
We choose the local splat size~$s_{p}$ as distance from $p$ to the farthest Voronoi vertex (Figure~\ref{fig:VoronoiCells}).
This ensures that all Delaunay triangles are still completely covered.
By choosing local splat sizes in this way, the visible deviation from the underlying geometry is reduced (Figure~\ref{fig:ChinesBowlIndividualSplat}).

\begin{figure*}[h!]
	\centering
	\newlength{\SplitImages}
	\setlength{\SplitImages}{0.119\linewidth}
	\begin{subfigure}[t]{\SplitImages}
		\centering
		\includegraphics[width=\textwidth]{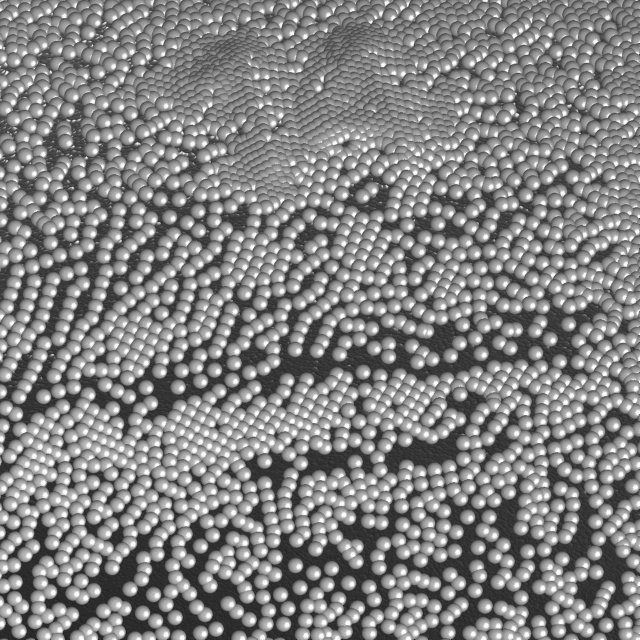}
		\captionsetup{labelformat=empty}
		\caption{Input.}
		\label{fig:ShampooBottlePoints}
	\end{subfigure}
	\begin{subfigure}[t]{\SplitImages}
		\centering
		\includegraphics[width=\textwidth]{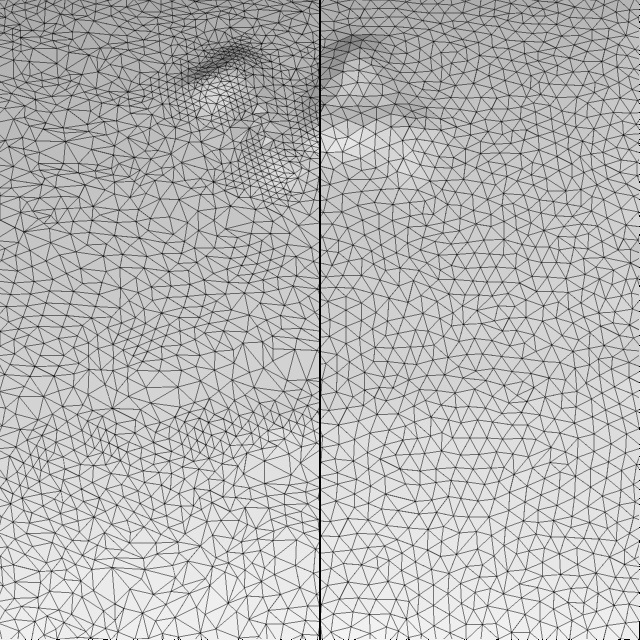}
		\captionsetup{labelformat=empty}
		\caption{Adv. Front.}
		\label{fig:AdvancingFront}
	\end{subfigure}
	\begin{subfigure}[t]{\SplitImages}
		\centering
		\includegraphics[width=\textwidth]{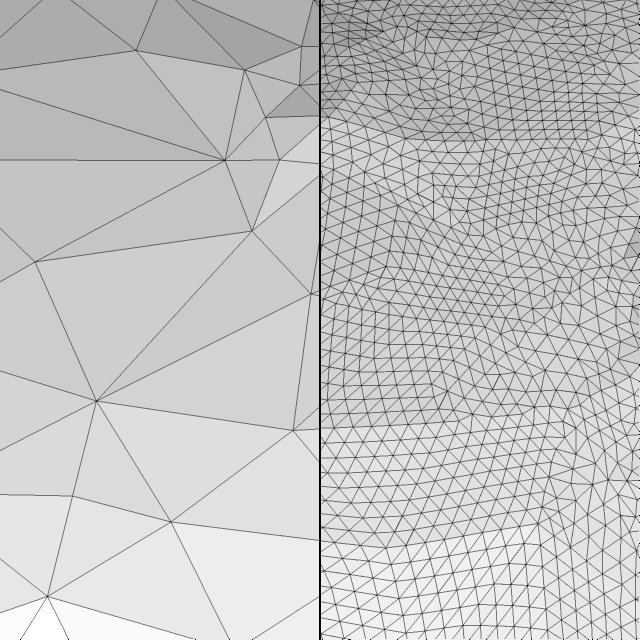}
		\captionsetup{labelformat=empty}
		\caption{Poisson.}
		\label{fig:Poisson}
	\end{subfigure}
	\begin{subfigure}[t]{\SplitImages}
		\centering
		\includegraphics[width=\textwidth]{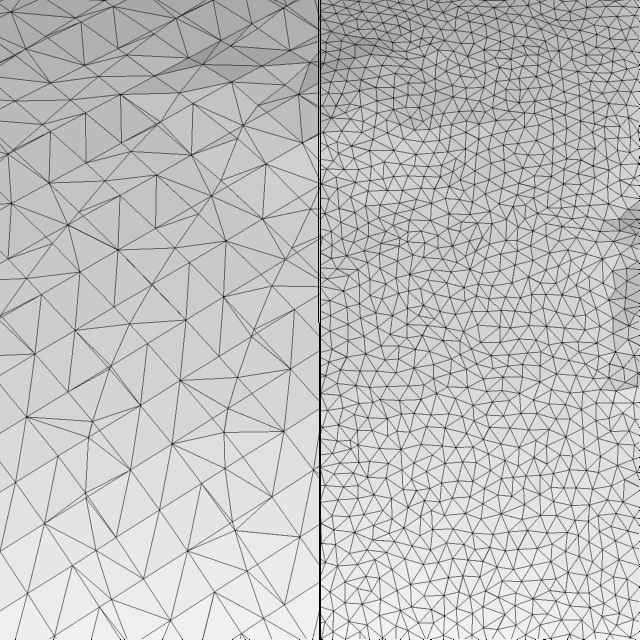}
		\captionsetup{labelformat=empty}
		\caption{Poisson MG.}
		\label{fig:PoissonGeo}
	\end{subfigure}
	\begin{subfigure}[t]{\SplitImages}
		\centering
		\includegraphics[width=\textwidth]{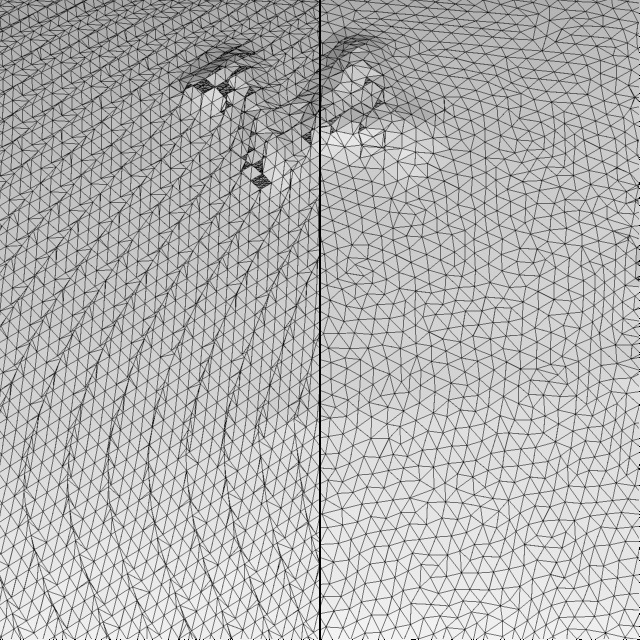}
		\captionsetup{labelformat=empty}
		\caption{RIMLS.}
		\label{fig:RIMLS}
	\end{subfigure}
	\begin{subfigure}[t]{\SplitImages}
		\centering
		\includegraphics[width=\textwidth]{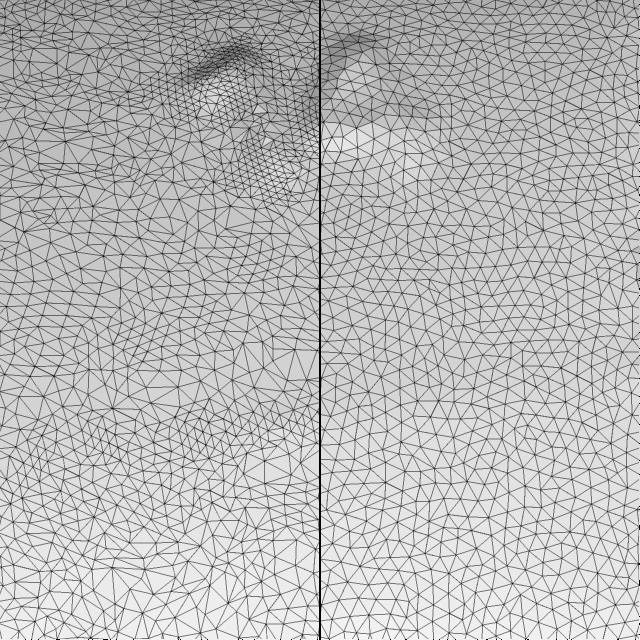}
		\captionsetup{labelformat=empty}
		\caption{Scale Space.}
		\label{fig:ScaleSpace}
	\end{subfigure}
	\begin{subfigure}[t]{\SplitImages}
		\centering
		\includegraphics[width=\textwidth]{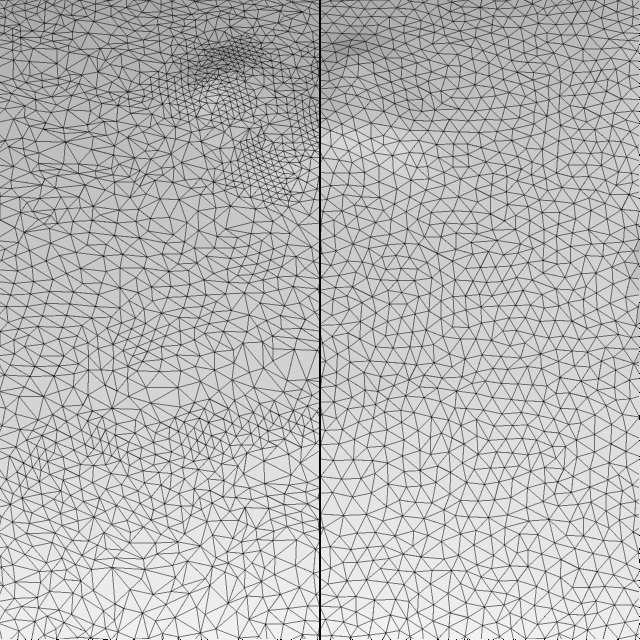}
		\captionsetup{labelformat=empty}
		\caption{Voronoi.}
		\label{fig:co3ne}
	\end{subfigure}	
	\begin{subfigure}[t]{\SplitImages}
		\centering
		\includegraphics[width=\textwidth]{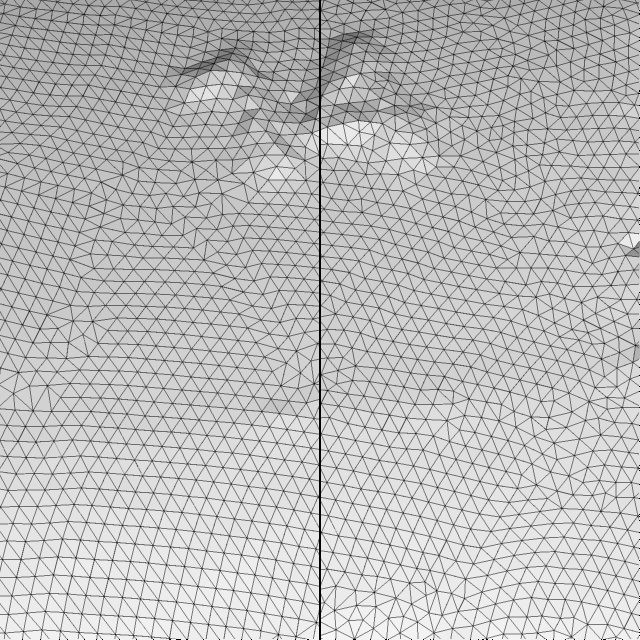}
		\captionsetup{labelformat=empty}
		\caption{Ours.}
		\label{fig:SphereMeshing}
	\end{subfigure}
	\caption{Qualitative comparison of named algorithms without remeshing (left) and with remeshing (right).}
	\label{fig:SplitMeshes}
\end{figure*}

\subsection{Processing Vertex Candidates}
\label{sec:ProcessingVertexCandidates}

\noindent The processing of vertex candidates, following Section~\ref{sec:DiskGrowingToCreateGraphAndRegions}, consists of the following steps: popping a vertex candidate from the priority queue, checking feasibility of the candidate, adding a suitable candidate as well as its edges to~$\mathcal{G}$, and adding new vertex candidates to the priority queue.

Because of the window size~$w$, there is a finite number of priorities, as given in Section~\ref{sec:PoppingAVertexCandidate}.
Each of these priorities is handled via its own queue that follows a strict first-in-first-out strategy, which enables popping of candidates in constant time~\cite[Chapter~2.4]{sedgewick2011algorithms}.

If a vertex still has correct priority, checking for conflict with existing vertices and performing the projection check from Figure~\ref{fig:RoofProjection} both requires access to nearby vertices.
This is a constant-time operation because of the box data structure that holds all relevant vertices.
Furthermore, the number of vertices within distance~$2d$ is, by construction, bounded from above by the densest sphere packing in space, which is a constant.

Once a new vertex~$v_{\text{new}}$ is created, we compute new vertex candidates having~$v_{\text{new}}$ as parent vertex.
Therefore, we need the set of all splats intersecting the ball of radius~$d$ centered at~$v_{\text{new}}$.
This is a subset of the set of those splats associated to the box containing~$v_\text{new}$.
To efficiently access potential second parent vertices, we maintain for each splat~$S$ a list of all vertices within distance~$d$ to $S$  (Figures~\ref{fig:CollectingSplatsTopView} and~\ref{fig:CollectingSplats}).

\section{Experiments}
\label{sec:Experiments}

\noindent This section is devoted to different experimental settings.
As described in the beginning, we aim for the reconstruction of real-world scan data.
When performing the comparison of different algorithms, we do so based on a quantitative analysis of the obtained triangle mesh~$\mathcal{T}$.
For this, given a triangle~${t\in\mathcal{T}}$, we denote the lengths of its edges by~$\ell_{t,1}$,~$\ell_{t,2}$, and~$\ell_{t,3}$.
The area of the triangle will be called~$A_t$.

Following the approach in~\cite[p.~307, Eq.~(13)]{ma2017guaranteed}, we measure the quality~$Q_t$ of a single triangle as
\begin{align*}
	Q_t = \frac{4\sqrt{3}A_t}{\ell_{t,1}^2 + \ell_{t,2}^2 + \ell_{t,3}^2}.
\end{align*}
This measure corresponds to a scaled version of the \emph{scale-invariant (smooth)} conditioning quality measure discussed by Shewchuk~\cite[Table~3]{shewchuk2002what}.
Based on the local, triangle-based measure~$Q_t$, further following~\cite{ma2017guaranteed}, we present a global metric for the entire triangle mesh~$\mathcal{T}$ as average over the quality of the triangles, that is
\begin{align*}
	Q_{\text{avg}} = \frac{1}{\left|\mathcal{T}\right|}\sum_{t\in\mathcal{T}}  Q_t.
\end{align*}
Note that the factors normalize this quality metric to be~$1$ for equilateral triangles and close to~$0$ for very narrow slivers.
Finally, we compute the root mean square deviation in percent~$Q_{\text{RMS}}$ as
\begin{align*}
	&Q_{\text{RMS}} = \frac{100}{Q_{\text{avg}}}\sqrt{\frac{1}{|\mathcal{T}|}\sum_{t\in\mathcal{T}}\left(Q_t-Q_{\text{avg}}\right)^2}.
\end{align*}
See Section~3 of~\cite{shewchuk2002what} for a relation of this quality measure to the stiffness matrix.
Furthermore, from the set of all edges in the triangulation, we consider the average edge length~$E_{\text{avg}}$ as well as the corresponding root mean square deviation~$E_{\text{RMS}}$, also in percent.

In order to demonstrate the quality of the meshes achieved by our algorithm, we turn to~20 scanned objects provided as part of a surface reconstruction benchmark~\cite{huang2022surface}.
Here, we concentrate on high-resolution scans obtained by an \emph{OKIO~5M} scanning device, resulting in~330k to~2,000k points per surface after~20 shots.
The shots are registered and do come with a normal field.
Out of the~20 point clouds, we used~19 as they are provided in the repository.
The scan of a remote control had a clear registration artifact, since one of the buttons of the remote was registered into the remote, pointing down, not up.
This, we corrected manually by removing the wrongly registered points.
Here, we compare our results to those made by various widely used algorithms from the field.
Then, we add different levels of noise to the data and investigate the stability of our algorithm.

As mentioned in Section~\ref{sec:Implementation}, there is a set of parameters which has to be chosen by the user.
For our experiments, we made the following choices.
The sphere diameter~$d$ was set to be~$0.2$, while the maximal border length~$\partial_{\max}$ was equal to~$40$.
For each model, the initial splat size~$s$ was chosen between~$0.2$ and~$0.4$ individually, depending on the considered point cloud.

\subsection{Experimental Comparison for Point Cloud Meshing}
\label{sec:ExperimentalResultsForPointCloudMeshing}

\begin{table*}
	\def\arraystretch{1.05}
	\begin{minipage}{.45\textwidth}
		\centering
		\tiny{
			\begin{tabular}{l|rrrrr}
				Algorithm & $|\mathcal{T}|$ & $E_{\text{avg}}$ & $E_{\text{RMS}}$ & $Q_{\text{avg}}$ & $Q_{\text{RMS}}$\\
				\hline
				Adv.~Front & 1.209,546 & 0.1799 & 39.6 & 0.8247 & 16.0 \\
				Adv.~Front (Re) & 928,850 & 0.2028 & 15.3 &  0.9416 & 6.1\\
				Poisson & 
				16,280 & 1.2946 & 74.8 & 0.8760 & 12.3 \\
				Poisson (Re) & 
				498,140 &  0.2657 & 38.6 & 0.9251 & 7.5\\
				Poisson MG & 
				150,770 & 0.5318 & 35.7 & 0.7204 & 33.7 \\
				Poisson MG (Re) &
				952,830 & \textbf{0.2015} & 16.3 & 0.9330 & 7.0 \\
				RIMLS & 
				1,907,781 & 0.1499 & 35.8 & 0.7055 & 35.1 \\
				RIMLS (Re) & 
				1,054,438 & 0.1905 & 19.3 & 0.9117 & 11.5 \\
				Scale Space & 
				1,209,093 & 0.1798 & 39.1 & 0.8248 & 16.0 \\
				Scale Space (Re) & 
				926,828 & 0.2028 & 15.2 &  0.9417 & 6.0 \\
				Voronoi & 
				1,209,792 & 0.1799 & 52.3 & 0.8241 & 16.1 \\
				Voronoi (Re) &
				923,476 & 0.2044 & 20.8 & 0.9407 & 6.8 \\
				Ours & 
				840,453 & 0.2131 & \textbf{11.2} & \textbf{0.9577} & \textbf{4.5}\\
				\rowcolor{grey1}
				Ours (Re) & 
				854,257 & 0.2098 & \textbf{10.4} & \textbf{0.9701} & \textbf{3.8}
			\end{tabular}
		}
		\vspace{-.2cm}
		\caption{\emph{Bottle Shampoo} (604,903 input points).}
		\vspace{.2cm}
		\label{tab:ExperimentalResultsShampoo}
	\end{minipage}
	\hspace{0.05\textwidth}
	\begin{minipage}{.45\textwidth}
		\centering
		\tiny{
			\begin{tabular}{l|rrrrr}
				Algorithm & $|\mathcal{T}|$ & $E_{\text{avg}}$ & $E_{\text{RMS}}$ & $Q_{\text{avg}}$ & $Q_{\text{RMS}}$\\
				\hline
				Adv.~Front & 
				1,212,636 & 0.2920 & 38.2 & 0.8045 & 18.6 \\
				Adv.~Front (Re) &
				2,407,002 & 0.2038 & 15.4 & 0.9405 & 6.2 \\
				Poisson & 
				13,584 & 2.3850 & 63.2 & 0.8845 & 11.7 \\
				Poisson (Re) & 
				637,488 & 0.3732 & 40.8 & 0.9301 & 6.9 \\
				Poisson MG &
				503,458 & 0.4710 & 39.8 & 0.7062 & 37.1 \\
				Poisson MG (Re) &
				2,409,076 & 0.2050 & 17.7 & 0.9223 & 7.9  \\
				RIMLS & 
				6,458,589 & 0.1331 & 40.4 & 0.6877 & 39.6 \\
				RIMLS (Re) & 
				2,441,143 & 0.2023 & 15.4 & 0.9394 & 6.3 \\
				Scale Space & 
				1,093,339 & 0.2779 & 34.9 & 0.8054 & 18.7  \\
				Scale Space (Re) & 
				1,947,592 & \textbf{0.2006} & 16.1 & 0.9351 & 7.3 \\
				Voronoi & 
				1,212,636 & 0.2916 & 38.4 & 0.8042 & 18.7 \\
				Voronoi (Re) & 
				2,398,584 & 0.2039 & 15.3 & 0.9405 & \textbf{6.1} \\
				Ours & 
				2,137,650 & 0.2167 & \textbf{14.8} & \textbf{0.9485} & 6.2 \\
				\rowcolor{grey1}
				Ours (Re) & 2,246,434 & 0.2093 & \textbf{11.4} & \textbf{0.9665} & \textbf{4.3}
			\end{tabular}
		}
		\vspace{-.2cm}
		\caption{\emph{Bowl Chinese} (606,320 input points).}
		\vspace{.2cm}
		\label{tab:ExperimentalResultsBowl}
	\end{minipage}
	
	\begin{minipage}{.45\textwidth}
		\centering
		\tiny{
			\begin{tabular}{l|rrrrr}
				Algorithm & $|\mathcal{T}|$ & $E_{\text{avg}}$ & $E_{\text{RMS}}$ & $Q_{\text{avg}}$ & $Q_{\text{RMS}}$\\
				\hline
				Adv.~Front & 2,037,574 & 0.1839 & 40.1 & 0.8143 & 17.3 \\
				Adv.~Front (Re) &
				1,739,214 & 0.1965 & 19.4 & 0.9179 & 9.5 \\
				Poisson & 
				147,940 & 0.6300 & 44.3 & 0.8805 & 12.0 \\
				Poisson (Re) & 
				1,488,112 & 0.2068 & 17.5 & 0.9311 & 7.2\\
				Poisson MG & 
				419,614 & 0.4086 & 38.5 & 0.7160 & 36.0 \\
				Poisson MG (Re) &
				1,463,018 & 0.2093 & 18.7 & 0.9154 & 8.8 \\
				RIMLS & 
				5,878,521 & 0.1154 & 39.9 & 0.6919 & 38.9 \\
				RIMLS (Re) &
				1,728,371 & \textbf{0.1978} & 20.1 & 0.9143 & 12.7 \\
				Scale Space & 
				2,036,816 & 0.1839 & 40.0 & 0.8139 & 17.4 \\
				Scale Space (Re) & 
				1,735,814 & 0.1965 & 19.3 & 0.9179 & 9.5 \\
				Voronoi & 
				2,037,270 & 0.1767 & 41.8 & 0.8067 & 18.1 \\
				Voronoi (Re) &
				1,514,160 & 0.2027 & \textbf{15.4} & 0.9407 & \textbf{6.3} \\
				Ours & 
				1,435,604 & 0.2181 & 15.7 & \textbf{0.9454} & 6.6 \\
				\rowcolor{grey1}
				Ours (Re) & 1,535,058 & 0.2089 & \textbf{12.5} & \textbf{0.9592} & \textbf{4.8}
			\end{tabular}
		}
		\vspace{-.2cm}
		\caption{\emph{Cloth Duck} (1,018,891 input points).}
		\label{tab:ExperimentalResultsClothDuck}
	\end{minipage}
	\hspace{0.05\textwidth}
	\begin{minipage}{.45\textwidth}
		\centering
		\tiny{
			\begin{tabular}{l|rrrrr}
				Algorithm & $|\mathcal{T}|$ & $E_{\text{avg}}$ & $E_{\text{RMS}}$ & $Q_{\text{avg}}$ & $Q_{\text{RMS}}$\\
				\hline
				Adv.~Front & 1,214,998 & 0.1474 & 36.3 & 0.8474 & 13.9 \\
				Adv.~Front (Re) &
				629,138 & 0.2024 & 15.2 & 0.9418 & 6.0 \\
				Poisson &
				20,134 &  1.0381 & 54.7 & 0.8882 & 11.7 \\
				Poisson (Re) &
				530,374 & 0.2193 & 22.8 & 0.9293 & 7.3 \\
				Poisson MG &
				432,268 & 0.2585 & 39.5 & 0.2623 & 37.1 \\
				Poisson MG (Re) &
				629,508 & \textbf{0.2021} & 15.0 & 0.9436 & 6.1 \\
				RIMLS &
				5,548,226 & 0.0730 & 40.0 & 0.6910 & 39.3 \\
				RIMLS (Re) &
				618,531 & 0.2049 & 16.2 & 0.9322 & 6.8 \\
				Scale Space &
				1,214,990 & 0.1474 & 36.3 & 0.8474 & 13.9 \\
				Scale Space (Re) &
				628,848 & 0.2025 & 15.2 & 0.9417 & 6.0 \\
				Voronoi &
				1,214,996 & 0.1471 & 36.5 & 0.8471 & 13.9 \\
				Voronoi (Re) &
				616,160 &  0.2041 & 15.0 & 0.9427 & 5.9 \\
				Ours &
				555,490 & 0.2159 & \textbf{13.5} & \textbf{0.9499} & \textbf{5.6} \\
				\rowcolor{grey1}
				Ours (Re) & 578,730 & 0.2096 & \textbf{11.5} & \textbf{0.9657} & \textbf{4.3}
			\end{tabular}
		}
		\vspace{-.2cm}
		\caption{\emph{Toy Bear} (607,501 input points).}
		\label{tab:ExperimentalResultsToyBear}
	\end{minipage}
\end{table*}

From the algorithms listed in Section~\ref{sec:RelatedWork}, Poisson~\cite{kazhdan2006poisson}, advancing front~\cite{CohenSteiner2004AGD}, and scale space~\cite{digne2011scale} are run with the standard parameters as implemented in~\cite{cgal2022user} except for the cleaning steps, which were unnecessary because of the high-quality input.
Multigrid Poisson~\cite{kazhdan2019multi} and Voronoi reconstruction~\cite{boltcheva2017surface} are run with the standard parameters as implemented in~\cite{levy2023geogram}.
RIMLS~\cite{oztireli2009feature} is run with the standard parameters from~\cite{cignoni2008meshlab}, using a smoothness of~2 and a grid resolution of~1000.

We aim for an algorithm that provides high-quality triangulations out-of-the-box, right after reconstruction.
However, as the comparison algorithms do not necessarily optimize for a uniform edge length, we take their respective results and process them with the ``Isotropic Explicit Remeshing'' filter of MeshLab~\cite{cignoni2008meshlab}.
This filter repeatedly applies edge flip, collapse, relax, and refine operations~\cite{hoppe1993Mesh}.
We run three iterations with a target edge length of~$0.2$ in absolute world units for the input~\cite{huang2022surface}.

In Tables~\ref{tab:ExperimentalResultsShampoo} to~\ref{tab:ExperimentalResultsToyBear}, we report both the results of the comparison algorithms and the result after these have been remeshed, indicated by~``(Re)''.
These tables include representative models.
A full report with data for all~20 models can be found in the supplementary material. 
We chose the \emph{Bottle Shampoo} and the \emph{Bowl Chinese} because of their features, as explored in Figures~\ref{fig:VaryingSplatSizes} and~\ref{fig:SplitMeshes}.
The \emph{Cloth Duck} is one of two models where the competing methods had the largest gain on our algorithm when measured by~$E_{\text{avg}}$ (see supplementary material for the \emph{Mug}).

A first thing to notice when regarding the results presented in Tables~\ref{tab:ExperimentalResultsShampoo} to~\ref{tab:ExperimentalResultsToyBear} is that our algorithm achieves the best, that is, highest values for~$Q_{\text{avg}}$ on all models.
This holds consistently across all~20 models from the repository. 
That is, our method produces the highest quality meshes, even when compared with the remeshed results of the other algorithms.
For comparison, we also add the remeshed version of our algorithm, which generally improves the quality metrics slightly while destroying the minimum edge length guarantee.
The goal of this paper is not to compare different remeshing approaches, but to present a method that can provide high-quality triangle meshes right after reconstruction, without remeshing.
Hence the remeshed version of our algorithm is set apart in gray and carries bold font if it causes an improvement on the previously best result.
In this setting, the comparison to the remeshed results just serves to place our results in a broader setting.

On most of the models, the deviation~$Q_{\text{RMS}}$ has also the lowest percentages for our algorithm.
Notable exceptions are the \emph{Bowl Chinese} (Table~\ref{tab:ExperimentalResultsBowl}) and the \emph{Cloth Duck} (Table~\ref{tab:ExperimentalResultsClothDuck}).
However, across all models, the lowest deviation~$Q_{\text{RMS}}$ is at most~$0.6\%$ better than ours, cf.~supplementary material.

Regarding the second metric, note that by construction, all edges produced by our algorithm are of length~$\geq0.2$.
Therefore, the average edge length is also always greater than~$0.2$, which places the remeshed output of other methods in the lead regarding the metric~$E_{\text{avg}}$.
However, the largest average edge length across all models is~$0.2181$ for our algorithm, attained on the \emph{Cloth Duck} (Table~\ref{tab:ExperimentalResultsClothDuck}), which is still very close to the target edge length.

Also, for almost all models, the width of the distribution of edge lengths, measured by~$E_{\text{RMS}}$, is the lowest for our algorithm.
That is, the triangulations produced are almost uniform.
As a final observation regarding the quality metrics, note that those comparison algorithms that provide better metrics on the models do so only after an additional remeshing step.
This shows that our algorithm does attain the goal of providing high-quality meshes immediately after reconstruction as it beats all comparison algorithms in this regard.

When inspecting the models visually, it is clear that, at least after remeshing, the triangulations are of high quality (Figure~\ref{fig:SplitMeshes}).
Note how some algorithms are not able to reproduce small details---for instance, a number~14 on the \emph{Bottle Shampoo}.
Even in the remeshed version, line-like artifacts are still visible for some of the comparison algorithms.
Our algorithm creates a mesh close to uniformity while retaining the details.

\begin{figure}
	\centering
	\begin{subfigure}[t]{.8\linewidth}
		\centering
		\includegraphics[width=\textwidth]{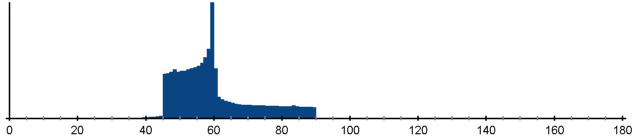}
		\caption{Angle distribution, target=$60^\circ$.}
	\end{subfigure}
	\begin{subfigure}[t]{.8\linewidth}
		\centering
		\includegraphics[width=\textwidth]{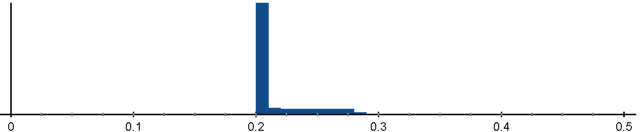}
		\caption{Edge lengths distribution, target=0.2.}
	\end{subfigure}
	\begin{subfigure}[t]{.8\linewidth}
		\centering
		\includegraphics[width=\textwidth]{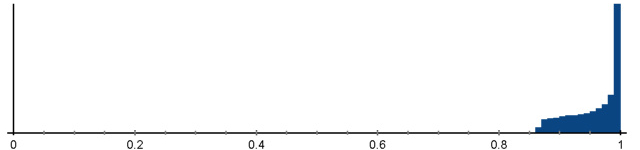}
		\caption{Distribution of quality $Q_t$, target=1.0.}
	\end{subfigure}
	\caption{
		Distributions of the \emph{Bottle Shampoo} as obtained by our algorithm (without remeshing).
	}
	\label{fig:BottleShampooHistogram}
\end{figure}

This uniformity can be observed by plotting histograms on the distribution of angles, edge lengths, and quality measures for a triangulation obtained by our algorithm.
See Figure~\ref{fig:BottleShampooHistogram} for a corresponding set of plots for the \emph{Bottle Shampoo} and find histograms for the other models in the supplementary material.
The histogram confirms that the angles of the triangles are centered around~$60^\circ$, indicating a strong tendency towards equilateral triangles.
Also, we see that the edge lengths are indeed starting from the set minimum of~0.2, with most edges actually attain this value.
Finally, the histogram of the triangle quality reveals that there are many equilateral triangles (corresponding to~${Q_t=1}$), with the distribution skewed towards this highest quality value.

\begin{figure}
	\includegraphics[width=1.\linewidth]{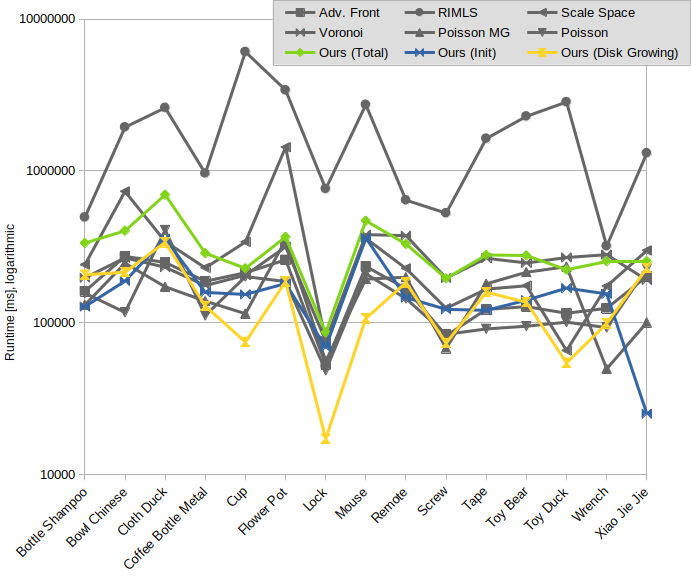}
	\caption{
		Log of the run time of the algorithms on several models. 
		Ours is additionally split into initialization and disk growing.
	}
	\label{fig:RunTimes}
\end{figure}

Unlike some competitors and the remeshing step, our algorithm is not iterative but produces the output in a single sweep over the input.
Run times for several models are given in Figure~\ref{fig:RunTimes}, where the competitors are reported including the remeshing time.
All experiments were run on a machine with an Intel$\textsuperscript{\textregistered}$ Core$^{\text{TM}}$ i7-5600U CPU 2.60GHz with four cores and 16GB of RAM.
Five of the models did not fit the RAM of this comparison machine.
Thus, we only provide timings for 15 models, while the qualitative data for the remaining five was acquired on another machine.
Note that our algorithm performs similarly to most of the competitors.
\begin{figure}
	\centering
	\includegraphics[width=.9\linewidth]{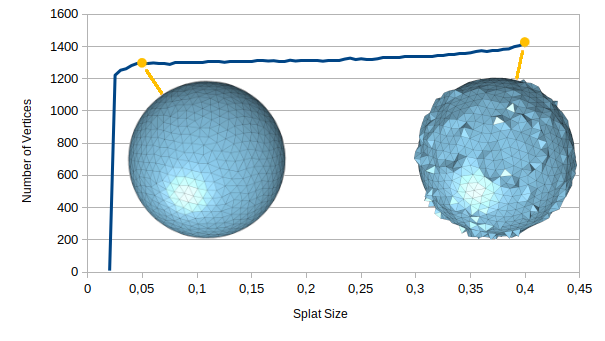}
	\includegraphics[width=.9\linewidth]{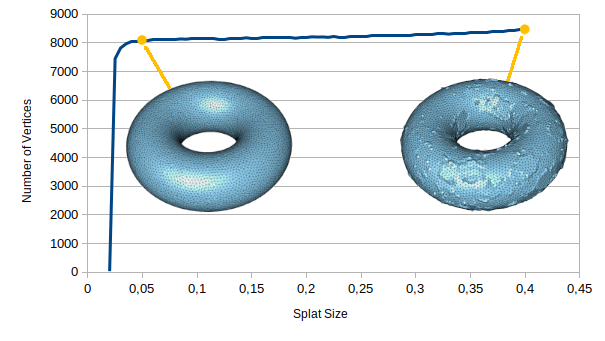}
	\includegraphics[width=.47\linewidth]{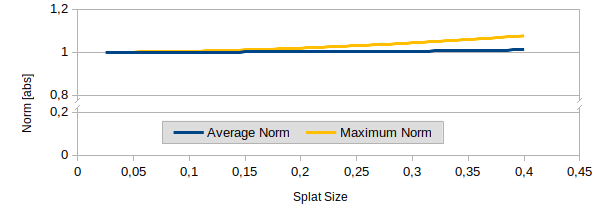}
	\hfill
	\includegraphics[width=.47\linewidth]{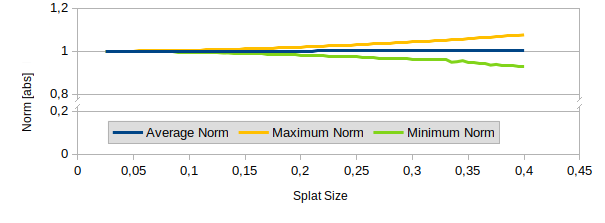}
	\caption{Measuring the reconstruction quality.}
	\label{fig:TorusReconstruction}
\end{figure}

\begin{figure*}
	\newlength{\StartImages}
	\setlength{\StartImages}{0.105\linewidth}
	\centering
	\begin{subfigure}[t]{\StartImages}
		\centering
		\includegraphics[width=\textwidth]{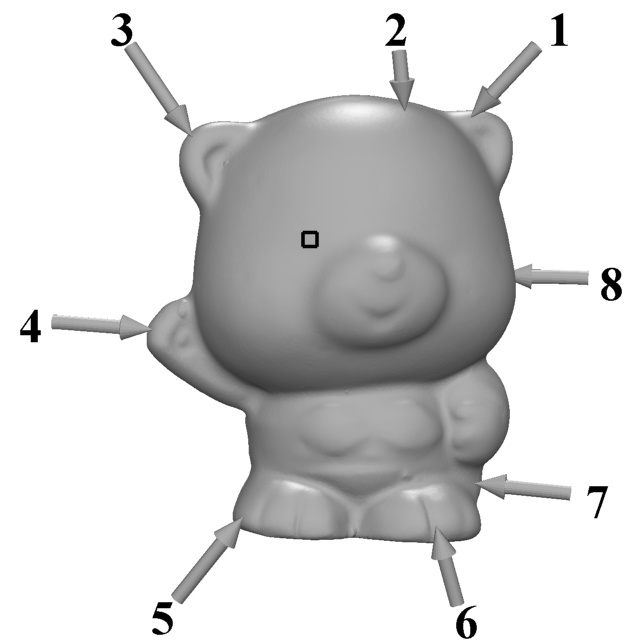}
	\end{subfigure}
	\begin{subfigure}[t]{\StartImages}
		\centering
		\includegraphics[width=\textwidth]{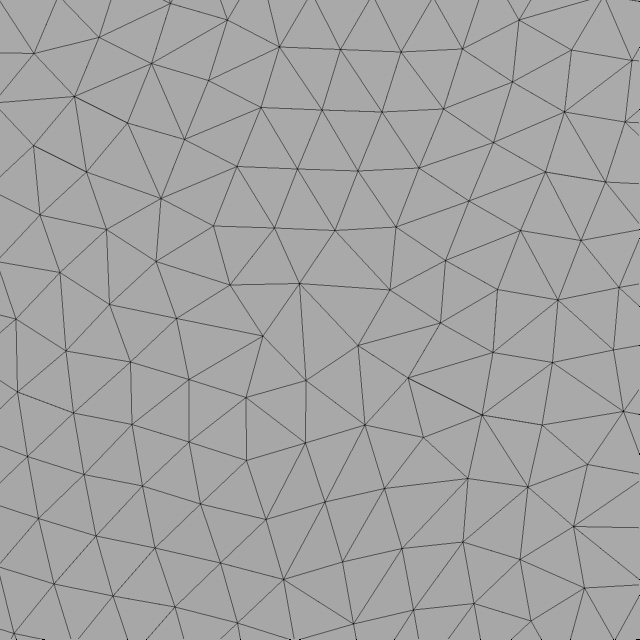}
	\end{subfigure}
	\begin{subfigure}[t]{\StartImages}
		\centering
		\includegraphics[width=\textwidth]{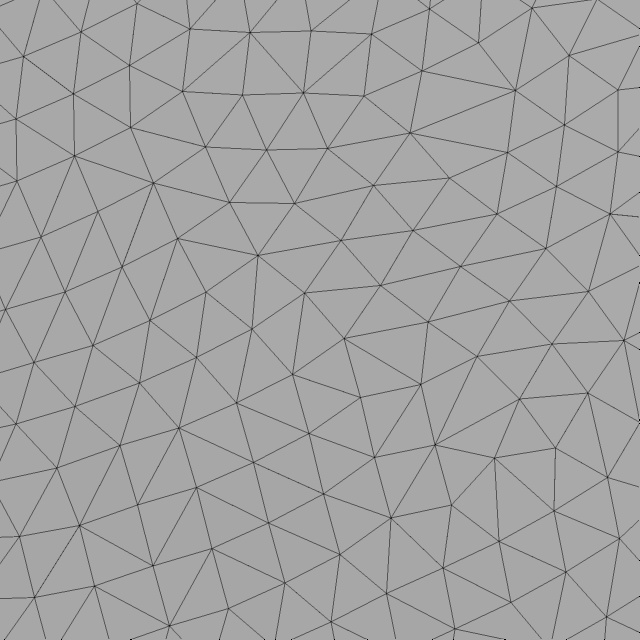}
	\end{subfigure}
	\begin{subfigure}[t]{\StartImages}
		\centering
		\includegraphics[width=\textwidth]{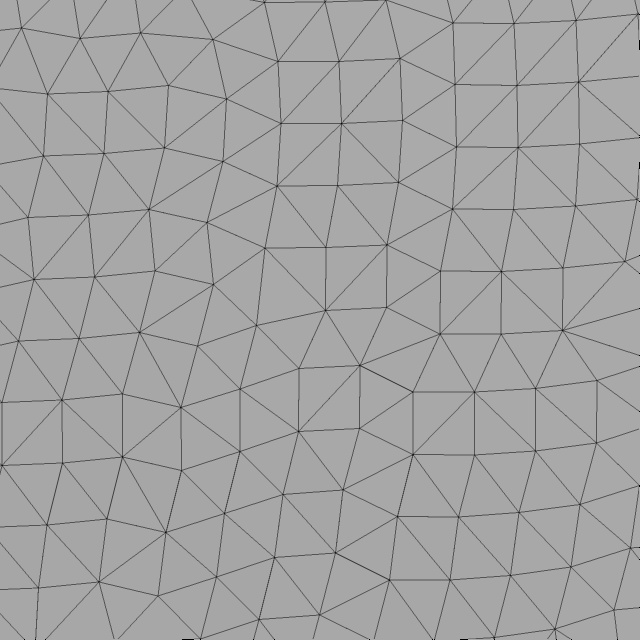}
	\end{subfigure}
	\begin{subfigure}[t]{\StartImages}
		\centering
		\includegraphics[width=\textwidth]{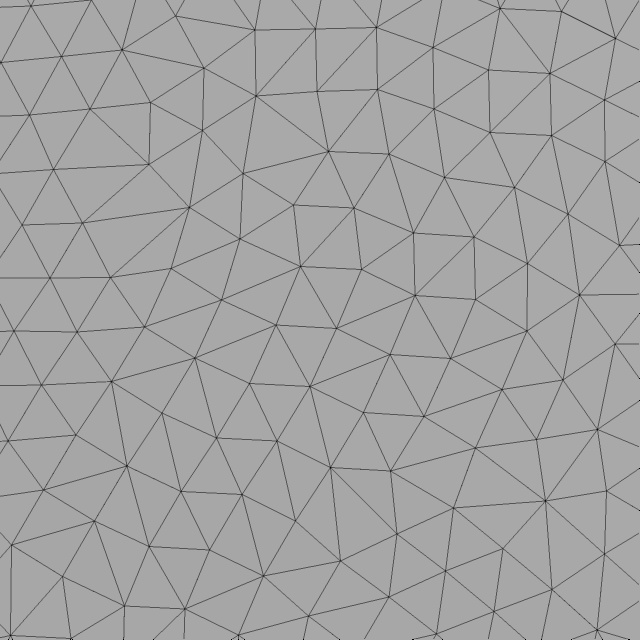}
	\end{subfigure}
	\begin{subfigure}[t]{\StartImages}
		\centering
		\includegraphics[width=\textwidth]{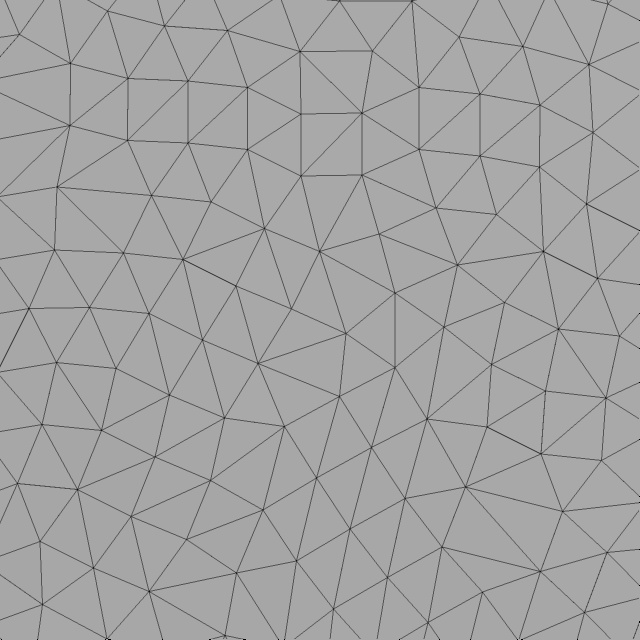}
	\end{subfigure}
	\begin{subfigure}[t]{\StartImages}
		\centering
		\includegraphics[width=\textwidth]{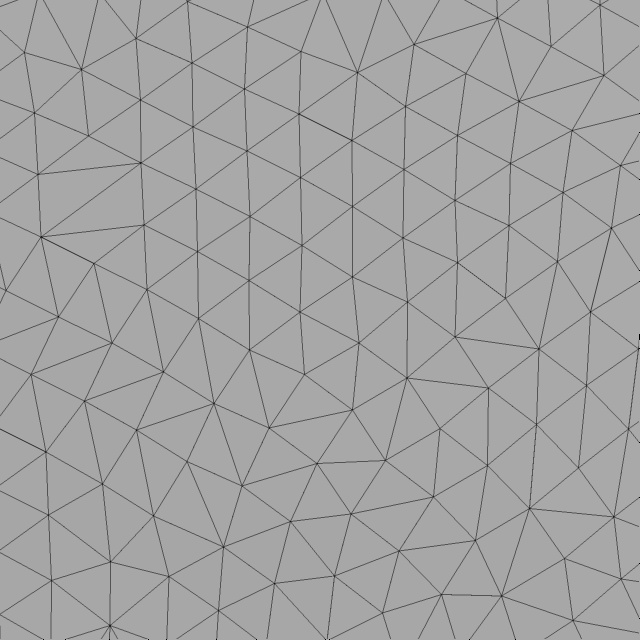}
	\end{subfigure}
	\begin{subfigure}[t]{\StartImages}
		\centering
		\includegraphics[width=\textwidth]{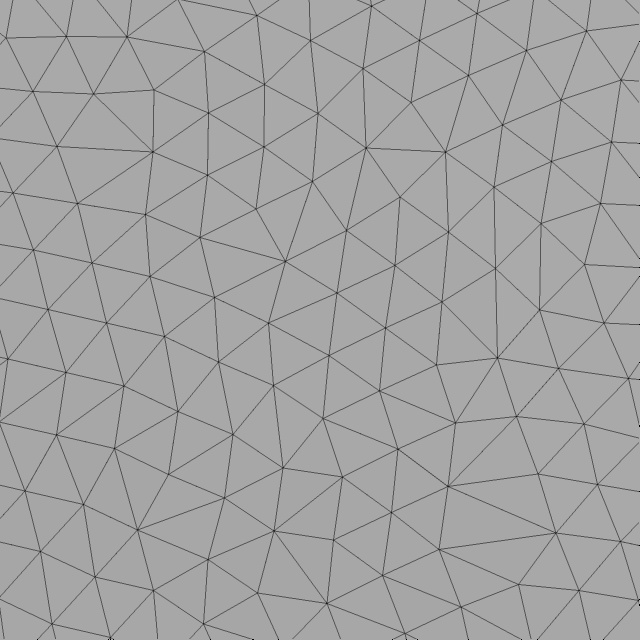}
	\end{subfigure}
	\begin{subfigure}[t]{\StartImages}
		\centering
		\includegraphics[width=\textwidth]{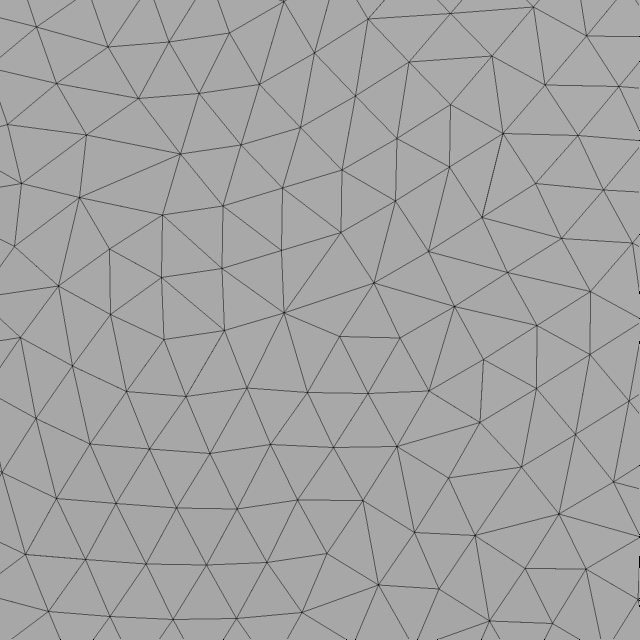}
	\end{subfigure}
	\caption{Results with various starting vertices. From left to right: The \emph{Toy Bear} with placements positions for starting vertex pairs, close up for pairs 1 to 8, showing the area at one eye emphasized in the first image.}
	\label{fig:VariousStartingVertices}
\end{figure*}
\subsection{Robustness to User Input}
\label{sec:Robustness}
As stated in Section~\ref{sec:Initialization}, the user is asked to provide two starting vertices to run the algorithm.
To investigate whether the quality of the obtained mesh is independent of the chosen starting vertices, we selected the \emph{Toy Bear} because of its various differently curved regions.
\reb{Further models promoting this observation are included in the supplementary material.}
As illustrated in Figure~\ref{fig:VariousStartingVertices}, we chose eight different regions to place the starting vertices in.
The results of these experiments show that the quality of the output is not sensitive to the choice of starting vertices.
For the eight resulting triangle meshes, the average edge length varies from~$0.2158$ to~$0.2159$, as does the average quality: from~$0.9499$ to~$0.9504$.
In all cases,~$E_{\text{RMS}}$ is equal to~$13.5$ while~$Q_{\text{RMS}}$ equals~$5.6$.

Next, we investigate the robustness of surface reconstruction depending on the splat size.
As the models discussed so far are real-world scans, there is no ground truth to compare the reconstruction with.
For this experiment, we turn to two models that satisfy all assumptions made in Section~\ref{sec:AssumptionsAndTheory} and that have an explicit mathematical parametrization to evaluate the reconstruction: the unit-sphere and a torus parametrized as a unit circle swept around a circle of radius~2.
We sample both models randomly, the sphere with~10,000 and the torus with~60,000 points, resulting in a similar density on the models.
The norm is a direct measure of the reconstruction quality.
For the sphere model, vertices with norm~$1$ lie directly on the sampled sphere.
For the torus model, we measure the norm as the distance to the circle of rotation, hence, a vertex with norm~$1$ lies directly on the sampled torus.
In this scenario, the sphere diameter~$d$ was chosen as~$0.1$ while a global splat size was chosen between~$0.02$ and~$0.4$.

For both models, given too small splat sizes, the algorithm fails to cover the entire model, resulting in a very small number of vertices.
Once a splat size is reached for which the entire model is covered, both the number of vertices and the reconstruction quality are stable until larger splat sizes are reached, which causes visible distortion in the reconstructed models (Figure~\ref{fig:TorusReconstruction}).
This shows that for splat sizes, just large enough to cover the geometry, our algorithm achieves close to optimal reconstruction results.
For the sphere model, all points created are on or outside the sphere, placing the closest vertex at a norm of~$1$ directly on the sphere.
On the torus model, points are lying both in and outside of the torus.
Even for the largest splat size of~$0.4$, which creates visible reconstruction artifacts, the reconstructed models are still manifold, in line with our guarantees from Section~\ref{sec:AssumptionsAndTheory}.

\subsection{Robustness to Noise}
\label{sec:ExperimentalResultsForMeshingNoisyPointClouds}
In order to investigate the robustness of our algorithm with respect to noisy data, we equip a selection of the high-quality models~\cite{huang2022surface} with different levels of noise~${\nu \in \mathbb{R}_{\geq 0}}$.
Here, we use those models that allow for placing moderate noise, for instance, the \emph{Bottle Shampoo} or the \emph{Bowl Chinese}, whereas we ignore those that already have details and elements that hinder manifold reconstruction even for tiny levels of noise.
Thereby, each input point is moved by a uniformly distributed random vector of length smaller or equal to~$\nu$.
This moves the noisy points within a bound of~${\pm\nu}$ around the ground truth.
To measure the quality of the output, for each level~$\nu$ of noise, we computed a triangulation based on the same parameter choices as above (Section~\ref{sec:Experiments}).
We find that the level of noise directly influences the number of vertices, similar to the observations made while increasing the splat size (Section~\ref{sec:Robustness}).
That is, with increasing noise level, the number of vertices of the output increases as well.
Depending on the model, we experience that for values ${\nu \in \left[0.06,0.1\right]}$, the output begins not to be manifold anymore.
That is, manifoldness is lost from~$2\nu$ between 60\% to 100\% of~$d$.
For the user, this experiment suggests that for a geometry with known or estimated noise level~$\nu$, choosing~$d\geq2\nu$ yields the best results.

\begin{table}
	\begin{minipage}{.47\textwidth}
		\centering
		\tiny{
			\begin{tabular}{l|ccccccccccc}
				Name $\backslash$ $\nu$ & \rotatebox[origin=c]{90}{0.00} & \rotatebox[origin=c]{90}{0.01} & \rotatebox[origin=c]{90}{0.02} & \rotatebox[origin=c]{90}{0.03} & \rotatebox[origin=c]{90}{0.04} & \rotatebox[origin=c]{90}{0.05} & \rotatebox[origin=c]{90}{0.06} & \rotatebox[origin=c]{90}{0.07} & \rotatebox[origin=c]{90}{0.08} & \rotatebox[origin=c]{90}{0.09} & \rotatebox[origin=c]{90}{0.1}\\
				\hline
				\emph{Bottle Shampoo} & \yes & \yes & \yes & \yes & \yes & \yes & \yes & \yes & \yes & \no  & \no \\
				\emph{Bowl Chinese}   & \yes & \yes & \yes & \yes & \yes & \yes & \yes & \yes & \yes & \yes & \no \\
				\emph{Cup} 			  & \yes & \yes & \yes & \yes & \yes & \yes & \yes & \yes & \no  & \no  & \no \\
				\emph{Flower Pot 2}   & \yes & \yes & \yes & \yes & \yes & \yes & \no  & \no  & \no  & \no  & \no \\
				\emph{Toy Bear} 	  & \yes & \yes & \yes & \yes & \yes & \yes & \yes & \yes & \no  & \no  & \no \\
				\emph{Toy Duck}		  & \yes & \yes & \yes & \yes & \yes & \yes & \yes & \yes & \no  & \no  & \no \\
			\end{tabular}
		}
		\caption{Noise levels~$\nu$ for which the reconstruction is (\yes) or is not (\no) manifold.}
		\label{tab:NoiseExperiments}
	\end{minipage}
\end{table}

\section{Extension to Surface Remeshing}
\label{sec:ExtensionToSurfaceRemeshing}

\noindent Here, we modify the ansatz presented in Sections~\ref{sec:TheoryAndMethodology} and~\ref{sec:Implementation} to remesh polyhedral surfaces to achieve an isotropic triangular mesh with guaranteed smallest edge length.
To guarantee the theoretical results collected in Section~\ref{sec:TheoryAndMethodology}, we assume the input polyhedral surface~$\mathcal{F}$ to represent an orientable, closed, and compact~$\mathcal{C}^2$-manifold embedded into~$\mathbb{R}^3$, which is of finite, positive reach~$\rho$.
The strict definition from Section~\ref{sec:AssumptionsAndTheory} would give a reach of~${\rho=0}$ for polyhedral surfaces.
However, several methods are available to compute an approximation of an idealized surface that the polyhedral input is assumed to represent, where a user-given parameter steers how closely the input should be taken into account~\cite{foskey2003efficient}.

In Section~\ref{sec:Methodology}, we equipped the input point cloud with circular splats on which the output surface is built on.
Now, turning to polyhedral surfaces, we can skip the splats and place the spheres on the faces of the polyhedral surface directly.
Consequently, we use the face normals instead of splat normals for further calculations.
This also removes the need for finding individual splat radii.
To build the box data structure described in Section~\ref{sec:AuxiliaryBoxDataStructure}, to each box~$b_j$, we associate all faces of the input with distance less than~$d$ to~$b_j$.
Afterwards, we run the algorithm as described in Sections~\ref{sec:Initialization} to~\ref{sec:TriangulatingTheResultingMesh} on the faces of the input geometry.
Hence, the resulting surface interpolates the input one.

\begin{table}[h!]
	\centering
	\tiny{
	\begin{tabular}{llllrrrr}
		& $d$ & $| \mathcal{V} |$ & $| \mathcal{T} |$ & $\alpha_{\min}$ & $\alpha_{\max}$& $\alpha_{\text{avg}}$\hspace{-4mm} 
		\\
		\hline
		input & --- &10,000 & 20,000 & 7.6919$^{\circ}$ & 153.3379$^{\circ}$ & 60$^{\circ}$\hspace{-4mm} 
		\\
		remesh & 0.02 & 17,673 & 35,346 & 30.2108$^{\circ}$ & 112.2197$^{\circ}$ & 60$^{\circ}$\hspace{-4mm} 
		\\
		remesh & 0.03 & 7,837 & 15,674 & 28.3165$^{\circ}$ & 115.9218$^{\circ}$ & 60$^{\circ}$\hspace{-4mm} 
		\\
		& & & & & & & \\
		& $E_{\min}$ & $E_{\text{max}}$ & $E_{\text{avg}}$ 
		& $Q_{\text{min}}$ & $Q_{\text{max}}$ & $Q_{\text{avg}}$\hspace{-4mm} 
		\\
		\hline
		& 0.0039 & 0.1119 & 0.0287 
		& 0.2296 & 0.9999 & 0.8438\hspace{-4mm} 
		\\
		& 0.02 & 0.0391 & 0.0213 
		& 0.6742 & 1.0000 & 0.9558\hspace{-4mm} 
		\\
		& 0.03 & 0.0613 & 0.0320 
		& 0.6391 & 1.0000 & 0.9552\hspace{-4mm} 
	\end{tabular}
}
	\caption{Experimental results for remeshing \emph{Kitten}.}
	\label{tab:ExperimentalResultsKitten}
\end{table}

\begin{figure}[h!]
		\centering
	\begin{subfigure}[t]{0.32\columnwidth}
		\includegraphics[width=\textwidth]{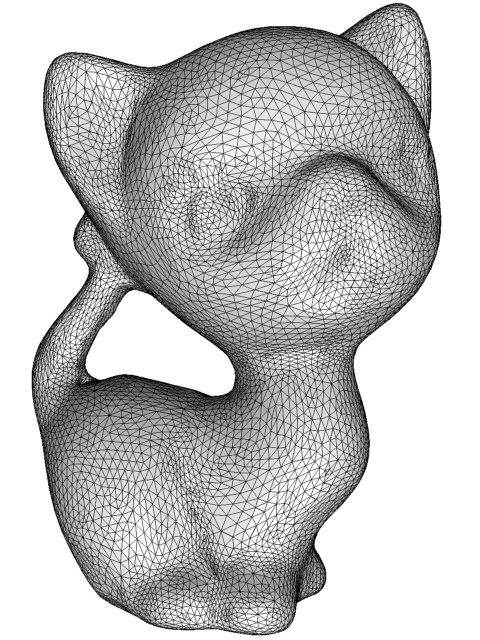}
	\end{subfigure}
	\begin{subfigure}[t]{0.32\columnwidth}
		\includegraphics[width=\textwidth]{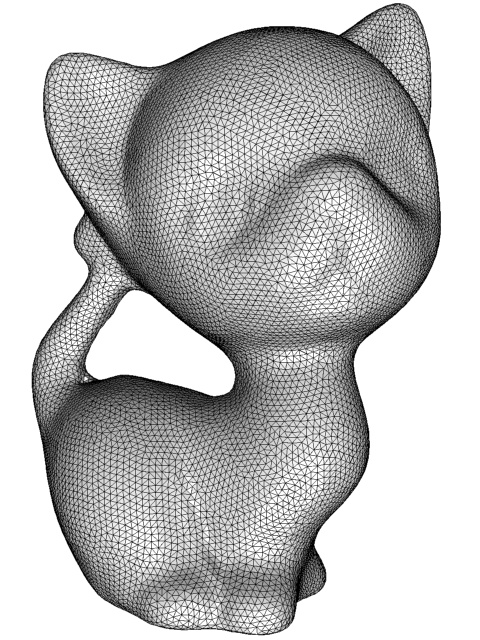}
	\end{subfigure}
	\begin{subfigure}[t]{0.32\columnwidth}
		\includegraphics[width=\textwidth]{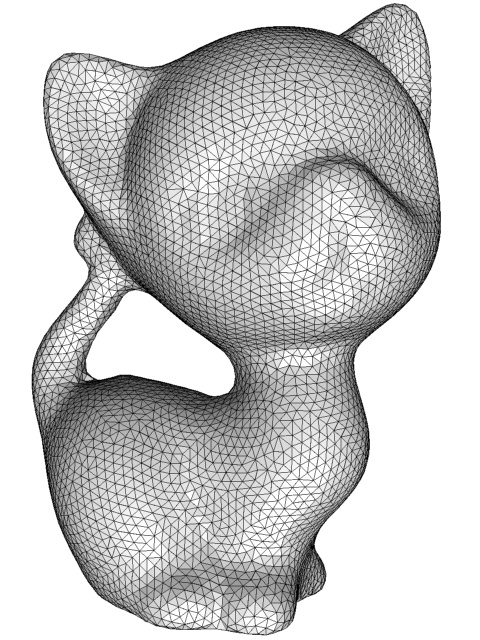}
	\end{subfigure}
	\caption{
		Result of the algorithm run on \emph{Kitten} model.
		Left: input geometry.
		Middle: input geometry remeshed with target edge length~$0.02$.
		Right: input geometry remeshed with target edge length~$0.03$.
	}
	\label{fig:Result_RemeshingKitten}
\end{figure}

We illustrate the applicability of our algorithm to remesh a freeform mesh by running it on the \emph{Kitten} model.
Here, we run our algorithm with target edge length~$d$ varying between~0.01 and~0.08.
Since the input consists of~10,000 vertices and has an average edge length of~$0.0287$, we consider the results achieved for~$d = 0.02$ and~$d = 0.03$ in more detail.
As shown in Table~\ref{tab:ExperimentalResultsKitten}, for both target edge lengths, the mesh quality is improved.
Both, the resulting edge lengths as well as the resulting angles are less widely distributed than provided by the input mesh.
Further, the model remeshed has a better average quality~$Q_{\text{avg}}$ equal to~0.9558 for~$d = 0.02$ or to~0.9552 for~$d = 0.03$, respectively, in comparison to~0.8438 of the input model.
The meshes resulting for the mentioned target edge lengths are shown in Figure~\ref{fig:Result_RemeshingKitten}.
The complete data set can be found in the supplementary material.

\begin{figure}[h!]
	\centering
	\begin{subfigure}[t]{0.49\columnwidth}
		\includegraphics[width=\textwidth]{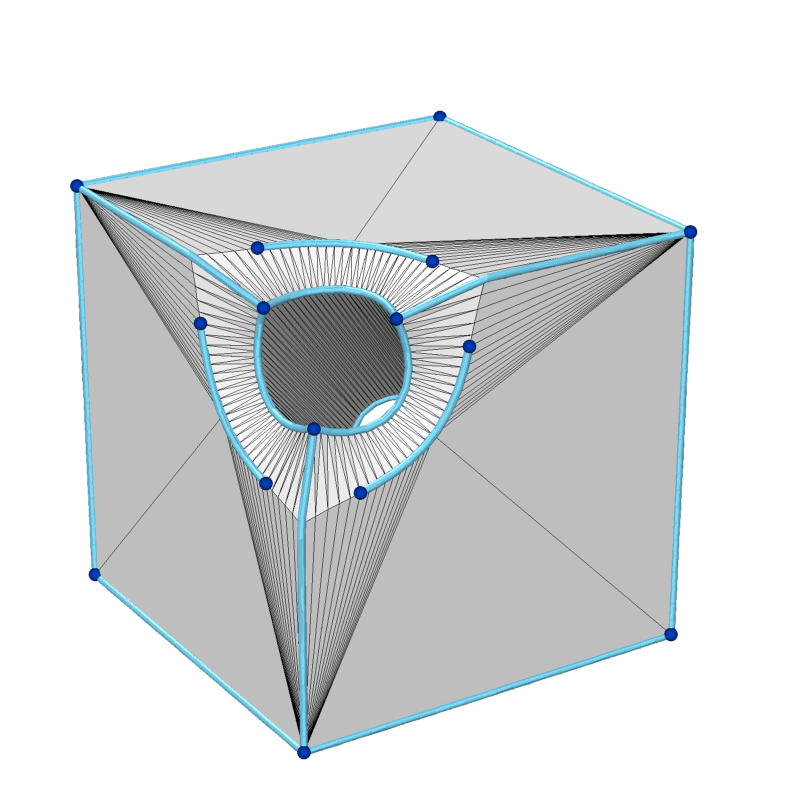}
	\end{subfigure}
	\begin{subfigure}[t]{0.49\columnwidth}
		\includegraphics[width=\textwidth]{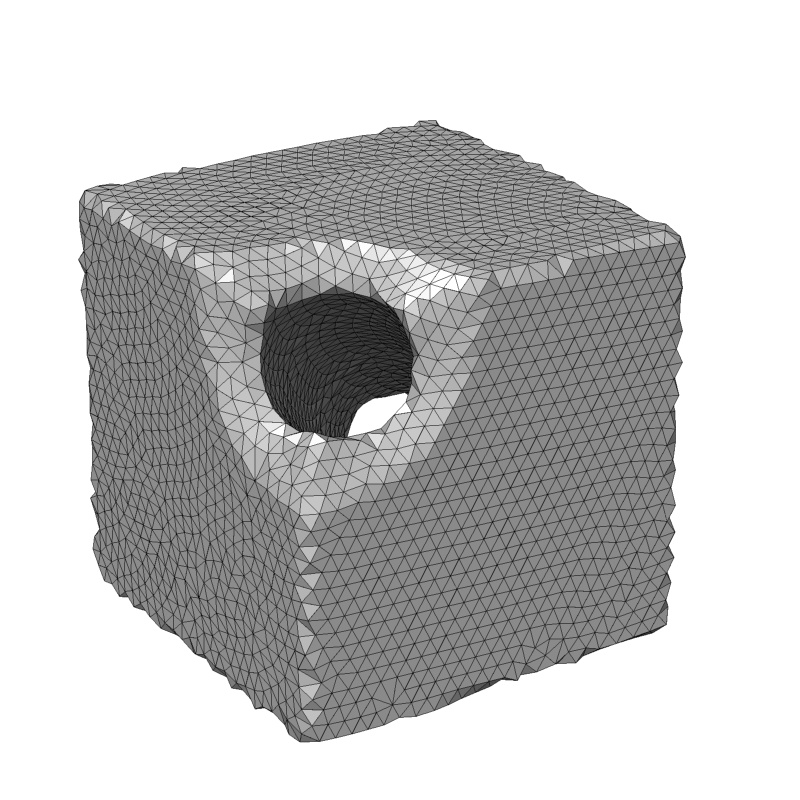}
	\end{subfigure}
	\caption{
		Result of the algorithm run on CAD model.
		Left: the input geometry with detected features (blue).
		Right: the remeshed output with more nearly regular triangles, but also a loss of features.
	}
	\label{fig:Result_SurfaceRemeshing}
\end{figure}

\begin{figure*}[h!]
	\centering
	\begin{subfigure}[t]{.15\textwidth}
		\centering
		\includegraphics[width=\textwidth]{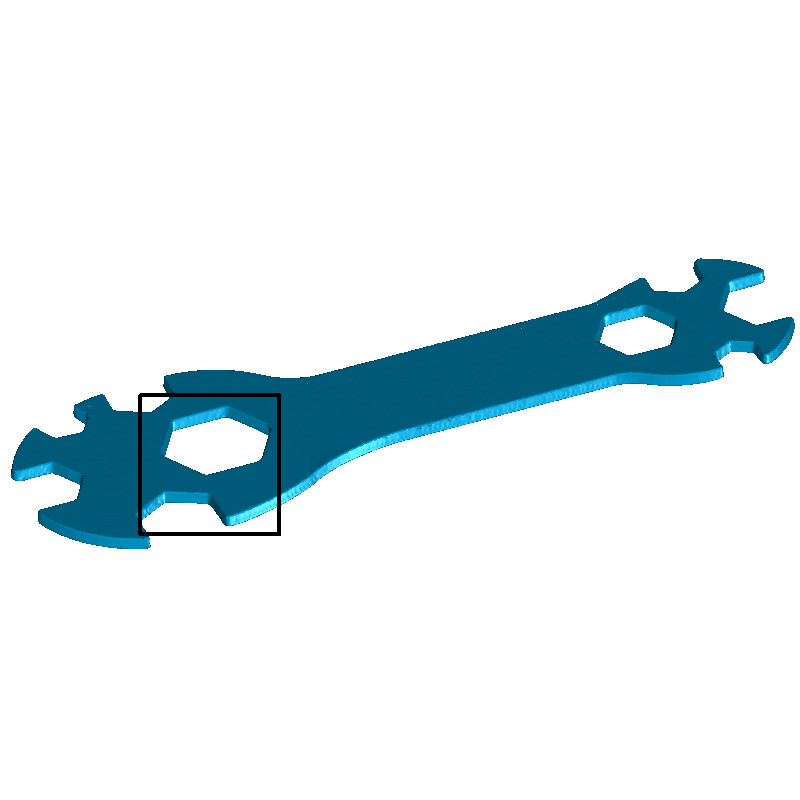}
		\caption{Input point cloud with highlighted sector.}
		\label{fig:InputHighlight_Wrench}
	\end{subfigure}
	~
	\begin{subfigure}[t]{.15\textwidth}
		\centering
		\includegraphics[width=\textwidth]{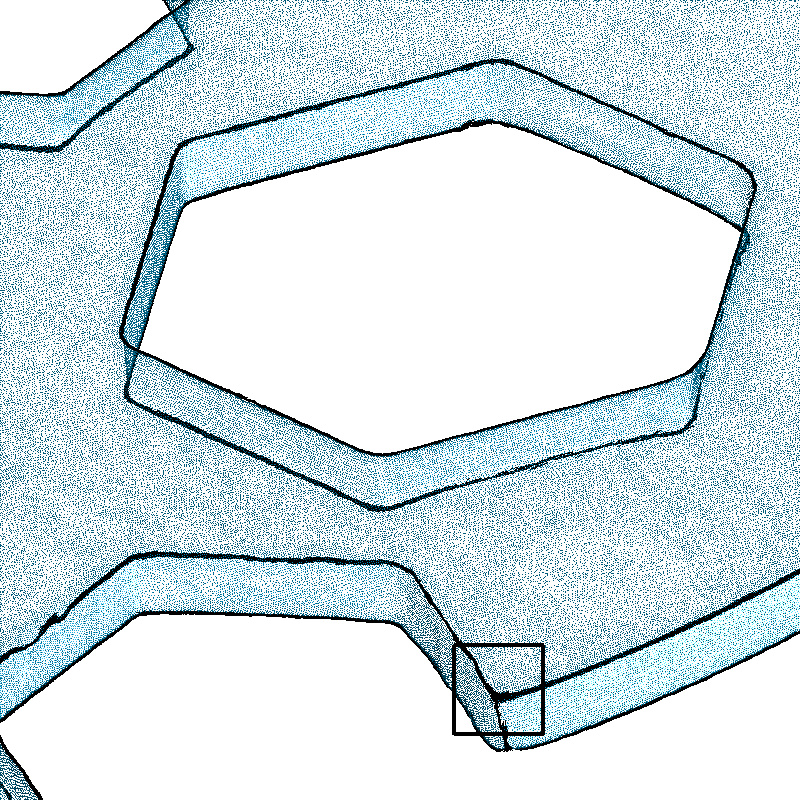}
		\caption{Collection of feature edges (black), $\vartheta=60^{\circ}$.}
		\label{fig:FeaturePointCloud_edges}
	\end{subfigure}
	~
	\begin{subfigure}[t]{.15\textwidth}
		\centering
		\includegraphics[width=\textwidth]{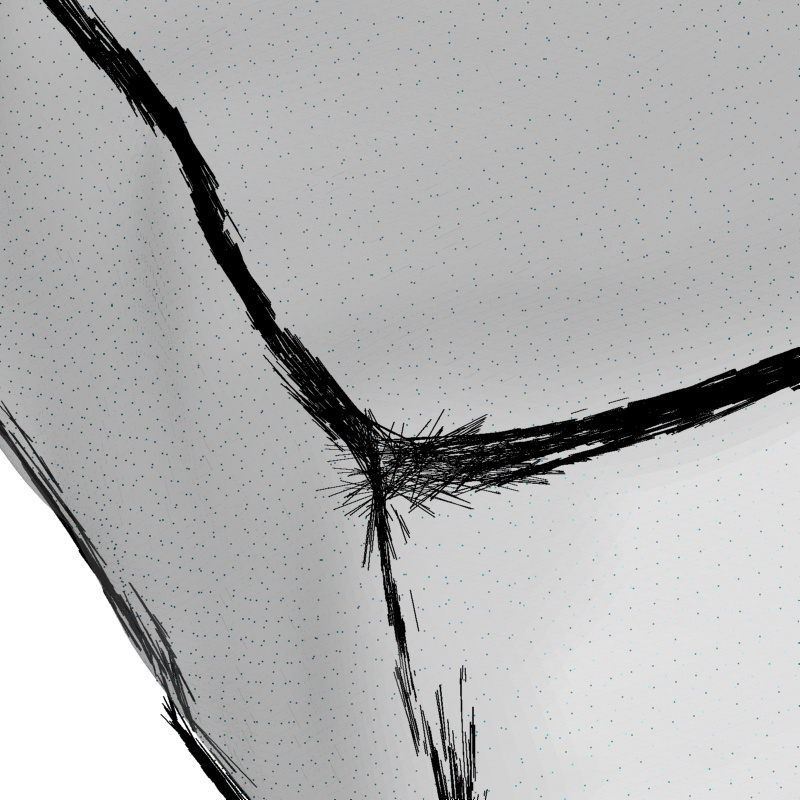}
		\caption{Collection of feature edges (close up).}
		\label{fig:FeaturePointCloud_edgesCloseUp}
	\end{subfigure}
	~
	\begin{subfigure}[t]{.15\textwidth}
		\centering
		\includegraphics[width=\textwidth]{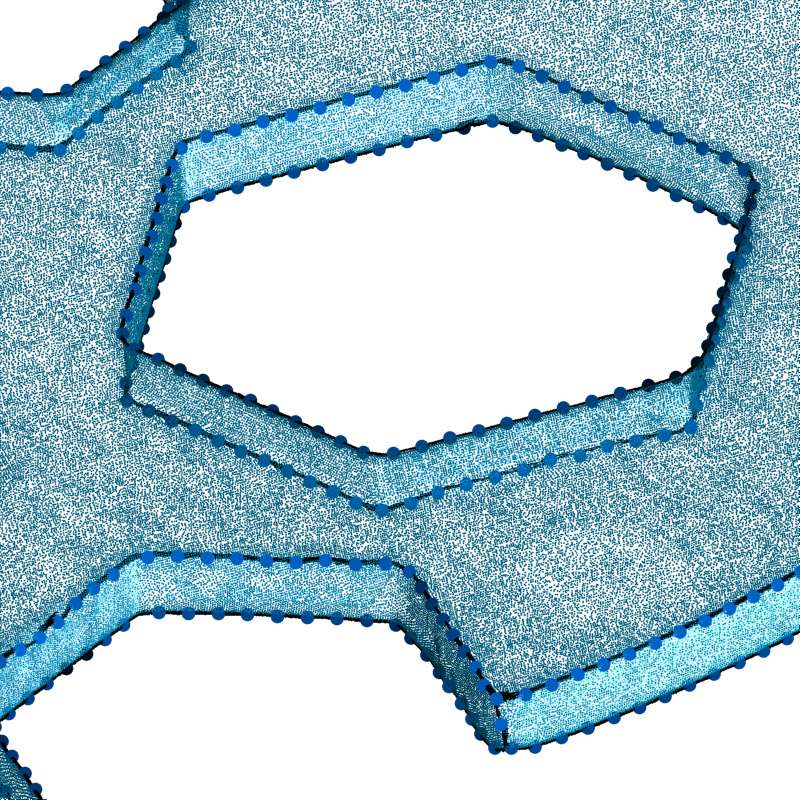}
		\caption{Feature vertices distributed on feature segments.}
		\label{fig:FeaturePointCloud_VerticesOnEdges}
	\end{subfigure}
	~
	\begin{subfigure}[t]{.15\textwidth}
		\centering
		\includegraphics[width=\textwidth]{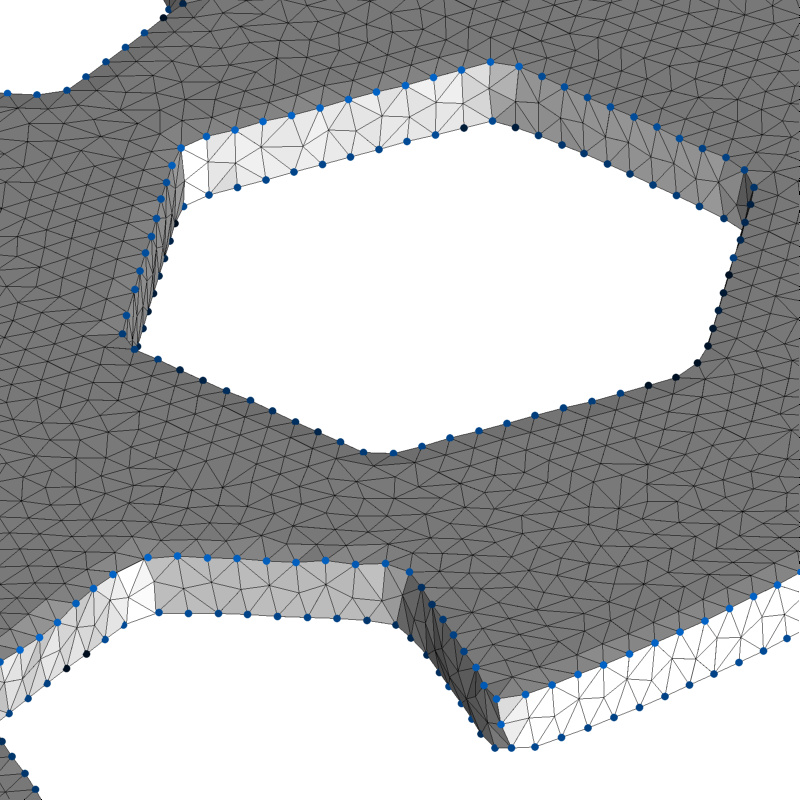}
		\caption{Meshed point cloud with feature detection.}
		\label{fig:FeaturePointCloud_meshed}
	\end{subfigure}
	~
	\begin{subfigure}[t]{.15\textwidth}
		\centering
		\includegraphics[width=\textwidth]{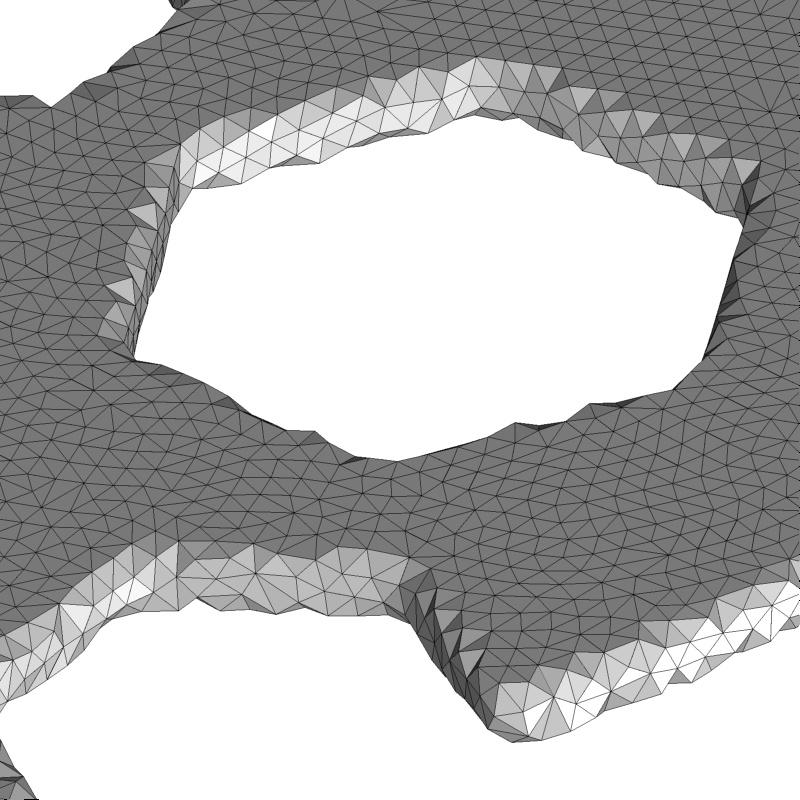}
		\caption{Meshed point cloud without feature detection.}
		\label{fig:FeaturePointCloud_withoutFeatureDetection}
	\end{subfigure}
	\caption{
		Feature detection on the \emph{Wrench} model.
	}
	\label{fig:FeaturesPointClouds}
\end{figure*}

Also for the CAD model shown in Figure~\ref{fig:Result_SurfaceRemeshing}, the modified algorithm returns a remeshed triangular polyhedral surface.
The result has more triangles than the input and the resulting triangles are closer to regular triangles than those of the input geometry.
The improved triangle quality does come with a caveat.
Since we place the spheres such that a newly introduced one touches two spheres already placed, the sphere centers---which later form the vertex set of the resulting mesh---do not necessarily coincide with input vertices nor are they likely to lie on edges of the input surface.
Hence, features like ridges, as shown in Figure~\ref{fig:Result_SurfaceRemeshing}, are worn off.
In the next section, we therefore discuss how to further modify the algorithm to maintain features on both polyhedral surfaces and point clouds.

\section{Feature Detection}
\label{sec:FeatureDetection}

In this section, we focus on retaining features on point clouds and polyhedral surfaces throughout our algorithm.
To define and to detect features, we use the dihedral angle~$\alpha$ formed by the normals of intersecting splats introduced in Section~\ref{sec:Methodology}.
In case of a polyhedral surface input, we use the dihedral angle~$\alpha$ formed by adjacent faces.
In either case, we additionally introduce an angle threshold~$\vartheta$ chosen by the user.
This threshold defines a lower bound such that every intersection with~${\alpha>\vartheta}$ will be considered to be a feature and thus to be retained throughout the remeshing.

\subsection{Feature Detection on Point Clouds}
\label{sec:FeatureDetectionOnPointClouds}

\begin{figure}[h!]
	\centering
	\includegraphics[width=\columnwidth]{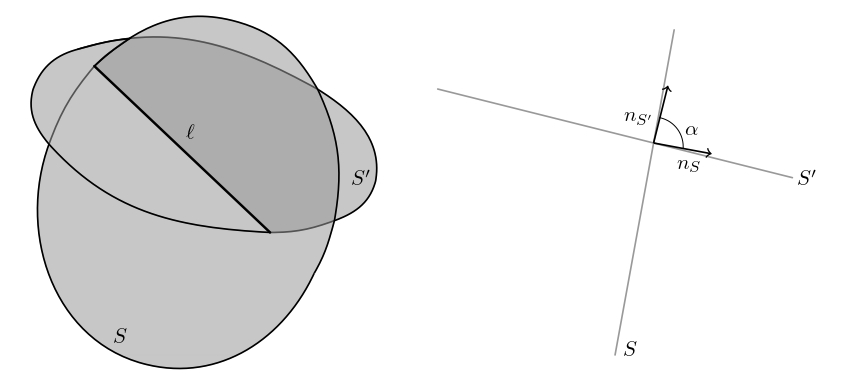}
	\caption{
		Intersection of two splats~$S$ and~$S'$ and resulting feature segment~$\ell$.
		Side view of intersecting splats.
	}
	\label{fig:FeatureSegment}
\end{figure}

We first turn to input point clouds and will discuss polyhedral surfaces as input subsequently.
Consider two splats~$S$ and~$S'$ intersecting in a straight line segment~$\ell$, see Figure~\ref{fig:FeatureSegment}.
We say that~$\ell$ is a \emph{feature segment}, if~$\alpha$ is larger than the threshold~$\vartheta$ (see Figure~\ref{fig:FeatureSegment}).
In Section~\ref{sec:ExperimentsAndEvaluation_PointClouds}, we will illustrate the connection between the input geometry and~$\vartheta$ experimentally.

For two splats, the length of the resulting feature segment~$\ell$ is at most equal to the diameter of the splat with smallest radius, depending on the individual splat sizes.
To avoid such---possibly very short---feature segments, we use the global splat diameter during feature segment detection, in contrast to Section~\ref{sec:DiscussionOfSplatSize}.
This leads to an increasing number of intersecting splats as larger splats are more likely to intersect other splats.
The features detected by a given splat size do depend on various properties such as, for instance, the noise level, the chosen value of~$\alpha$, or the feature scales of the input geometry.
Therefore, choosing a suitable splat size is highly dependent on the input settings at hand.

As depicted in Figures~\ref{fig:FeaturePointCloud_edges} and~\ref{fig:FeaturePointCloud_edgesCloseUp}, after collecting all feature segments, regions of aligned feature segments occur.
Similar to the disconnected splats representing the input point cloud, the feature segments are a disconnected representation of features of the surface underlying the input.
To structure this further, we collect a set of \emph{feature vertex candidates}, which is formed by all intersection points of three splats.

For initialization of our algorithm, we construct the graph~$\mathcal{G}$ based on the detected features of the input geometry.
First, we sequentially add all feature vertex candidates to~$\mathcal{G}$.
However, we only add a feature vertex candidate, if it is at least distance~$d$ away from previously added ones to maintain the minimal edge lengths, otherwise, the feature vertex candidate is discarded.
Next, we grow feature lines by iteratively looking for a position on a feature segment at distance~$d$ from an already existing feature vertex~$v$.
Again, these positions have to be at least distance~$d$ away from all other previously placed feature vertices.
As long as we find such positions, we place a feature vertex there and connect it to the parent vertex~$v$ to which it has distance~$d$.
The connecting edges between two feature vertices are called \emph{feature edges}.
Hence, in contrast to the initialization described in Section~\ref{sec:Initialization},~$\mathcal{G}$ does not contain solely two starting vertices, but vertices and edges representing the features of the input geometry.
In this way, feature lines can be grown similar to the disk growing process.

\begin{figure*}
	\centering
	\begin{subfigure}[t]{.13\textwidth}
		\centering
		\includegraphics[width=\textwidth]{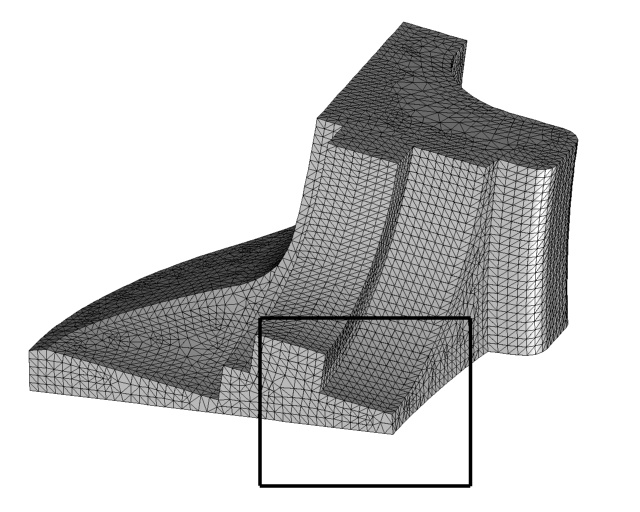}
		\label{fig:FeatureAlignmentInput}
	\end{subfigure}
	\begin{subfigure}[t]{.13\textwidth}
		\centering
		\includegraphics[width=\textwidth]{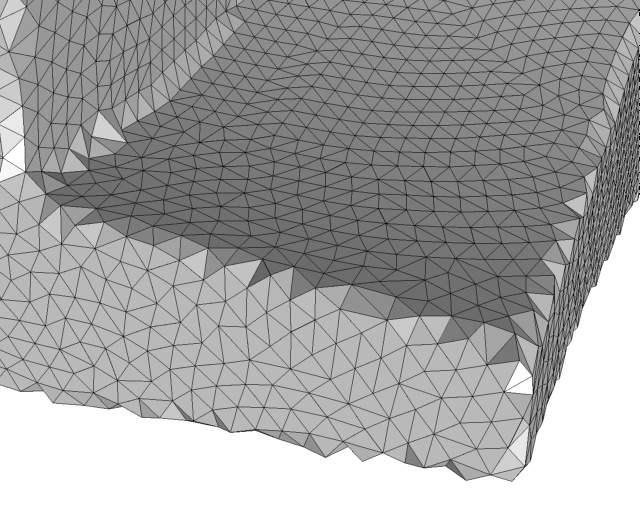}
		\label{fig:FeatureAlignmentNoInitialization}
	\end{subfigure}
	\begin{subfigure}[t]{.13\textwidth}
		\centering
		\includegraphics[width=\textwidth]{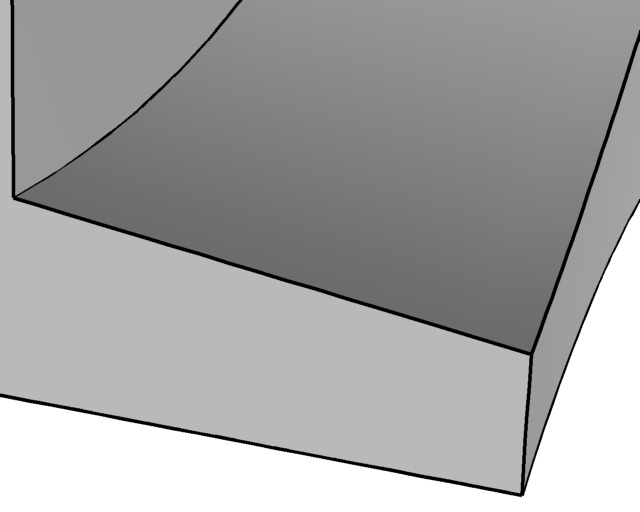}
		\label{fig:FeatureAlignmentFeatureLines}
	\end{subfigure}
	\begin{subfigure}[t]{.13\textwidth}
		\centering
		\includegraphics[width=\textwidth]{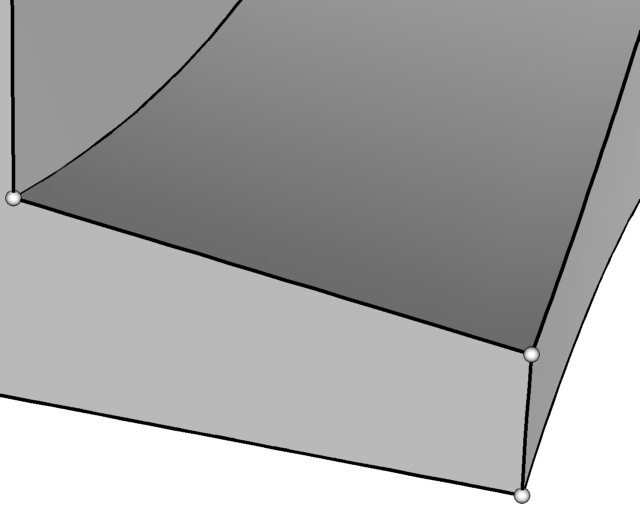}
		\label{fig:FeatureAlignmentLineEnds}
	\end{subfigure}
	\begin{subfigure}[t]{.13\textwidth}
		\centering
		\includegraphics[width=\textwidth]{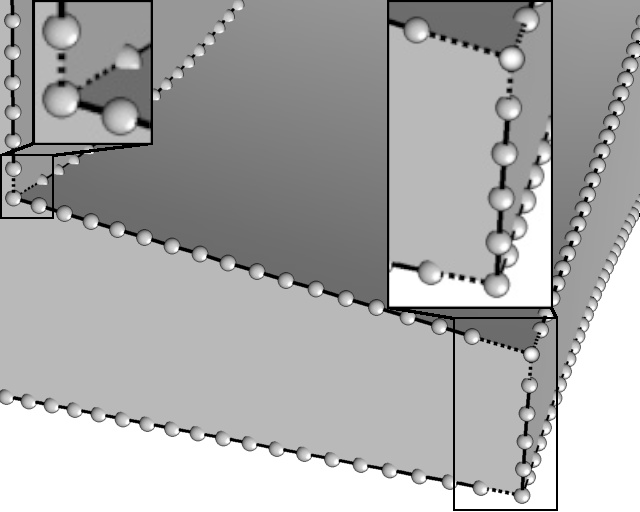}
		\label{fig:FeatureAlignmentSpherePlacement}
	\end{subfigure}
	\begin{subfigure}[t]{.13\textwidth}
		\centering
		\includegraphics[width=\textwidth]{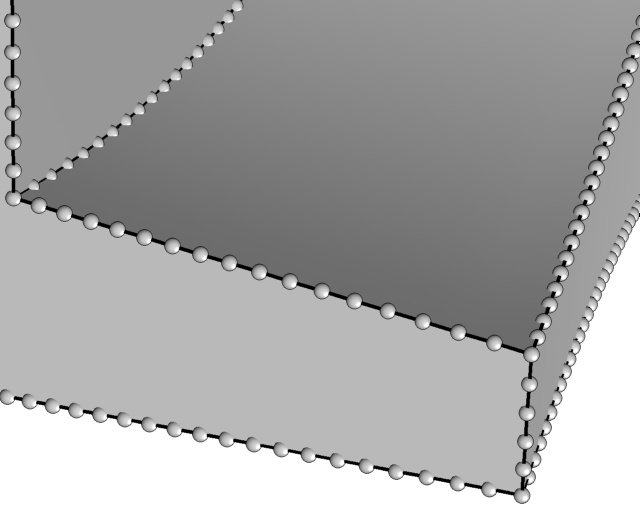}
		\label{fig:FeatureAlignmentRedistribution}
	\end{subfigure}
	\begin{subfigure}[t]{.13\textwidth}
		\centering
		\includegraphics[width=\textwidth]{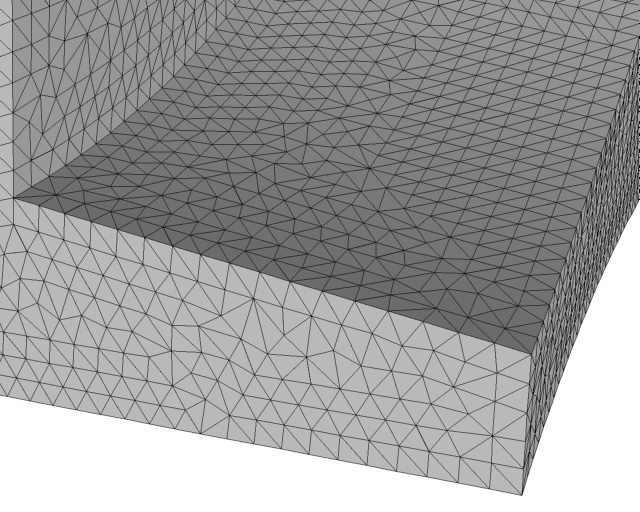}
		\label{fig:RemeshedGeometryAfterFeatureDetection}
	\end{subfigure}
	\caption{
		Feature detection on polyhedral surface (\emph{Fandisk} model).
		From left to right: Input with highlighted section; algorithm applied without feature detection; detected feature edges; detected feature edges and feature vertices; end of feature lines where no further vertex can be placed; feature vertices regularly distributed along feature lines; remeshed geometry with feature detection.
		}
	\label{fig:FeatureAlignmentAndPreservation}
\end{figure*}

On a point cloud, not all feature segments might be close enough to a feature line evolving from a previously placed feature vertex such that spheres are placed on them being connected to a feature vertex.
Therefore, after the growing process described above, we iterate through all feature lines that we have not yet placed a feature vertex on.
If we can still do so, according to the distance criterion to all other vertices, we place a feature vertex and continue the growing process from there.
Once all feature segments are sampled by vertices, we look for pairs of vertices with valence~$1$ in the graph, which have a distance smaller than~$2d$ to each other, and connect them by an edge.
This closes gaps in~$\mathcal{G}$, where vertices were added iteratively starting from both ends on a feature line of the geometry.
The positions of the two vertices connected by such a closing edge are moved slightly towards the center of the closing edge to avoid edges on feature lines with length close to~$2d$.
Finally, after having processed all feature lines, we perform disk growing based on~$\mathcal{G}$ as described in Section~\ref{sec:DiskGrowingToCreateGraphAndRegions}.

\subsection{Feature Detection on Polyhedral Surfaces}
\label{sec:FeatureDetectionOnPolyhedralSurfaces}

\begin{figure}[h!]
	\centering
	\begin{subfigure}[t]{0.32\columnwidth}
		\includegraphics[width=\textwidth]{cubeWithHole_0p1_noFeatureRemesh}
	\end{subfigure}
	\begin{subfigure}[t]{0.32\columnwidth}
		\includegraphics[width=\textwidth]{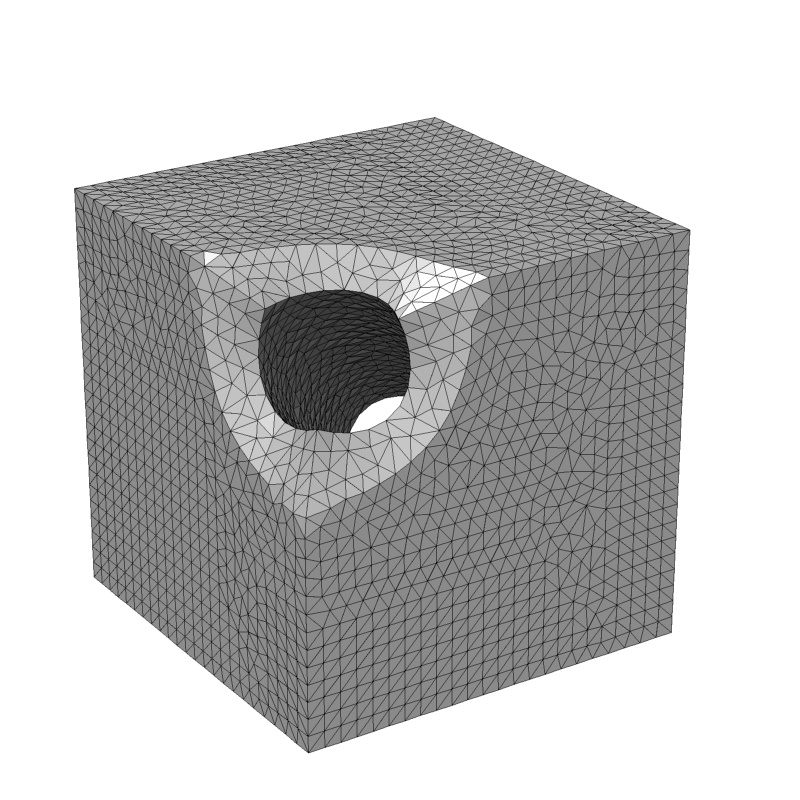}
	\end{subfigure}
	\begin{subfigure}[t]{0.32\columnwidth}
		\includegraphics[width=\textwidth]{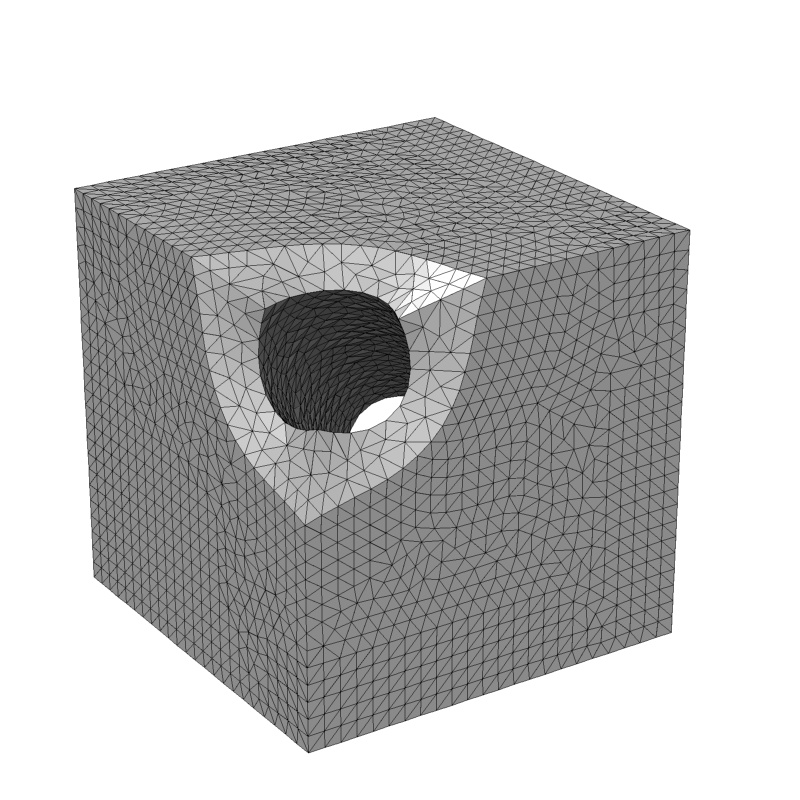}
	\end{subfigure}
	\caption{
		Result of the algorithm run on \emph{Cube} model.
		Left: input geometry remeshed without feature detection (see Figure~\ref{fig:Result_SurfaceRemeshing}).
		Middle: input geometry remeshed with~$\vartheta = 40^{\circ}$ and target edge length~$0.1$.
		Right: input geometry remeshed with~$\vartheta = 24^{\circ}$ and target edge length~$0.1$.
	}
	\label{fig:Result_SurfaceRemeshingWithFeatures}
\end{figure}

In contrast to an input point cloud, on a polyhedral surface input, we already have a set of edges.
As stated above, an edge~$e$ in the input is called a \emph{feature segment} if~${\alpha > \vartheta}$.
Similarly, a vertex of the input is said to be a \emph{feature vertex} if it is incident to either one feature segment or to at least three.
In Figure~\ref{fig:Result_SurfaceRemeshing}, the features detected for~$\vartheta = 40^{\circ}$ are shown.
Feature vertices are depicted as dark blue dots.
All the other end points of feature segments are incident to exactly one other feature segment.
Hence, the feature segments form feature lines consisting of consecutive feature segments connecting the feature vertices.
In case, a feature segment is not yet contained in such a feature line, it is part of a closed feature line.
This allows for their reconstruction as shown in Figure~\ref{fig:FeatureFandisk}.
On each feature line, we distribute feature vertices with distance~$d$ until no further vertex can be placed.
In order to distribute the sphere centers more regularly, feature vertices are moved iteratively on the feature lines.
Finally, after having processed all feature lines, we perform disk growing based on~$\mathcal{G}$ as described in Section~\ref{sec:ExtensionToSurfaceRemeshing}.
Figures~\ref{fig:FeatureAlignmentAndPreservation} and~\ref{fig:Result_SurfaceRemeshingWithFeatures} show the results achieved after running the algorithm on a CAD model.
Further evaluations of the output achieved by the algorithm presented here are included in the supplementary material and  \href{https://graphics.tudelft.nl/guaranteed-smallest-edge-length-manifold-meshing}{available online}.
Here, complete data sets in addition to graphical representations of the models used in this paper as well as models not presented so far can be found.

\section{Experiments on Point Clouds}
\label{sec:ExperimentsAndEvaluation_PointClouds}

\begin{figure*}
	\centering
	\begin{subfigure}[t]{.125\textwidth}
		\centering
		\includegraphics[width=\textwidth]{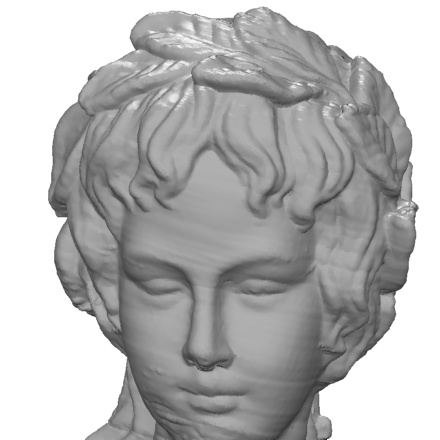}
	\end{subfigure}
	~
	\begin{subfigure}[t]{.125\textwidth}
		\centering
		\includegraphics[width=\textwidth]{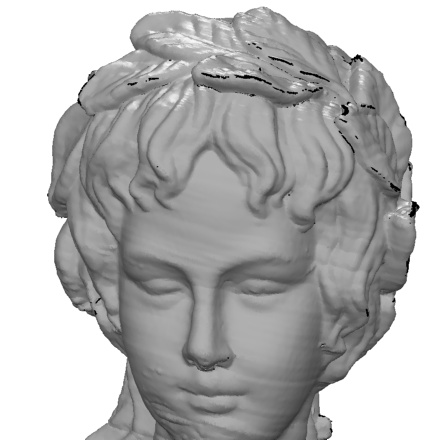}
	\end{subfigure}
	~
	\begin{subfigure}[t]{.125\textwidth}
		\centering
		\includegraphics[width=\textwidth]{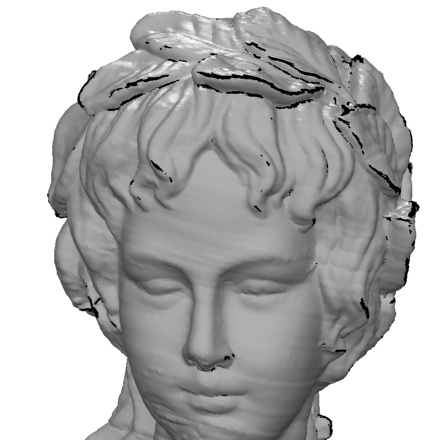}
	\end{subfigure}
	~
	\begin{subfigure}[t]{.125\textwidth}
		\centering
		\includegraphics[width=\textwidth]{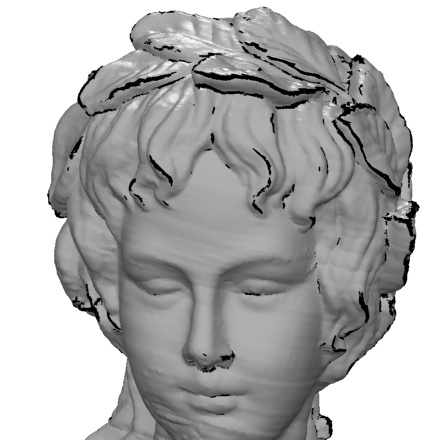}
	\end{subfigure}
	~
	\begin{subfigure}[t]{.125\textwidth}
		\centering
		\includegraphics[width=\textwidth]{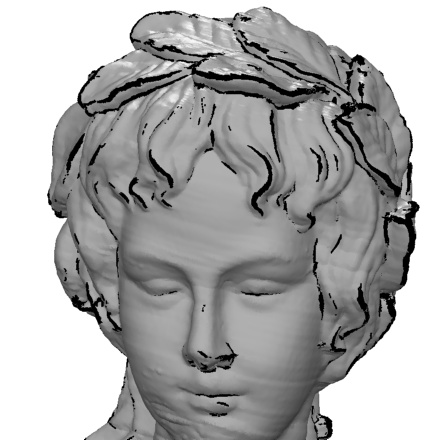}
	\end{subfigure}
	~
	\begin{subfigure}[t]{.125\textwidth}
		\centering
		\includegraphics[width=\textwidth]{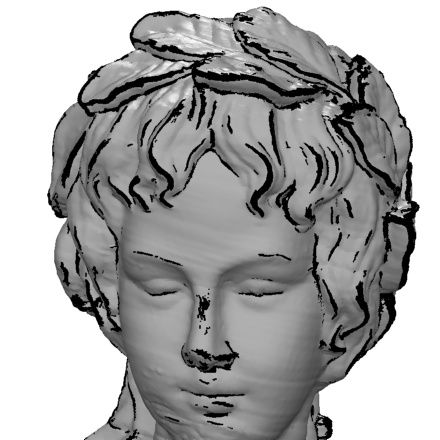}
	\end{subfigure}
	~
	\begin{subfigure}[t]{.125\textwidth}
		\centering
		\includegraphics[width=\textwidth]{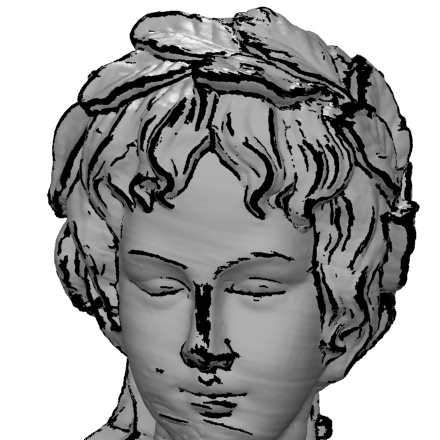}
	\end{subfigure}
	
	\vspace{2mm}
	
	\begin{subfigure}[t]{.125\textwidth}
		\centering
		\includegraphics[width=\textwidth]{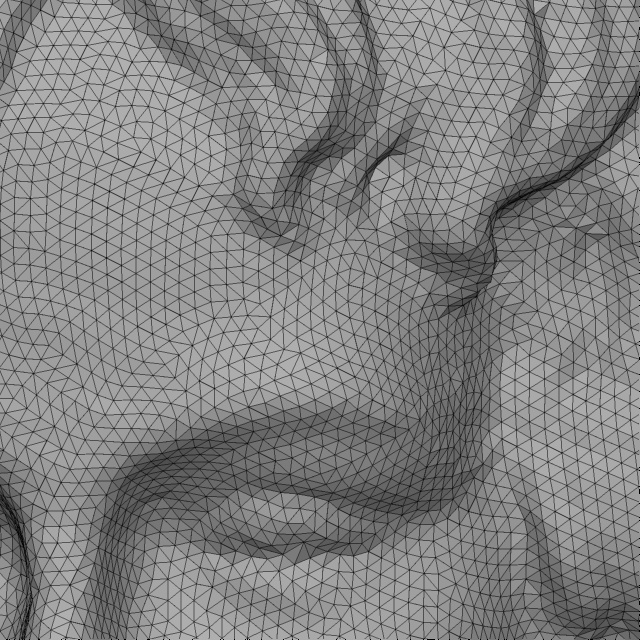}
	\end{subfigure}
	~
	\begin{subfigure}[t]{.125\textwidth}
		\centering
		\includegraphics[width=\textwidth]{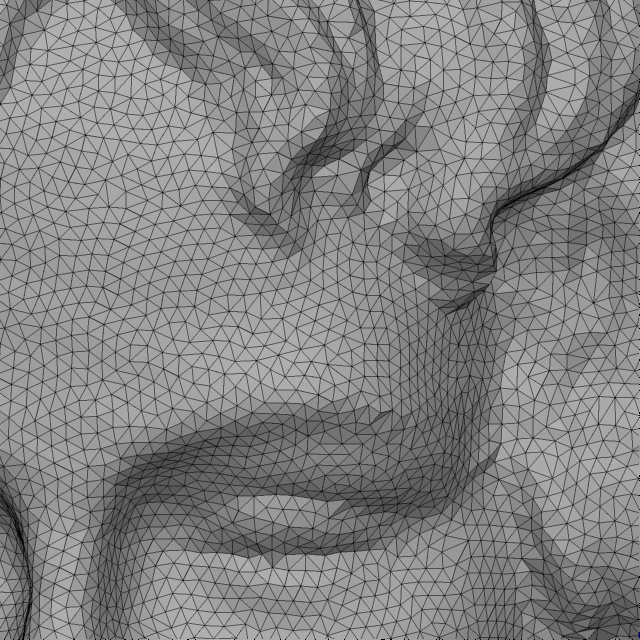}
	\end{subfigure}
	~
	\begin{subfigure}[t]{.125\textwidth}
		\centering
		\includegraphics[width=\textwidth]{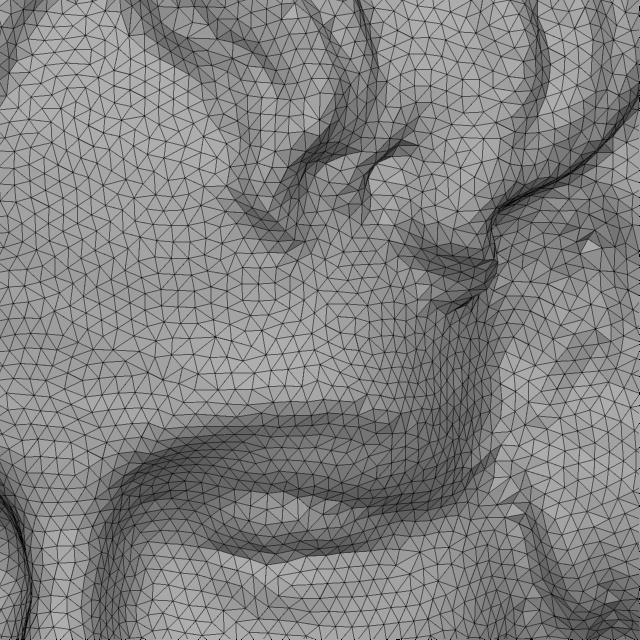}
	\end{subfigure}
	~
	\begin{subfigure}[t]{.125\textwidth}
		\centering
		\includegraphics[width=\textwidth]{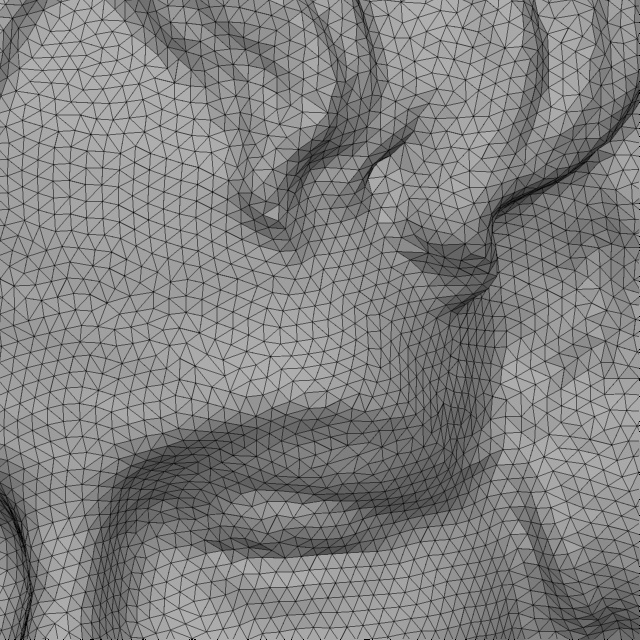}
	\end{subfigure}
	~
	\begin{subfigure}[t]{.125\textwidth}
		\centering
		\includegraphics[width=\textwidth]{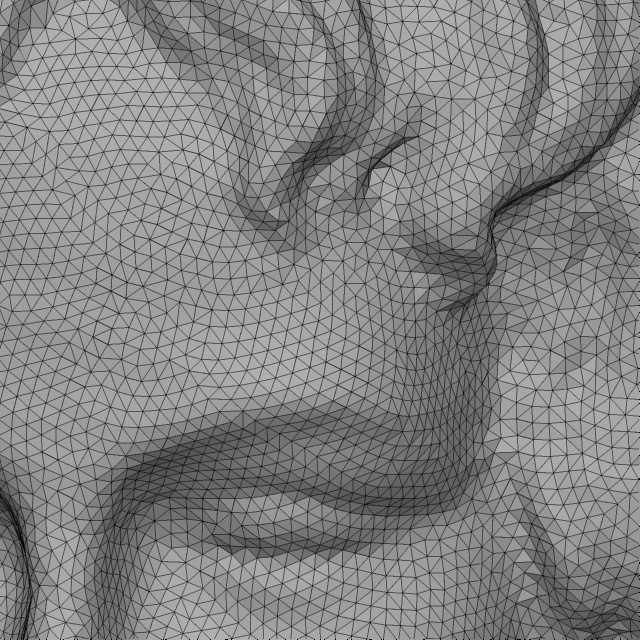}
	\end{subfigure}
	~
	\begin{subfigure}[t]{.125\textwidth}
		\centering
		\includegraphics[width=\textwidth]{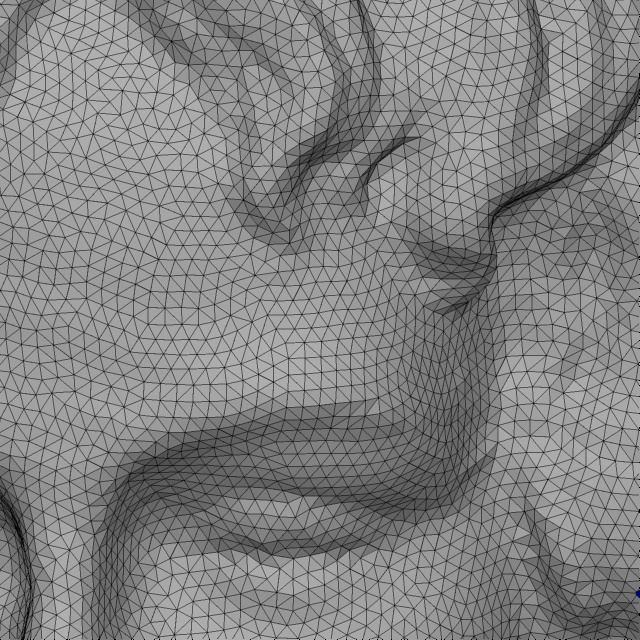}
	\end{subfigure}
	~
	\begin{subfigure}[t]{.125\textwidth}
		\centering
		\includegraphics[width=\textwidth]{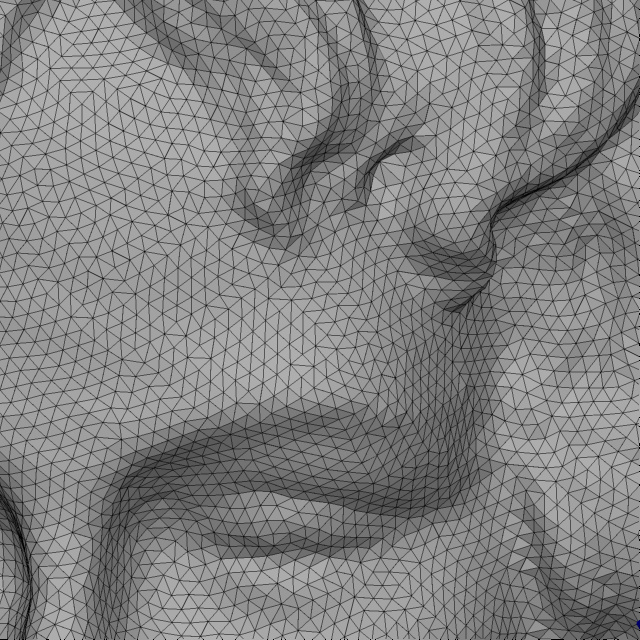}
	\end{subfigure}
	\caption{Feature detection on point clouds, illustrated on the \emph{Xiao Jie Jie} model. Upper row: point cloud and features detected.
		Lower row: close-up of left eye.
		Both rows, 	left to right: without feature detection, detected features for~$\vartheta = 90^{\circ}, 80^{\circ}, 70^{\circ}, 60^{\circ},50^{\circ},40^{\circ}$.}
	\label{fig:FeatureDetectionVaryingAngles}
\end{figure*}

\noindent From the set of models provided in~\cite{huang2022surface}, we run the modified version of the algorithm on those models possessing feature ridges like the \emph{Wrench}, the \emph{Screw}, and the \emph{Xiao Jie Jie}.
Complete data sets of the models listed and others are provided in Table~\ref{tab:Qualities_PointCloudMesh}.
As illustrated in Figure~\ref{fig:FeaturesPointClouds}, there is a visually striking improvement made.
The \emph{Wrench} model with feature ridges shown in Figure~\ref{fig:FeaturePointCloud_meshed} gives the impression of being a CADed version of the result shown in Figure~\ref{fig:FeaturePointCloud_withoutFeatureDetection}.

To quantify the results, we measure the level of interpolation achieved by the triangulation in comparison to the input point cloud.
Therefore, we calculate the shortest distance between an input point~${p \in \mathcal{P}}$ and the output mesh~$\mathcal{T}$,
\begin{align*}
	d(p, \mathcal{T}) = \min_{p' \in \mathcal{T}} \left\| p - p' \right\|,
\end{align*}
where~$p'$ denotes a point on the output mesh.
We use~$d(p,\mathcal{T})$ to determine the one-sided Hausdorff distance between~$\mathcal{P}$ and~$\mathcal{T}$ as 
\begin{align*}
	d_{\max} = \max_{p \in \mathcal{P}}\,d(p, \mathcal{T}).
\end{align*}
Next, we determine the average distance between~$\mathcal{P}$ and~$\mathcal{T}$ as
\begin{align*}
	d_{\text{avg}} = \frac{1}{\left\vert \mathcal{P} \right\vert}\, \sum_{p\in\mathcal{P}}\,d(p,\mathcal{T})
\end{align*}
and the corresponding root mean square deviation in percent~$d_{\text{RMS}}$
\begin{align*}
	d_{\text{RMS}} = \frac{100}{d_{\text{avg}}}\sqrt{\frac{1}{|\mathcal{P}|}\sum_{p\in\mathcal{P}}\left(d \left(p, \mathcal{T} \right)-d_{\text{avg}}\right)^2}.
\end{align*}
We use the implementation of the one-sided Hausdorff distance~\cite{cignoni1998metro} in MeshLab~\cite{cignoni2008meshlab} to compute these values practically.
In this sample-based approach, we always sample on vertices, edges, and faces.
Further, we utilize ten times the number of samples suggested by MeshLab.

As shown in Table~\ref{tab:Qualities_PointCloudMesh}, applying feature detection does not significantly change the number of vertices and edges in the output geometry.
The maximal distance between point cloud and output geometry, the average distance, and the root mean square error are reduced.

\begin{table}
	\centering
	\tiny{
		\begin{tabular}{l|rrrrrrr}
			Model & $d$\hspace{-1mm} & $\vartheta$ & $\vert \mathcal{V} \vert$ & $\vert \mathcal{T} \vert$ & $d_{\max}$ & $d_{\text{avg}}$ & $d_{\text{RMS}}$\\
			\hline
			\emph{Wrench} & 1.0\hspace{-1mm} & --- & 8946 & 17,896 & 0.6630 & 0.0360 & 7.4933 \\
			with f.d. & 1.0\hspace{-1mm} & 60$^{\circ}$ & 8871 & 17,746 & 0.4353 & 0.0112 & 2.1373 \\
			\hline
			\emph{Screw} & 0.6\hspace{-1mm} & --- & 13,411 & 26,842 & 0.4605 & 0.0438 & 6.0548 \\
			with f.d. & 0.6\hspace{-1mm} & 90$^{\circ}$ & 13,472 & 26,988 & 0.3595 & 0.0190 & 2.3354 \\
			\hline
			\emph{Xiao Jie Jie} & 0.8\hspace{-1mm} & --- & 22,584 & 45,164 & 0.7918 & 0.0312 & 4.0479 \\
			with f.d. & 0.8\hspace{-1mm} & 70$^{\circ}$ & 22,180 & 44,364 & 0.6131 & 0.0269 & 3.1905\\
			\hline
			\emph{Lock} & 0.8\hspace{-1mm} & --- & 5,951 & 11,902 & 0.4609 & 0.0167 & 3.2285 \\
			with f.d.   & 0.8\hspace{-1mm} & 60$^{\circ}$ & 5,837 & 11,674 & 0.2963 & 0.0113 & 1.6142 \\
			\hline
			\emph{Remote} & 0.8\hspace{-1mm} & --- & 22,595 & 45,186 & 0.7816 & 0.0296 & 5.6993 \\
			with f.d. & 0.8\hspace{-1mm} & 50$^{\circ}$ & 22,239 & 44,474 & 0.4880 & 0.0179 & 2.9222
		\end{tabular}
	}
	\vspace{-.2cm}
	\caption{
		Evaluation of one-sided Hausdorff distance from point cloud to approximating surface, without and with feature detection (f.d.) in comparison to input point cloud.
	}
	\label{tab:Qualities_PointCloudMesh}
\end{table}

In Section~\ref{sec:FeatureDetection}, we introduced the threshold~$\vartheta$ for feature detection.
Experiments for decreasing values of~$\vartheta$ showed an increase in the number and density of feature segments as depicted in Figure~\ref{fig:FeatureDetectionVaryingAngles}.
A suitable choice for~$\vartheta$ depends on the input as well as on the remeshing goals of the user.
Choosing~${\vartheta = 90^{\circ}}$ can be taken as a recommendation to start the search for a reasonable value of~$\vartheta$ with.
That is, since this allows for the detection of geometry parts intersecting under an angle equal to~$90^{\circ}$ like those found on the \emph{Wrench} model as illustrated in Figure~\ref{fig:FeaturePointCloud_edgesCloseUp}.
Table~\ref{tab:Qualities_PointCloudMesh} lists the angle~$\vartheta$ used per model, which was chosen to preserve the majority of features present in the model.

Figure~\ref{fig:FeatureDetectionVaryingAngles} illustrates the increase in the number of feature segments with shrinking values of~$\vartheta$ on the \emph{Xiao Jie Jie} model.
Since for decreasing angle~$\vartheta$ the number of feature segments increases, regions in the point cloud occur being densely covered with feature segments.
In contrast to a single feature segment or a small number of nearly parallel ones, the features are blurred.
The quality of the meshes does not differ significantly from the result achieved without feature detection.
For target edge length~$0.4$ and~${\vartheta = 40^{\circ}}$, we achieve the worst value for~$Q_{\text{avg}}$, which is~$0.9420$, while without feature detection, the corresponding value of~$Q_{\text{avg}}$ is equal to~$0.9552$ determined for the same target edge length.
An overview on all data collected for this model is contained in the supplementary material, Table~25.
Based on the results achieved, the feature angle to choose depends on the geometry.

\subsection{Varying Sampling Density}
\label{sec:VaryingSamplingDensity}

\begin{figure}[h!]
	\begin{subfigure}[t]{.48\columnwidth}
		\includegraphics[width=\columnwidth]{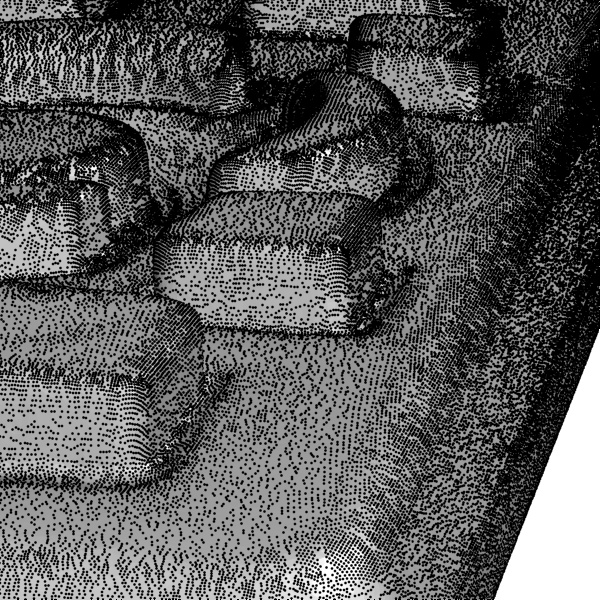}
	\end{subfigure}
	~
	\begin{subfigure}[t]{.48\columnwidth}
		\includegraphics[width=\columnwidth]{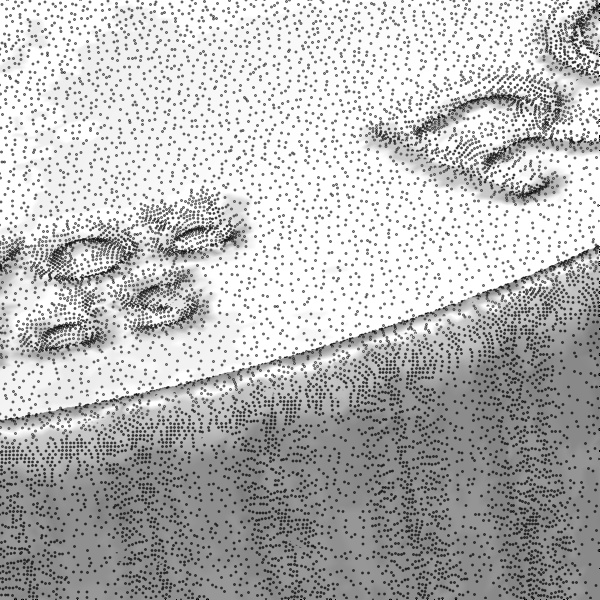}
	\end{subfigure}

	\vspace{3mm}

	\begin{subfigure}[t]{.48\columnwidth}
		\includegraphics[width=\columnwidth]{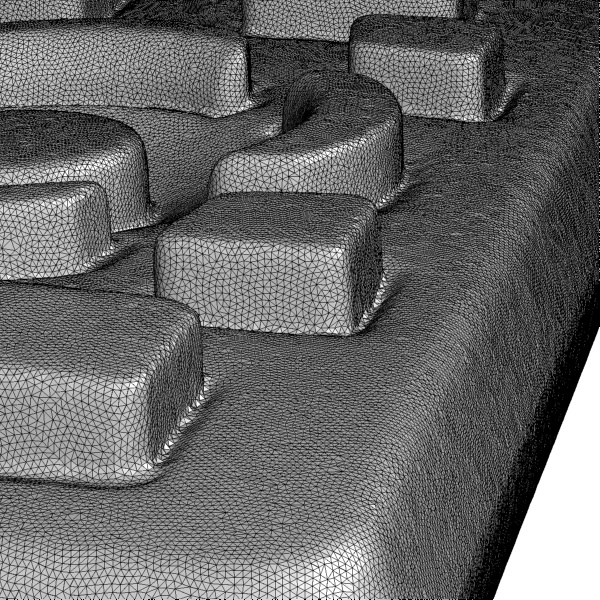}
	\end{subfigure}
	~
	\begin{subfigure}[t]{.48\columnwidth}
		\includegraphics[width=\columnwidth]{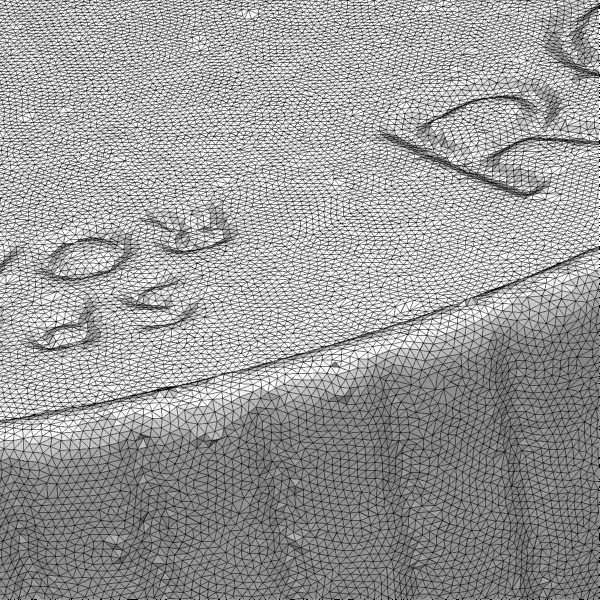}
	\end{subfigure}
	\caption{
		Upper row: varying sample density, detected on the \emph{Remote} model and on the \emph{Coffee Bottle Plastic} model.
		Lower row: resulting isotropic meshes derived from the point samplings shown above.
	}
	\label{fig:VaryingSampleDensity}
\end{figure}

Depending on the scanning process, the derived sampling points may be distributed in varying density over the geometry.
In the upper row of Figure~\ref{fig:VaryingSampleDensity}, close-ups of two models taken from~\cite{huang2022surface} are depicted. 
Both show a higher density of sampling points in areas of high curvature than in low-curvature areas.
In the algorithm presented here, the individual splat sizes are chosen based on the distribution of sample points.
Hence, areas with low curvature are well represented by a few splats of larger radii while a higher number of splats of smaller radii cover regions of higher curvature.
The images in the lower row of Figure~\ref{fig:VaryingSampleDensity} illustrate that the quality of the resulting meshes is not influenced by the variation of the sampling density of the input.

\subsection{Handling Reach Criterion}
\label{sec:HandlingReachCriterion}

\begin{figure}[h!]
	\begin{subfigure}[t]{.9\columnwidth}
		\includegraphics[width=\columnwidth]{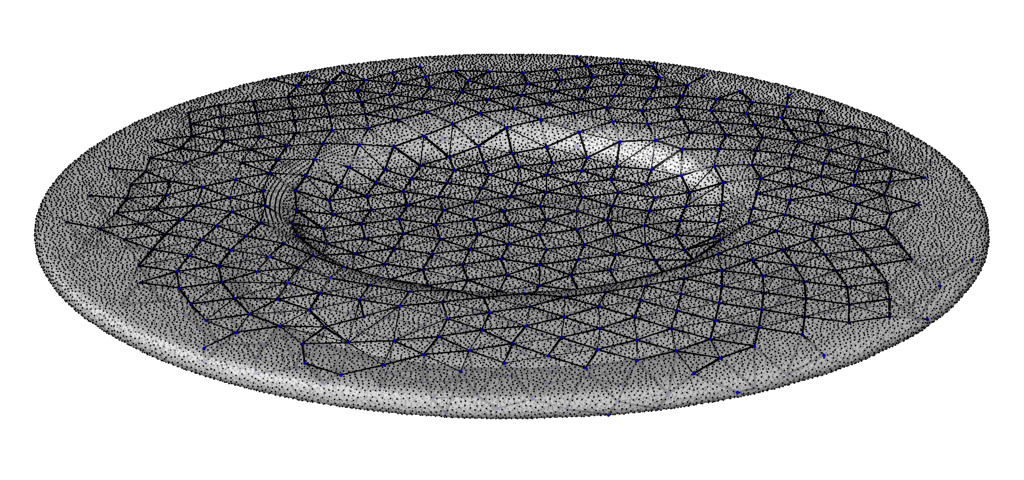}
	\end{subfigure}
	\begin{subfigure}[t]{.9\columnwidth}
		\includegraphics[width=\columnwidth]{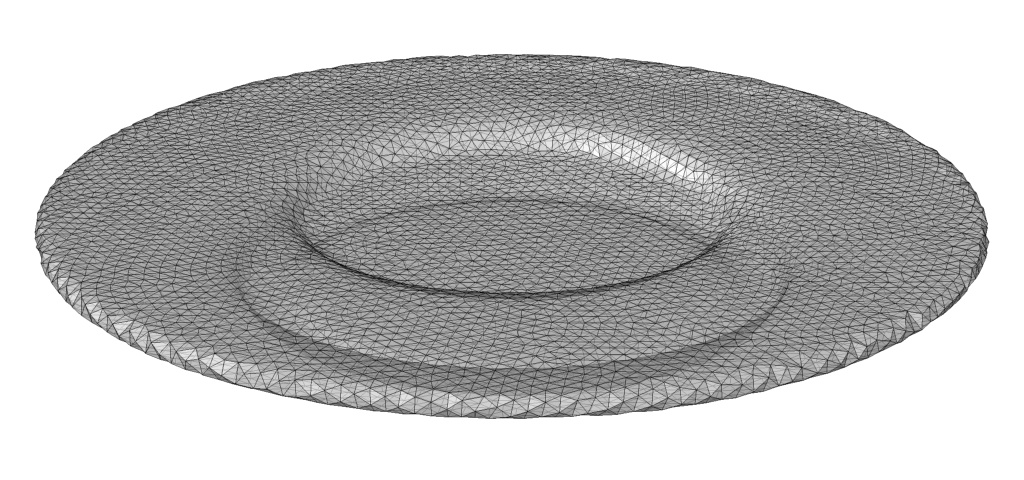}
	\end{subfigure}
	\caption{
		Results achieved by processing the \emph{Plate} model.
		Above: failed attempt based on~$d$ chosen too big.
		Below: successful application.
	}
	\label{fig:ViolatedReachCriterion}
\end{figure}

As mentioned in Section~\ref{sec:AuxiliaryBoxDataStructure}, the parameter~$d$ has to be chosen by the user, depending on the reach of the input.
In case~$d$ is smaller or equal to the reach~$\rho$, the output consists of an isotropic triangular mesh, as discussed in Section~\ref{sec:TheoryAndMethodology}.
However, choosing~$d$ larger than the reach might lead to issues in the growing process.
Both choices are illustrated in Figure~\ref{fig:ViolatedReachCriterion}.
For~${d > \rho}$, the growing process was started on the top side of the \emph{Plate} model (available online~\cite{reitebuch2025CADmodels}) and failed to grow over the rim to the lower side.
For any~${d < \rho}$, the \emph{Plate} model is meshed successfully, as illustrated in the lower image of Figure~\ref{fig:ViolatedReachCriterion}.

\section{Experiments on Polyhedral Surfaces}
\label{sec:ExperimentsOnPolyhedralSurfaces}

\noindent To further illustrate the effectiveness of the modifications to the algorithm presented in Sections~\ref{sec:ExtensionToSurfaceRemeshing} and~\ref{sec:FeatureDetection}, we run the algorithm on a broad choice of polyhedral surfaces, with a varying number of sharp features as well as an increasing level of topology.
The models we remesh in the following are either taken from an online repository of commonly used meshes~\cite{jacobson2025repository} or self-made (available online~\cite{reitebuch2025CADmodels}).

We compare the results achieved by our algorithm to the \emph{Isometric Explicit Remeshing}~\cite{hoppe1993Mesh} pipeline of MeshLab~\cite{cignoni2008meshlab} and to the \emph{Remeshing}~\cite{botsch2004remeshing} module of the PMP library~\cite{pmp23}.
In all experiments, for both algorithms, we use the standard number ten of iterations.
Furthermore, we do not use adaptive remeshing for either implementation as it conflicts the goal of obtaining a uniform edge length.
All other parameters are left in their standard configuration of the respective implementation, except for using world-length coordinates in PMP---in the standard configuration, lengths are given relative to the bounding box.
In Table~\ref{tab:DataCADModels}, a representative choice of CAD models processed with said routines is listed.
The derived data show that our single-sweep algorithm produces comparable mesh quality with and without feature detection.
A larger variety of models with more detailed data sets can be found in the supplementary material.

\subsection{Influence of Feature Detection on Hausdorff Distance}
\label{sec:InfluenceOfFeatureDetectionOnHausdorffDistance}

In Figure~\ref{fig:Result_SurfaceRemeshingWithFeatures}, we observe the striking differences between the results without feature detection and with feature detection.
While all features in the result achieved without feature detection are lost, varying the threshold angle~$\vartheta$ shows an increase of features maintained.
In addition to investigating the one-sided Hausdorff distance~$d_{\max}$ as introduced in Section~\ref{sec:ExperimentsAndEvaluation_PointClouds}, we will consider the \emph{relative error}~$\frac{d_{\max}}{d}$ as well.
Remeshing the \emph{Cube} with~$\vartheta = 40^{\circ}$ does not allow for a complete maintenance of all feature lines and vertices while choosing~$\vartheta = 24^{\circ}$ does as illustrated in Table~\ref{tab:EvaluationHausdorffCubeFandisk} and depicted in Figure~\ref{fig:Result_SurfaceRemeshingWithFeatures}.
Further, we can observe the same behavior on the \emph{Fandisk} as well.

\begin{table}[h!]
	\centering
	\tiny{
	\begin{tabular}{l|rrrrr}
		
	\end{tabular}
}
\end{table} 

\begin{table}[h!]
	\centering
	\tiny{
	\begin{tabular}{l|l|rcrrrr}
		 & Algo. & $d$ & $\vartheta$ & $E_{\text{avg}}$ & $E_{\text{RMS}}$ & $Q_{\text{avg}}$ & $Q_{\text{RMS}}$\\
		\hline
		\parbox[t]{2mm}{\multirow{6}{*}{\rotatebox[origin=c]{90}{\emph{Cube}}}} & ML & 0.04 & $24^{\circ}$ & 0.0408 & 12.8 & 0.9600 & 4.5\\
		& PMP & 0.04 & $24^{\circ}$ & 0.0364 & 12.2 & 0.9636 & 3.5\\
		& ours & 0.04 & $24^{\circ}$ & 0.0436 & 14.1 & 0.9295 & 5.8 \\
		& ML & 0.04 & -- & 0.0404 & 14.5 & 0.9525 & 7.3 \\
		& PMP & 0.04 & -- & 0.0364 & 11.8 & 0.9663 & 3.1 \\
		& ours & 0.04 & -- & 0.0426 & 11.3 & 0.9584 & 4.5 \\
		\hline
		\parbox[t]{2mm}{\multirow{6}{*}{\rotatebox[origin=c]{90}{\emph{Fandisk}}}} & ML & 0.012 & $60^{\circ}$ & 0.0124 & 18.9 & 0.9526 & 5.0 \\
		& PMP & 0.012 & $60^{\circ}$ & 0.0112 & 12.3 & 0.9713 & 2.9 \\
		& ours & 0.012 & $60^{\circ}$ & 0.0131 & 14.2 & 0.9221 & 5.3 \\
		& ML & 0.012 & -- & 0.0124 & 12.9 & 0.9531 & 4.9 \\
		& PMP & 0.012 & -- & 0.0125 & 12.2 & 0.9720 & 2.8 \\
		& ours & 0.012 & -- & 0.0127 & 11.1 & 0.9592 & 4.4 \\
		\hline
		\parbox[t]{2mm}{\multirow{6}{*}{\rotatebox[origin=c]{90}{\emph{Flange}}}} & ML & 1.0 & $80^{\circ}$ & 1.0241 & 13.2 & 0.9562 & 5.1 \\
		& PMP & 1.0 & $80^{\circ}$ & 0.9240 & 12.6 & 0.9601 & 3.8 \\
		& ours & 1.0 & $80^{\circ}$ & 1.0645 & 11.8 & 0.9585 & 4.9 \\
		& ML & 1.0 & -- & 0.9361 & 27.7 & 0.8823 & 21.4 \\
		& PMP & 1.0 & -- & 0.9269 & 12.2 & 0.9626 & 3.5 \\
		& ours & 1.0 & -- & 1.0666 & 11.4 & 0.9572 & 4.5 \\
		\hline
		\parbox[t]{2mm}{\multirow{6}{*}{\rotatebox[origin=c]{90}{\emph{Oloid}}}} & ML & 0.02 & $80^{\circ}$ & 0.202 & 15.8 & 0.9455 & 9.4 \\
		& PMP & 0.02 & $80^{\circ}$ & 0.0183 & 12.3 & 0.9624 & 3.4 \\
		& ours & 0.02 & $80^{\circ}$ & 0.0214 & 11.8 & 0.9508 & 4.5 \\
		& ML & 0.02 & -- & 0.0201 & 17.3 & 0.9373 & 11.2 \\
		& PMP & 0.02 & -- & 0.0183 & 12.3 & 0.9625 & 3.4 \\
		& ours & 0.02 & -- & 0.0212 & 10.9 & 0.9606 & 4.4
	\end{tabular}
	}
	\caption{
		Selection of CAD models remeshed with and without feature detection using MeshLab (ML), Polygon Mesh Processing (PMP), and our algorithm.
		A more detailed evaluation is presented in the supplementary material.
	}
	\label{tab:DataCADModels}
\end{table}

\begin{table}[h!]
	\centering
	\tiny{
\begin{tabular}{l|c|llrr|llrr}
	 & Algo.& $d$\hspace{-2mm} & $\vartheta$\hspace{-2mm} & $d_{\max}$\hspace{-2mm} & $\frac{d_{\max}}{d}$ & $d$\hspace{-2mm} & $\vartheta$\hspace{-2mm} & $d_{\max}$\hspace{-2mm} & $\frac{d_{\max}}{d}$\\
	\hline
	\parbox[t]{2mm}{\multirow{9}{*}{\rotatebox[origin=c]{90}{\emph{Cube}}}} & ML & 0.04\hspace{-2mm} & ---\hspace{-2mm} & 0.0347\hspace{-2mm} & 0.8687 & 0.1\hspace{-2mm} & ---\hspace{-2mm} & 0.0808\hspace{-2mm} & 0.8082\\
	& PMP & 0.04\hspace{-2mm} & ---\hspace{-2mm} & 0.0334\hspace{-2mm} & 0.8353 & 0.1\hspace{-2mm} & ---\hspace{-2mm} & 0.2781\hspace{-2mm} & 0.6623\\
	& ours & 0.04\hspace{-2mm} & ---\hspace{-2mm} & 0.0324\hspace{-2mm} & 0.8122 & 0.1\hspace{-2mm} & ---\hspace{-2mm} & 0.0776\hspace{-2mm} & 0.7765\\
	& ML & 0.04\hspace{-2mm} & 24$^{\circ}$\hspace{-2mm} & 0.0043\hspace{-2mm} & 0,1078 & 0.1\hspace{-2mm} & 24$^{\circ}$\hspace{-2mm} & 0.0062\hspace{-2mm} & 0.0620\\
	& PMP & 0.04\hspace{-2mm} & 24$^{\circ}$\hspace{-2mm} & 0.0038\hspace{-2mm} & 0.0951 & 0.1\hspace{-2mm} & 24$^{\circ}$\hspace{-2mm} & 0.0061\hspace{-2mm} & 0.0614\\
	& ours & 0.04\hspace{-2mm} & 24$^{\circ}$\hspace{-2mm} & 0.0042\hspace{-2mm} & 0.1059 & 0.1\hspace{-2mm} & 24$^{\circ}$\hspace{-2mm} & 0.0084\hspace{-2mm} & 0.0843\\
	& ML & 0.04\hspace{-2mm} & 40$^{\circ}$\hspace{-2mm} & 0.0043\hspace{-2mm} & 0.1033 & 0.1\hspace{-2mm} & 40$^{\circ}$\hspace{-2mm} & 0.0190\hspace{-2mm} & 0.1908\\
	& PMP & 0.04\hspace{-2mm} & 40$^{\circ}$\hspace{-2mm} & 0.0065\hspace{-2mm} & 0.1637 & 0.1\hspace{-2mm} & 40$^{\circ}$\hspace{-2mm} & 0.0196\hspace{-2mm} & 0.1961\\
	& ours & 0.04\hspace{-2mm} & 40$^{\circ}$\hspace{-2mm} & 0.0097\hspace{-2mm} & 0.2427 & 0.1\hspace{-2mm} & 40$^{\circ}$\hspace{-2mm} & 0.0213\hspace{-2mm} & 0.2134\\
	\hline
	\parbox[t]{2mm}{\multirow{12}{*}{\rotatebox[origin=c]{90}{\emph{Fandisk}}}} & ML & 0.012\hspace{-2mm} & ---\hspace{-2mm} & 0.0111\hspace{-2mm} & 0.9302 & 0.024\hspace{-2mm} & ---\hspace{-2mm} & 0.0216\hspace{-2mm} & 0.9039\\
	& PMP & 0.012\hspace{-2mm} & ---\hspace{-2mm} & 0.0080\hspace{-2mm} & 0.6721 & 0.024\hspace{-2mm} & ---\hspace{-2mm} & 0.0180\hspace{-2mm} & 0.7524\\
	& ours & 0.012\hspace{-2mm} & ---\hspace{-2mm} & 0.0097\hspace{-2mm} & 0.8147 & 0.024\hspace{-2mm} & ---\hspace{-2mm} & 0.0184\hspace{-2mm} & 0.7679 \\
	& ML & 0.012\hspace{-2mm} & 60$^{\circ}$\hspace{-2mm} & 0.0035\hspace{-2mm} & 0.2929 & 0.024\hspace{-2mm} & 60$^{\circ}$\hspace{-2mm} & 0.0061\hspace{-2mm} & 0.2561\\
	& PMP & 0.012\hspace{-2mm} & 60$^{\circ}$\hspace{-2mm} & 0.0027\hspace{-2mm} & 0.2331 & 0.024\hspace{-2mm} & 60$^{\circ}$\hspace{-2mm} & 0.0055\hspace{-2mm} & 0.2302\\
	& ours & 0.012\hspace{-2mm} & 60$^{\circ}$\hspace{-2mm} & 0.0034\hspace{-2mm} & 0.2835 & 0.024\hspace{-2mm} & 60$^{\circ}$\hspace{-2mm} & 0.0072\hspace{-2mm} & 0.2983 \\
	& ML & 0.036\hspace{-2mm} & ---\hspace{-2mm} & 0.0318\hspace{-2mm} & 0.8840 & 0.048\hspace{-2mm} & ---\hspace{-2mm} & 0.0461\hspace{-2mm} & 0.9618\\
	& PMP & 0.036\hspace{-2mm} & ---\hspace{-2mm} & 0.0261\hspace{-2mm} & 0.7252 & 0.048\hspace{-2mm} & ---\hspace{-2mm} & 0.0334\hspace{-2mm} & 0.6976\\
	& ours & 0.036\hspace{-2mm} & ---\hspace{-2mm} & 0.0302\hspace{-2mm} & 0.8376 & 0.048\hspace{-2mm} & ---\hspace{-2mm} & 0.0337\hspace{-2mm} & 0.7025 \\
	& ML & 0.036\hspace{-2mm} & 60$^{\circ}$\hspace{-2mm} & 0.0060\hspace{-2mm} & 0.1690 & 0.048\hspace{-2mm} & 60$^{\circ}$\hspace{-2mm} & 0.0115\hspace{-2mm} & 0.2405\\
	& PMP & 0.036\hspace{-2mm} & 60$^{\circ}$\hspace{-2mm} & 0.0088\hspace{-2mm} & 0.2454 & 0.048\hspace{-2mm} & 60$^{\circ}$\hspace{-2mm} & 0.0125\hspace{-2mm} & 0.2612\\
	& ours & 0.036\hspace{-2mm} & 60$^{\circ}$\hspace{-2mm} & 0.0085\hspace{-2mm} & 0.2367 & 0.048\hspace{-2mm} & 60$^{\circ}$\hspace{-2mm} & 0.0134\hspace{-2mm} & 0.2806
\end{tabular}
}
\caption{Evaluation of one-sided Hausdorff distance of CAD models (\emph{Cube} and \emph{Fandisk}), comparing to the out put of MeshLab (ML) and Polygon Mesh Processing (PMP).}
\label{tab:EvaluationHausdorffCubeFandisk}
\end{table}

\subsection{Influence of Feature Detection on Edge Length and Angle Distribution}
\label{sec:InfluenceOfFeatureDetectionOnEdgeLengthAndAngleDistribution}

\begin{figure}[h!]
	\centering
	\begin{subfigure}[t]{.8\linewidth}
		\centering
		\includegraphics[width=\textwidth]{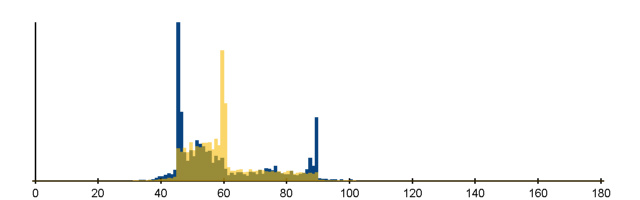}
		\caption{Angle distribution, target=$60^\circ$.}
	\end{subfigure}
	\begin{subfigure}[t]{.8\linewidth}
		\centering
		\includegraphics[width=\textwidth]{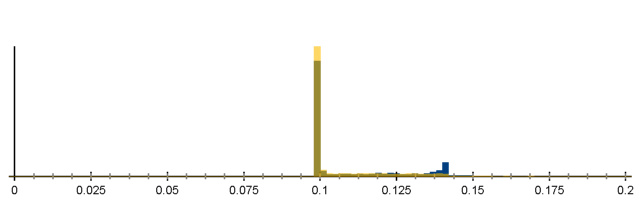}
		\caption{Edge lengths distribution, target=$0.1$.}
	\end{subfigure}
	\begin{subfigure}[t]{.8\linewidth}
		\centering
		\includegraphics[width=\textwidth]{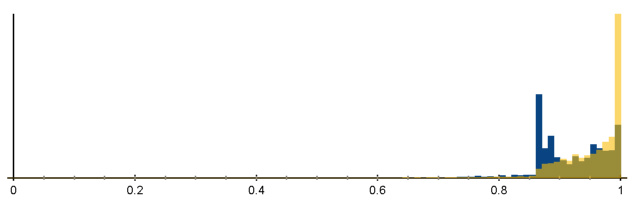}
		\caption{Distribution of quality $Q_t$, target=1.0.}
	\end{subfigure}
	\caption{
		Distributions obtained by our algorithm applied to the \emph{Cube} model with feature detection (shown in blue) for $\vartheta = 24^{\circ}$ and without feature detection (shown in yellow).
	}
	\label{fig:CubeHistogram}
\end{figure}

\begin{figure}[h!]
	\centering
	\begin{subfigure}[t]{.8\linewidth}
		\centering
		\includegraphics[width=\textwidth]{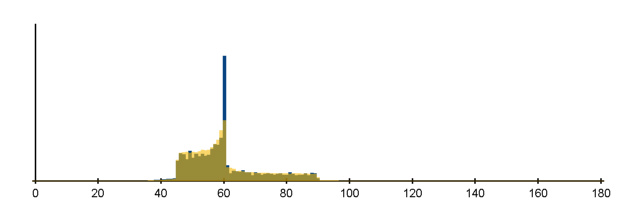}
		\caption{Angle distribution, target=$60^\circ$.}
	\end{subfigure}
	\begin{subfigure}[t]{.8\linewidth}
		\centering
		\includegraphics[width=\textwidth]{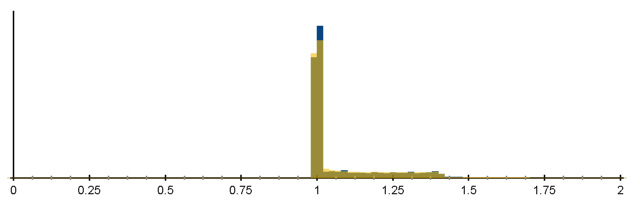}
		\caption{Edge lengths distribution, target=$1.0$.}
	\end{subfigure}
	\begin{subfigure}[t]{.8\linewidth}
		\centering
		\includegraphics[width=\textwidth]{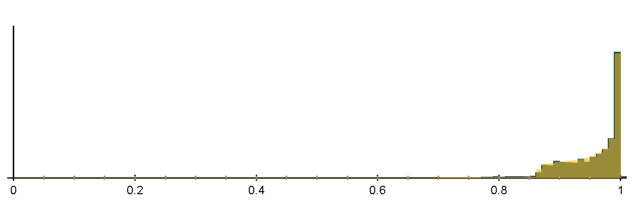}
		\caption{Distribution of quality $Q_t$, target=1.0.}
	\end{subfigure}
	\caption{
		Distributions obtained by our algorithm applied to the \emph{Flange} model with feature detection (shown in blue) for $\vartheta = 80^{\circ}$ and without feature detection (shown in yellow).
	}
	\label{fig:FlangeHistogram}
\end{figure}

In Figure~\ref{fig:CubeHistogram}, the angle distribution, the edge length distribution, and the distribution of quality are shown, evaluated on a remeshing of the \emph{Cube} model with and without feature detection.
Respecting features of a geometry results in a broader distribution for each of the three quantities evaluated.
While the edge length distribution does not change drastically, the angle distribution has a peak at a smaller angle.
The quality~$Q_t$ is still good, but does not contain as many ideal triangles as the remeshing without feature detection.
In contrast to the results achieved by the algorithms mentioned above, we perform better with respect to the minimum angle as well as to the minimum triangle quality.
The \emph{Cube} model possesses several feature lines and corners easily to be detected visually.
Most angles at points detected as feature vertices of higher valence by our algorithm are close to~$90^{\circ}$.
This leads to a larger number of nearly isosceles triangles with angles close to~$45^{\circ}$ and~$90^{\circ}$, which can be seen in the histograms in Figures~\ref{fig:CubeHistogram}.
In case the sharp features of the input geometry do not meet at an angle close to~$90^{\circ}$, the angle distribution of the remeshed geometry varies less, comparing the results achieved with and without feature detection as depicted in Figure~\ref{fig:FlangeHistogram} on the \emph{Flange}.

\subsection{Sharp Angles}
\label{sec:SharpAngles}

\begin{figure}[h!]
	\begin{subfigure}[t]{.48\columnwidth}
		\includegraphics[width=\columnwidth]{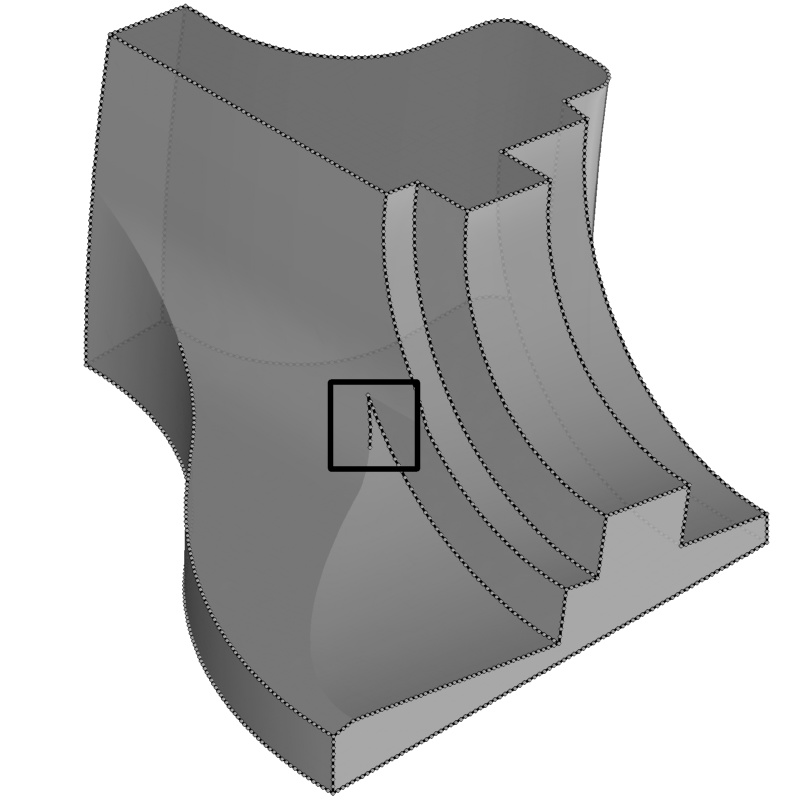}
	\end{subfigure}
~
	\begin{subfigure}[t]{.48\columnwidth}
		\includegraphics[width=\columnwidth]{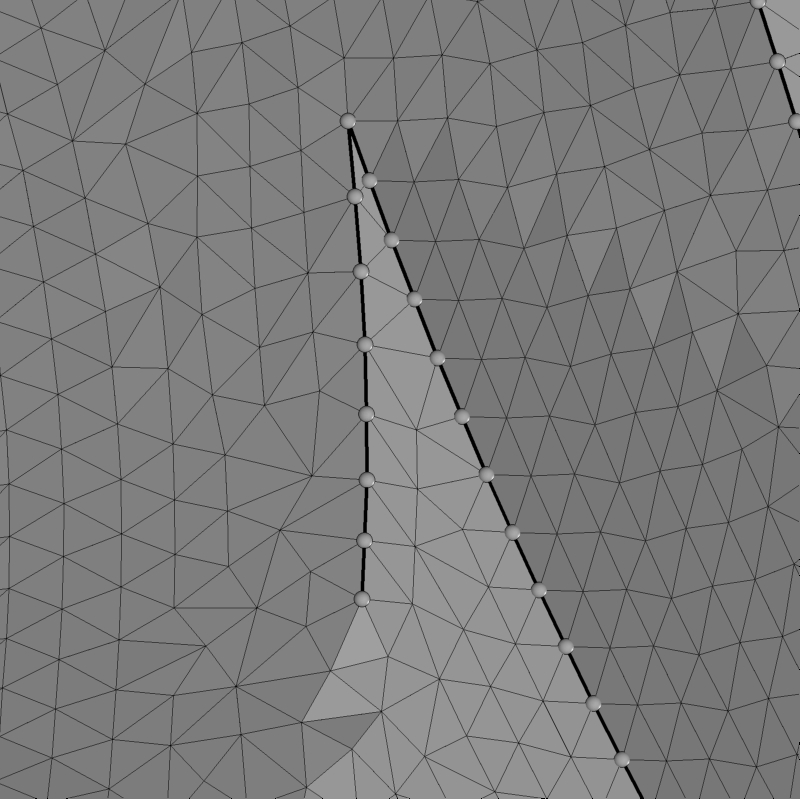}
	\end{subfigure}
	\caption{Feature lines meeting at angle~$< 60^{\circ}$.
	Left: \emph{Fandisk} model with emphasized feature lines meeting.
	Right: triangulation of surface between meeting features, containing edges shorter in length than the preset target edge length.}
	\label{fig:FeatureFandisk}
\end{figure}

In Figure~\ref{fig:FeatureFandisk}, two feature lines are shown, meeting under an angle~$\gamma$ less than~$60^{\circ}$.
Triangulating the area between the feature lines may introduce edges within the triangulation step shorter than the target edge length.
Since the length of the third side of the triangle filling the space at the meeting point of the feature lines depends on~$\gamma$, there are two different ways to go.
The first one consists of maintaining the features detected and accept the occurrence of (a comparably small number of) edges shorter than the target edge length, while the second one maintains the target edge length, but loses one of the features.
Here, we decided to follow the first possibility, which is illustrated in Figure~\ref{fig:FeatureFandisk}.

\begin{figure}[h!]
	\begin{subfigure}[t]{.48\columnwidth}
		\includegraphics[width=\columnwidth]{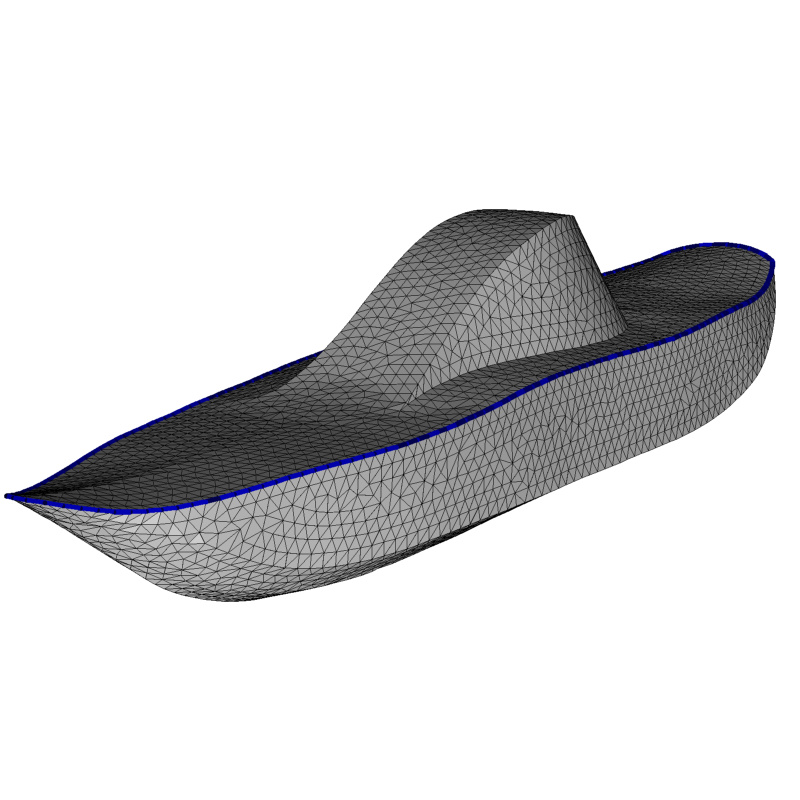}
	\end{subfigure}
~
	\begin{subfigure}[t]{.48\columnwidth}
		\includegraphics[width=\columnwidth]{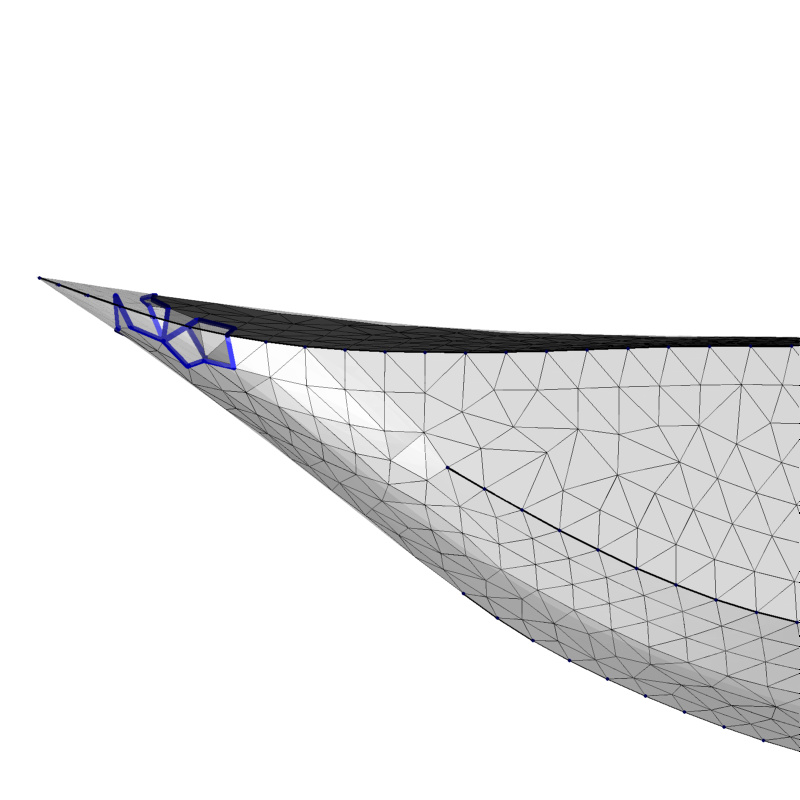}
	\end{subfigure}
	\caption{Sharp feature ridges on \emph{Boat} model.
	Left: \emph{Boat} model remeshed with features.
	Right: close up of \emph{Boat} model showing faulty edges.}
	\label{fig:FeatureBoat}
\end{figure}

A similar obstacle occurs in case the angle between the two parts of the surfaces joint along a single feature line is less or equal to~$60^{\circ}$.
Here, the disk growing step is restricted to one side solely, since no sphere can be placed on the other side of the ridge because it is too close to spheres already placed.
Hence, internal connections between the two sides may occur as well as a non-manifold result since the lack of possibilities to place spheres cuts a hole into the surface as illustrated in Figure~\ref{fig:FeatureBoat}.
In case such a feature line forms a closed curve on the geometry, as it does on the \emph{Boat} model, the geometry can be cut into two parts along the feature curve.
Since the boundaries of both parts are already covered with spheres placed in the feature detection step, we use them for initialization.
Subsequently, we place further spheres in a disk growing process applied to each part of the geometry individually.
Hence, we prevent the spheres placed to intersect the other part of the geometry.

The ansatz we described above to mesh feature lines, is applicable to geometries with boundaries as well.
Here, the spheres placed on the boundaries directly are used for initialization.
In case, a feature line does not close, but makes an angle less than~$60^{\circ}$, the geometry has to be segmented in order to prevent the spheres to intersect the geometry several times.
To find such a segmentation is left as future work.

\section{Conclusion and Future Work}

\noindent In this paper, we built on a surface representation by equally sized spheres~\cite{lipschutz2021single} to provide a feature-aware algorithm for meshing of point clouds and remeshing of polyhedral surfaces.
A prior publication of the algorithm introduced the meshing of point clouds with a guaranteed smallest edge length~\cite{lipschutz2024manifold}, see Sections~\ref{sec:TheoryAndMethodology} to~\ref{sec:Experiments}.
This extension enables the algorithm to remesh polyhedral surfaces with and without sharp features, see Sections~\ref{sec:ExtensionToSurfaceRemeshing} to~\ref{sec:ExperimentsOnPolyhedralSurfaces}.
The algorithm still only needs a single, greedy sweep across the input geometry to obtain the resulting mesh.
Respecting features of the input point cloud or surface mesh makes it possible to process a broader variety of input geometries in a way that is suitable for a variety of follow-up applications.
The results achieved remain guaranteed to be manifold, based on the theory discussed in Section~\ref{sec:TheoryAndMethodology}.

The \emph{Fandisk} model and the \emph{Boat} model discussed in Section~\ref{sec:SharpAngles} already hinted at some shortcomings of our algorithm.
These arise when the model demands for very sharp angles in the geometry that cannot be modeled while maintaining the desired minimum edge length.
While this can be interpreted as a feature of the algorithm, in these cases---especially with the tip of the \emph{Boat} model---it would be preferential to let the user decide to violate the distance requirement between non-connected vertices to obtain a faithful reconstruction of the input geometry.
A similar case could be made for the \emph{Fandisk} model, where the angle made between features might be worthwhile preserving over the cost of having few badly shaped triangles.
Highlighting these cases and letting the user decide how to handle them is left as future work.

On a different note, none of the surfaces handled here are equipped with boundary.
Technically, handling boundary is not very different from handling features: If the boundary is given as a polygon, it can be used for initialization.
Then, when triangulating the regions following Section~\ref{sec:TriangulatingTheResultingMesh}, regions solely bounded by boundary edges are omitted.
The tricky element here is to identify a boundary polygon for point cloud input.
While corresponding approaches exist in the literature~\cite{mineo2019novel}, implementing such in our context is left as future work.

Finally, this paper used the uniform sphere representation of surfaces~\cite{lipschutz2021single} in the context of remeshing.
Fundamentally, the representation is not limited to surfaces and could also be applied to represent volumes by a collection of single-sized spheres.
Similar to the remeshing of surfaces, a volume representation could thus be used for tetrahedral meshing and remeshing of volumetric objects.
Like the previous considerations, this is left as future work.

\bibliographystyle{abbrv}
\bibliography{mybibfile}

\newpage
\newpage
\newpage 

\onecolumn

\appendix

\section*{Feature-aware Manifold Meshing and Remeshing of Point Clouds and
	Polyhedral Surfaces with Guaranteed Smallest Edge Length}
\subsection*{Supplementary Material}

\begin{figure}[b]
	\includegraphics[width=0.095\textwidth]{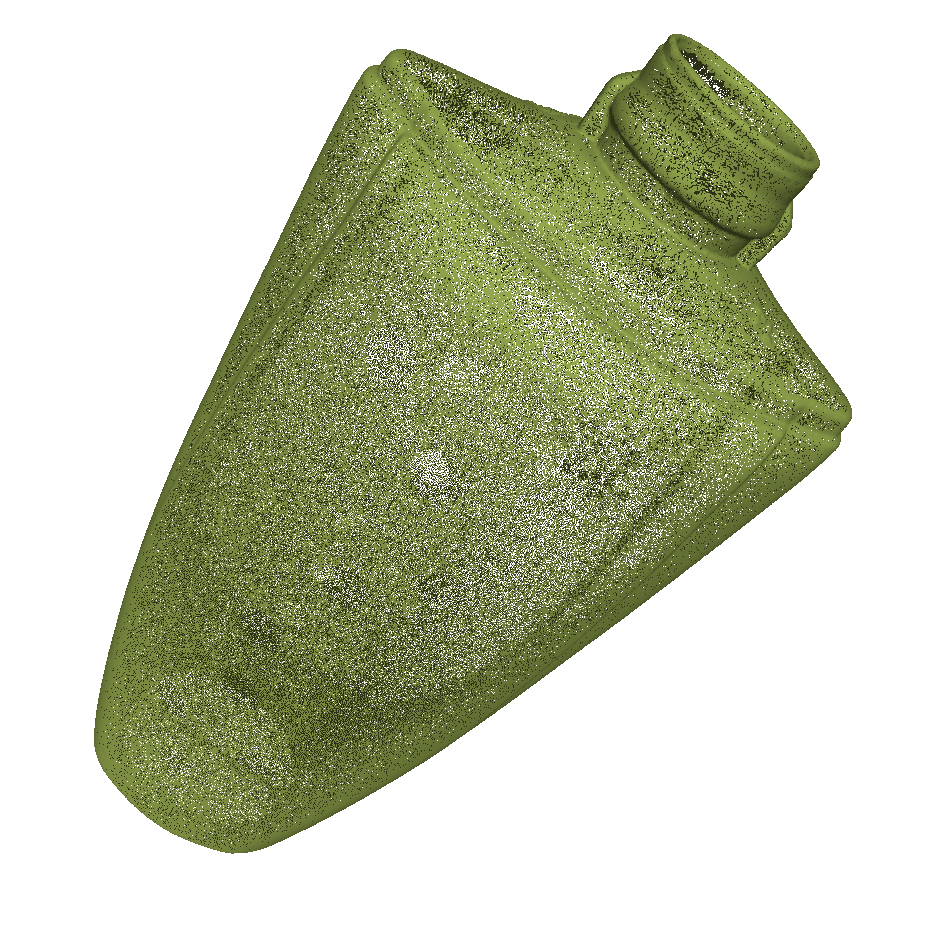}
	\includegraphics[width=0.095\textwidth]{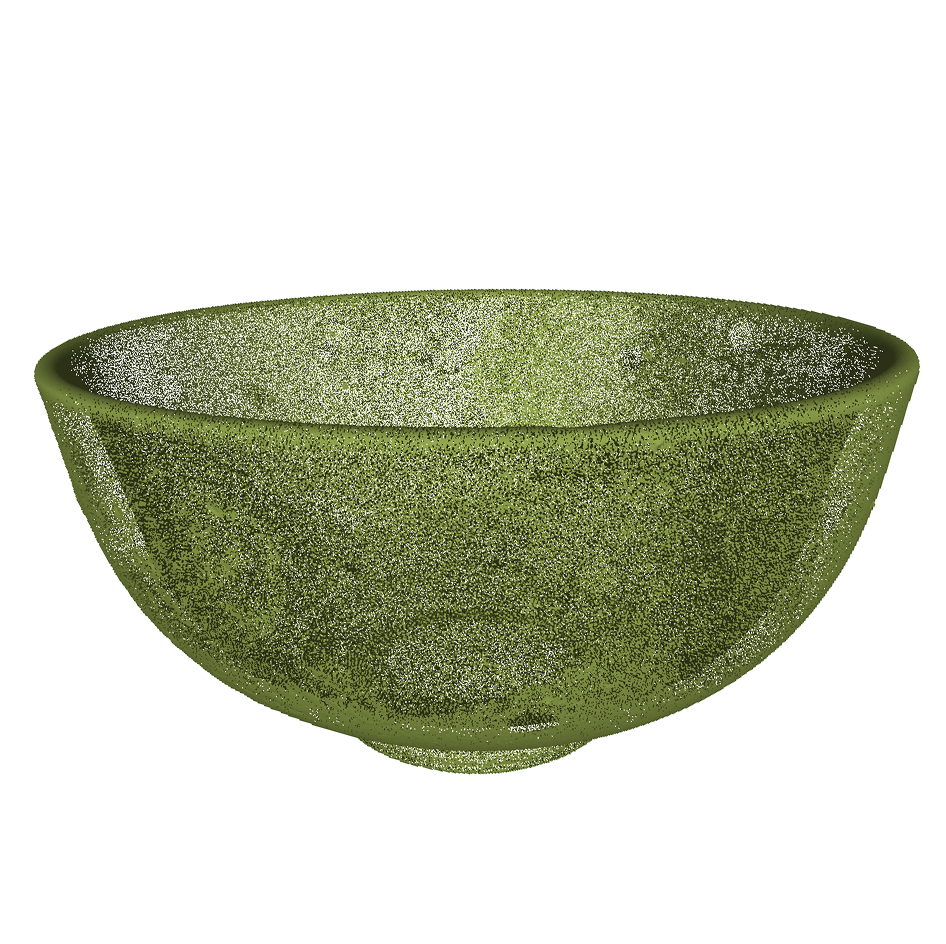}
	\includegraphics[width=0.095\textwidth]{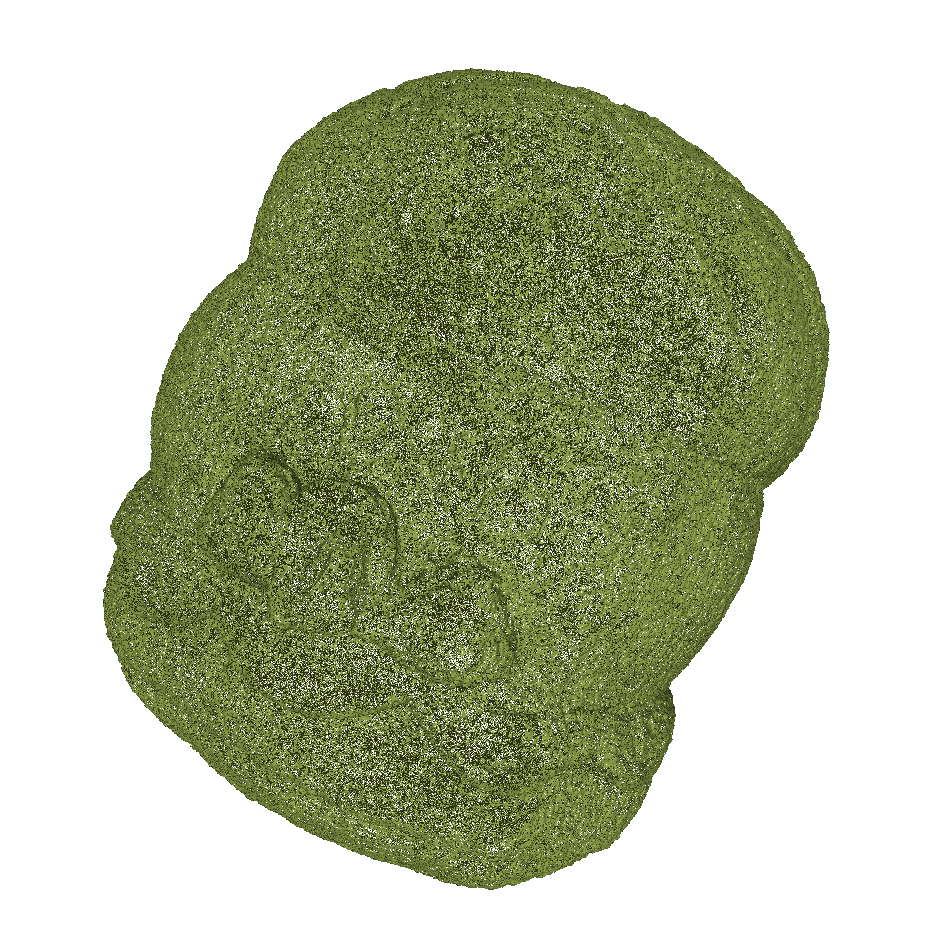}
	\includegraphics[width=0.095\textwidth]{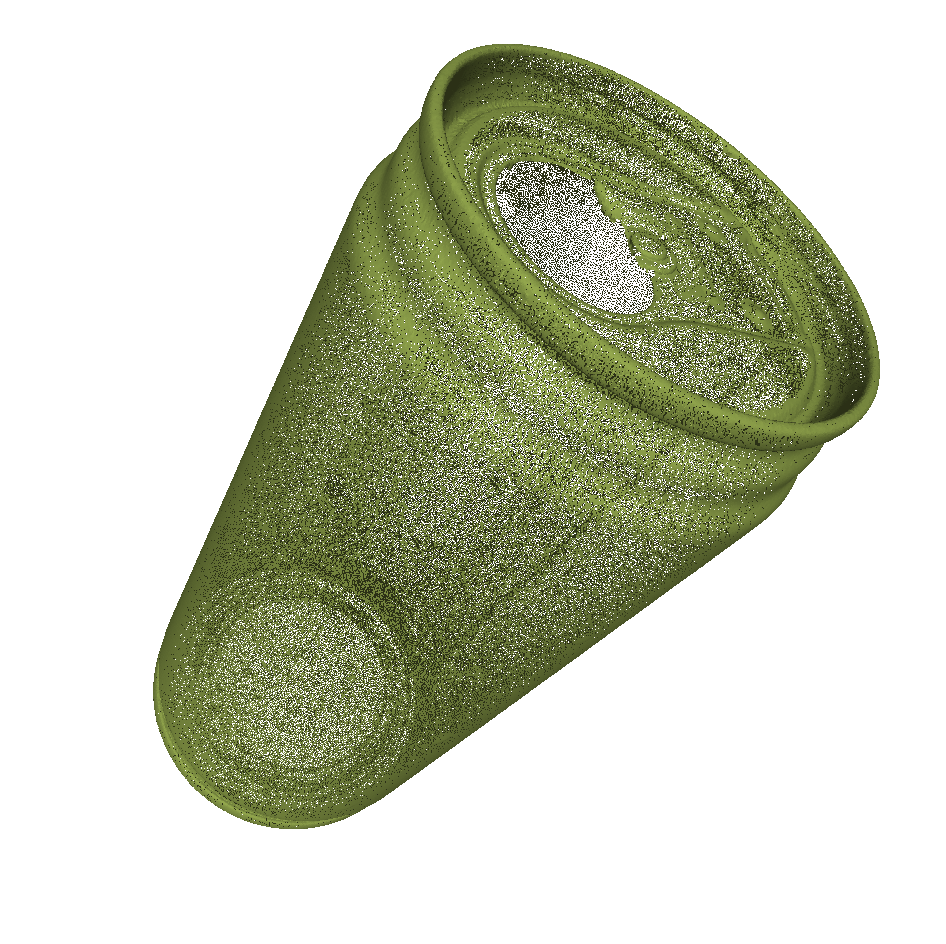}
	\includegraphics[width=0.095\textwidth]{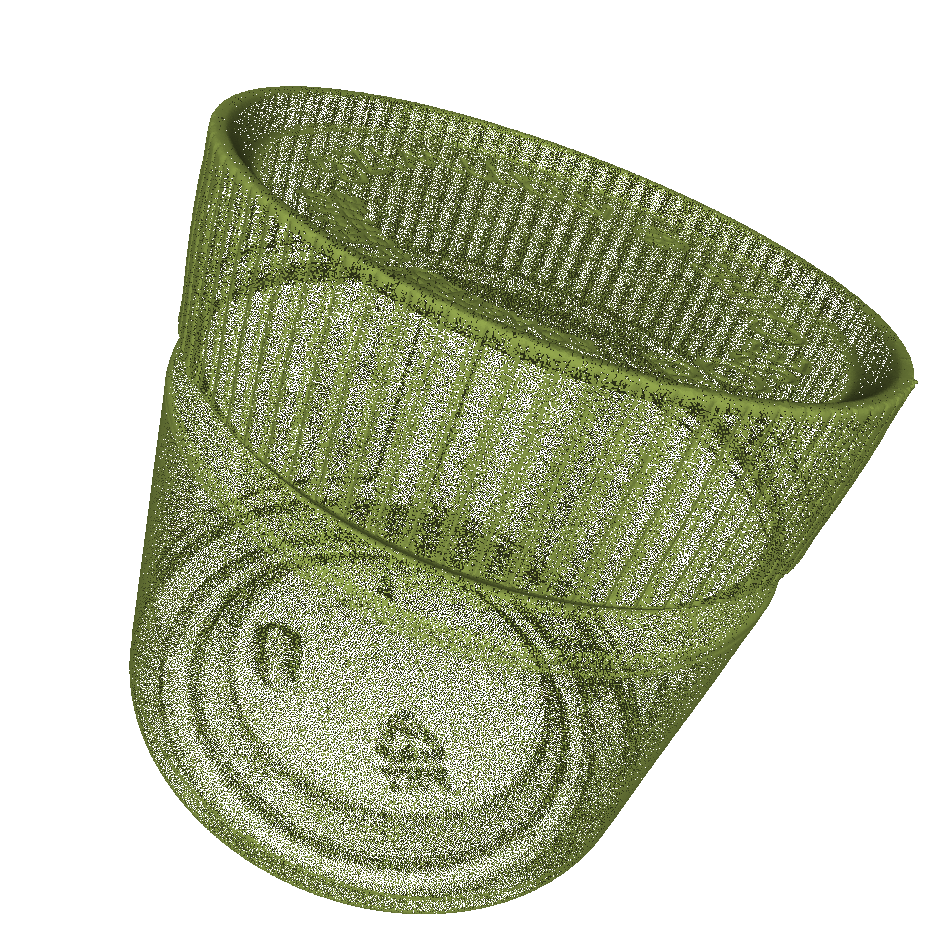}
	\includegraphics[width=0.095\textwidth]{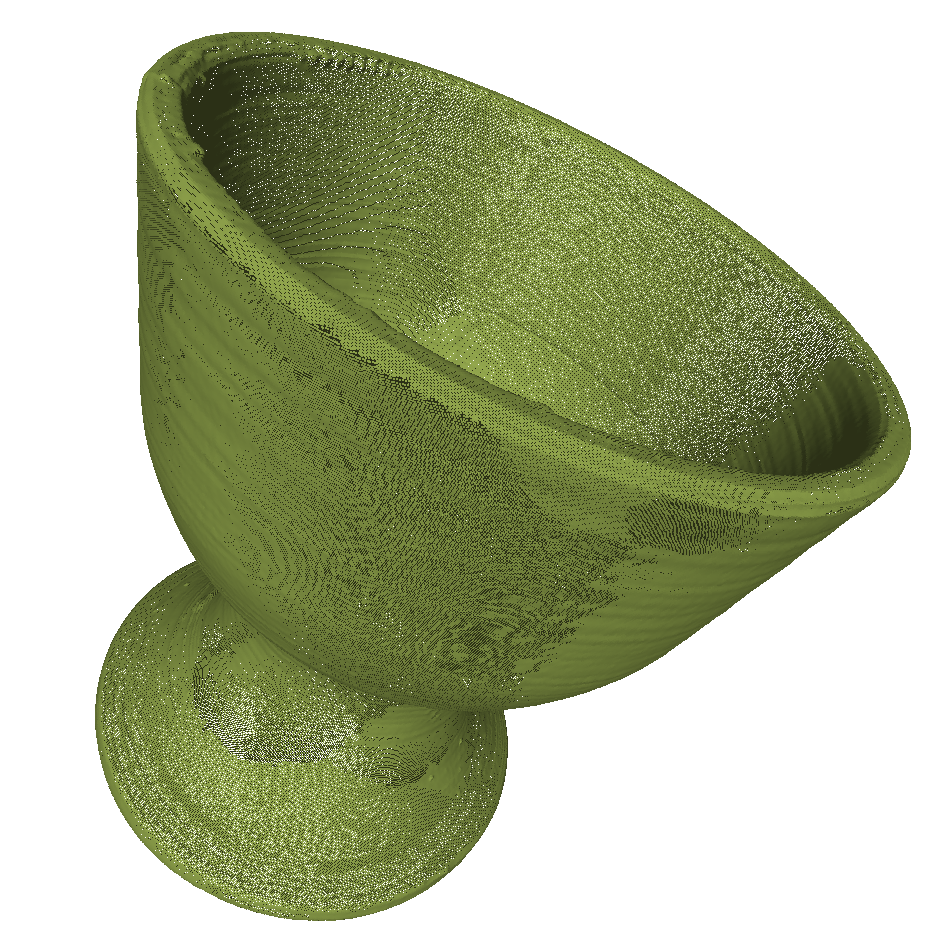}
	\includegraphics[width=0.095\textwidth]{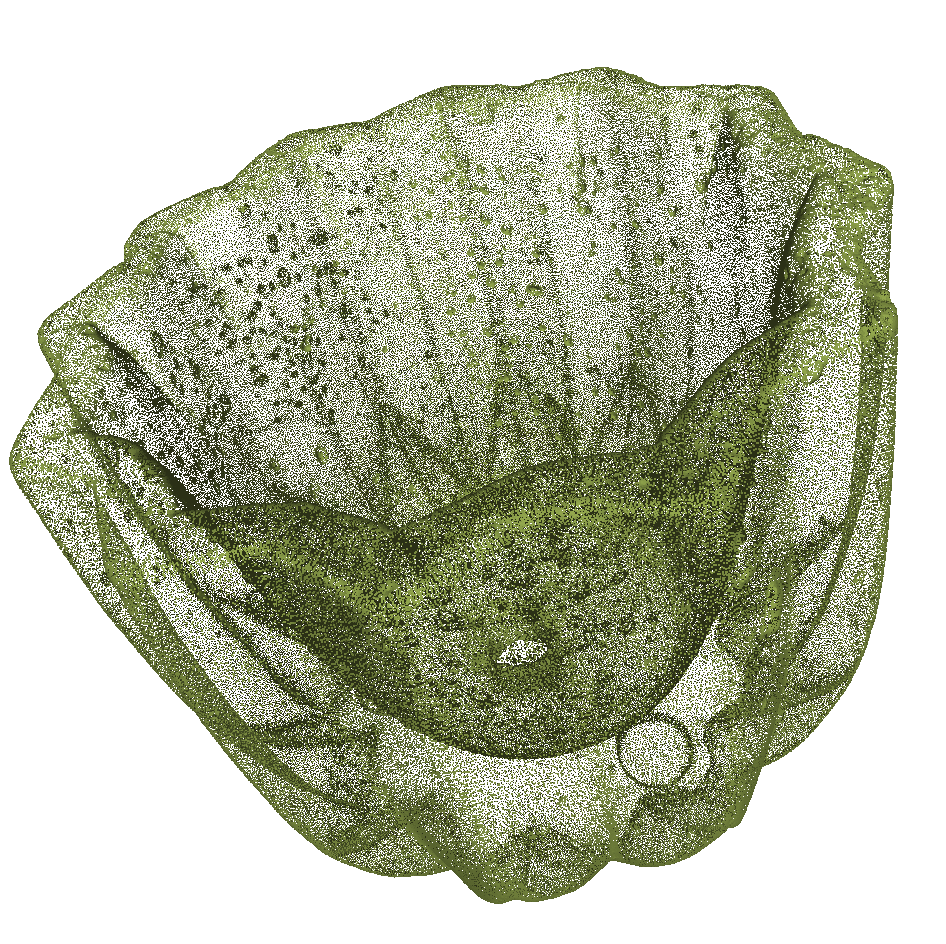}
	\includegraphics[width=0.095\textwidth]{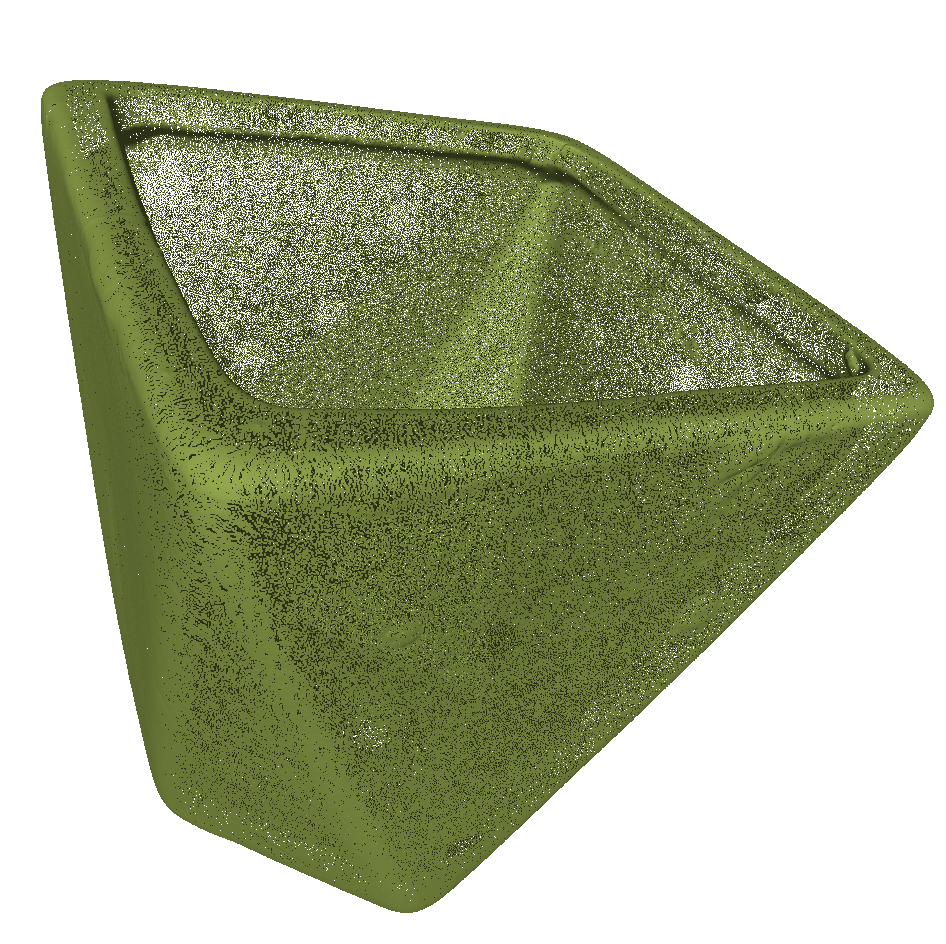}
	\includegraphics[width=0.095\textwidth]{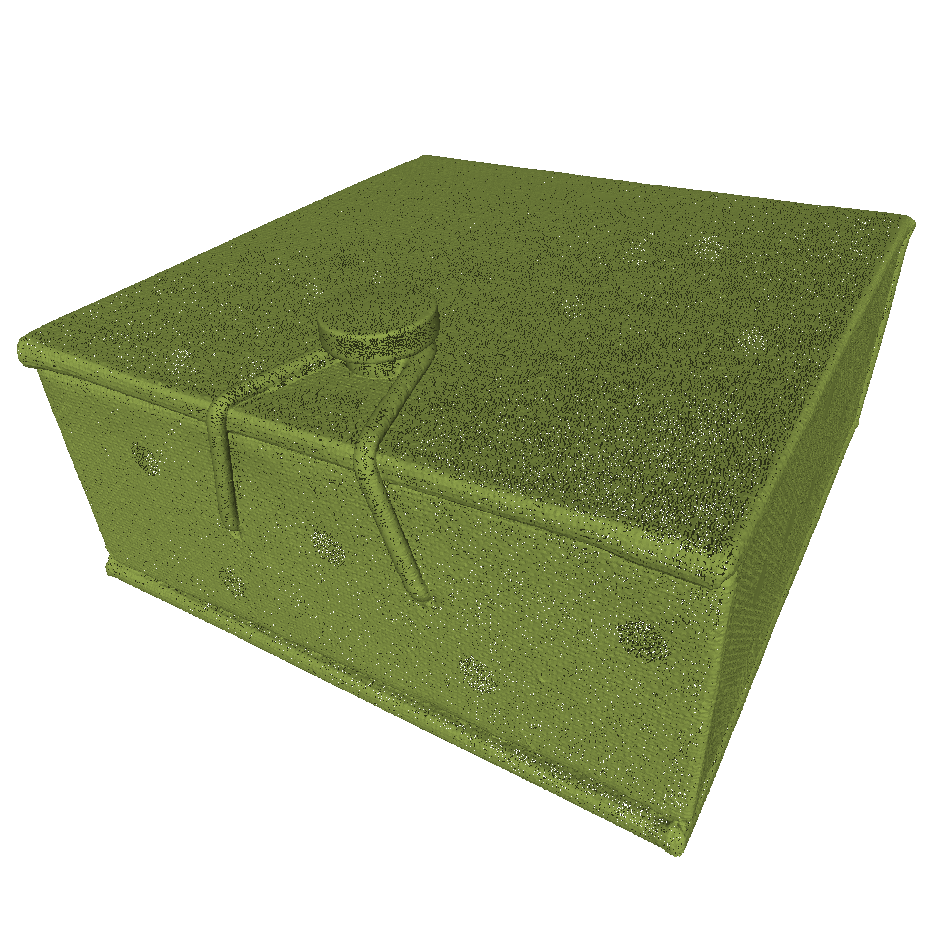}
	\includegraphics[width=0.095\textwidth]{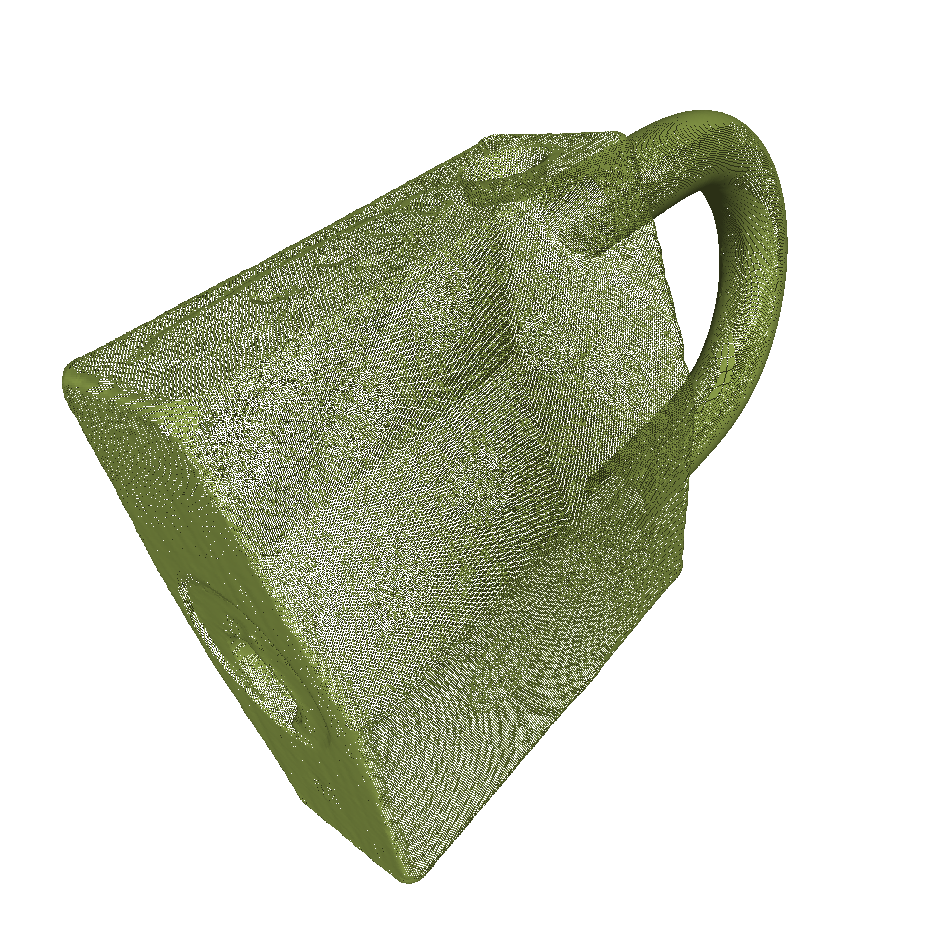}
	\includegraphics[width=0.095\textwidth]{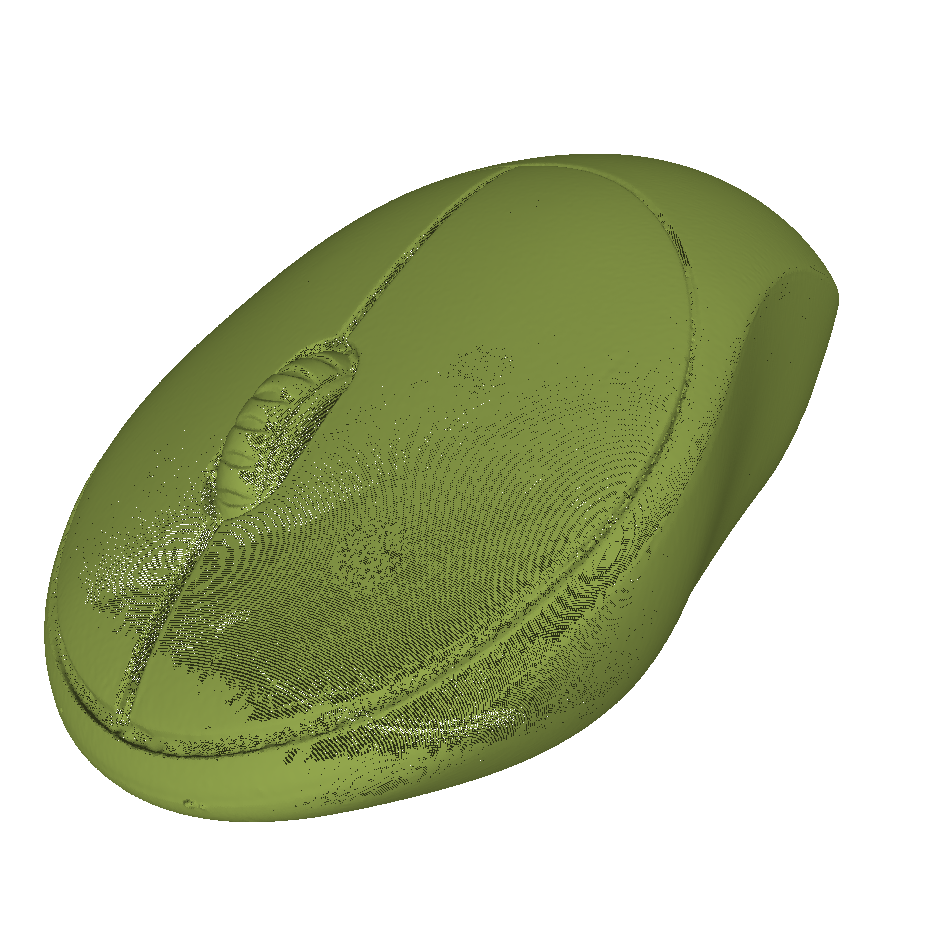}
	\includegraphics[width=0.095\textwidth]{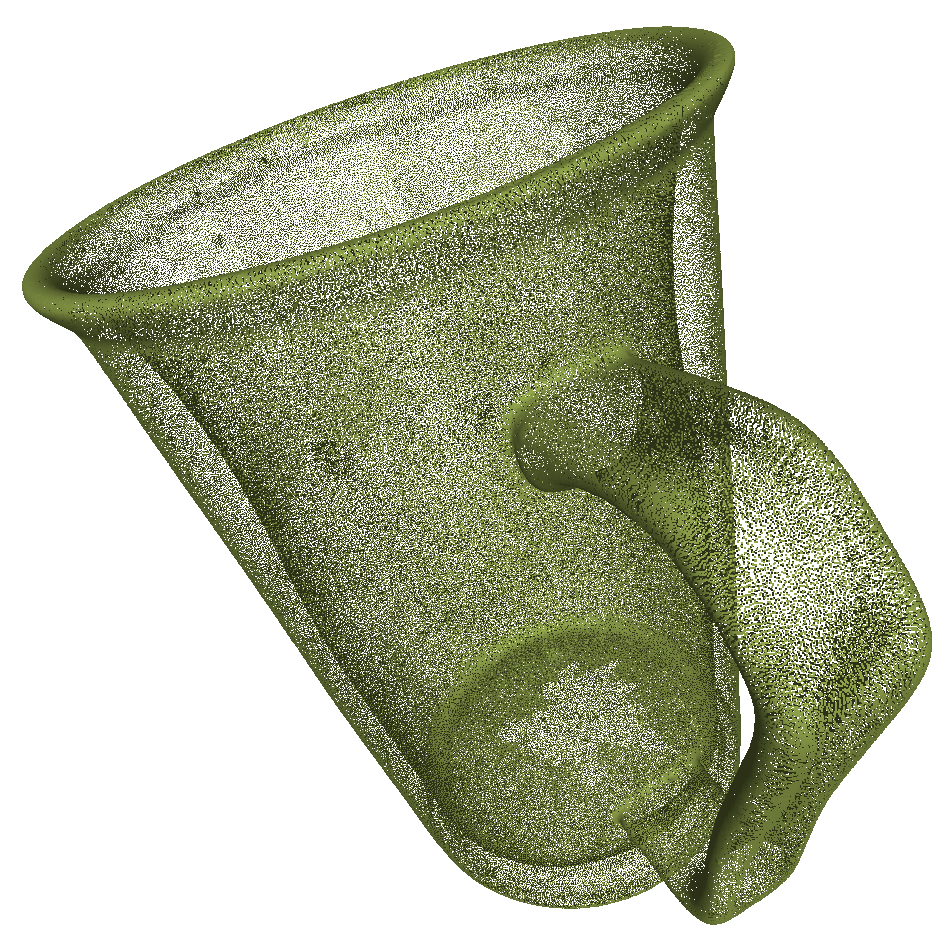}
	\includegraphics[width=0.095\textwidth]{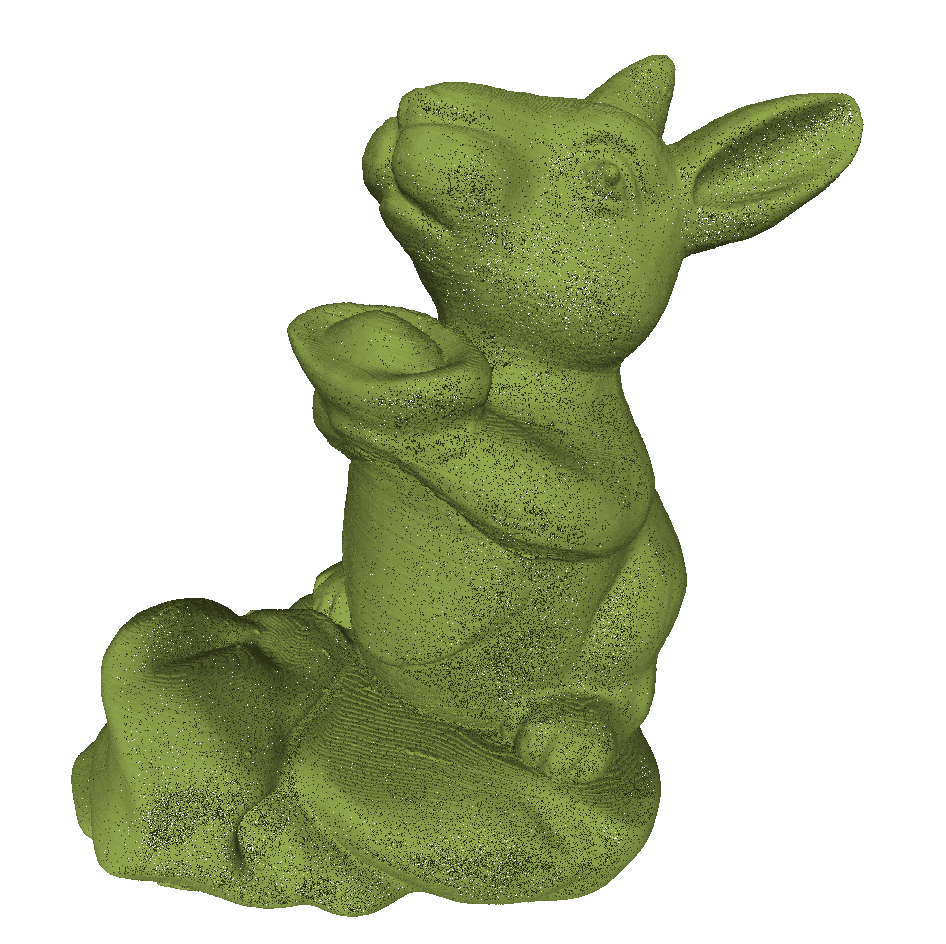}
	\includegraphics[width=0.095\textwidth]{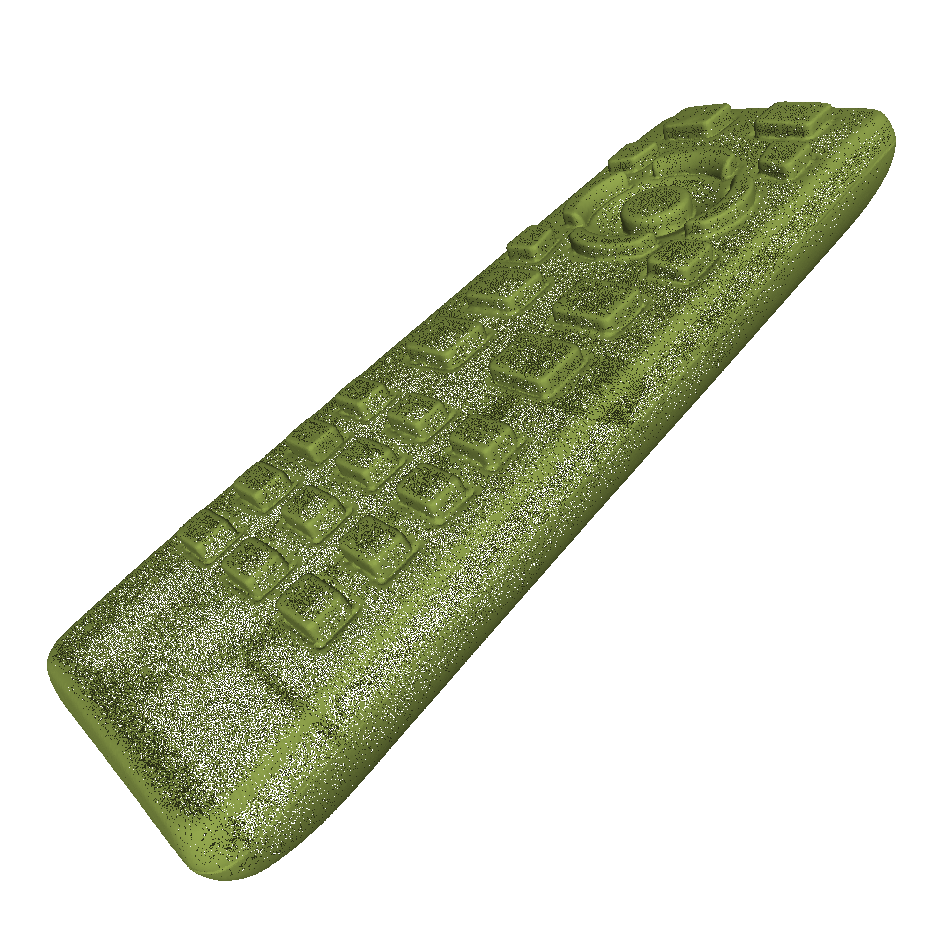}
	\includegraphics[width=0.095\textwidth]{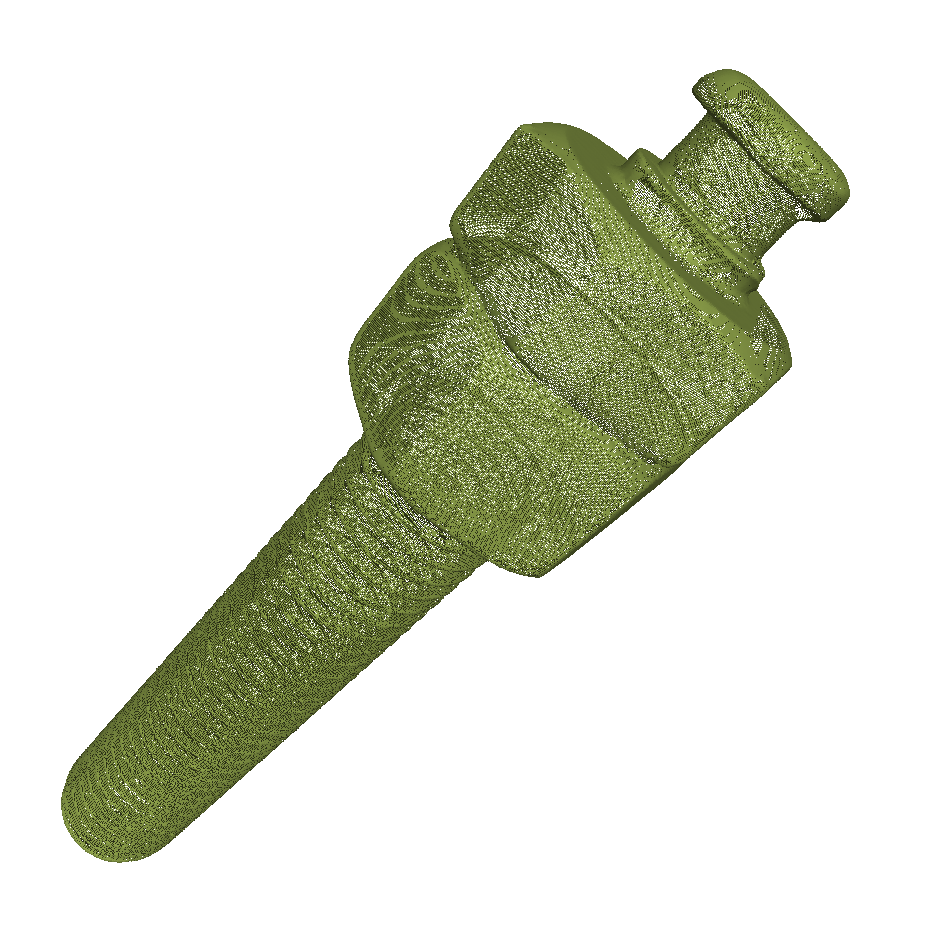}
	\includegraphics[width=0.095\textwidth]{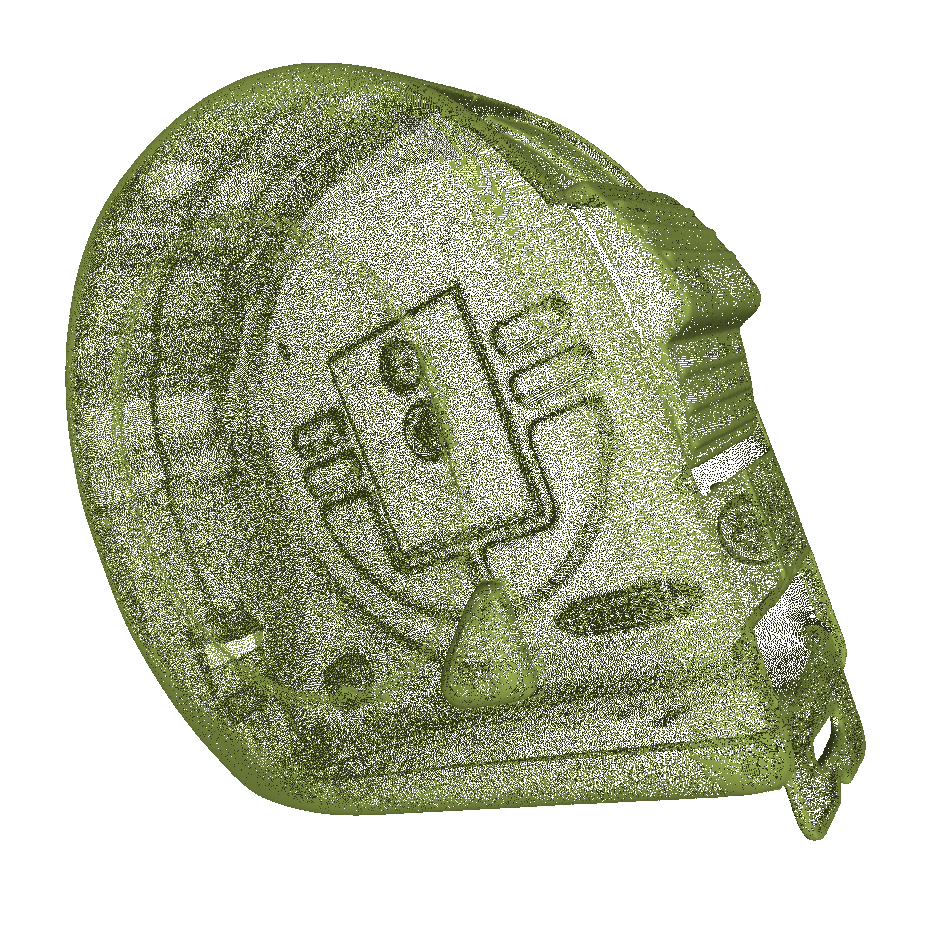}
	\includegraphics[width=0.095\textwidth]{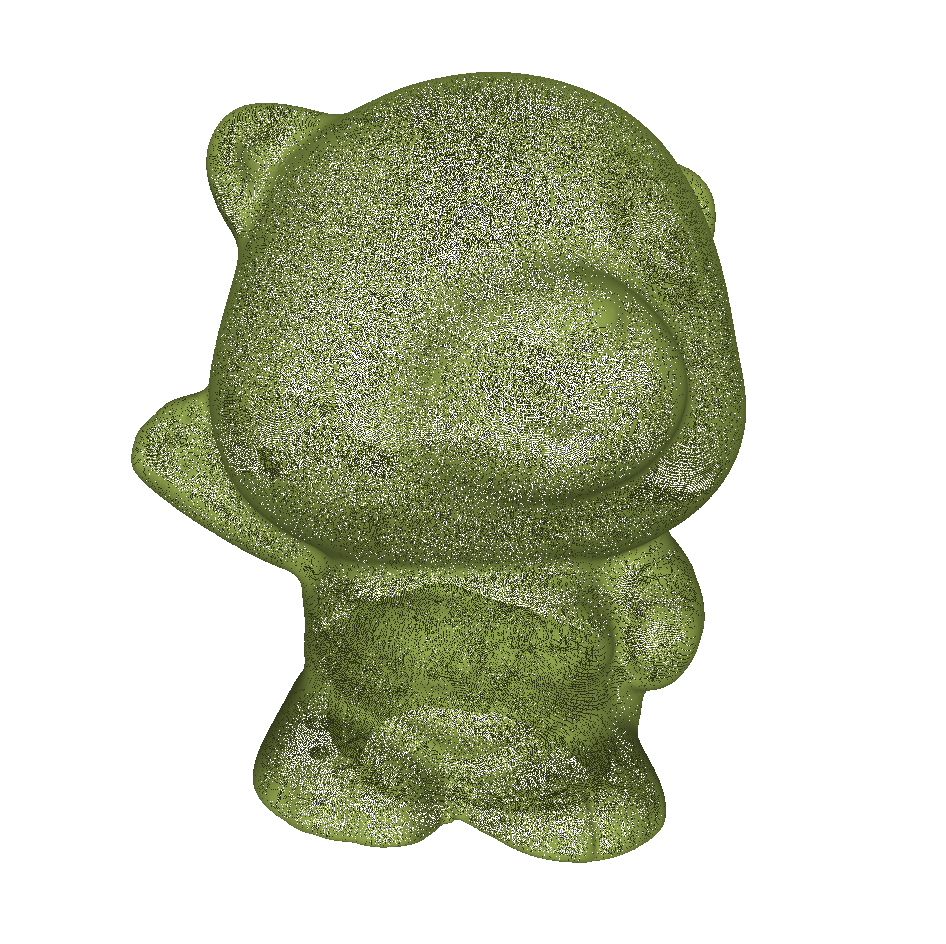}
	\includegraphics[width=0.095\textwidth]{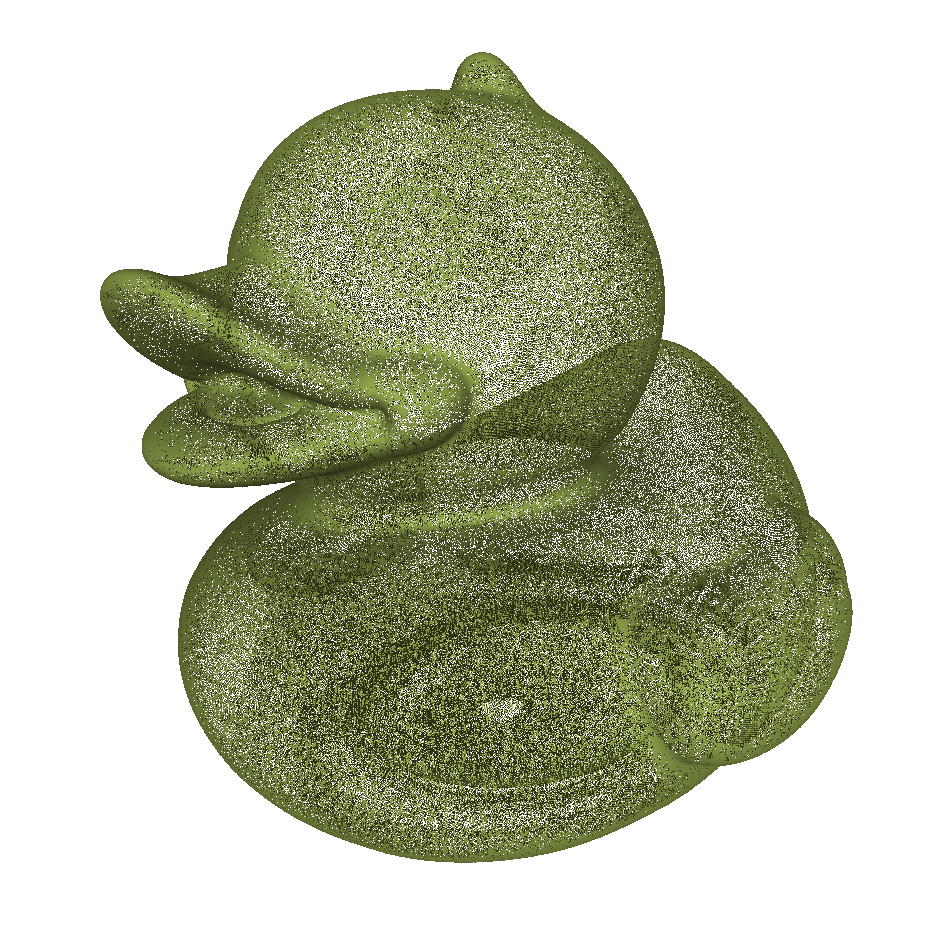}
	\includegraphics[width=0.095\textwidth]{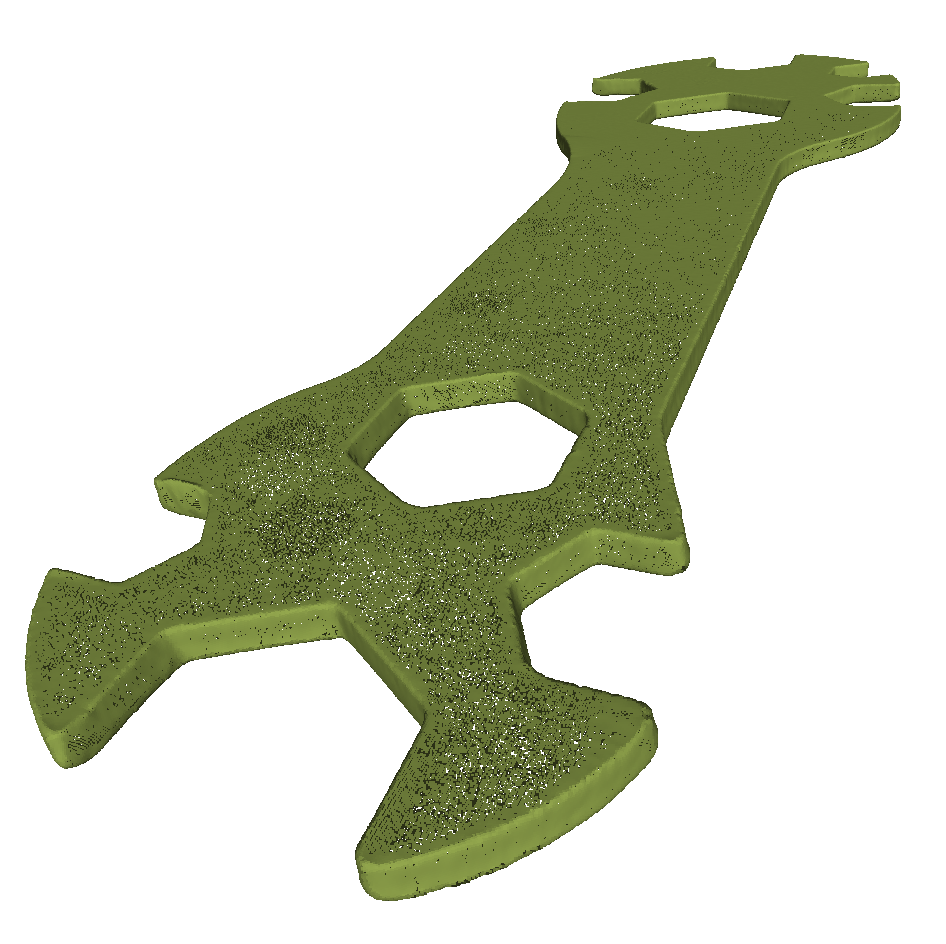}
	\includegraphics[width=0.095\textwidth]{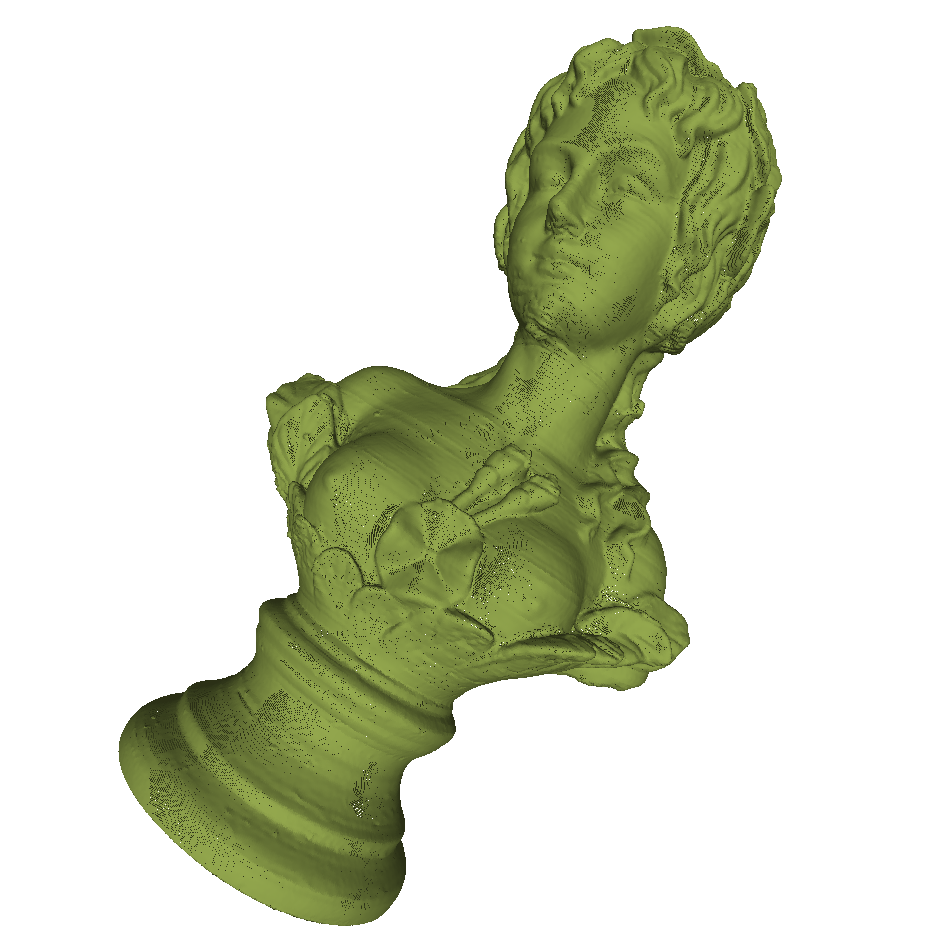}
	\caption{Real-world scanned models as point clouds as presented in~\cite{huang2022surface}. From left to right: \emph{Bottle Shampoo}, \emph{Bowl Chinese}, \emph{Cloth Duck}, \emph{Coffee Bottle Metal}, \emph{Coffee Bottle Plastic}, \emph{Cup}, \emph{Flower Pot}, \emph{Flower Pot 2}, \emph{Gift Box}, \emph{Lock}, \emph{Mouse}, \emph{Mug}, \emph{Rabbit}, \emph{Remote}, \emph{Screw}, \emph{Tape}, \emph{Toy Bear}, \emph{Toy Duck}, \emph{Wrench}, \emph{Xiao Jie Jie}.}
	\label{fig:SetOfModels}
\end{figure}

\noindent In this supplementary material, we present a pseudo code segment summarizing Sections~3.3 to~3.7 and results of the algorithm introduced in our submission, applied to the complete set of twenty models as provided by the authors of~\cite{huang2022surface}.
See Figure~\ref{fig:SetOfModels} for all models.
In order to emphasize the overall simple structure and to allow for easy reproduction of the algorithm presented here, we provide the following pseudo code fragment.

\begin{algorithm}
	\caption{Isotropic Point Cloud Meshing using unit Spheres}\label{alg:Pseudocode}
	\begin{algorithmic}
		\Require point cloud $\mathcal{P}$, normal field $\mathcal{N}$, target edge length $d$, splat size $s$, starting vertices $q,q'$, window size $w$, maximum border length $\partial_{\max}$
		\Ensure triangle mesh $\mathcal{T}$ with edge lengths close to uniformity
		\State \noindent\rule[0.25ex]{\linewidth}{0.5pt}
		\State Build box grid, register splats in all boxes up to distance $d$ (possibly with individual splat sizes), filter points according to the average normal, compute box normals. 
		\State Project $q$ and $q'$ to their closest splats, start a graph $\mathcal{G}$ by adding the projections as vertices. 
		\State Compute initial vertex candidates and their priority, add candidates to the queue.
		\While{candidate vertex $v_c$ exists in the queue} 
		\If{$\mathcal{G}$ has vertex~$v$ s.t. $\left\|v-v_c\right\|_2<d$ or $v_c$ fails the projection check}
		\State Discard $v_c$.
		\ElsIf{priority of $v$ is not correct}
		\State Correct priority by pushing $v_c$ back to the queue.
		\Else
		\State Add $v_c$ and edges to its parent vertices to $\mathcal{G}$, update region borders.
		Compute new vertex candidates around $v_c$ and their priorities, add them to the queue.
		\EndIf
		\EndWhile
		\For{each region $R\in\mathcal{R}$} 
		\While{length of region border $\partial R$ is $\geq3$ and $<\partial_{\max}$}
		\State Cut triangle $t$ at the smallest border angle, add $t$ to~$\mathcal{T}$.
		\EndWhile
		\EndFor
	\end{algorithmic}
\end{algorithm}

The experiments run on said repository are subdivided into three sections.
In Section~\ref{sec:WindowSize}, we investigate the window size~$w$ by running the algorithm on the \emph{Bottle Shampoo} with varying values for~$w$.
This motivates our standard parameter choice~$w=8$ as presented in the paper.
Then, in Section~\ref{sec:TriangulationOfBorders}, we discuss and illustrate how the triangulation of borders is done in detail.
In Section~\ref{sec:CollectionOfModels}, we present results for each model.
Section~\ref{sec:robustness} is dedicated to the investigation whether the output is sensitive to the choice of starting vertices.
This includes: 
\begin{compactitem}
	\item an image of the meshed model, as obtained by our algorithm,
	\item histograms showing the angle distribution, edge lengths distribution, and distribution of the triangle quality~$Q_t$ for the model,
	\item and the number of triangles and aggregated quality measures in tables.
\end{compactitem}
See Section~5 of the paper for an explanation of these quantities.
Note that the images and histograms displayed in Section~\ref{sec:CollectionOfModels} come from the output of our algorithm without applying a remeshing step.
As in the paper, the best values achieved for the root mean square deviation~$E_{\text{RMS}}$, for the quality~$Q_t$ its average~$Q_{\text{avg}}$, and the root mean square deviation~$Q_{\text{RMS}}$ per model are highlighted.
The table also shows values for a remeshed version of our results for comparison.
Values are highlighted here if they improve those shown above.

Note that our algorithm produces a triangulation of edge lengths close to uniformity.
This is reflected in the second histogram presented for each model, i.e., in the edge lengths distribution. 
The target edge length for all models is~$0.2$, which---by construction---is also the lowest edge length possible in the output of our algorithm.
Larger edges are possible, but there is a significant spike at the target edge length~$0.2$.
This insurance of a minimal edge length is the main reason why our algorithm does not perform best, but very close to best among the compared algorithms, when comparing average edge lengths in Tables~1 to~4 of the main paper.
Because of the strict minimum at~0.2, there are no shorter edges that can bring the average closer to~0.2 again, cf.~the discussion in Section~5 of the paper.

As the edges are all of very uniform length, the triangles are almost equilateral. 
This is noticeable by the large spike at~$60^\circ$ in the first histogram of each model, showing the angle distribution.

Note that a significant number of triangles~$t$ of the resulting triangulations actually achieve very high up to highest quality possible, i.e.,~${0.99<Q_t=1}$, and are thus sorted into the last bin of the respective histogram.
The distribution of this quality is very tight across all models.
See Section~5 of the paper for the definition of this quality measure.

In Section~\ref{sec:robustness}, we include additional data sets for a reasonable selection of models to illustrate the robustness of the algorithm.
Therefore, we chose the \emph{Wrench}, the \emph{Lock}, and the \emph{Bottle Shampoo} because of the features in contrast to the comparibly smooth \emph{Toy Bear} shown in the paper and run the algorithm with eight different pairs of starting vertices.
Further, we included the data sets corresponding to the \emph{Toy Bear} for completeness.
The data sets consist of significant characteristics including the total number of generated triangles~$\left| \mathcal{T} \right|$, the maximum edge length~$E_{\text{max}}$, the average edge length~$E_{\text{avg}}$, the root mean square deviation~$E_{\text{RMS}}$, the minimum triangle quality~$Q_{\text{min}}$, the maximum triangle quality~$Q_{\text{max}}$, the average triangle quality~$Q_{\text{avg}}$, and the root mean square deviation~$Q_{\text{RMS}}$.

Finally, Sections~\ref{sec:CollectionOfPointCloudsForFeatureDetection} and~\ref{sec:CADModelsForFeatureDetection} display a variety of point clouds as well as CAD models used to illustrate our advances in feature detection and preservation.
The point clouds belong to the repository named above, while the CAD models are based on models priviously available at AIM@SHAPE.
Their reconstructions are made by U. Reitebuch and they can be downloaded at~\cite{reitebuch2025CADmodels}.
This supplementary material contains all information not included in the paper for completeness.
Next to histograms showing the distributions of angles, edge lengths, and triangle quality, we illustrate our findings with two different resolutions for the target edge length~$d$.
For point clouds, a table lists information like target edge length, feature angle (if chosen), number of vertices, number of triangles, as well as~$d_{\text{avg}}$, $d_{\max}$, and $d_{\text{RMS}}$ as defined in the original paper.
In case of feature detection, we additionally list the number of detected feature segments and feature vertices.
The CAD models come with a table listing the target edge length, the feature angle chosen by the user, the resulting number of vertices~$| \mathcal{V} |$ and of triangles~$| \mathcal{T} |$, the minimal angle~$\alpha_{\min}$ achieved, the maximal angle~$\alpha_{\max}$, the average angle~$\alpha_{\text{avg}}$, and the corresponding RMS value~$\alpha_{\text{RMS}}$.
Further, the table contains the maximal edge length~$E_{\max}$, the average edge length~$E_{\text{avg}}$ as well as the corresponsing RMS value~$E_{\text{RMS}}$.
In addition, we evaluate the quality measure introduced in the original article and collect the resulting values for $Q_{\min}$, $Q_{\max}$, $Q_{\text{avg}}$, and $Q_{\text{RMS}}$.
As mentioned in the main article, we compare results achieved by running our algorithm to those achieved by running a certain routine implemented in MeshLab.
For two models, the \emph{Oloid} model and the \emph{Body of Constant Width} model, we increased the number of iterations until there were no overfoldings occuring anymore.
These overfoldings are not visible in the tables shown in Sections~\ref{sec:BodyOfConstantWidth} and~\ref{sec:Oloid}.

\newpage

\section{Window Size}
\label{sec:WindowSize}

\begin{figure}[h!]
	\centering
	\includegraphics[width=.8\textwidth]{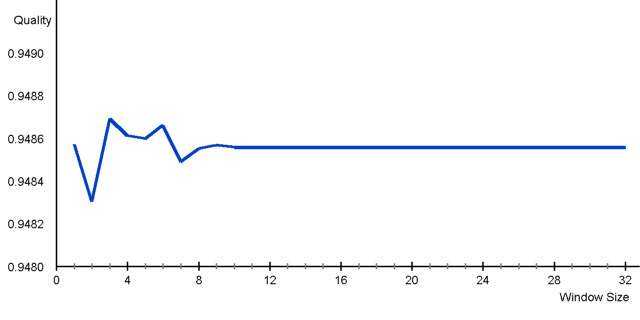}
	\caption{Quality~$Q_t$ for varying window sizes.}
	\label{fig:WindowSize_Quality}
\end{figure}

\begin{figure}[h!]
	\centering
	\includegraphics[width=.8\textwidth]{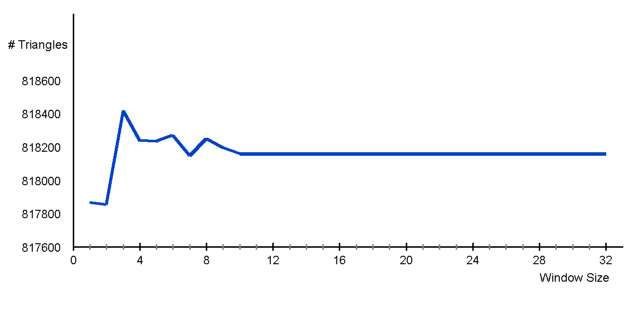}
	\caption{Number of triangles for varying window sizes.}
	\label{fig:WindowSize_NumTriangles}
\end{figure}

As mentioned in the paper, the algorithm asks for a window size~$w$, which is used to limit tracing along borders in order to place the next vertex candidate.
Experiments run with different window sizes show that the triangulation quality and the number of triangles obtain plateau from~$w=8$ onward.
This makes~$w=8$ a reasonable choice for a reliable output with respect to the occurring lengths of borders before triangulating the created regions.
See the result of the experiments in Figures~\ref{fig:WindowSize_Quality} and~\ref{fig:WindowSize_NumTriangles}.
All experiments were run on the \emph{Bottle Shampoo}.


On this model, the effect of changing the window sizes within this comparably small regime ([0,32]) is not measurable when considering the run time.
That is on the one hand side because all these look-ups are constant-time as opposed to considering the entire border, which can grow towards an expected time of~$\mathcal{O}(\sqrt{n})$ for~$n$ vertices placed.
On the other hand size, traversing a border for a small number of vertices already suffices to find that the traversed points are on the same border region.
Thus, an increased window size does not have any effect on these specific queries.


\newpage

%
%


\noindent The window size~$w$ influences the border length of the arising regions.
As illustrated in Figure~7 in the article, a very small window size, i.e.,~$w = 0$, results in borders up to length~$93$ while increasing the window size to~$w = 8$ reduces the border length to~$11$.

In contrast to the target edge length~$d$ chosen to illustrate the effect on the \emph{Bottle Shampoo} in the paper, here, the target edge length is equal to~$0.2$ to conincide with the other experiments presented in the supplementary material.
As shown in Figure~\ref{fig:HistogramDifferentWindowSizes}, a bigger window size~$w$ prevents longer borders.
Here, for $w = 0$, the longest border is of length~$80$ while for~$w = 8$, there is no region of border length bigger than~$20$.
Most resulting regions have a border length settled between~$3$ and~$12$ for both choices for~$w$.
Note that the histogram in Figure~\ref{fig:HistogramDifferentWindowSizes} has a log-scale~$y$-axis. 
The obtained triangle quality stems to the largest extend from the placement of the vertices, rather than from triangulating regions with borders larger than~$3$.
Even though changing the window size does not influence the run time significantly, it influences the visual aspects of the output, but the mesh quality after trigangulating the regions remains untouched as aforementioned.

\begin{figure}[h!]
	\includegraphics[width=\textwidth]{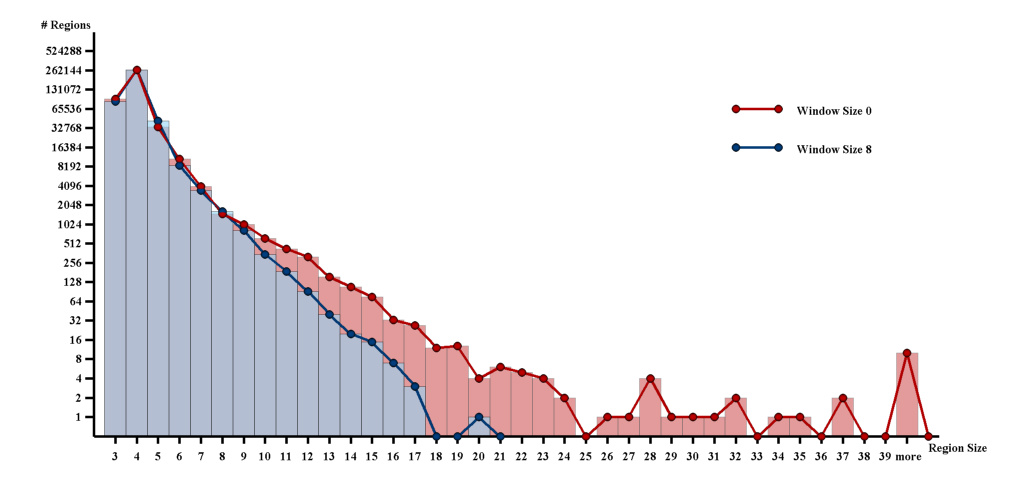}
	\caption{Region sizes for window sizes $w = 0$ (shown in blue) and $w = 8$ (shown in red) apllied to the \emph{Bottle Shampoo}.}
	\label{fig:HistogramDifferentWindowSizes}
\end{figure}

\newpage

\section{Triangulation of Borders}\label{sec:TriangulationOfBorders}

\noindent After the algorithm was run on a geometry, the final graph may contain region borders longer than~$3$.
In order to obtain a triangular mesh, these have to be triangulated.
During application of our algorithm, we assign to every vertex the normal~$n_b$ of the box the considered point is associated to as its vertex normal.
Additionally, this box normal gives us a best fitting local approximation of the resconstructed surface.
Hence, for each point~$p$ on a boundary~$\partial R$, we project the pair of edges adjacent to~$p$ belonging to~$\partial R$ to the tangent plane~$T_p$ defined by the corresponding box normal~$n_b$.
From the set of oriented angles~$\left\lbrace \alpha_p \mid p \in \partial R \right\rbrace$ made by the projected pairs of edges adjacent to~$p$, we choose the smallest one and connect the two points~$p_{\ell}$ and~$p_r$ adjacent to~$p$ by an edge.
We repeat this process now with the new border~$\left(\partial R \setminus \left\lbrace \left( p_{\ell}, p\right), \left(p, p_r\right) \right\rbrace \right) \cup \left\lbrace \left( p_{\ell}, p_r\right) \right\rbrace$ until the remaining border is of length~$3$.

\begin{figure}[b!]
	\centering
	\begin{minipage}{\textwidth}
		\centering
		\scalebox{.85}{\input{sm_borderToTriangulate}}
	\end{minipage}
	\begin{minipage}{\textwidth}
		\centering
		\scalebox{.75}{\input{sm_cuttingSmallestAngle}}
	\end{minipage}
	\caption{Top: Border~$\partial R$ to be triangulated. Bottom: Projection of border edges to tangent plane.}
\end{figure}


\newpage

\section{Handling of Noisy Data}\label{sec:NoisyData}

In this section, we collect several observations made during the consideration of noisy data.
For given reasons, the geometry sampled by a point cloud~$\mathcal{P}$ with normal field~$\mathcal{N}$ is a plane~$P$.
Without loss of generality, we assume this plane to be the~$x$-$y$-plane.
In the following, we analyze three different settings.
In the first one, we solely assume the points in~$\mathcal{P}$ to be noisy.
Next, we assume the points of~$\mathcal{P}$ to lie in~$P$ and the normals to be noisy.
In the third case, we combine the two situations considered before.

Even though~$P$ is not a proper example fulfilling the properties listed in Section~3.1 in the paper, it allows us to demonstrate several thoughts because of its simple structure and well known properties.
Since the reach~$\rho$ of~$P$ is undefined, the target edge length~$d$ may be chosen depending on the level of noise
and the initial splat size~$s$ may be equal to~$d$.

\subsubsection*{Noisy Points, Correct Normal Field}

Here, we assume the points in~$\mathcal{P}$ to come with some noise while the normal field is assumed to be correct.
Hence, based on the assumptions made above, all normals are equal to~$(0,0,1)^T \in \mathbb{R}^3$ and hence~$\langle n_i, n_j \rangle = 1 > 0$.
Set~$\nu = \max_{i \in \lbrace 1, \dots, n \rbrace} \left| \left( p_i \right)_z \right|$, where~$\left( . \right)_z$ denotes the~$z$-coordinate of~$p_i \in \mathcal{P}$.
Then, the two planes parallel to~$P$ at distance~$\nu$ enclose~$\mathcal{P}$.
By construction, all generated splats lie between these two planes as well as all generated vertices.
Especially, if~$d > 2 \nu$, no edges cross already existing ones under projection to the tangent plane, which is here the plane~$P$ itself, as described in Section~3.6 in the paper.
Hence, after triangulating the occuring regions, the result is a graph above~$P$ and hence topologically correct.


\subsubsection*{Correct Points, Noisy Normals}

Now, we assume the points in~$\mathcal{P}$ to lie in~$P$ while the normals deviate by some angle~$\alpha_i \in \left\lbrack 0, \frac{\pi}{2} \right\lbrack$ from the correct normal~$(0,0,1)^T$.
Let~$\alpha_{\max} = \max_{i \in \lbrace 1, \dots, n \rbrace} \alpha_i$ denote the biggest deviation angle.
Since the elements of~$\mathcal{P}$ lie in~$P$, but the normals deviate from the correct ones, the splats do not lie in~$P$ but stick out.
Hence, in~$z$-direction, the height of such a splat is bounded by~$\nu_{\alpha} = s\, \sin (\alpha_{\max})$ and similar to the scenario described before, the splats are enclosed by two parallel planes determined by~$\alpha_{\max}$.
Additionally, the box normals do not deviate more from the correct normal than the noisy ones.
For~$d = 2 \nu$, problems may occur while projecting the edges to the respective tangent plane.
In order to prevent these faulty projections, choose~$d > d_{\min} = 2 \nu_{\alpha} \cos \left( \alpha_{\max} \right)$.

\begin{figure}[h!]
	\centering
	\begin{tikzpicture}
		\draw (1.5,0) -- (10,0);
		\draw[dotted] (1.5,1.5) -- (10,1.5);
		\draw[dotted] (1.5,-1.5) -- (10,-1.5);
		\node (P) at (1.25,0){$P$};
		\draw[fill=black] (4,0) circle(.05);
		\node (p) at (4.25,-.25){$p$};
		\draw[very thick] (2,-1.5) -- (6,1.5);
		\node (sp) at(1.8,-1.25){$S$};
		\draw (4.5,0) arc(0:36.87:.5);
		\node (alpha) at (4.9,.2){$\alpha_{\max}$};
		\begin{scope}[shift={(4,0)}]
			\begin{scope}[rotate=36.87]
				\draw[->] (0,0) -- (0,1);
			\end{scope}
		\end{scope}
		\node (n) at (3.2,.8){$n$};
		\draw[dashed] (4,-1.75) -- (4,1.75);
		\draw[very thin,<->] (9,1.5) -- (9,-1.5);
		\node (nu) at (9.4,-1.3){$2 \nu_{\alpha}$};
		\draw (6,1.5) -- (8.25,-1.5);
		\node (dmin) at (7.7,-1.25){$d_{\min}$};
	\end{tikzpicture}
\end{figure}

%
%

\subsubsection*{Noisy Points, Noisy Normals}

Now, assuming both the points and the normals to be noisy, can be visualized as adding some offset in form of a translation into~$z$-direction.
This offset is given as~$dz = \max_{i \in \lbrace 1, \dots, n \rbrace} \left| \left( p_i \right)_z \right|$.
Therefore, the distance of the bounding planes to~$P$ is given by~$dz = \nu + \nu_{\alpha}$ and choose~$d > 2 dz \cos \left( \alpha_{\max} \right)$.

\begin{figure}[h!]
	\centering
	\begin{tikzpicture}
		\draw (0,0) -- (8,0);
		\draw[dotted] (0,1) -- (8,1);
		\draw[dotted] (0,-1) -- (8,-1);
		\begin{scope}[shift={(4,1)}]
			\begin{scope}[rotate=-20]
				\draw[very thick] (-1,0) -- (1,0);
				\draw[thick,->] (0,0) -- (0,1);
				\draw[fill=black] (0,0) circle(.05);
			\end{scope}
		\end{scope}
		\draw[thin,<->] (7.5,0) -- (7.5,-1);
		\node (dz) at (7.75,-.5){$\nu$};
		\draw[very thin,dashed] (0,1.34) -- (8,1.34);
		\draw[very thin,dashed] (0,-1.34) -- (8,-1.34);
		\draw[thin,<->] (7.5,-1) -- (7.5,-1.34);
		\node(nu) at (7.75,-1.171){$\nu_{\alpha}$};
	\end{tikzpicture}
\end{figure}

\newpage

\section{Collection of Models}
\label{sec:CollectionOfModels}

\subsection{\emph{Bottle Shampoo}}

\begin{figure}[h!]
	\centering
	\begin{minipage}{0.45\textwidth}
		\begin{subfigure}{\textwidth}
			\includegraphics[width=1.\textwidth]{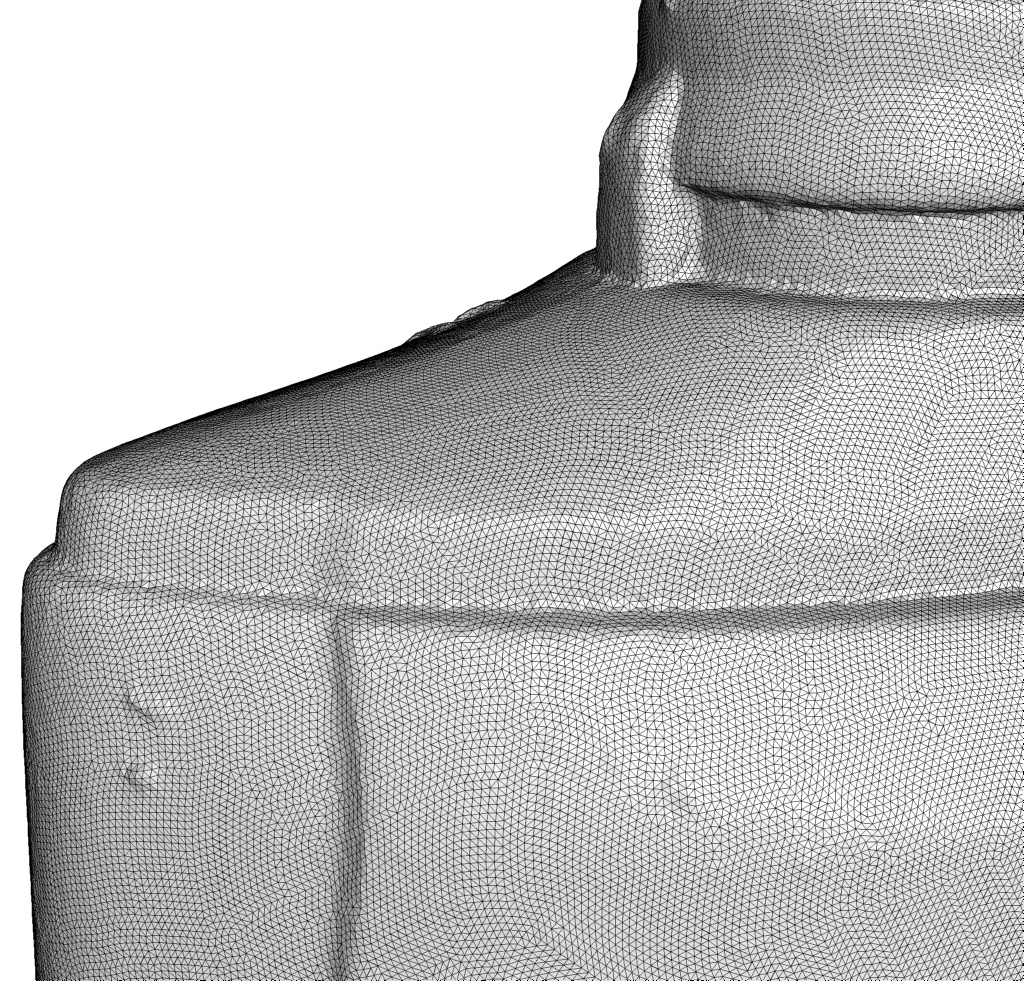}
		\end{subfigure}
	\end{minipage}
	\hfill
	\begin{minipage}{0.5\textwidth}
		\begin{subfigure}[t]{\Histogram}
			\centering
			\includegraphics[width=\textwidth]{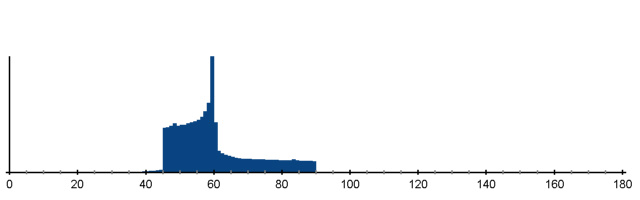}
			\caption{Angle distribution, target=$60^\circ$.}
		\end{subfigure}
		\begin{subfigure}[t]{\Histogram}
			\centering
			\includegraphics[width=\textwidth]{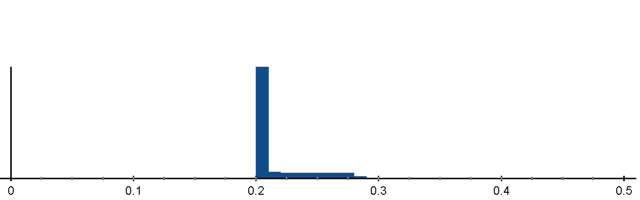}
			\caption{Edge lengths distribution, target=0.2.}
		\end{subfigure}
		\begin{subfigure}[t]{\Histogram}
			\centering
			\includegraphics[width=\textwidth]{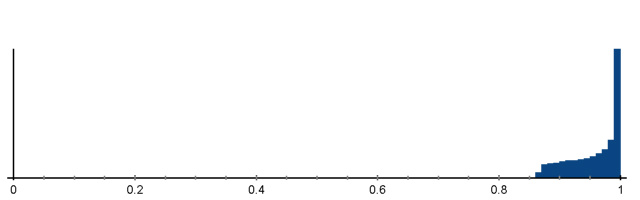}
			\caption{Distribution of quality $Q_t$, target=1.0.}
		\end{subfigure}
	\end{minipage}
	\caption{
		\emph{Bottle Shampoo}.
	}
\end{figure}

\begin{table}[b!]
	\def\arraystretch{1.05}
	\centering
	\begin{tabular}{l|rrrrr}
		Algorithm & $|\mathcal{T}|$ & $E_{\text{avg}}$ & $E_{\text{RMS}}$ & $Q_{\text{avg}}$ & $Q_{\text{RMS}}$\\
		\hline
		Adv.~Front &  1.209,546 & 0.1799 & 39.6 & 0.8247 & 16.0\\
		Adv.~Front (Re) & 928,850 & 0.2028 & 15.3 &  0.9416 & 6.1\\
		Poisson &  16,280 & 1.2946 & 74.8 & 0.8760 & 12.3\\
		Poisson (Re) & 498,140 &  0.2657 & 38.6 & 0.9251 & 7.5\\
		Poisson MG & 150,770 & 0.5318 & 35.7 & 0.7204 & 33.7\\
		Poisson MG (Re) & 952,830 & \textbf{0.2015} & 16.3 & 0.9330 & 7.0\\
		RIMLS & 1,907,781 & 0.1499 & 35.8 & 0.7055 & 35.1\\
		RIMLS (Re) & 1,054,438 & 0.1905 & 19.3 & 0.9117 & 11.5\\
		Scale Space & 1,209,093 & 0.1798 & 39.1 & 0.8248 & 16.0\\
		Scale Space (Re) & 926,828 & 0.2028 & 15.2 &  0.9417 & 6.0\\
		Voronoi & 1,209,792 & 0.1799 & 52.3 & 0.8241 & 16.1 \\
		Voronoi (Re) & 923,476 & 0.2044 & 20.8 & 0.9407 & 6.8\\
		Ours & 840,453 & 0.2131 & \textbf{11.2} & \textbf{0.9577} & \textbf{4.5} \\
		\rowcolor{grey1}
		Ours (Re) & 854,257 & 0.2098 & \textbf{10.4} & \textbf{0.9701} & \textbf{3.8}
	\end{tabular}
	\caption{Experimental results for the \emph{Bottle Shampoo} from~\cite{huang2022surface} (604,903 points).}
	\label{tab:BottleShampoo}
\end{table}

\newpage

\subsection{\emph{Bowl Chinese} \textcolor{sowaswieweiss}{y}}

\begin{figure}[h!]
	\centering
	\begin{minipage}{0.45\textwidth}
		\begin{subfigure}{\textwidth}
			\includegraphics[width=1.\textwidth]{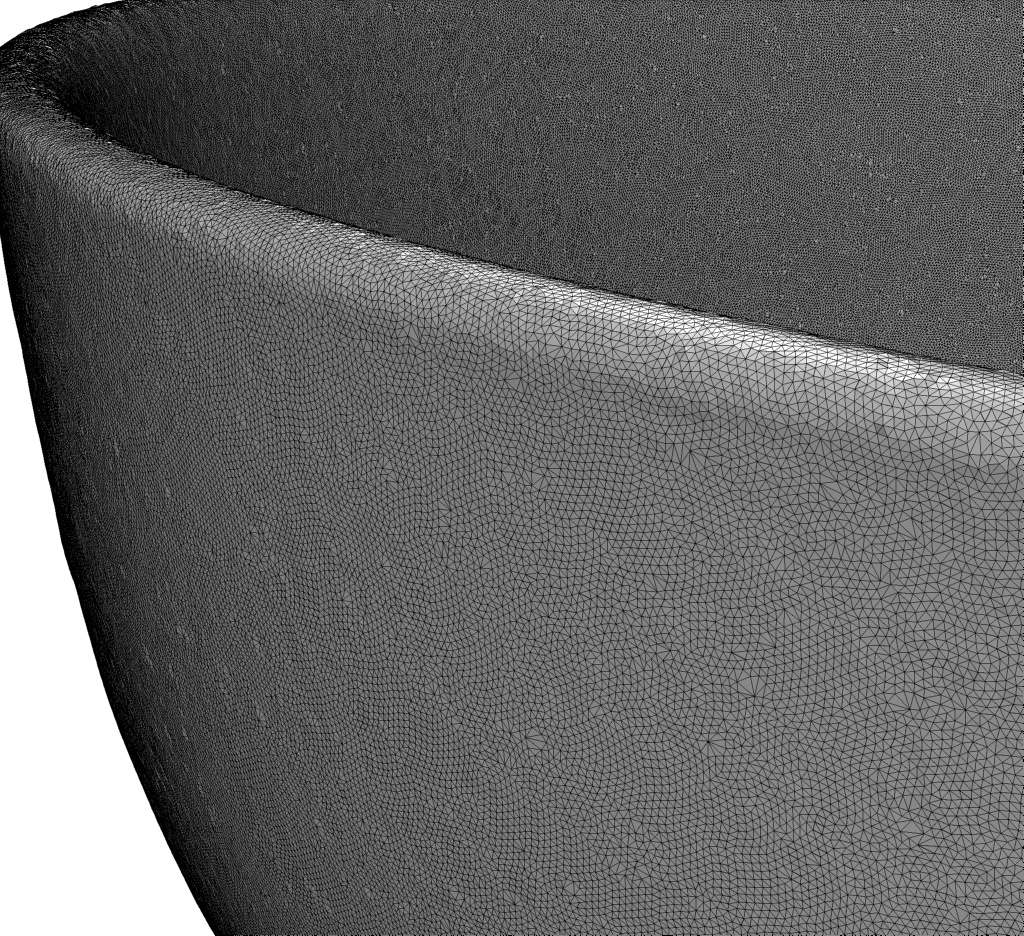}
		\end{subfigure}
	\end{minipage}
	\hfill
	\begin{minipage}{0.5\textwidth}
		\begin{subfigure}[t]{\Histogram}
			\centering
			\includegraphics[width=\textwidth]{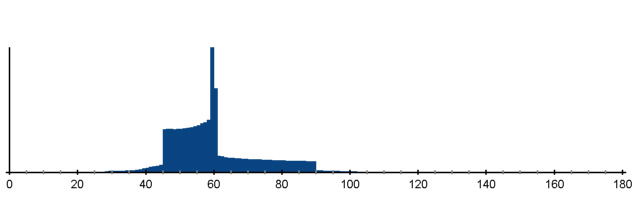}
			\caption{Angle distribution, target=$60^\circ$.}
		\end{subfigure}
		\begin{subfigure}[t]{\Histogram}
			\centering
			\includegraphics[width=\textwidth]{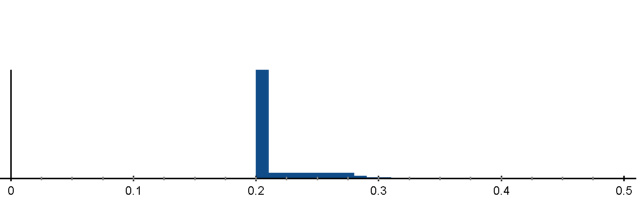}
			\caption{Edge lengths distribution, target=0.2.}
		\end{subfigure}
		\begin{subfigure}[t]{\Histogram}
			\centering
			\includegraphics[width=\textwidth]{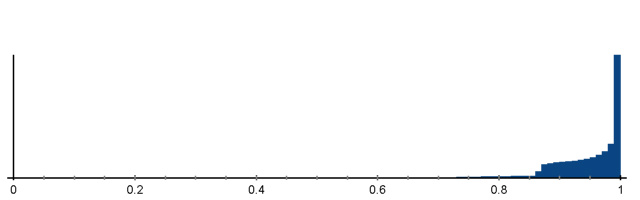}
			\caption{Distribution of quality $Q_t$, target=1.0.}
		\end{subfigure}
	\end{minipage}
	\caption{
		\emph{Bowl Chinese}.
	}
\end{figure}

\begin{table}[b!]
	\def\arraystretch{1.05}
	\centering
	\begin{tabular}{l|rrrrr}
		Algorithm & $|\mathcal{T}|$ & $E_{\text{avg}}$ & $E_{\text{RMS}}$ & $Q_{\text{avg}}$ & $Q_{\text{RMS}}$\\
		\hline
		Adv.~Front &  1,212,636 & 0.2920 & 38.2 & 0.8045 & 18.6 \\
		Adv.~Front (Re) & 2,407,002 & 0.2038 & 15.4 & 0.9405 & 6.2 \\
		Poisson &  13,584 & 2.3850 & 63.2 & 0.8845 & 11.7 \\
		Poisson (Re) & 637,488 & 0.3732 & 40.8 & 0.9301 & 6.9 \\
		Poisson MG &  503,458 & 0.4710 & 39.8 & 0.7062 & 37.1 \\
		Poisson MG (Re) & 2,409,076 & 0.2050 & 17.7 & 0.9223 & 7.9 \\
		RIMLS & 6,458,589 & 0.1331 & 40.4 & 0.6877 & 39.6 \\
		RIMLS (Re) & 2,441,143 & 0.2023 & 15.4 & 0.9394 & 6.3 \\
		Scale Space & 1,093,339 & 0.2779 & 34.9 & 0.8054 & 18.7\\
		Scale Space (Re) & 1,947,592 & \textbf{0.2006} & 16.1 & 0.9351 & 7.3 \\
		Voronoi & 1,212,636 & 0.2916 & 38.4 & 0.8042 & 18.7 \\
		Voronoi (Re) & 2,398,584 & 0.2039 & 15.3 & 0.9405 & \textbf{6.1} \\
		Ours & 2,137,650 & 0.2167 & \textbf{14.8} & \textbf{0.9485} & 6.2 \\
		\rowcolor{grey1}
		Ours (Re) & 2,246,434 & 0.2093 & \textbf{11.4} & \textbf{0.9665} & \textbf{4.3}
	\end{tabular}
	\caption{Experimental results for the \emph{Bowl Chinese} from~\cite{huang2022surface} (606,320 points).}
	\label{tab:BowlChinese}
\end{table}

\newpage

\subsection{\emph{Cloth Duck} \textcolor{sowaswieweiss}{y}}

\begin{figure}[h!]
	\centering
	\begin{minipage}{0.45\textwidth}
		\begin{subfigure}{\textwidth}
			\includegraphics[width=1.\textwidth]{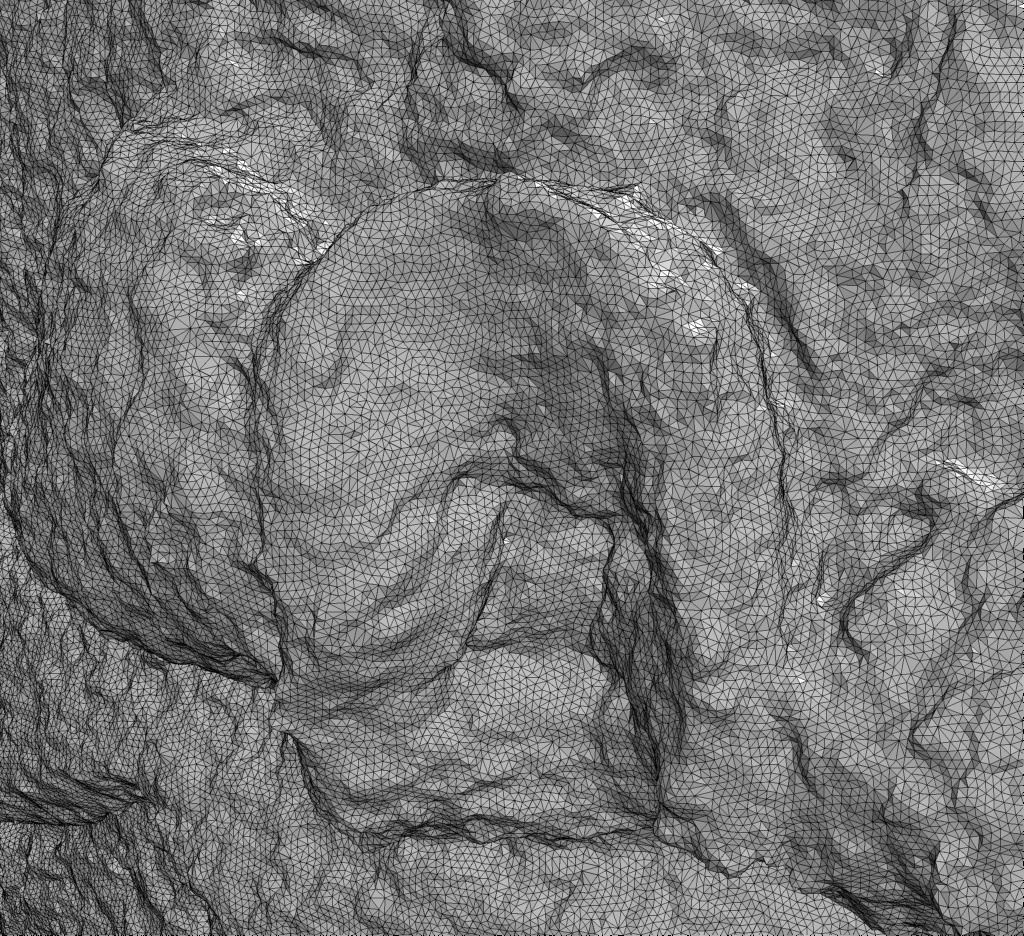}
		\end{subfigure}
	\end{minipage}
	\hfill
	\begin{minipage}{0.5\textwidth}
		\begin{subfigure}[t]{\Histogram}
			\centering
			\includegraphics[width=\textwidth]{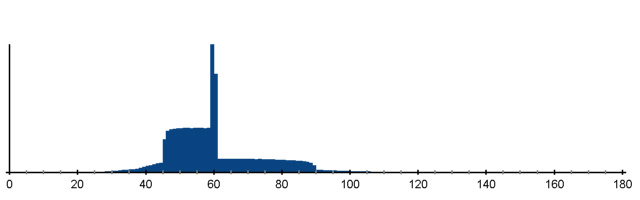}
			\caption{Angle distribution, target=$60^\circ$.}
		\end{subfigure}
		\begin{subfigure}[t]{\Histogram}
			\centering
			\includegraphics[width=\textwidth]{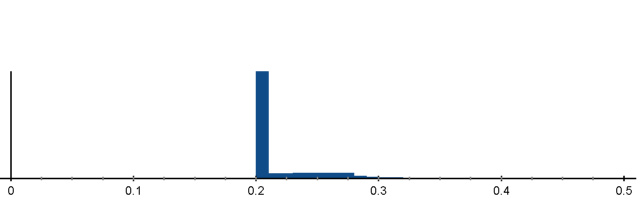}
			\caption{Edge lengths distribution, target=0.2.}
		\end{subfigure}
		\begin{subfigure}[t]{\Histogram}
			\centering
			\includegraphics[width=\textwidth]{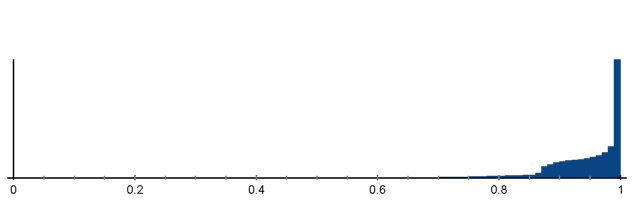}
			\caption{Distribution of quality $Q_t$, target=1.0.}
		\end{subfigure}
	\end{minipage}
	\caption{
		\emph{Cloth Duck}.
	}
\end{figure}

\begin{table}[b!]
	\def\arraystretch{1.05}
	\centering
	\begin{tabular}{l|rrrrr}
		Algorithm & $|\mathcal{T}|$ & $E_{\text{avg}}$ & $E_{\text{RMS}}$ & $Q_{\text{avg}}$ & $Q_{\text{RMS}}$\\
		\hline
		Adv.~Front &  2,037,574 & 0.1839 & 40.1 & 0.8143 & 17.3\\
		Adv.~Front (Re) & 1,739,214 & 0.1965 & 19.4 & 0.9179 & 9.5 \\
		Poisson &  147,940 & 0.6300 & 44.3 & 0.8805 & 12.0 \\
		Poisson (Re) & 1,488,112 & 0.2068 & 17.5 & 0.9311 & 7.2\\
		Poisson MG &  419,614 & 0.4086 & 38.5 & 0.7160 & 36.0 \\
		Poisson MG (Re) & 1,463,018 & 0.2093 & 18.7 & 0.9154 & 8.8 \\
		RIMLS & 5,878,521 & 0.1154 & 39.9 & 0.6919 & 38.9 \\
		RIMLS (Re) & 1,728,371 & \textbf{0.1978} & 20.1 & 0.9143 & 12.7\\
		Scale Space & 2,036,816 & 0.1839 & 40.0 & 0.8139 & 17.4 \\
		Scale Space (Re) & 1,735,814 & 0.1965 & 19.3 & 0.9179 & 9.5 \\
		Voronoi & 2,037,270 & 0.1767 & 41.8 & 0.8067 & 18.1 \\
		Voronoi (Re) & 1,514,160 & 0.2027 & \textbf{15.4} & 0.9407 & \textbf{6.3}\\
		Ours & 1,435,604 & 0.2181 & 15.7 & \textbf{0.9454} & 6.6 \\
		\rowcolor{grey1}
		Ours (Re) & 1,535,058 & 0.2089 & \textbf{12.5} & \textbf{0.9592} & \textbf{4.8}
	\end{tabular}
	\caption{Experimental results for the \emph{Cloth Duck} from~\cite{huang2022surface} (1,018,891 points).}
	\label{tab:ClothDuck}
\end{table}

\newpage

\subsection{\emph{Coffee Bottle Metal}}

\begin{figure}[h!]
	\centering
	\begin{minipage}{0.45\textwidth}
		\begin{subfigure}{\textwidth}
			\includegraphics[width=1.\textwidth]{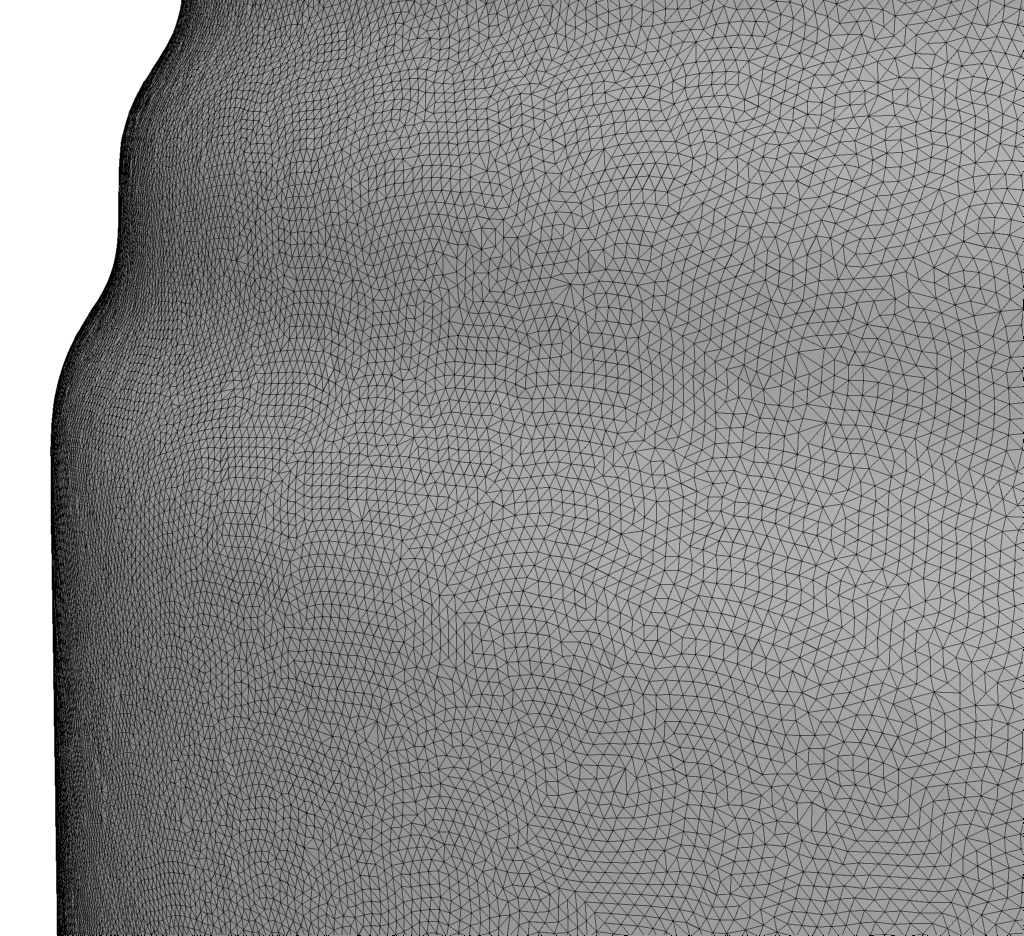}
		\end{subfigure}
	\end{minipage}
	\hfill
	\begin{minipage}{0.5\textwidth}
		\begin{subfigure}[t]{\Histogram}
			\centering
			\includegraphics[width=\textwidth]{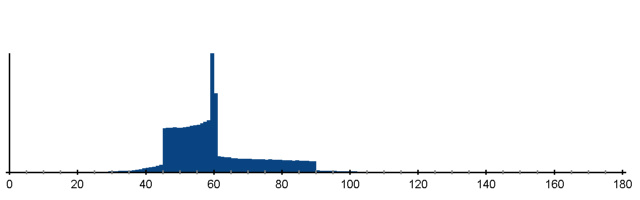}
			\caption{Angle distribution, target=$60^\circ$.}
		\end{subfigure}
		\begin{subfigure}[t]{\Histogram}
			\centering
			\includegraphics[width=\textwidth]{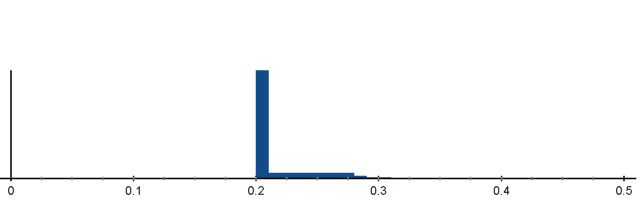}
			\caption{Edge lengths distribution, target=0.2.}
		\end{subfigure}
		\begin{subfigure}[t]{\Histogram}
			\centering
			\includegraphics[width=\textwidth]{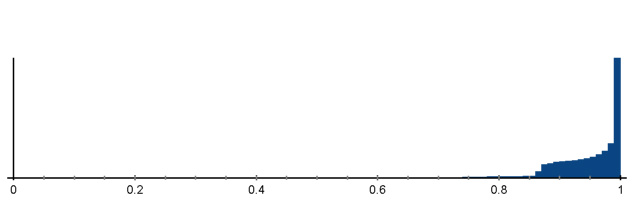}
			\caption{Distribution of quality $Q_t$, target=1.0.}
		\end{subfigure}
	\end{minipage}
	\caption{
		\emph{Coffee Bottle Metal}.
	}
\end{figure}

\begin{table}[b!]
	\def\arraystretch{1.05}
	\centering
	\begin{tabular}{l|rrrrr}
		Algorithm & $|\mathcal{T}|$ & $E_{\text{avg}}$ & $E_{\text{RMS}}$ & $Q_{\text{avg}}$ & $Q_{\text{RMS}}$\\
		\hline
		Adv.~Front &  1,211,290 & 0.2204 & 44.2 & 0.8036 & 18.4 \\
		Adv.~Front (Re) & 1,423,429 & 0.2029 & 15.2 & 0.9416 & 6.1 \\
		Poisson & 25,842 & 1.2170 & 84.0 & 0.8743 & 12.4  \\
		Poisson (Re) & 670,954 & 0.2805 & 41.6 & 0.9254 & 7.5 \\
		Poisson MG &  249,103 & 0.5440 & 46.4 & 0.7265 & 33.0 \\
		Poisson MG (Re) & 1,657,210 &  0.2097 & 20.8 & 0.9143 & 8.0 \\
		RIMLS & 3,024,537 & 0.1477 & 35.4 & 0.7107 & 34.4 \\
		RIMLS (Re) & 1,489,387 & \textbf{0.1986} & 18.0 & 0.9229 & 10.1\\
		Scale Space & 1,197,246 & 0.2188 & 43.4 & 0.8039 & 18.5 \\
		Scale Space (Re) & 1,379,802 & 0.2021 & 15.5 & 0.9397 & 6.5 \\
		Voronoi & 1,211,618 & 0.2198 & 52.6 & 0.8021 & 18.6 \\
		Voronoi (Re) & 1,412,630 & 0.2043 & 21.1 & 0.9411 & 6.6 \\
		Ours & 1,260,935 & 0.2162 & \textbf{14.0} & \textbf{0.9494} & \textbf{5.9} \\
		\rowcolor{grey1}
		Ours (Re) & 1,317,181 & 0.2095 & \textbf{11.4} & \textbf{0.9661} & \textbf{4.3}
	\end{tabular}
	\caption{Experimental results for the \emph{Coffee Bottle Metal} from~\cite{huang2022surface} (605,826 points).}
	\label{tab:CoffeeBottleMetal}
\end{table}

\newpage

\subsection{\emph{Coffee Bottle Plastic}}

\begin{figure}[h!]
	\centering
	\begin{minipage}{0.45\textwidth}
		\begin{subfigure}{\textwidth}
			\includegraphics[width=1.\textwidth]{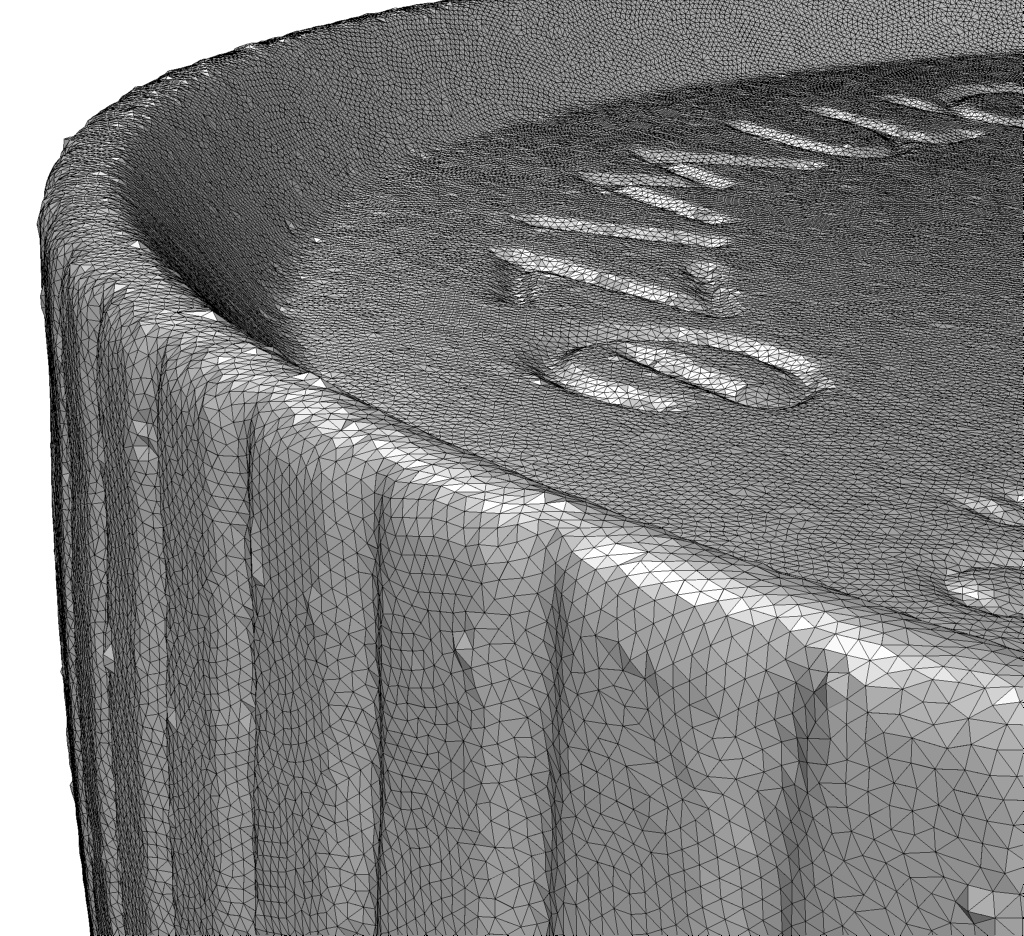}
		\end{subfigure}
	\end{minipage}
	\hfill
	\begin{minipage}{0.5\textwidth}
		\begin{subfigure}[t]{\Histogram}
			\centering
			\includegraphics[width=\textwidth]{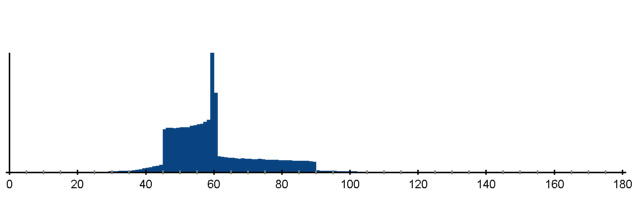}
			\caption{Angle distribution, target=$60^\circ$.}
		\end{subfigure}
		\begin{subfigure}[t]{\Histogram}
			\centering
			\includegraphics[width=\textwidth]{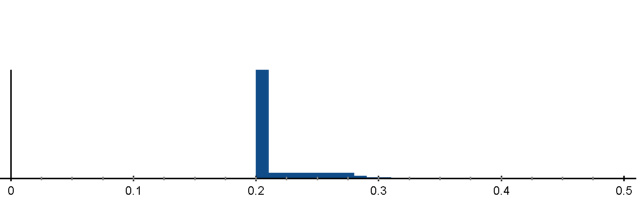}
			\caption{Edge lengths distribution, target=0.2.}
		\end{subfigure}
		\begin{subfigure}[t]{\Histogram}
			\centering
			\includegraphics[width=\textwidth]{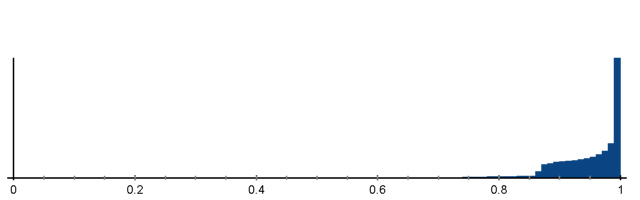}
			\caption{Distribution of quality $Q_t$, target=1.0.}
		\end{subfigure}
	\end{minipage}
	\caption{
		\emph{Coffee Bottle Plastic}.
	}
\end{figure}

\begin{table}[b!]
	\def\arraystretch{1.05}
	\centering
	\begin{tabular}{l|rrrrr}
		Algorithm & $|\mathcal{T}|$ & $E_{\text{avg}}$ & $E_{\text{RMS}}$ & $Q_{\text{avg}}$ & $Q_{\text{RMS}}$\\
		\hline
		Adv.~Front &  1,218,482 & 0.2128 & 45.7 & 0.8049 & 18.2\\
		Adv.~Front (Re) & 1,359,038 & 0.2027 & 15.2 & 0.9417 & 6.1 \\
		Poisson &  29,206 & 1.1586 & 76.5 & 0.8636 & 13.2 \\
		Poisson (Re) & 746,200 & 0.2557 & 44.2 & 0.9205 & 7.8 \\
		Poisson MG &  576,442 & 0.3273 & 35.4 & 0.7267 & 33.2 \\
		Poisson MG (Re) & 1,530,906 & 0.1931 & 16.1 & 0.9134 & \textbf{5.8} \\
		RIMLS & 7,438,789 & 0.0922 & 35.9 & 0.7121 & 34.8 \\
		RIMLS (Re) & 1,308,171 & 0.2070 & 15.5 & 0.9373 & 6.5 \\
		Scale Space & 1,204,794 & 0.2112 & 45.0 & 0.8050 & 18.2 \\
		Scale Space (Re) & 1,316,070 & \textbf{0.2019} & 15.5 & 0.9398 & 6.4 \\
		Voronoi & 1,218,482 & 0.2114 & 46.4 & 0.8028 & 18.4 \\
		Voronoi (Re) & 1,341,154 & 0.2032 & 15.1 & 0.9422 & 6.0 \\
		Ours & 1,203,156 & 0.2162 & \textbf{14.0} & \textbf{0.9495} & \textbf{5.8} \\
		\rowcolor{grey1}
		Ours (Re) & 1,257,141 & 0.2094 & \textbf{11.4} & \textbf{0.9659} & \textbf{4.3}
	\end{tabular}
	\caption{Experimental results for the \emph{Coffee Bottle Plastic} from~\cite{huang2022surface} (609,243 points).}
	\label{tab:CoffeeBottlePlastic}
\end{table}

\newpage

\subsection{\emph{Cup}}

\begin{figure}[h!]
	\centering
	\begin{minipage}{0.45\textwidth}
		\begin{subfigure}{\textwidth}
			\includegraphics[width=1.\textwidth]{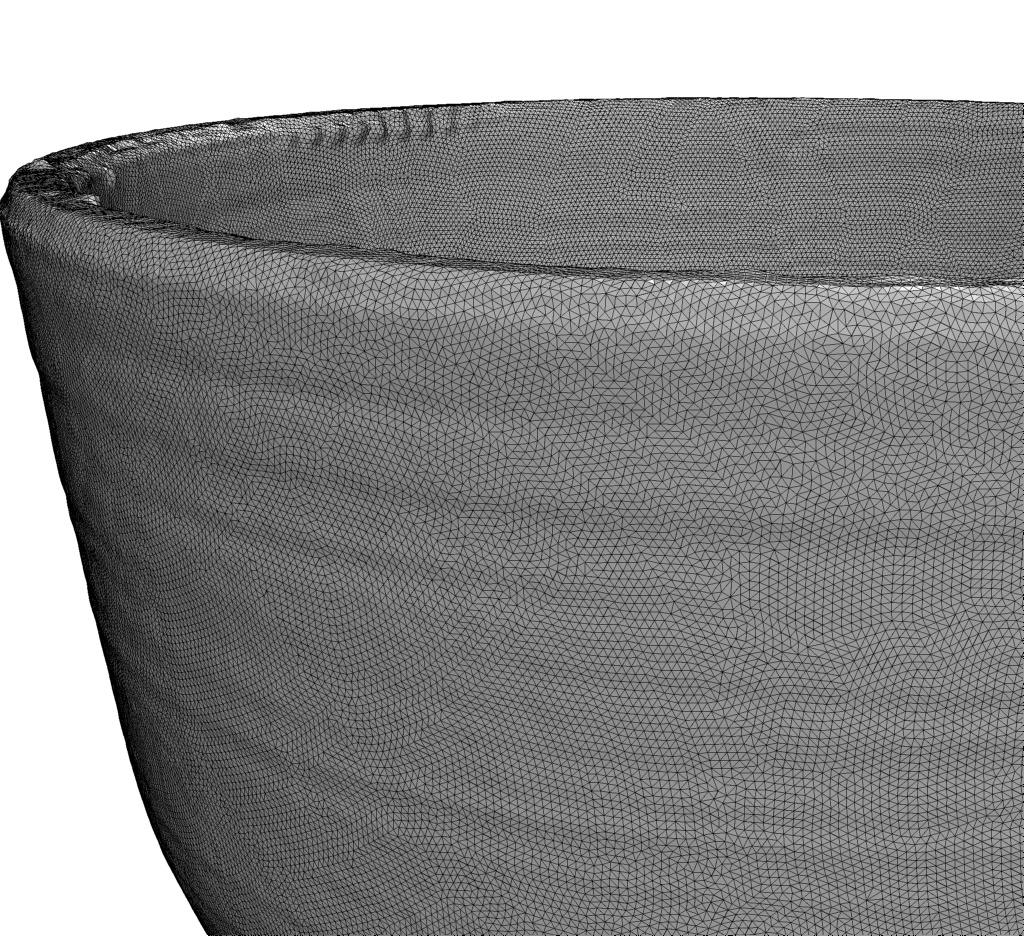}
		\end{subfigure}
	\end{minipage}
	\hfill
	\begin{minipage}{0.5\textwidth}
		\begin{subfigure}[t]{\Histogram}
			\centering
			\includegraphics[width=\textwidth]{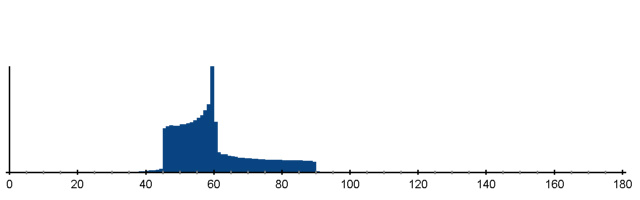}
			\caption{Angle distribution, target=$60^\circ$.}
		\end{subfigure}
		\begin{subfigure}[t]{\Histogram}
			\centering
			\includegraphics[width=\textwidth]{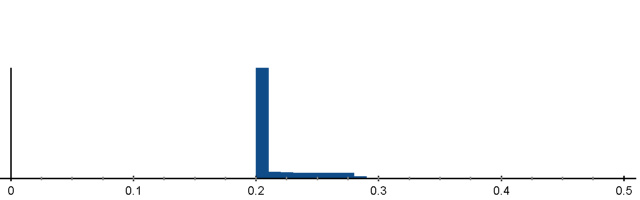}
			\caption{Edge lengths distribution, target=0.2.}
		\end{subfigure}
		\begin{subfigure}[t]{\Histogram}
			\centering
			\includegraphics[width=\textwidth]{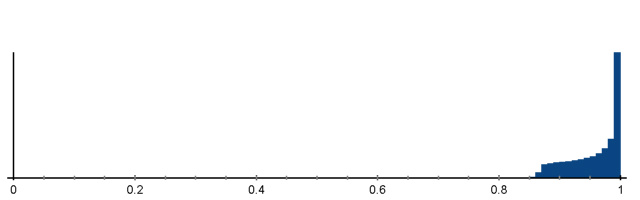}
			\caption{Distribution of quality $Q_t$, target=1.0.}
		\end{subfigure}
	\end{minipage}
	\caption{
		\emph{Cup}.
	}
\end{figure}

\begin{table}[b!]
	\def\arraystretch{1.05}
	\centering
	\begin{tabular}{l|rrrrr}
		Algorithm & $|\mathcal{T}|$ & $E_{\text{avg}}$ & $E_{\text{RMS}}$ & $Q_{\text{avg}}$ & $Q_{\text{RMS}}$\\
		\hline
		Adv.~Front & 2,820,618 & 0.1038 & 21.5 & 0.8978 & 6.1 \\
		Adv.~Front (Re) & 711,716 & 0.2032 & 15.0 & 0.9430 & 5.9 \\
		Poisson &  31,450 & 0.8540 & 63.3 & 0.8744 &12.5 \\
		Poisson (Re) & 660,134 & 0.2110 & 19.5 & 0.9297 & 7.2 \\
		Poisson MG & 605,996 & 0.2330 & 39.6 & 0.7050 & 37.2 \\
		Poisson MG (Re) & 692,170 & 0.2064 & 17.1 & 0.9323 & 7.0 \\
		RIMLS & 7,815,640 & 0.0658 & 40.2 & 0.6880 & 39.4 \\
		RIMLS (Re) & 713,396 & 0.2040 & 16.5 & 0.9295 & 7.1 \\
		Scale Space & 2,820,618 & 0.1038 & 21.5 &0.8978 & 6.1 \\
		Scale Space (Re) & 712,010 & \textbf{0.2031} & 15.0 &0.9430 & 5.9 \\
		Voronoi & 2,820,618 & 0.1035 & 21.6 & 0.8976 & 6.2 \\
		Voronoi (Re) & 664,994 & 0.2097 & 14.8 & 0.9420 & 6.1 \\
		Ours & 645,670 & 0.2133 & \textbf{11.3} & \textbf{0.9570} & \textbf{4.5} \\
		\rowcolor{grey1}
		Ours (Re) & 657,232 & 0.2099 & \textbf{10.6} & \textbf{0.9694} & \textbf{3.9}
	\end{tabular}
	\caption{Experimental results for the \emph{Cup} from~\cite{huang2022surface} (1,410,309 points).}
	\label{tab:CupModel}
\end{table}

\newpage

\subsection{\emph{Flower Pot} \textcolor{sowaswieweiss}{y}}

\begin{figure}[h!]
	\centering
	\begin{minipage}{0.45\textwidth}
		\begin{subfigure}{\textwidth}
			\includegraphics[width=1.\textwidth]{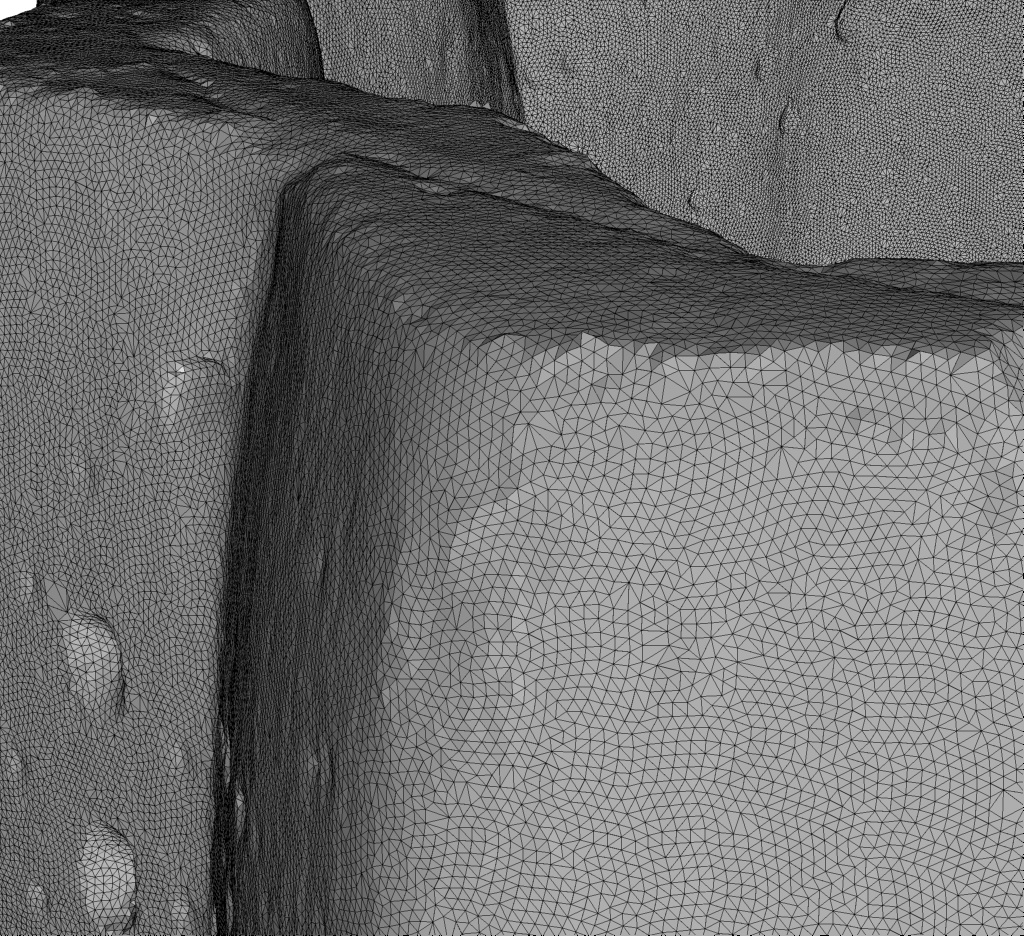}
		\end{subfigure}
	\end{minipage}
	\hfill
	\begin{minipage}{0.5\textwidth}
		\begin{subfigure}[t]{\Histogram}
			\centering
			\includegraphics[width=\textwidth]{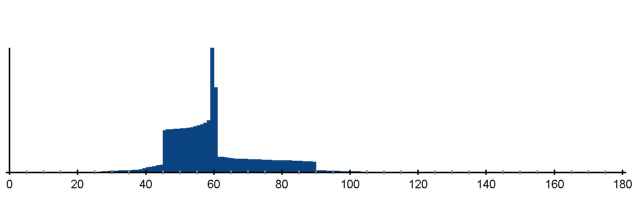}
			\caption{Angle distribution, target=$60^\circ$.}
		\end{subfigure}
		\begin{subfigure}[t]{\Histogram}
			\centering
			\includegraphics[width=\textwidth]{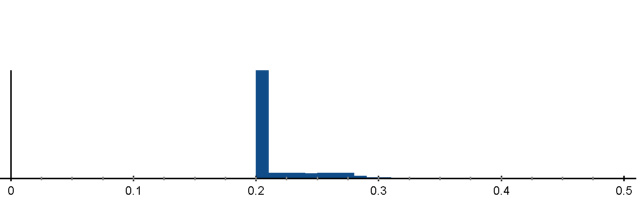}
			\caption{Edge lengths distribution, target=0.2.}
		\end{subfigure}
		\begin{subfigure}[t]{\Histogram}
			\centering
			\includegraphics[width=\textwidth]{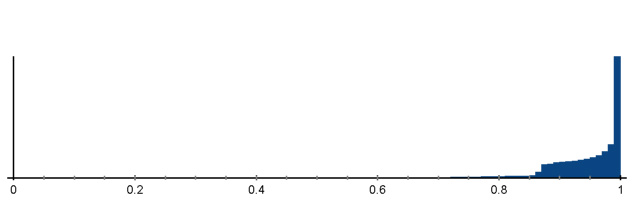}
			\caption{Distribution of quality $Q_t$, target=1.0.}
		\end{subfigure}
	\end{minipage}
	\caption{
		\emph{Flower Pot}.
	}
\end{figure}

\begin{table}[b!]
	\def\arraystretch{1.05}
	\centering
	\begin{tabular}{l|rrrrr}
		Algorithm & $|\mathcal{T}|$ & $E_{\text{avg}}$ & $E_{\text{RMS}}$ & $Q_{\text{avg}}$ & $Q_{\text{RMS}}$\\
		\hline
		Adv.~Front &  1,226,042 & 0.2780 & 46.1 & 0.8010 &19.0 \\
		Adv.~Front (Re) & 2,304,698 & 0.2043 & 15.7 & 0.9376 & 6.5 \\
		Poisson &  30,656 & 1.5601 & 67.1 & 0.8589 & 13.6 \\
		Poisson (Re) & 1,164,952 & 0.2722 & 41.4 & 0.9157 & 8.1 \\
		Poisson MG &  747,300 & 0.3767 & 36.6 & 0.7235 & 33.9 \\
		Poisson MG (Re) & 2,296,376 & 0.2052 & 16.7 & 0.9247 & 8.2 \\
		RIMLS & 9,625,452 & 0.1063 & 37.2 & 0.7050 & 36.1 \\
		RIMLS (Re) & 2,252,713 & 0.2064 & \textbf{15.1} & 0.9410 & \textbf{6.2} \\
		Scale Space & 1,057,169 & 0.2555 & 42.0 & 0.8017 & 19.1 \\
		Scale Space (Re) & 1,644,152 & \textbf{0.1997} & 16.8 & 0.9288 & 8.3 \\
		Voronoi & 1,226,042 & 0.2764 & 46.6 & 0.7996 & 19.2 \\
		Voronoi (Re) & 2,284,266 & 0.2045 & 15.7 & 0.9378 & 6.4 \\
		Ours & 2,052,950 & 0.2170 & \textbf{15.1} & \textbf{0.9476} & 6.4 \\
		\rowcolor{grey1}
		Ours (Re) & 2,161,017 & 0.2095 & \textbf{11.7} & \textbf{0.9642} & \textbf{4.6}
	\end{tabular}
	\caption{Experimental results for the \emph{Flower Pot} from~\cite{huang2022surface} (613,021 points).}
	\label{tab:FlowerPot}
\end{table}

\newpage

\subsection{\emph{Flower Pot 2} \textcolor{sowaswieweiss}{y}}

\begin{figure}[h!]
	\centering
	\begin{minipage}{0.45\textwidth}
		\begin{subfigure}{\textwidth}
			\includegraphics[width=1.\textwidth]{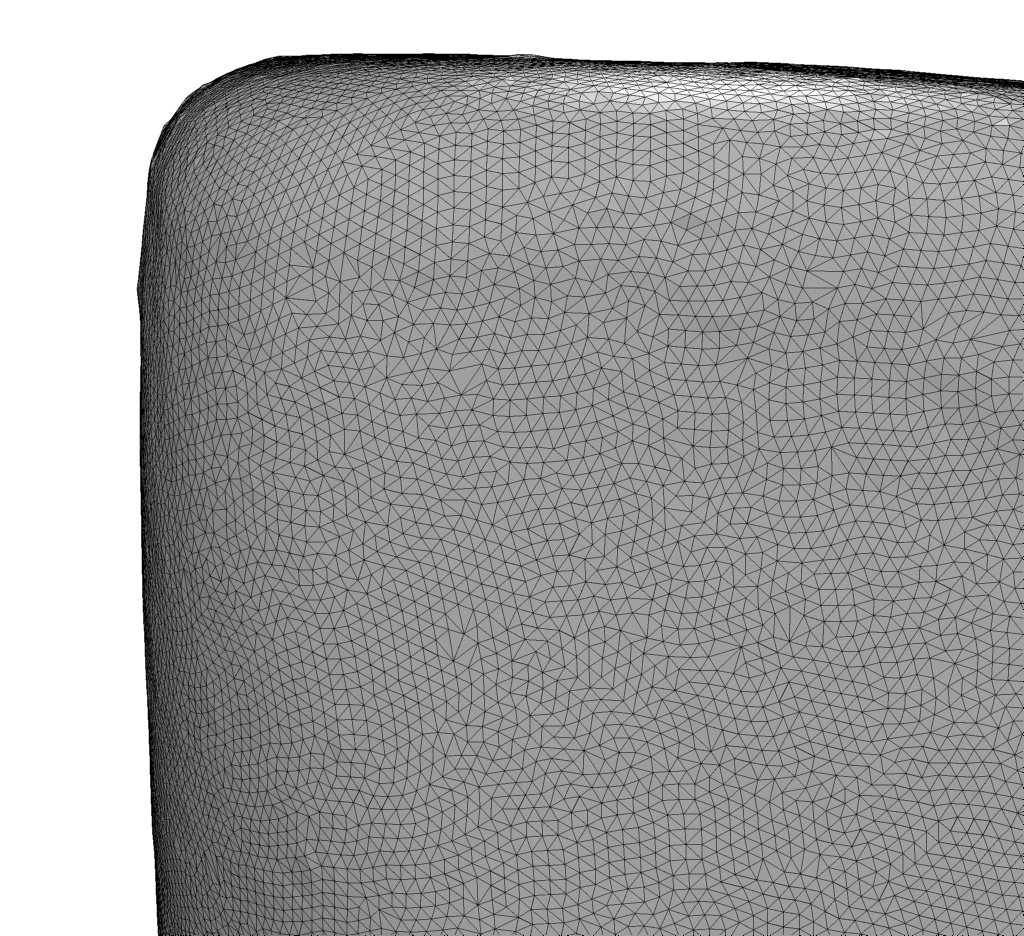}
		\end{subfigure}
	\end{minipage}
	\hfill
	\begin{minipage}{0.5\textwidth}
		\begin{subfigure}[t]{\Histogram}
			\centering
			\includegraphics[width=\textwidth]{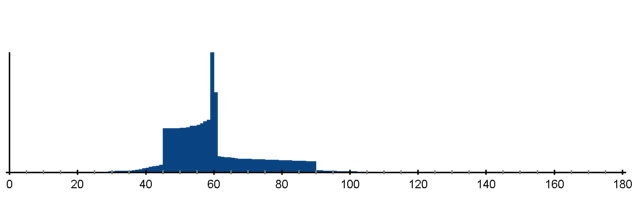}
			\caption{Angle distribution, target=$60^\circ$.}
		\end{subfigure}
		\begin{subfigure}[t]{\Histogram}
			\centering
			\includegraphics[width=\textwidth]{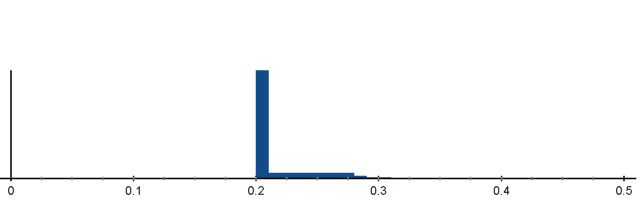}
			\caption{Edge lengths distribution, target=0.2.}
		\end{subfigure}
		\begin{subfigure}[t]{\Histogram}
			\centering
			\includegraphics[width=\textwidth]{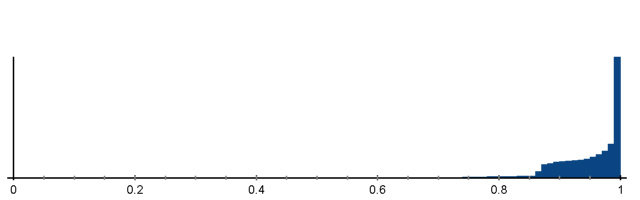}
			\caption{Distribution of quality $Q_t$, target=1.0.}
		\end{subfigure}
	\end{minipage}
	\caption{
		\emph{Flower Pot 2}.
	}
\end{figure}

\begin{table}[b!]
	\def\arraystretch{1.05}
	\centering
	\begin{tabular}{l|rrrrr}
		Algorithm & $|\mathcal{T}|$ & $E_{\text{avg}}$ & $E_{\text{RMS}}$ & $Q_{\text{avg}}$ & $Q_{\text{RMS}}$\\
		\hline
		Adv.~Front &  2,826,864 & 0.2135 & 38.1 & 0.8136 & 17.3 \\
		Adv.~Front (Re) & 3,001,322 & 0.2031 & 15.1 & 0.9425 & \textbf{5.9} \\
		Poisson &  26,436 & 1.7814 & 80.3 & 0.8668 & 13.0 \\
		Poisson (Re) & 1,062,612 & 0.3053 & 56.0 & 0.9198 & 7.7 \\
		Poisson MG &  640,912 & 0.4605 & 36.4 & 0.7400 & 32.7 \\
		Poisson MG (Re) & 3,016,388 & 0.2043 & 18.7 & 0.9180 & 8.4 \\
		RIMLS & 8,225,140 & 0.1302 & 37.0 & 0.7205 & 35.3 \\
		RIMLS (Re) & 2,939,770 & 0.2053 & 15.4 & 0.9392 & 6.7 \\
		Scale Space & 2,815,621 & 0.2130 & 37.8 & 0.8138 & 17.2 \\
		Scale Space (Re) & 2,969,237 & \textbf{0.2028} & 15.2 & 0.9418 & 6.1 \\
		Voronoi & 2,826,862 & 0.2133 & 38.2 & 0.8133 & 17.3 \\
		Voronoi (Re) & 2,985,334 & 0.2034 & 15.0 & 0.9428 & \textbf{5.9} \\
		Ours & 2,661,776 & 0.2162 & \textbf{14.0} & \textbf{0.9494} & \textbf{5.9} \\
		\rowcolor{grey1}
		Ours (Re) & 2,782,563 & 0.2094 & \textbf{11.4} & \textbf{0.9662} & \textbf{4.3}
	\end{tabular}
	\caption{Experimental results for the \emph{Flower Pot 2} from~\cite{huang2022surface} (1,413,342 points).}
	\label{tab:FlowerPot2}
\end{table}

\newpage

\subsection{\emph{Gift Box}}

\begin{figure}[h!]
	\centering
	\begin{minipage}{0.45\textwidth}
		\begin{subfigure}{\textwidth}
			\includegraphics[width=1.\textwidth]{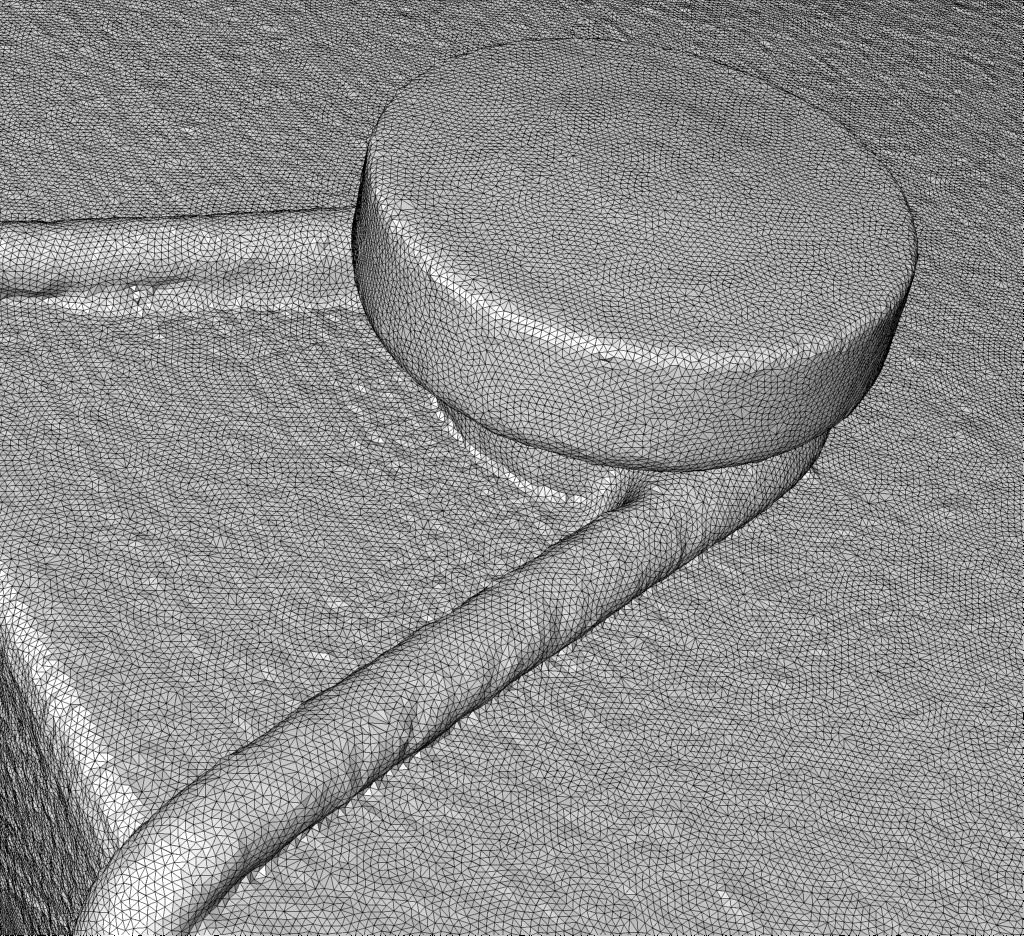}
		\end{subfigure}
	\end{minipage}
	\hfill
	\begin{minipage}{0.5\textwidth}
		\begin{subfigure}[t]{\Histogram}
			\centering
			\includegraphics[width=\textwidth]{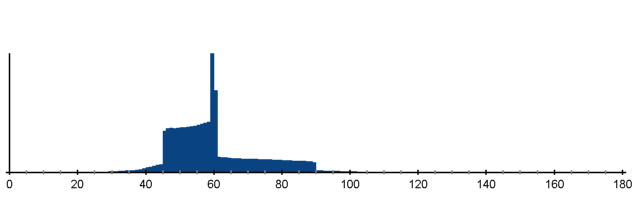}
			\caption{Angle distribution, target=$60^\circ$.}
		\end{subfigure}
		\begin{subfigure}[t]{\Histogram}
			\centering
			\includegraphics[width=\textwidth]{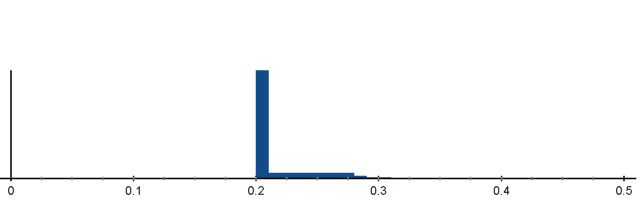}
			\caption{Edge lengths distribution, target=0.2.}
		\end{subfigure}
		\begin{subfigure}[t]{\Histogram}
			\centering
			\includegraphics[width=\textwidth]{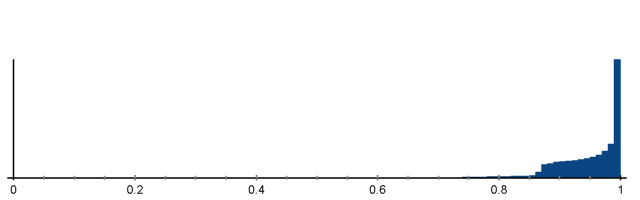}
			\caption{Distribution of quality $Q_t$, target=1.0.}
		\end{subfigure}
	\end{minipage}
	\caption{
		\emph{Gift Box}.
	}
\end{figure}

\begin{table}[b!]
	\def\arraystretch{1.05}
	\centering
	\begin{tabular}{l|rrrrr}
		Algorithm & $|\mathcal{T}|$ & $E_{\text{avg}}$ & $E_{\text{RMS}}$ & $Q_{\text{avg}}$ & $Q_{\text{RMS}}$\\
		\hline
		Adv.~Front &  3,063,655 &  0.1809 & 40.0 & 0.8313 & 15.7 \\
		Adv.~Front (Re) & 2,400,932 & 0.2022 & 15.5 & 0.9409 & 6.3 \\
		Poisson &  68,578 & 0.7763 & 130.2 & 0.8681 & 12.9 \\
		Poisson (Re) & 845,348 & 0.2990 & 62.8 & 0.9242 & 7.7 \\
		Poisson MG &  400,500 & 0.6026 & 61.4 & 0.7877 & 24.7 \\
		Poisson MG (Re) & 3,389,028 & 0.2185 & 31.1 & 0.9126 & 8.5 \\
		RIMLS & 4,649,233 & 0.1531 & 74.9 & 0.7641 & 28.0 \\
		RIMLS (Re) & 4,648,407 & 0.1522 & 75.2 & 0.7761 & 25.7 \\
		Scale Space & 3,059,551 & 0.1808 & 39.9 & 0.8313 & 15.7 \\
		Scale Space (Re) & 2,395,066 & \textbf{0.2020} & 15.6 & 0.9405 & 6.4 \\
		Voronoi & 3,063,361 & 0.1796 & 40.7 & 0.8297 & 15.9 \\
		Voronoi (Re) & 		2,360,337 & 0.2028 & 15.3 & 0.9415 & 6.3 \\
		Ours & 2,115,895 & 0.2164 & \textbf{14.2} & \textbf{0.9487} & \textbf{6.1} \\
		\rowcolor{grey1}
		Ours (Re) & 2,211,935 & 0.2096 & \textbf{11.8} & \textbf{0.9639} & \textbf{4.5}
	\end{tabular}
	\caption{Experimental results for the \emph{Gift Box} from~\cite{huang2022surface} (1,532,008 points).}
	\label{tab:}
\end{table}

\newpage

\subsection{\emph{Lock} \textcolor{sowaswieweiss}{y}}

\begin{figure}[h!]
	\centering
	\begin{minipage}{0.45\textwidth}
		\begin{subfigure}{\textwidth}
			\includegraphics[width=1.\textwidth]{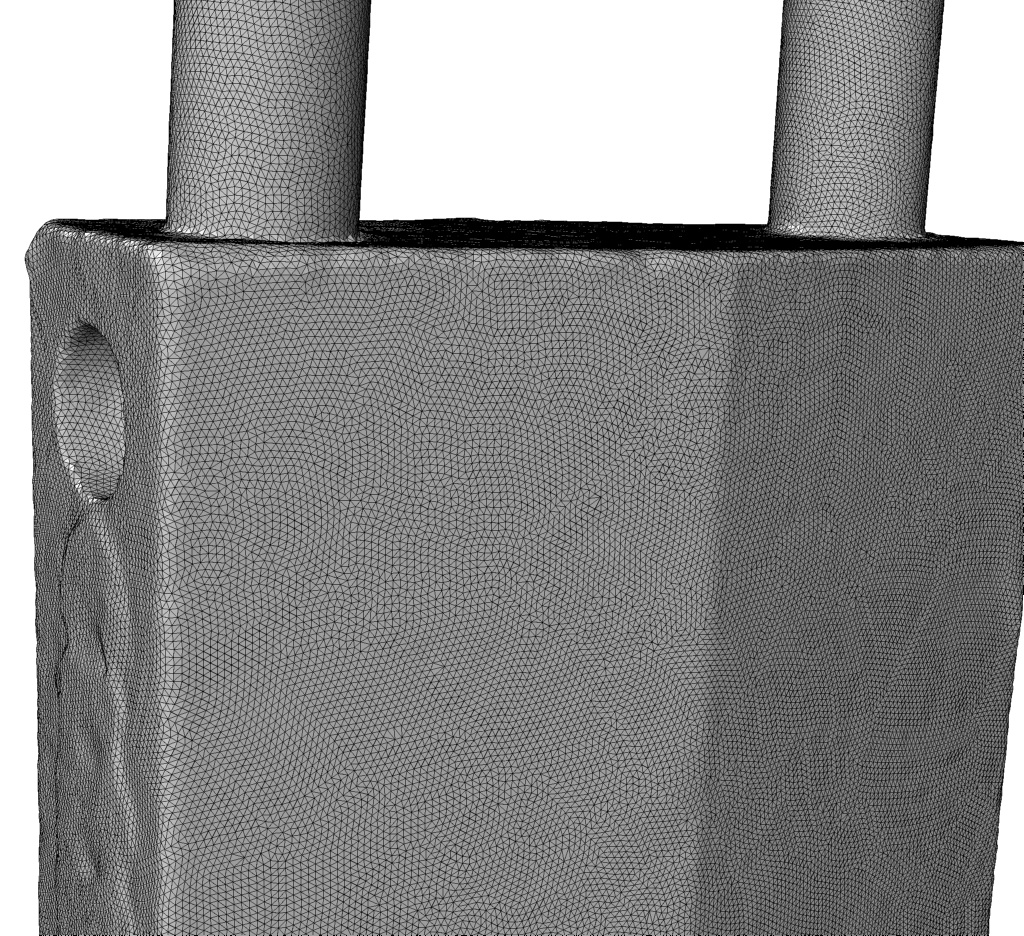}
		\end{subfigure}
	\end{minipage}
	\hfill
	\begin{minipage}{0.5\textwidth}
		\begin{subfigure}[t]{\Histogram}
			\centering
			\includegraphics[width=\textwidth]{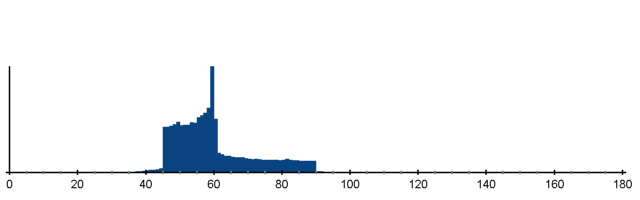}
			\caption{Angle distribution, target=$60^\circ$.}
		\end{subfigure}
		\begin{subfigure}[t]{\Histogram}
			\centering
			\includegraphics[width=\textwidth]{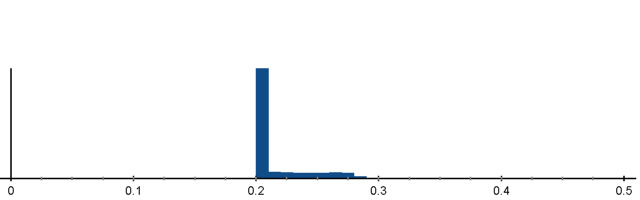}
			\caption{Edge lengths distribution, target=0.2.}
		\end{subfigure}
		\begin{subfigure}[t]{\Histogram}
			\centering
			\includegraphics[width=\textwidth]{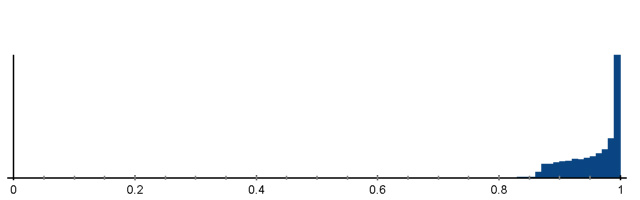}
			\caption{Distribution of quality $Q_t$, target=1.0.}
		\end{subfigure}
	\end{minipage}
	\caption{
		\emph{Lock}.
	}
\end{figure}

\begin{table}[b!]
	\def\arraystretch{1.05}
	\centering
	\begin{tabular}{l|rrrrr}
		Algorithm & $|\mathcal{T}|$ & $E_{\text{avg}}$ & $E_{\text{RMS}}$ & $Q_{\text{avg}}$ & $Q_{\text{RMS}}$\\
		\hline
		Adv.~Front &  668,388 & 0.1152 & 18.3 & 0.9227 & 6.6 \\
		Adv.~Front (Re) & 211,646 & \textbf{0.2035} & 15.0 & 0.9434 & 5.9 \\
		Poisson &  12,598 &  0.6966 & 79.2 & 0.8654 &13.1 \\
		Poisson (Re) & 153,344 & 0.2318 & 33.9 & 0.9241 & 7.7  \\
		Poisson MG & 201,226 & 0.2201 & 36.3 & 0.7118 & 34.4 \\
		Poisson MG (Re) & 204,718 & 0.2078 & 17.0 & 0.9319 & 6.9 \\
		RIMLS & 2,585,492 & 0.0620 & 36.7 & 0.6966 & 36.1 \\
		RIMLS (Re) & 212,924 & 0.2043 & 16.9 & 0.9251 & 7.4 \\
		Scale Space & 668,388 & 0.1152 & 18.3 & 0.9227 & 6.6 \\
		Scale Space (Re) & 211,572 & 0.2036 & 15.0 & 0.9434 & 5.8 \\
		Voronoi & 668,388 & 0.1147 & 18.5 & 0.9224 &6.7\\
		Voronoi (Re) & 201,296 & 0.2089 & 15.5 & 0.9313 & 7.4 \\
		Ours & 192,260 & 0.2136 & \textbf{11.6} & \textbf{0.9560} & \textbf{4.6} \\
		\rowcolor{grey1}
		Ours (Re) & 196,018 & 0.2101 & \textbf{11.0} & \textbf{0.9673} & \textbf{4.3}
	\end{tabular}
	\caption{Experimental results for the \emph{Lock} from~\cite{huang2022surface} (334,194 points).}
	\label{tab:Lock}
\end{table}

\newpage

\subsection{\emph{Mouse} \textcolor{sowaswieweiss}{y}}

\begin{figure}[h!]
	\centering
	\begin{minipage}{0.45\textwidth}
		\begin{subfigure}{\textwidth}
			\includegraphics[width=1.\textwidth]{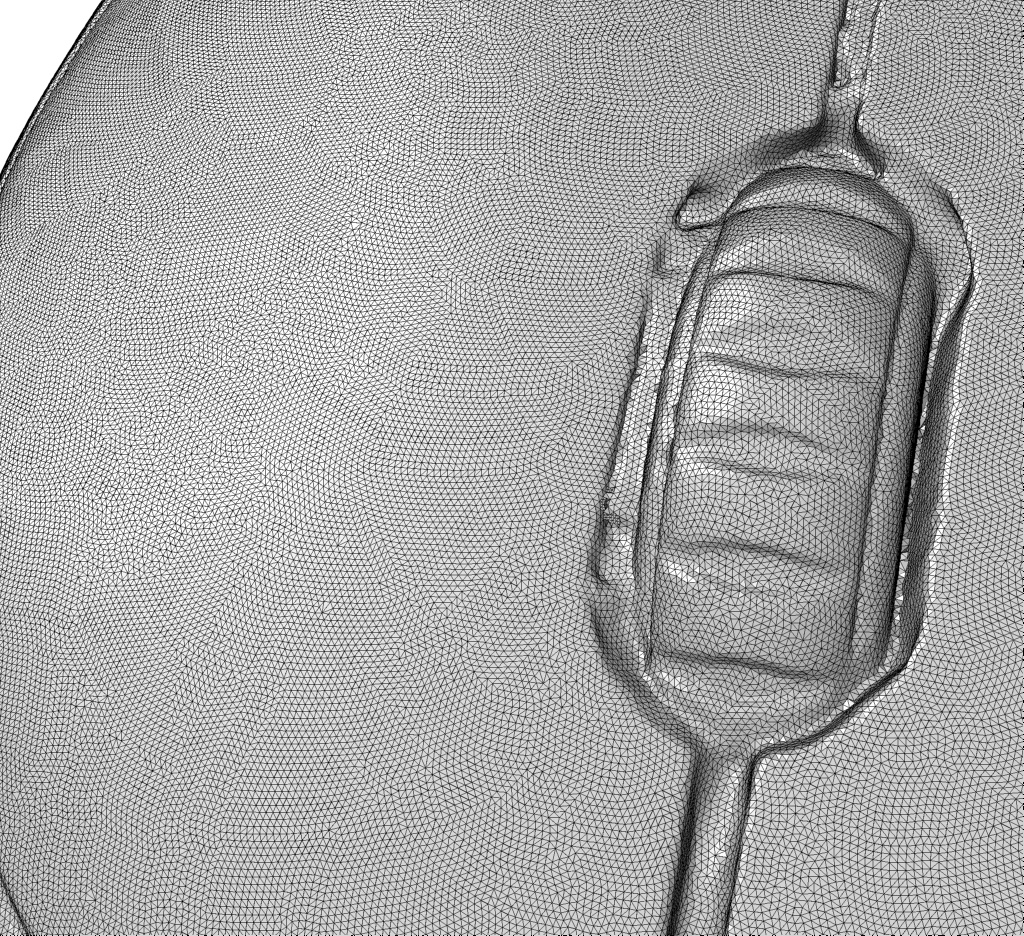}
		\end{subfigure}
	\end{minipage}
	\hfill
	\begin{minipage}{0.5\textwidth}
		\begin{subfigure}[t]{\Histogram}
			\centering
			\includegraphics[width=\textwidth]{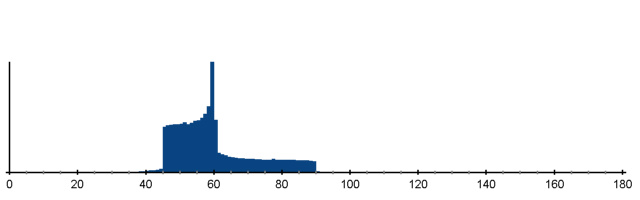}
			\caption{Angle distribution, target=$60^\circ$.}
		\end{subfigure}
		\begin{subfigure}[t]{\Histogram}
			\centering
			\includegraphics[width=\textwidth]{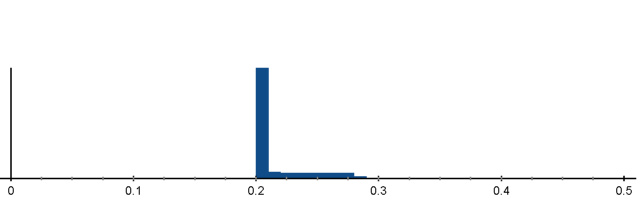}
			\caption{Edge lengths distribution, target=0.2.}
		\end{subfigure}
		\begin{subfigure}[t]{\Histogram}
			\centering
			\includegraphics[width=\textwidth]{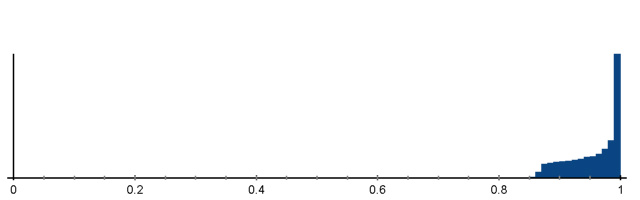}
			\caption{Distribution of quality $Q_t$, target=1.0.}
		\end{subfigure}
	\end{minipage}
	\caption{
		\emph{Mouse}.
	}
\end{figure}

\begin{table}[b!]
	\def\arraystretch{1.05}
	\centering
	\begin{tabular}{l|rrrrr}
		Algorithm & $|\mathcal{T}|$ & $E_{\text{avg}}$ & $E_{\text{RMS}}$ & $Q_{\text{avg}}$ & $Q_{\text{RMS}}$\\
		\hline
		Adv.~Front &  2,564,191 & 0.1265 & 26.6 & 0.8806 & 8.2 \\
		Adv.~Front (Re) & 947,782 & 0.2036 & 15.2 & 0.9421 & 6.2 \\
		Poisson &  33,934 & 0.8160 & 97.7 & 0.8711 & 12.9 \\
		Poisson (Re) & 585,622 & 0.2521 & 34.8 & 0.9298 & 7.3 \\
		Poisson MG &  265,630 & 0.4047 & 38.4 & 0.7125 & 35.7 \\
		Poisson MG (Re) & 894,136 & 0.2108 & 18.2 & 0.9190 & 8.5 \\
		RIMLS & 3,409,562 & 0.1145 & 39.1 & 0.6910 & 37.9 \\
		RIMLS (Re) & 979,361 & \textbf{0.1984} & 22.2 & 0.9083 & 17.9 \\
		Scale Space & 2,563,854 & 0.1265 & 26.6 & 0.8805 & 8.2 \\
		Scale Space (Re) & 948,202 & 0.2035 & 15.2 & 0.9420 & 6.3 \\
		Voronoi & 2,564,340 & 0.1262 & 26.9 & 0.8801 & 8.3 \\
		Voronoi (Re) & 881,846 & 0.2103 & 14.2 & 0.9464 & 6.2 \\
		Ours & 862,392 &  0.2134 & \textbf{11.5} & \textbf{0.9562} & \textbf{4.7} \\
		\rowcolor{grey1}
		Ours (Re) & 878,134 & 0.2100 & \textbf{10.9} & \textbf{0.9676} & \textbf{4.3}
	\end{tabular}
	\caption{Experimental results for the \emph{Mouse} from~\cite{huang2022surface} (1,282,234 points).}
	\label{tab:Mouse}
\end{table}

\newpage

\subsection{\emph{Mug}}

\begin{figure}[h!]
	\centering
	\begin{minipage}{0.45\textwidth}
		\begin{subfigure}{\textwidth}
			\includegraphics[width=1.\textwidth]{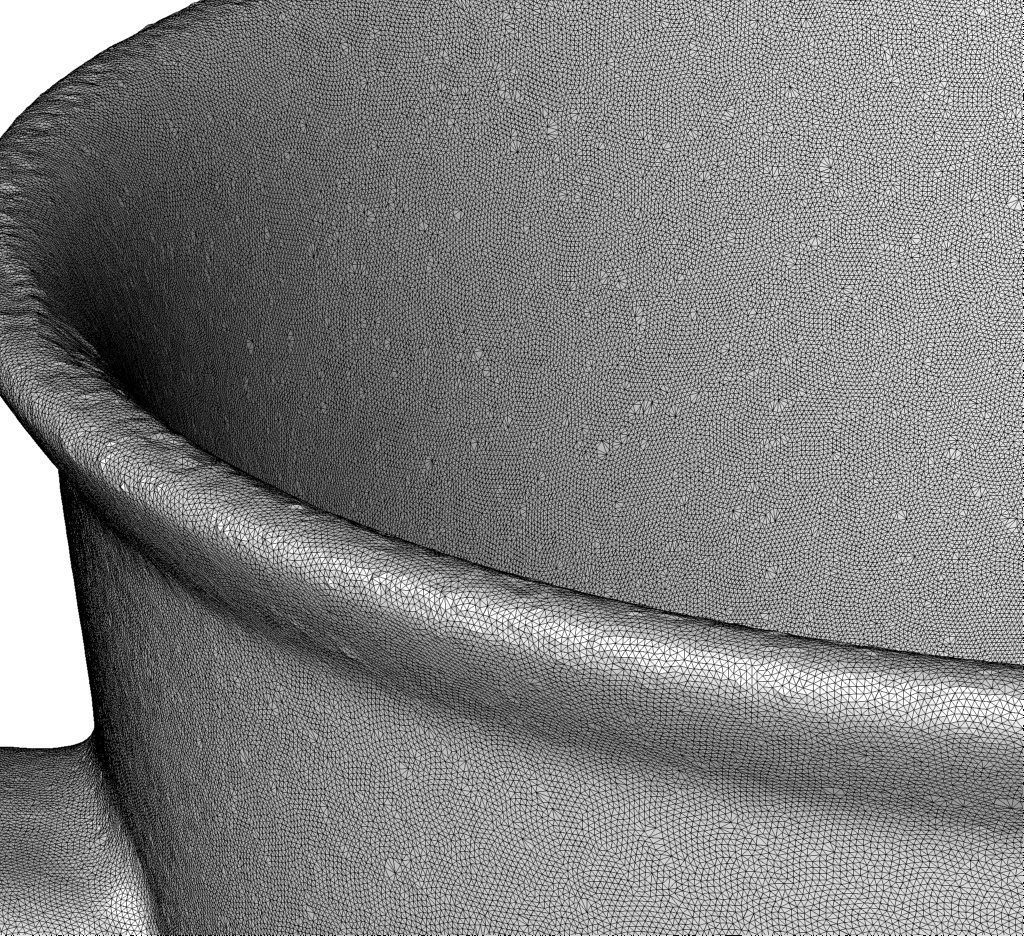}
		\end{subfigure}
	\end{minipage}
	\hfill
	\begin{minipage}{0.5\textwidth}
		\begin{subfigure}[t]{\Histogram}
			\centering
			\includegraphics[width=\textwidth]{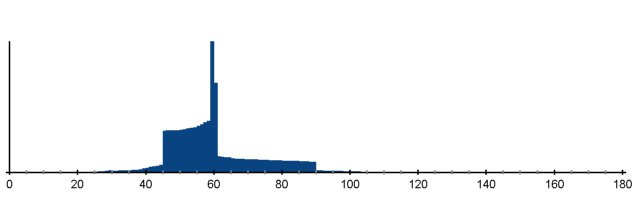}
			\caption{Angle distribution, target=$60^\circ$.}
		\end{subfigure}
		\begin{subfigure}[t]{\Histogram}
			\centering
			\includegraphics[width=\textwidth]{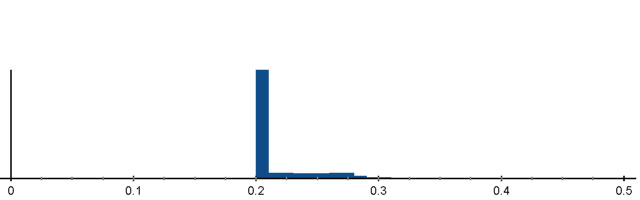}
			\caption{Edge lengths distribution, target=0.2.}
		\end{subfigure}
		\begin{subfigure}[t]{\Histogram}
			\centering
			\includegraphics[width=\textwidth]{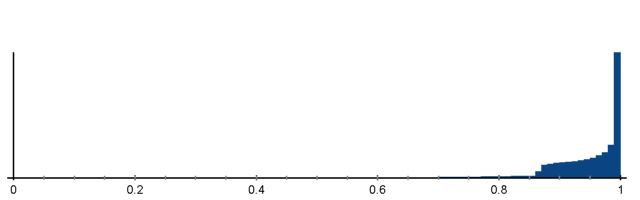}
			\caption{Distribution of quality $Q_t$, target=1.0.}
		\end{subfigure}
	\end{minipage}
	\caption{
		\emph{Mug}.
	}
\end{figure}

\begin{table}[b!]
	\def\arraystretch{1.05}
	\centering
	\begin{tabular}{l|rrrrr}
		Algorithm & $|\mathcal{T}|$ & $E_{\text{avg}}$ & $E_{\text{RMS}}$ & $Q_{\text{avg}}$ & $Q_{\text{RMS}}$\\
		\hline
		Adv.~Front &  1,212,626 & 0.3419 & 43.8 & 0.7999 & 19.1 \\
		Adv.~Front (Re) & 3,387,245 & 0.2049 & \textbf{15.9} & 0.9362 & 6.5 \\
		Poisson &  16,520 & 2.5528 & 65.7 & 0.8822 & 11.8 \\
		Poisson (Re) & 811,746 & 0.3845 & 47.9 & 0.9259 & 7.3 \\
		Poisson MG &  550,330 & 0.5340 & 35.8 & 0.7233 & 33.3 \\
		Poisson MG (Re) & 3,500,438 & 0.2016 & 16.1 & 0.9340 & 6.8 \\
		RIMLS & 7,059,114 & 0.1507 & 36.3 & 0.7062 & 35.3 \\
		RIMLS (Re) & 3,901,787 & 0.1919 & 16.8 & 0.9230 & \textbf{6.1} \\
		Scale Space & 840,041 & 0.2861 & 36.8 & 0.8014 & 19.4 \\
		Scale Space (Re) & 1,606,753 & \textbf{0.1985} & 17.3 & 0.9240 & 8.9 \\
		Voronoi & 1,212,628 & 0.3413 & 44.0 & 0.7995 & 19.2 \\
		Voronoi (Re) & 3,376,934 & 0.2050 & \textbf{15.9} & 0.9362 & 6.5 \\
		Ours & 3,011,551 & 0.2175 & 16.2 & \textbf{0.9470} & 6.7 \\
		\rowcolor{grey1}
		Ours (Re) & 3,190,711 & 0.2093 & \textbf{11.7} & \textbf{0.9646} & \textbf{4.6}
	\end{tabular}
	\caption{Experimental results for the \emph{Mug} from~\cite{huang2022surface} (606,322 points).}
	\label{tab:Mug}
\end{table}

\newpage

\subsection{\emph{Rabbit} \textcolor{sowaswieweiss}{y}}

\begin{figure}[h!]
	\centering
	\begin{minipage}{0.45\textwidth}
		\begin{subfigure}{\textwidth}
			\includegraphics[width=1.\textwidth]{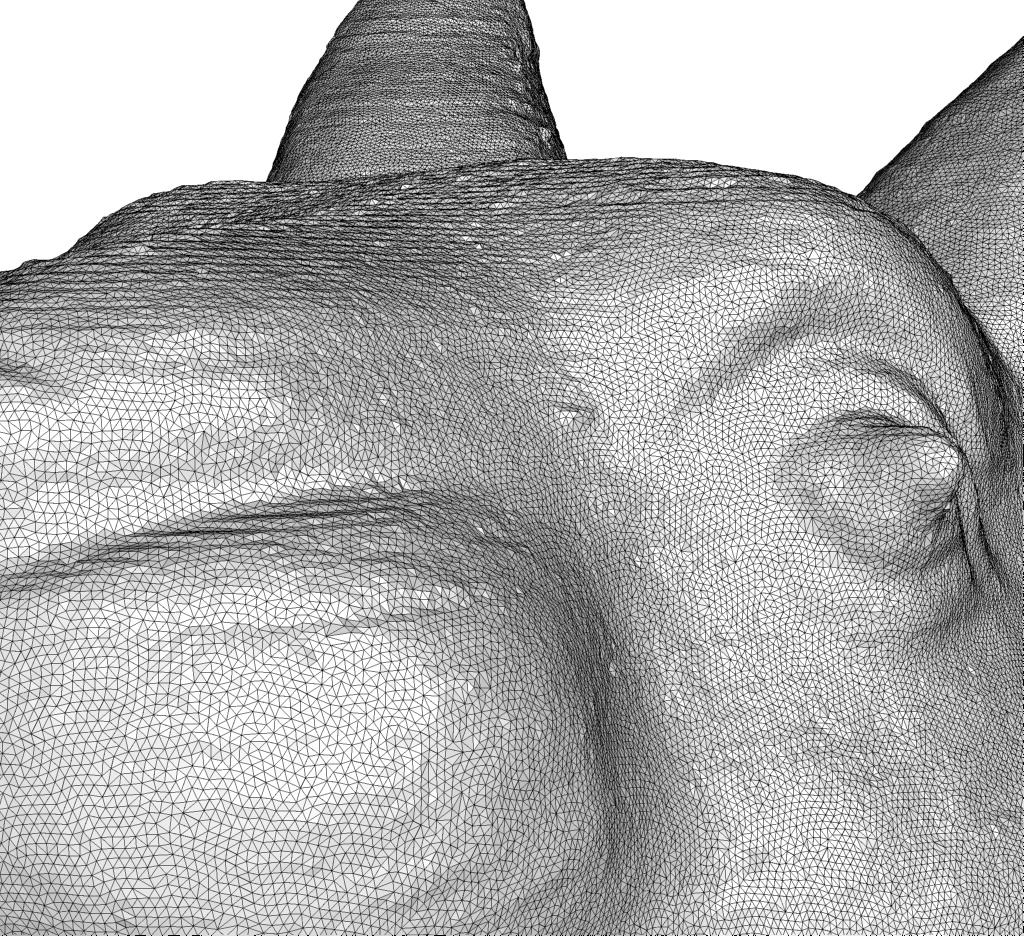}
		\end{subfigure}
	\end{minipage}
	\hfill
	\begin{minipage}{0.5\textwidth}
		\begin{subfigure}[t]{\Histogram}
			\centering
			\includegraphics[width=\textwidth]{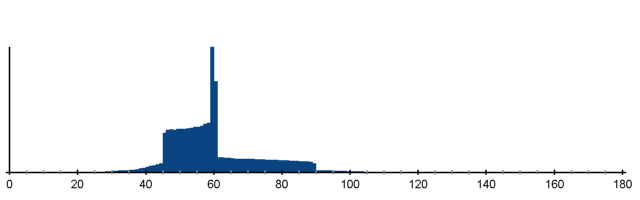}
			\caption{Angle distribution, target=$60^\circ$.}
		\end{subfigure}
		\begin{subfigure}[t]{\Histogram}
			\centering
			\includegraphics[width=\textwidth]{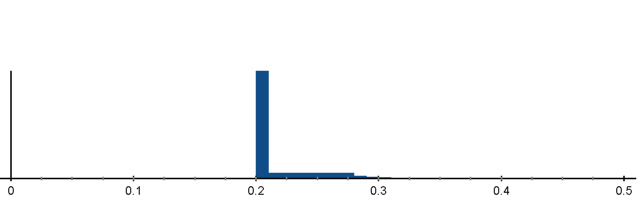}
			\caption{Edge lengths distribution, target=0.2.}
		\end{subfigure}
		\begin{subfigure}[t]{\Histogram}
			\centering
			\includegraphics[width=\textwidth]{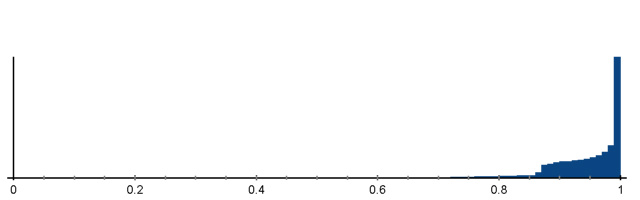}
			\caption{Distribution of quality $Q_t$, target=1.0.}
		\end{subfigure}
	\end{minipage}
	\caption{
		\emph{Rabbit}.
	}
\end{figure}

\begin{table}[b!]
	\def\arraystretch{1.05}
	\centering
	\begin{tabular}{l|rrrrr}
		Algorithm & $|\mathcal{T}|$ & $E_{\text{avg}}$ & $E_{\text{RMS}}$ & $Q_{\text{avg}}$ & $Q_{\text{RMS}}$\\
		\hline
		Adv.~Front &  4,046,178 & 0.1563 & 45.4 & 0.7782 & 19.9 \\
		Adv.~Front (Re) & 2,370,715 & 0.2006 & 16.5 & 0.9337 & 7.2 \\
		Poisson &  79,654 & 0.9804 & 60.4 & 0.8767 & 12.5 \\
		Poisson (Re) & 1,851,254 & 0.2215 & 29.1 & 0.9264 & 7.5 \\
		Poisson MG &  293,676 & 0.5974 & 38.4 & 0.7154 & 35.8 \\
		Poisson MG (Re) & 2,559,026 & 0.1931 & 16.6 & 0.9216 & \textbf{6.1} \\
		RIMLS & 3,573,629 & 0.1692 & 38.8 & 0.6973 & 38.1 \\
		RIMLS (Re) & 2,631,908 & 0.1868 & 27.0 & 0.8500 & 22.9 \\
		Scale Space & 4,041,938 & 0.1561 & 45.1 & 0.7782 & 19.9 \\
		Scale Space (Re) & 2,357,012 & \textbf{0.2005} & 16.6 & 0.9332 & 7.3 \\
		Voronoi & 4,046,157 & 0.1552 & 45.6 & 0.7767 & 20.0 \\
		Voronoi (Re) & 2,323,985 & 0.2014 & 15.9 & 0.9374 & 6.5 \\
		Ours & 2,039,974 & 0.2173 & \textbf{15.1} & \textbf{0.9470} & 6.4 \\
		\rowcolor{grey1}
		Ours (Re) & 2,152,626 & 0.2093 & \textbf{11.8} & \textbf{0.9636} & \textbf{4.5}
	\end{tabular}
	\caption{Experimental results for the \emph{Rabbit} from~\cite{huang2022surface} (2,023,131 points).}
	\label{tab:Rabbit}
\end{table}

\newpage

\subsection{\emph{Remote} \textcolor{sowaswieweiss}{y}}

\begin{figure}[h!]
	\centering
	\begin{minipage}{0.45\textwidth}
		\begin{subfigure}{\textwidth}
			\includegraphics[width=1.\textwidth]{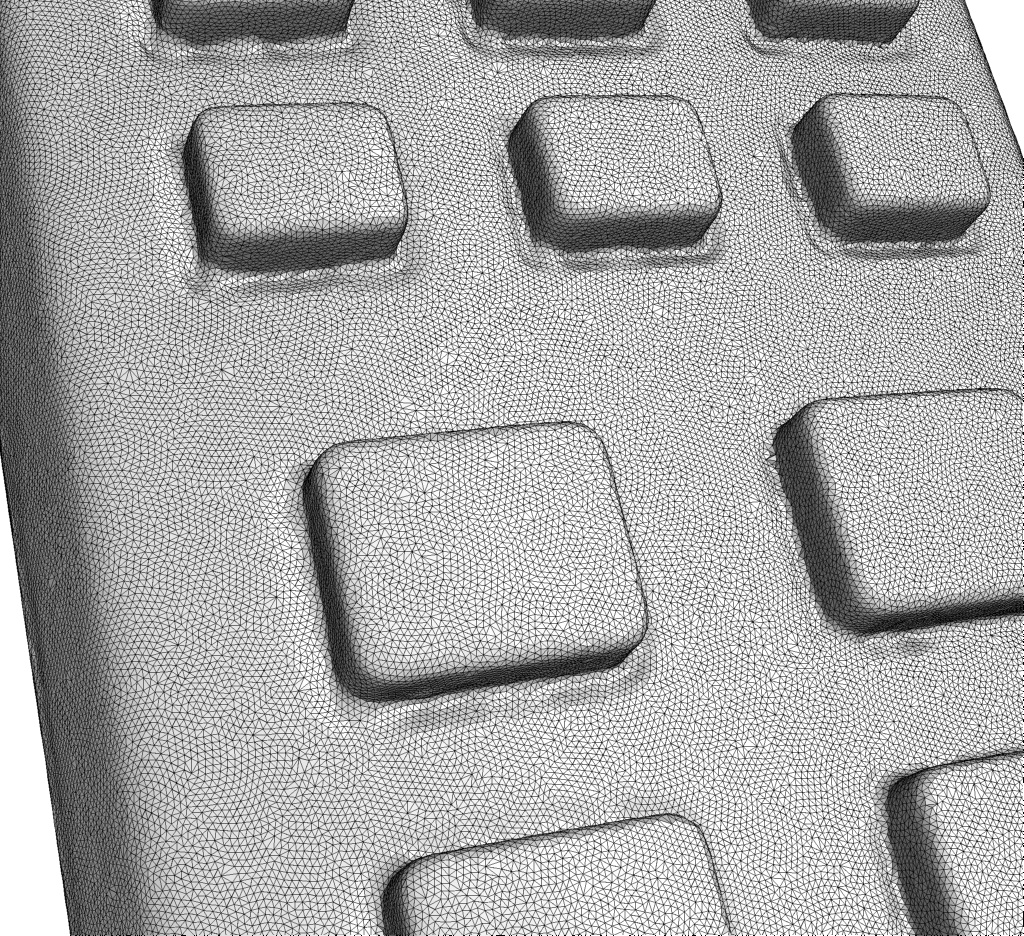}
		\end{subfigure}
	\end{minipage}
	\hfill
	\begin{minipage}{0.5\textwidth}
		\begin{subfigure}[t]{\Histogram}
			\centering
			\includegraphics[width=\textwidth]{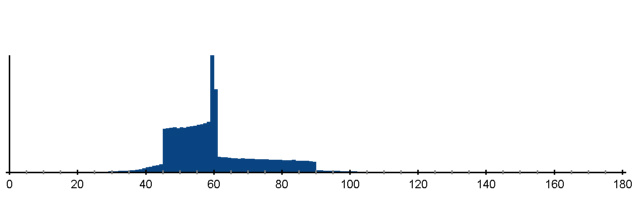}
			\caption{Angle distribution, target=$60^\circ$.}
		\end{subfigure}
		\begin{subfigure}[t]{\Histogram}
			\centering
			\includegraphics[width=\textwidth]{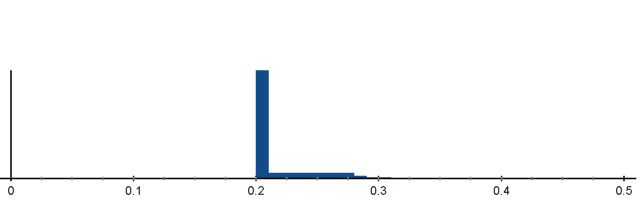}
			\caption{Edge lengths distribution, target=0.2.}
		\end{subfigure}
		\begin{subfigure}[t]{\Histogram}
			\centering
			\includegraphics[width=\textwidth]{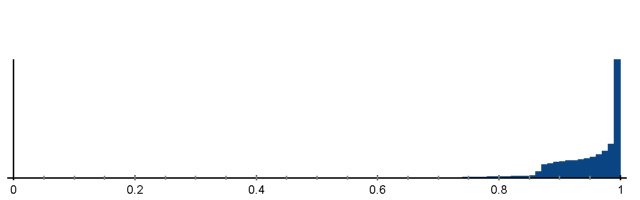}
			\caption{Distribution of quality $Q_t$, target=1.0.}
		\end{subfigure}
	\end{minipage}
	\caption{
		\emph{Romote}.
	}
\end{figure}

\begin{table}[b!]
	\def\arraystretch{1.05}
	\centering
	\begin{tabular}{l|rrrrr}
		Algorithm & $|\mathcal{T}|$ & $E_{\text{avg}}$ & $E_{\text{RMS}}$ & $Q_{\text{avg}}$ & $Q_{\text{RMS}}$\\
		\hline
		Adv.~Front & 1,211,452 & 0.1681 & 40.7 & 0.8298 & 15.9 \\
		Adv.~Front (Re) & 826,007 & 0.2023 & 15.5 & 0.9405 & 6.3 \\
		Poisson &  37,808 & 0.7858 & 78.8 & 0.8603 & 13.5 \\
		Poisson (Re) & 603,656 & 0.2306 & 31.2 & 0.9232 & 7.7 \\
		Poisson MG &  132,682 & 0.5261 & 37.1 & 0.7352 & 34.0 \\
		Poisson MG (Re) & 829,026 & \textbf{0.2014} & 16.5 & 0.9318 & 7.3 \\
		RIMLS & 1,691,586 & 0.1490 & 37.4 & 0.7162 & 36.6 \\
		RIMLS (Re) & 1,006,124 & 0.1834 & 22.0 & 0.8900 & 15.7 \\
		Scale Space & 1,211,373 & 0.1681 & 40.7 & 0.8298 & 15.9 \\
		Scale Space (Re) & 825,804 & 0.2023 & 15.5 & 0.9404 & 6.3 \\
		Voronoi & 1,211,428 & 0.1667 & 41.5 & 0.8275 & 16.1 \\
		Voronoi (Re) & 806,808 & 0.2034 & 15.3 & 0.9415 & 6.2 \\
		Ours & 726,550 & 0.2164 & \textbf{14.0} & \textbf{0.9488} & \textbf{5.9} \\
		\rowcolor{grey1}
		Ours (Re) & 760,465 & 0.2094 & \textbf{11.5} & \textbf{0.9654} & \textbf{4.4}
	\end{tabular}
	\caption{Experimental results for the \emph{Remote} from~\cite{huang2022surface} (608,874 points).}
	\label{tab:Remote}
\end{table}

\newpage

\subsection{\emph{Screw} \textcolor{sowaswieweiss}{y}}

\begin{figure}[h!]
	\centering
	\begin{minipage}{0.45\textwidth}
		\begin{subfigure}{\textwidth}
			\includegraphics[width=1.\textwidth]{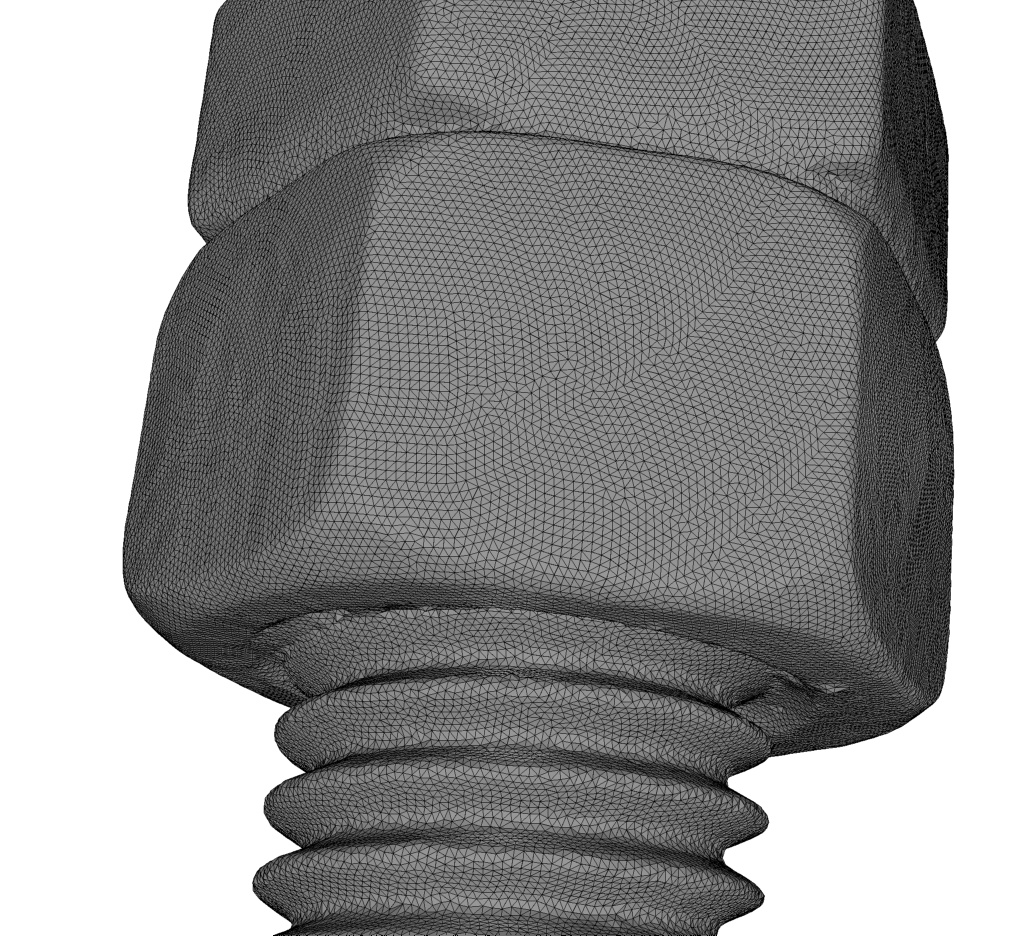}
		\end{subfigure}
	\end{minipage}
	\hfill
	\begin{minipage}{0.5\textwidth}
		\begin{subfigure}[t]{\Histogram}
			\centering
			\includegraphics[width=\textwidth]{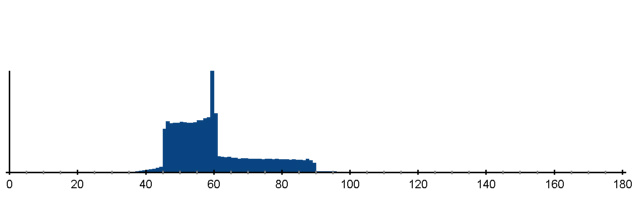}
			\caption{Angle distribution, target=$60^\circ$.}
		\end{subfigure}
		\begin{subfigure}[t]{\Histogram}
			\centering
			\includegraphics[width=\textwidth]{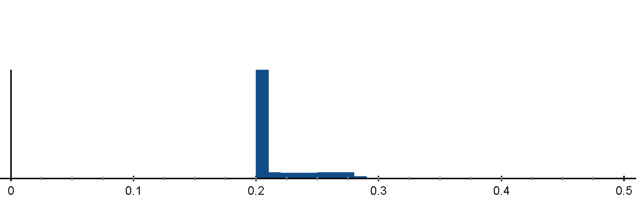}
			\caption{Edge lengths distribution, target=0.2.}
		\end{subfigure}
		\begin{subfigure}[t]{\Histogram}
			\centering
			\includegraphics[width=\textwidth]{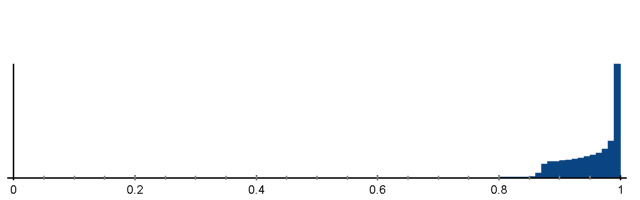}
			\caption{Distribution of quality $Q_t$, target=1.0.}
		\end{subfigure}
	\end{minipage}
	\caption{
		\emph{Screw}.
	}
\end{figure}

\begin{table}[b!]
	\def\arraystretch{1.05}
	\centering
	\begin{tabular}{l|rrrrr}
		Algorithm & $|\mathcal{T}|$ & $E_{\text{avg}}$ & $E_{\text{RMS}}$ & $Q_{\text{avg}}$ & $Q_{\text{RMS}}$\\
		\hline
		Adv.~Front &  990,307 & 0.1102 & 23.3 & 0.8999 & 8.9 \\
		Adv.~Front (Re) & 283,248 & 0.2028 & 15.7 & 0.9397 & 6.5 \\
		Poisson &  72,344 & 0.3693 & 56.8 & 0.8624 & 12.9 \\
		Poisson (Re) & 268,408 & 0.2069 & 18.9 & 0.9271 & 7.5 \\
		Poisson MG &  126,076 & 0.3162 & 39.7 & 0.7102 & 37.1 \\
		Poisson MG (Re) & 297,510 & 0.1962 & 17.0 & 0.9176 & 8.6 \\
		RIMLS & 1,687,736 & 0.0895 & 40.7 & 0.6873 & 40.0 \\
		RIMLS (Re) & 284,956 & \textbf{0.2023} & 19.1 & 0.9228 & 12.4 \\
		Scale Space & 990,296 & 0.1102 & 23.3 & 0.8999 & 8.9 \\
		Scale Space (Re) & 283,400 & 0.2027 & 15.7 & 0.9397 & 6.6 \\
		Voronoi & 990,314 & 0.1065 & 23.9 & 0.8970 & 9.3 \\
		Voronoi (Re) & 252,708 & 0.2079 & 15.4 & 0.9394 & 6.6 \\
		Ours & 254,936 & 0.2144 & \textbf{12.1} & \textbf{0.9531} & \textbf{5.0} \\
		\rowcolor{grey1}
		Ours (Re) & 261,268 & 0.2103 & \textbf{11.6} & \textbf{0.9633} & \textbf{4.4}
	\end{tabular}
	\caption{Experimental results for the \emph{Screw} from~\cite{huang2022surface} (495,182 points).}
	\label{tab:Screw}
\end{table}

\newpage

\subsection{\emph{Tape}}

\begin{figure}[h!]
	\centering
	\begin{minipage}{0.45\textwidth}
		\begin{subfigure}{\textwidth}
			\includegraphics[width=1.\textwidth]{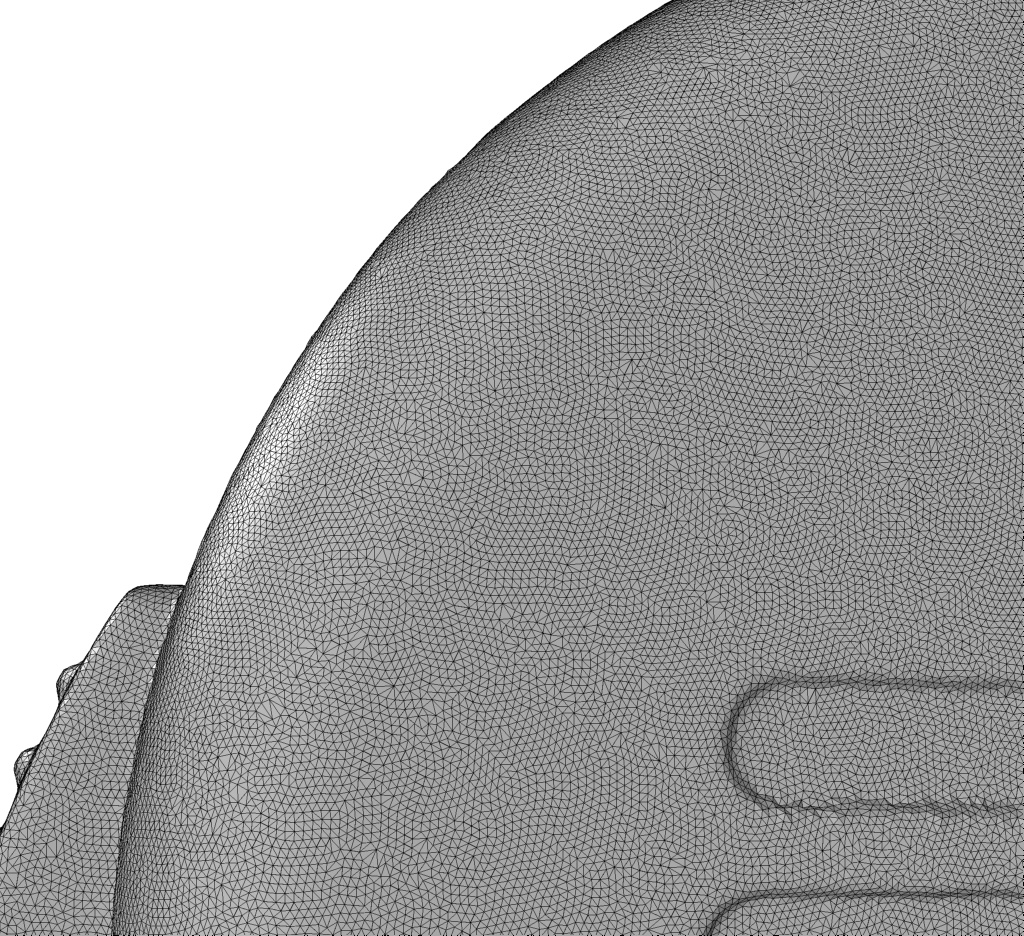}
		\end{subfigure}
	\end{minipage}
	\hfill
	\begin{minipage}{0.5\textwidth}
		\begin{subfigure}[t]{\Histogram}
			\centering
			\includegraphics[width=\textwidth]{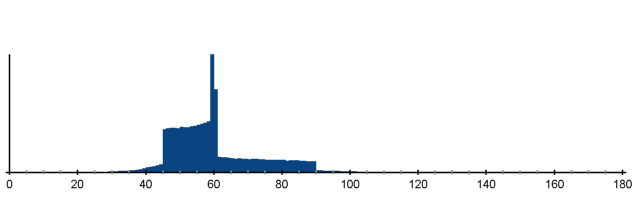}
			\caption{Angle distribution, target=$60^\circ$.}
		\end{subfigure}
		\begin{subfigure}[t]{\Histogram}
			\centering
			\includegraphics[width=\textwidth]{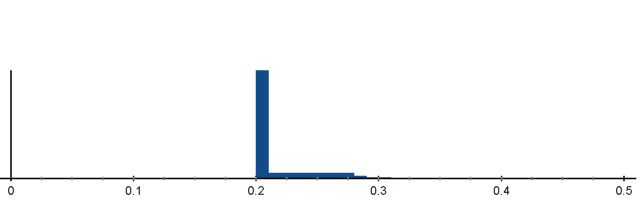}
			\caption{Edge lengths distribution, target=0.2.}
		\end{subfigure}
		\begin{subfigure}[t]{\Histogram}
			\centering
			\includegraphics[width=\textwidth]{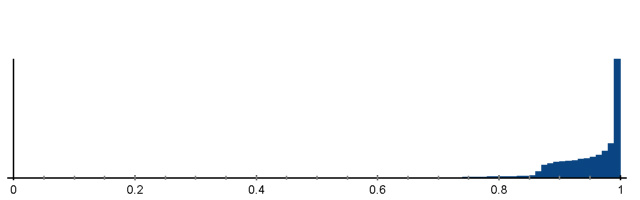}
			\caption{Distribution of quality $Q_t$, target=1.0.}
		\end{subfigure}
	\end{minipage}
	\caption{
		\emph{Tape}.
	}
\end{figure}

\begin{table}[b!]
	\def\arraystretch{1.05}
	\centering
	\begin{tabular}{l|rrrrr}
		Algorithm & $|\mathcal{T}|$ & $E_{\text{avg}}$ & $E_{\text{RMS}}$ & $Q_{\text{avg}}$ & $Q_{\text{RMS}}$\\
		\hline
		Adv.~Front &  1,022,926 & 0.1693 & 47.8 & 0.8250 & 16.4 \\
		Adv.~Front (Re) & 704,575 & 0.2028 & 16.2 & 0.9385 & 7.1 \\
		Poisson &  27,660 & 0.8333 & 83.3 & 0.8650 & 13.2 \\
		Poisson (Re) & 474,692 & 0.2364 & 38.2 & 0.9203 & 7.9 \\
		Poisson MG &  280,190 & 0.3369 & 35.6 & 0.7373 & 32.3 \\
		Poisson MG (Re) & 773,900 & 0.1962 & 16.2 & 0.9097 & 6.1 \\
		RIMLS & 3,577,797 & 0.0950 & 35.9 & 0.2486 & 34.5 \\
		RIMLS (Re) & 680,432 & 0.2055 & 16.2 & 0.9357 & 7.6 \\
		Scale Space & 1,021,944 & 0.1685 & 41.5 & 0.8255 & 16.2 \\
		Scale Space (Re) & 698,119 & \textbf{0.2025} & 15.5 & 0.9405 & 6.3 \\
		Voronoi & 1,023,090 & 0.1680 & 46.9 & 0.8234 & 16.5 \\
		Voronoi (Re) & 690,444 & 0.2042 & 16.5 & 0.9400 & 6.8 \\
		Ours & 616,091 & 0.2162 & \textbf{14.0} & \textbf{0.9491} & \textbf{5.8} \\
		\rowcolor{grey1}
		Ours (Re) & 643,856 & 0.2097 & \textbf{11.9} & \textbf{0.9633} & \textbf{4.7}
	\end{tabular}
	\caption{Experimental results for the \emph{Tape} from~\cite{huang2022surface} (511,569 points).}
	\label{tab:Tape}
\end{table}

\newpage

\subsection{\emph{Toy Bear}}

\begin{figure}[h!]
	\centering
	\begin{minipage}{0.45\textwidth}
		\begin{subfigure}{\textwidth}
			\includegraphics[width=1.\textwidth]{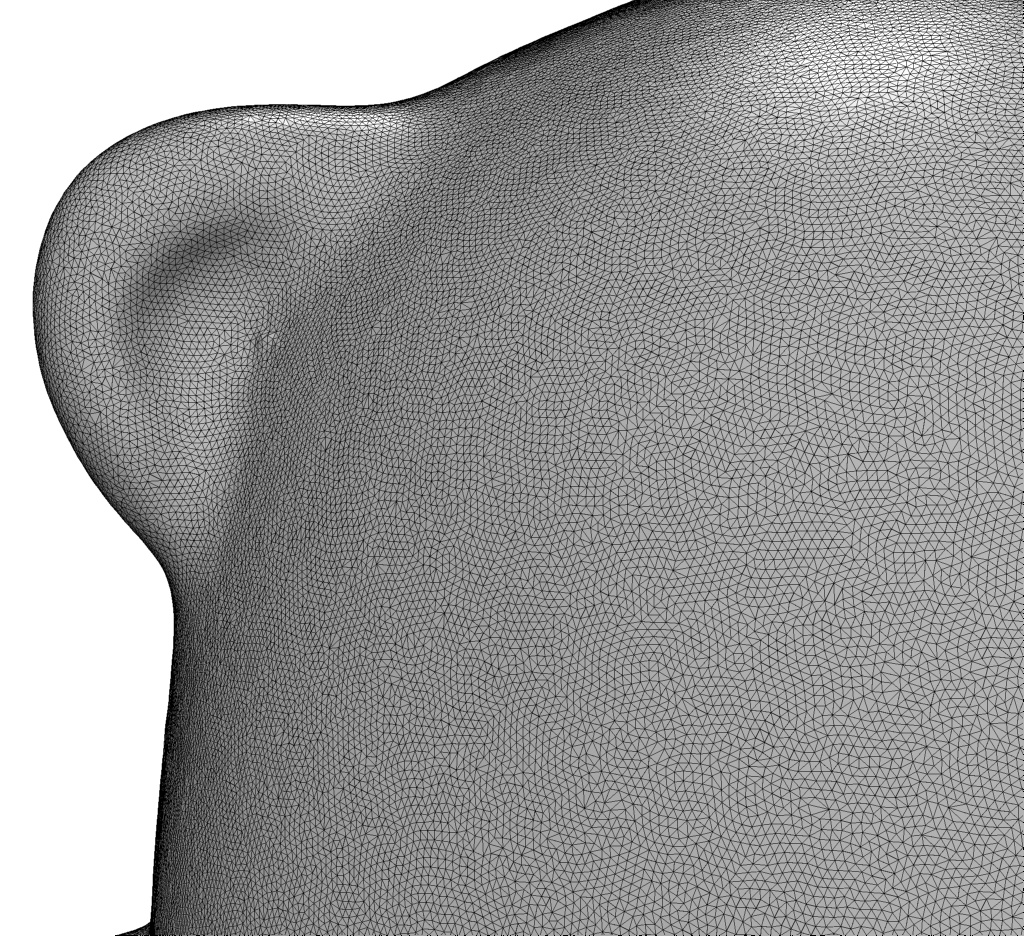}
		\end{subfigure}
	\end{minipage}
	\hfill
	\begin{minipage}{0.5\textwidth}
		\begin{subfigure}[t]{\Histogram}
			\centering
			\includegraphics[width=\textwidth]{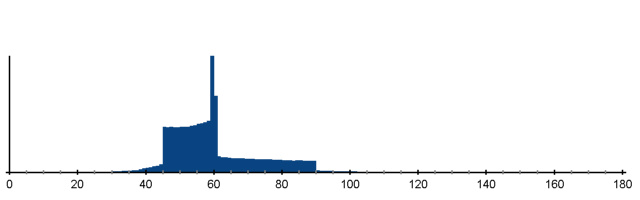}
			\caption{Angle distribution, target=$60^\circ$.}
		\end{subfigure}
		\begin{subfigure}[t]{\Histogram}
			\centering
			\includegraphics[width=\textwidth]{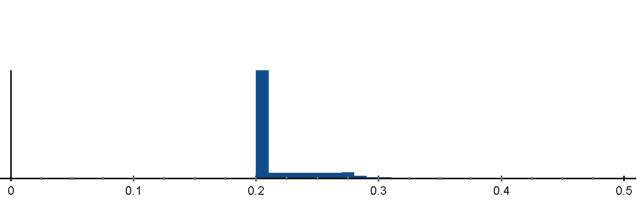}
			\caption{Edge lengths distribution, target=0.2.}
		\end{subfigure}
		\begin{subfigure}[t]{\Histogram}
			\centering
			\includegraphics[width=\textwidth]{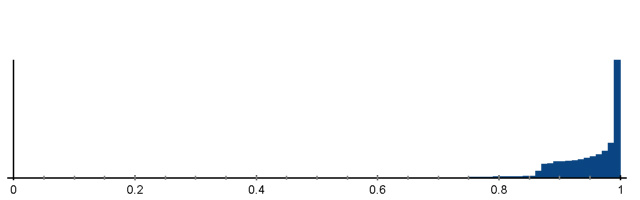}
			\caption{Distribution of quality $Q_t$, target=1.0.}
		\end{subfigure}
	\end{minipage}
	\caption{
		\emph{Toy Bear}.
	}
\end{figure}

\begin{table}[b!]
	\def\arraystretch{1.05}
	\centering
	\begin{tabular}{l|rrrrr}
		Algorithm & $|\mathcal{T}|$ & $E_{\text{avg}}$ & $E_{\text{RMS}}$ & $Q_{\text{avg}}$ & $Q_{\text{RMS}}$\\
		\hline
		Adv.~Front &  1,214,998 & 0.1474 & 36.3 & 0.8474 & 13.9 \\
		Adv.~Front (Re) & 629,138 & 0.2024 & 15.2 & 0.9418 & 6.0 \\
		Poisson &  20,134 &  1.0381 & 54.7 & 0.8882 & 11.7 \\
		Poisson (Re) & 530,374 & 0.2193 & 22.8 & 0.9293 & 7.3 \\
		Poisson MG &  432,268 & 0.2585 & 39.5 & 0.2623 & 37.1 \\
		Poisson MG (Re) & 629,508 & \textbf{0.2021} & 15.0 & 0.9436 & 6.1 \\
		RIMLS & 5,548,226 & 0.0730 & 40.0 & 0.6910 & 39.3 \\
		RIMLS (Re) & 618,531 & 0.2049 & 16.2 & 0.9322 & 6.8 \\
		Scale Space & 1,214,990 & 0.1474 & 36.3 & 0.8474 & 13.9 \\
		Scale Space (Re) & 628,848 & 0.2025 & 15.2 & 0.9417 & 6.0 \\
		Voronoi & 1,214,996 & 0.1471 & 36.5 & 0.8471 & 13.9 \\
		Voronoi (Re) & 616,160 &  0.2041 & 15.0 & 0.9427 & 5.9 \\
		Ours & 555,490 & 0.2159 & \textbf{13.5} & \textbf{0.9499} & \textbf{5.6} \\
		\rowcolor{grey1}
		Ours (Re) & 578,730 & 0.2096 & \textbf{11.5} & \textbf{0.9657} & \textbf{4.3}
	\end{tabular}
	\caption{Experimental results for the \emph{Toy Bear} from~\cite{huang2022surface} (607,501 points).}
	\label{tab:ToyBear}
\end{table}

\newpage

\subsection{\emph{Toy Duck}}

\begin{figure}[h!]
	\centering
	\begin{minipage}{0.45\textwidth}
		\begin{subfigure}{\textwidth}
			\includegraphics[width=1.\textwidth]{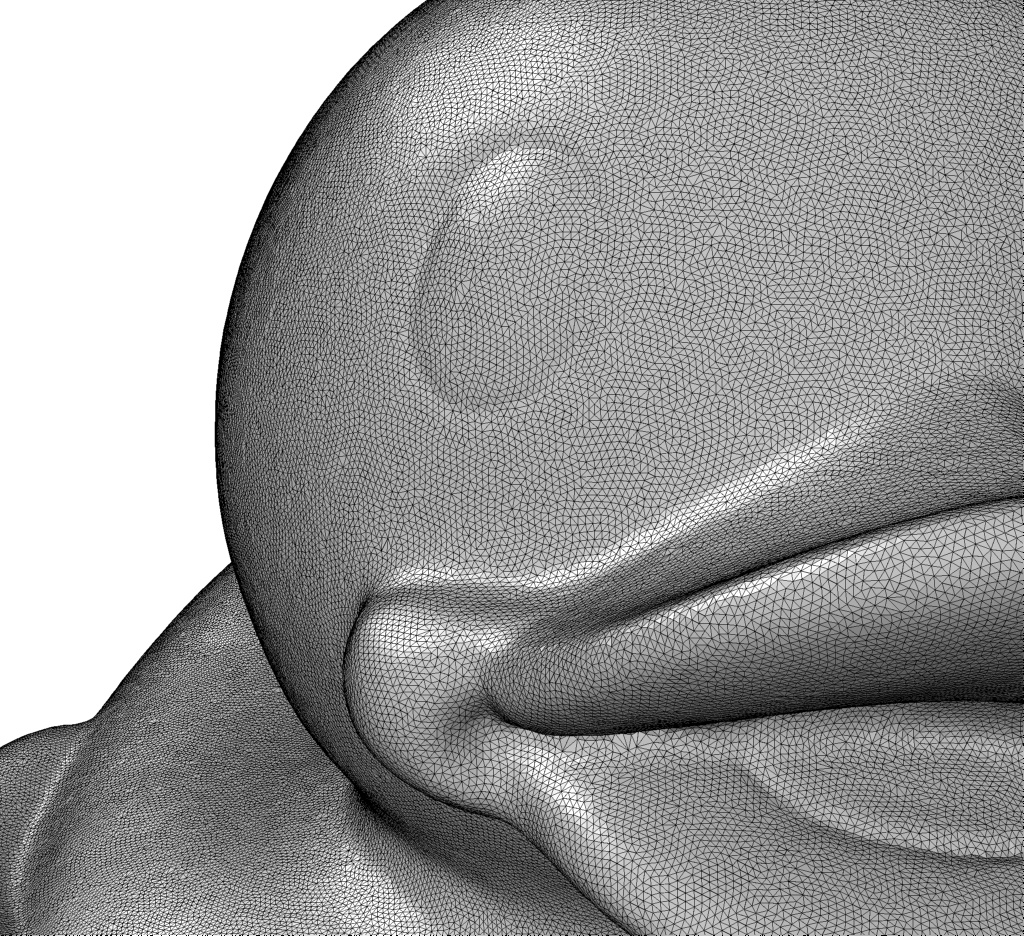}
		\end{subfigure}
	\end{minipage}
	\hfill
	\begin{minipage}{0.5\textwidth}
		\begin{subfigure}[t]{\Histogram}
			\centering
			\includegraphics[width=\textwidth]{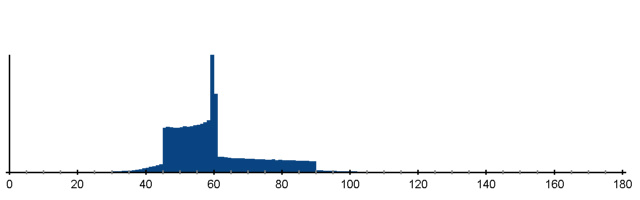}
			\caption{Angle distribution, target=$60^\circ$.}
		\end{subfigure}
		\begin{subfigure}[t]{\Histogram}
			\centering
			\includegraphics[width=\textwidth]{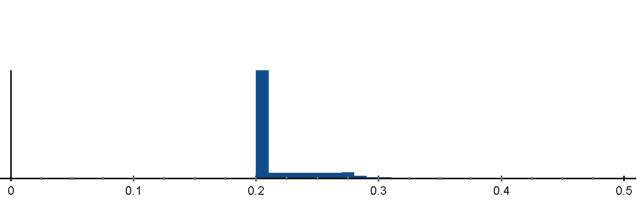}
			\caption{Edge lengths distribution, target=0.2.}
		\end{subfigure}
		\begin{subfigure}[t]{\Histogram}
			\centering
			\includegraphics[width=\textwidth]{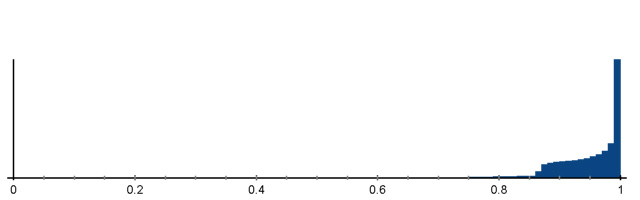}
			\caption{Distribution of quality $Q_t$, target=1.0.}
		\end{subfigure}
	\end{minipage}
	\caption{
		\emph{Toy Duck}.
	}
\end{figure}

\begin{table}[b!]
	\def\arraystretch{1.05}
	\centering
	\begin{tabular}{l|rrrrr}
		Algorithm & $|\mathcal{T}|$ & $E_{\text{avg}}$ & $E_{\text{RMS}}$ & $Q_{\text{avg}}$ & $Q_{\text{RMS}}$\\
		\hline
		Adv.~Front &  1,208,620 & 0.1523 & 36.7 & 0.8347 & 14.9 \\
		Adv.~Front (Re) & 664,788 & 0.\textbf{2023} & 15.3 & 0.9412 & 6.1 \\
		Poisson &  18,238 & 1.1320 & 51.1 & 0.8876 & 11.8 \\
		Poisson (Re) & 578,760 & 0.2158 & 20.9 & 0.9337 & 7.0 \\
		Poisson MG &  503,882 & 0.2456 & 39.4 & 0.7090 & 36.9 \\
		Poisson MG (Re) & 642,274 & 0.2064 & 16.8 & 0.9315 & 6.8 \\
		RIMLS & 6,470,657 & 0.0694 & 40.1 & 0.6910 & 39.4 \\
		RIMLS (Re) & 655,847 & 0.2044 & 16.4 & 0.9310 & 7.0 \\
		Scale Space & 1,208,618 & 0.1523 & 36.7 & 0.8347 & 14.9 \\
		Scale Space (Re) & 664,854 &  \textbf{0.2023} & 15.3 & 0.9412 & 6.1 \\
		Voronoi & 1,208,620 & 0.1520 & 36.8 & 0.8345 & 14.9 \\
		Voronoi (Re) & 654,288 & 0.2036 & 15.1 & 0.9422 & 6.0 \\
		Ours & 585,744 & 0.2160 & \textbf{13.6} & \textbf{0.9497} & \textbf{5.7} \\
		\rowcolor{grey1}
		Ours (Re) & 610,908 & 0.2095 & \textbf{11.5} & \textbf{0.9658} & \textbf{4.3}
	\end{tabular}
	\caption{Experimental results for the \emph{Toy Duck} from~\cite{huang2022surface} (604,312 points).}
	\label{tab:ToyDuck}
\end{table}

\newpage

\subsection{\emph{Wrench} \textcolor{sowaswieweiss}{y}}

\begin{figure}[h!]
	\centering
	\begin{minipage}{0.45\textwidth}
		\begin{subfigure}{\textwidth}
			\includegraphics[width=1.\textwidth]{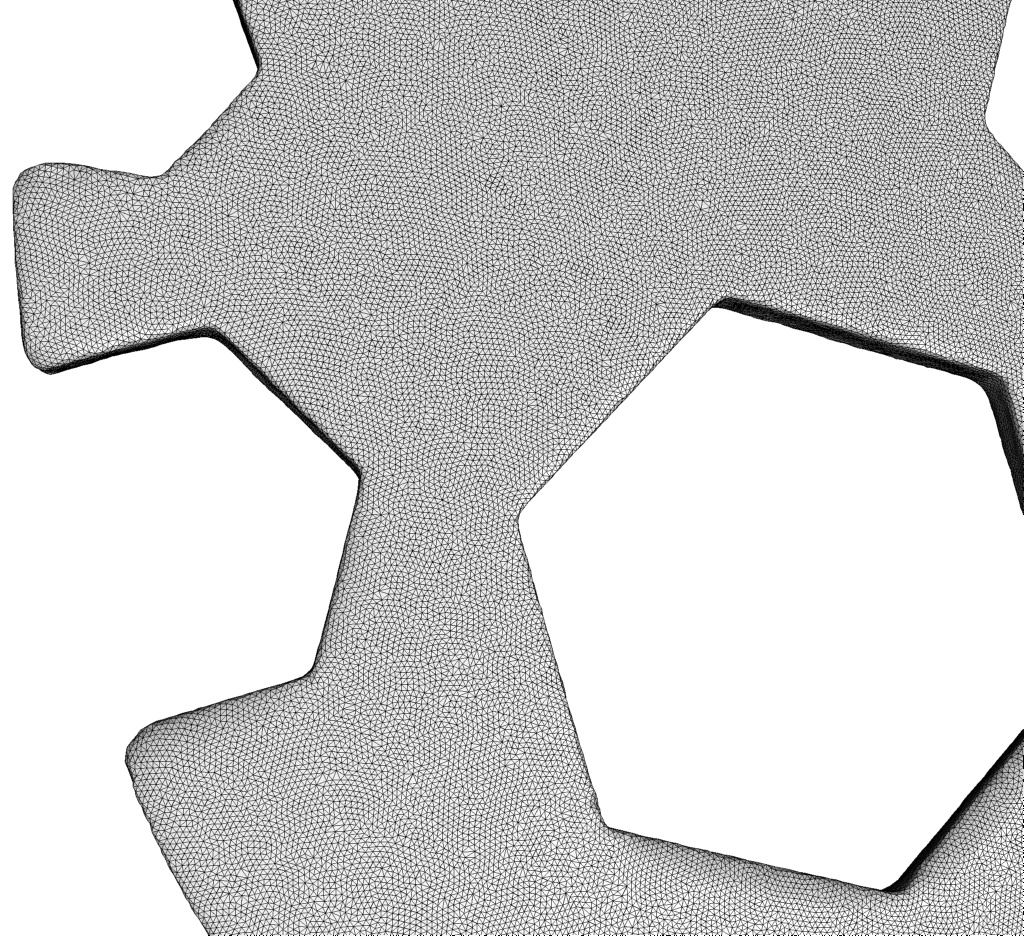}
		\end{subfigure}
	\end{minipage}
	\hfill
	\begin{minipage}{0.5\textwidth}
		\begin{subfigure}[t]{\Histogram}
			\centering
			\includegraphics[width=\textwidth]{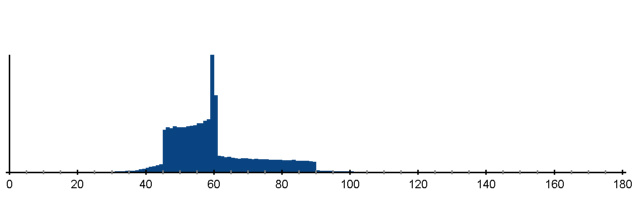}
			\caption{Angle distribution, target=$60^\circ$.}
		\end{subfigure}
		\begin{subfigure}[t]{\Histogram}
			\centering
			\includegraphics[width=\textwidth]{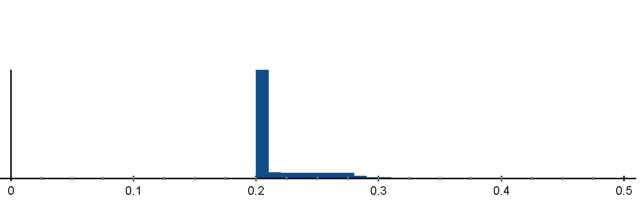}
			\caption{Edge lengths distribution, target=0.2.}
		\end{subfigure}
		\begin{subfigure}[t]{\Histogram}
			\centering
			\includegraphics[width=\textwidth]{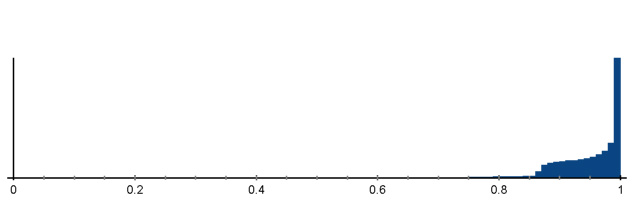}
			\caption{Distribution of quality $Q_t$, target=1.0.}
		\end{subfigure}
	\end{minipage}
	\caption{
		\emph{Wrench}.
	}
\end{figure}

\begin{table}[b!]
	\def\arraystretch{1.05}
	\centering
	\begin{tabular}{l|rrrrr}
		Algorithm & $|\mathcal{T}|$ & $E_{\text{avg}}$ & $E_{\text{RMS}}$ & $Q_{\text{avg}}$ & $Q_{\text{RMS}}$\\
		\hline
		Adv.~Front &  1,219,826 &  0.1335 & 33.7 & 0.8538 & 12.6 \\
		Adv.~Front (Re) & 516,056 & \textbf{0.2020} & 15.5 & 0.9401 & 6.2 \\
		Poisson &  26,428 &  0.6729 & 99.7 & 0.8469 & 14.2 \\
		Poisson (Re) & 300,868 & 0.2535 & 37.5 & 0.9190 & 7.9\\
		Poisson MG &  66,946 & 0.5781 & 31.9 & 0.7818 & 26.0 \\
		Poisson MG (Re) & 496,824 & 0.2054 & 15.1 & 0.9358 & 6.8 \\
		RIMLS & 801,420 & 0.1656 & 31.2 & 0.7667 & 28.2 \\
		RIMLS (Re) & 638,096 & 0.1808 & 20.6 & 0.8739 & 13.4 \\
		Scale Space & 1,219,826 & 0.1335 & 33.7 & 0.8538 & 12.6 \\
		Scale Space (Re) & 516,184 & \textbf{0.2020} & 15.4 & 0.9403 & 6.2 \\
		Voronoi & 1,219,826 & 0.1326 & 34.3 & 0.8543 & 12.7 \\
		Voronoi (Re) & 503,056 & 0.2037 & 15.4 & 0.9402 & 6.2 \\
		Ours & 454,492 &  0.2157 & \textbf{13.3} & \textbf{0.9508} & \textbf{5.5} \\
		\rowcolor{grey1}
		Ours (Re) & 472,946 & 0.2095 & \textbf{11.5} & \textbf{0.9655} & \textbf{4.4}
	\end{tabular}
	\caption{Experimental results for the \emph{Wrench} from~\cite{huang2022surface} (609,911 points).}
	\label{tab:Wrench}
\end{table}

\newpage

\subsection{\emph{Xiao Jie Jie} \textcolor{sowaswieweiss}{y}}

\begin{figure}[h!]
	\centering
	\begin{minipage}{0.45\textwidth}
		\begin{subfigure}{\textwidth}
			\includegraphics[width=1.\textwidth]{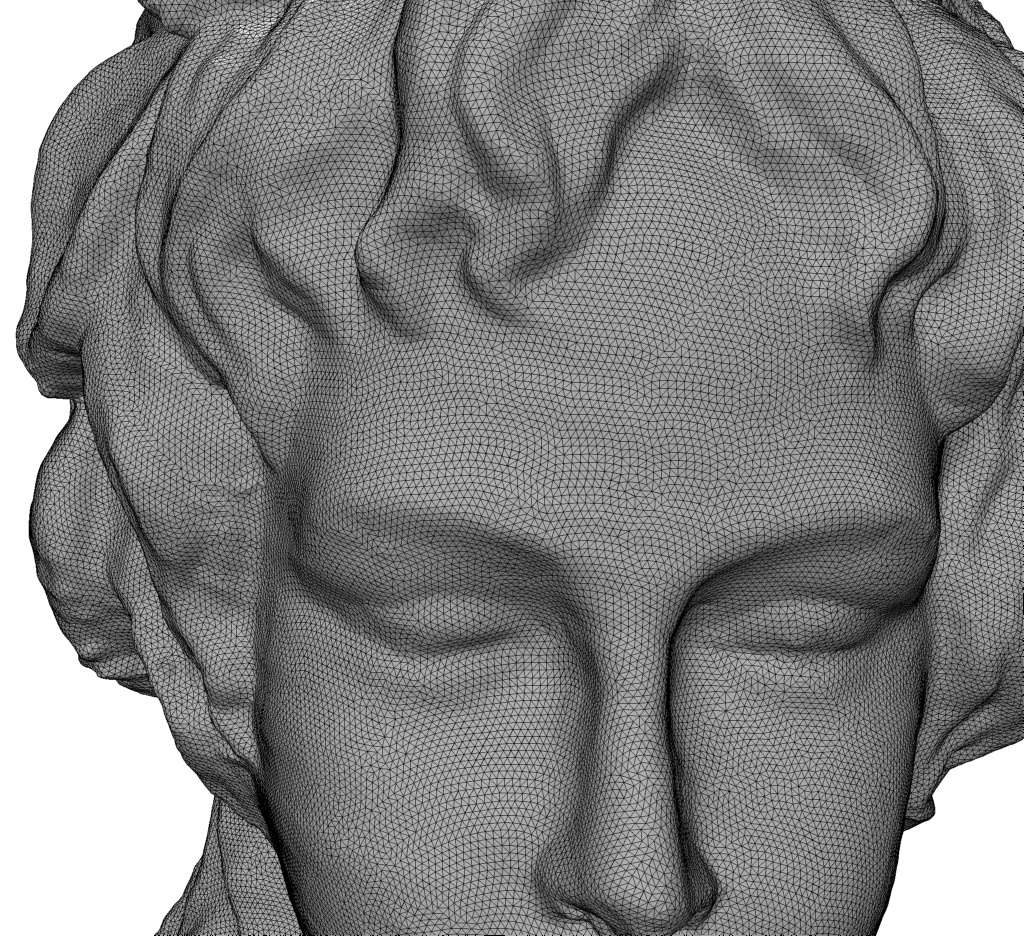}
		\end{subfigure}
	\end{minipage}
	\hfill
	\begin{minipage}{0.5\textwidth}
		\begin{subfigure}[t]{\Histogram}
			\centering
			\includegraphics[width=\textwidth]{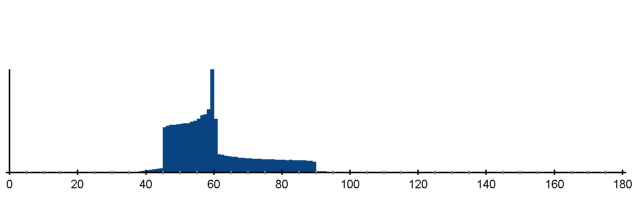}
			\caption{Angle distribution, target=$60^\circ$.}
		\end{subfigure}
		\begin{subfigure}[t]{\Histogram}
			\centering
			\includegraphics[width=\textwidth]{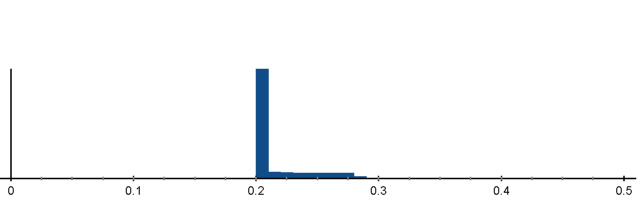}
			\caption{Edge lengths distribution, target=0.2.}
		\end{subfigure}
		\begin{subfigure}[t]{\Histogram}
			\centering
			\includegraphics[width=\textwidth]{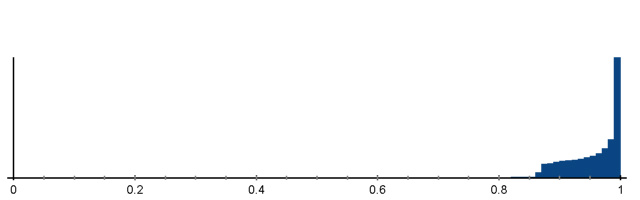}
			\caption{Distribution of quality $Q_t$, target=1.0.}
		\end{subfigure}
	\end{minipage}
	\caption{
		\emph{Xiao Jie Jie}.
	}
\end{figure}

\begin{table}[b!]
	\def\arraystretch{1.05}
	\centering
	\begin{tabular}{l|rrrrr}
		Algorithm & $|\mathcal{T}|$ & $E_{\text{avg}}$ & $E_{\text{RMS}}$ & $Q_{\text{avg}}$ & $Q_{\text{RMS}}$\\
		\hline
		Adv.~Front &  2,473,735 & 0.1198 & 27.2 & 0.8755 & 9.3 \\
		Adv.~Front (Re) & 811,986 & \textbf{0.2037} & 15.0 & 0.9433 & 6.0\\
		Poisson &  87,740 & 0.5677 & 55.9 & 0.8785 & 12.2 \\
		Poisson (Re) & 755,538 & 0.2097 & 21.0 & 0.9321 &7.1 \\
		Poisson MG &  200,446 & 0.4316 & 39.7 & 0.7043 & 37.7\\
		Poisson MG (Re) & 786,370 & 0.2073 & 18.7 & 0.9152 & 8.7\\
		RIMLS & 2,528,244 &  0.1219 & 40.2 & 0.6833 & 40.0 \\
		RIMLS (Re) & 1,066,017 & 0.1703 & 37.3 & 0.7956 & 34.3\\
		Scale Space & 2,473,712 &  0.1198 & 27.2 &0.8750 & 9.3 \\
		Scale Space (Re) & 812,416 & \textbf{0.2037} & 15.0 & 0.9432 & 6.0\\
		Voronoi & 2,473,692 & 0.1186 & 27.5 & 0.8751 & 9.4 \\
		Voronoi (Re) & 753,938 & 0.2094 & 14.4 & 0.9463 & 5.8 \\
		Ours & 739,325 &  0.2138 & \textbf{11.6} & \textbf{0.9557} & \textbf{4.6} \\
		\rowcolor{grey1}
		Ours (Re) & 754,086 & 0.2101 & \textbf{11.2} & \textbf{0.9663} & \textbf{4.4}
	\end{tabular}
	\caption{Experimental results for the \emph{Xiao Jie Jie} from~\cite{huang2022surface} (1,236,884 points).}
	\label{tab:XiaoJieJie}
\end{table}

\newpage

\section{Robustness of the Algorithm}\label{sec:robustness}

For all models, we chose the same target edge length~$d = 0.2$.

\subsection{\emph{Toy Bear}}

\begin{figure}[h!]
	\begin{minipage}[h!]{.24\linewidth}
		\centering
		\includegraphics[width=0.85\linewidth]{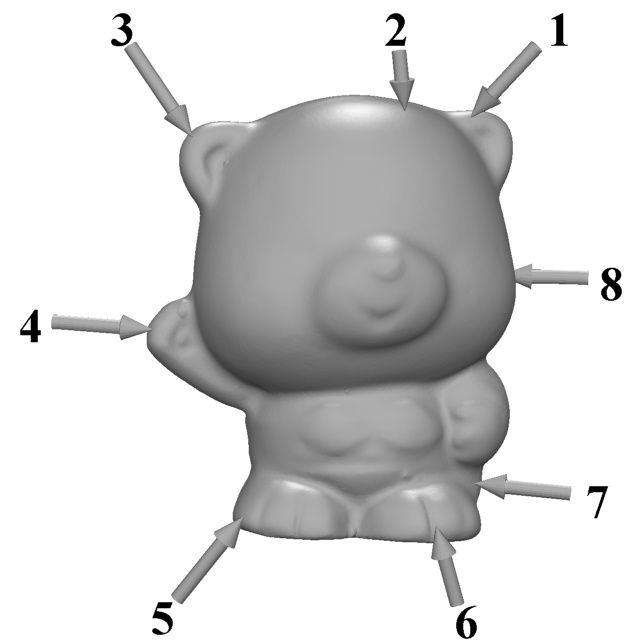}
	\end{minipage}\hfill
	\begin{minipage}[t]{.75\linewidth}
		\centering
		\begin{tabular}{l|rrrrrrrr}
			Pair & $| \mathcal{T} |$ & $E_{\text{max}}$ & $E_{\text{avg}}$ & $E_{\text{RMS}}$ & $Q_{\text{min}}$ & $Q_{\text{max}}$ & $Q_{\text{avg}}$ & $Q_{\text{RMS}}$ \\
			\hline
			1 & 555,750 & 0.5395 & 0.2157 & 13.5 & 0.4817 & 1.0 & 0.9504 & 5.6 \\
			2 & 555,768 & 0.5151 & 0.2157 & 13.5 & 0.4465 & 1.0 & 0.9504 & 5.6 \\
			3 & 555,508 & 0.5107 & 0.2158 & 13.5 & 0.3856 & 1.0 & 0.9499 & 5.6 \\
			4 & 555,736 & 0.5229 & 0.2158 & 13.5 & 0.4164 & 1.0 & 0.9503 & 5.6 \\
			5 & 555,504 & 0.5489 & 0.2158 & 13.5 & 0.3950 & 1.0 & 0.9501 & 5.6 \\
			6 & 555,396 & 0.5466 & 0.2158 & 13.5 & 0.4358 & 1.0 & 0.9500 & 5.6 \\
			7 & 555,554 & 0.6009 & 0.2158 & 13.5 & 0.4632 & 1.0 & 0.9502 & 5.6 \\
			8 & 555,592 & 0.5493 & 0.2158 & 13.5 & 0.4253 & 1.0 & 0.9501 & 5.6 
		\end{tabular}
	\end{minipage}
	\caption{\emph{Toy Bear} with eight different pairs of starting vertices and evaluation of results achieved.}
\end{figure}

\subsection{\emph{Wrench}} \textcolor{sowaswieweiss}{y}

\begin{figure}[h!]
	\begin{minipage}[h!]{.24\linewidth}
		\centering
		\includegraphics[width=0.85\linewidth]{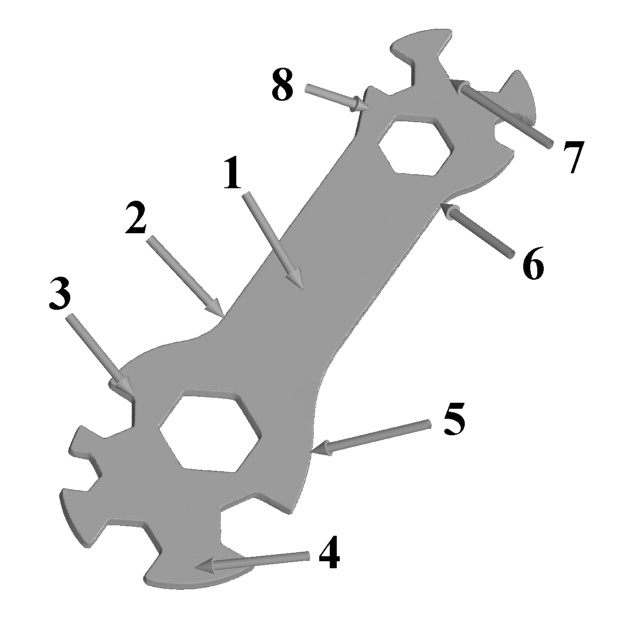}
	\end{minipage}\hfill
	\begin{minipage}[t]{.75\linewidth}
		\centering
		\begin{tabular}{l|rrrrrrrr}
			Pair & $| \mathcal{T} |$ & $E_{\text{max}}$ & $E_{\text{avg}}$ & $E_{\text{RMS}}$ & $Q_{\text{min}}$ & $Q_{\text{max}}$ & $Q_{\text{avg}}$ & $Q_{\text{RMS}}$ \\
			\hline
			1 & 454,510 & 0.4911 & 0.2156 & 13.3 & 0.4748 & 1.0 & 0.9507 & 5.5 \\
			2 & 454,630 & 0.4905 & 0.2156 & 13.3 & 0.4528 & 1.0 & 0.9507 & 5.5 \\
			3 & 454,550 & 0.4724 & 0.2156 & 13.3 & 0.4645 & 1.0 & 0.9506 & 5.5 \\
			4 & 454,520 & 0.5160 & 0.2156 & 13.3 & 0.4499 & 1.0 & 0.9505 & 5.5 \\
			5 & 454,618 & 0.4747 & 0.2156 & 13.3 & 0.4817 & 1.0 & 0.9507 & 5.5 \\
			6 & 454,514 & 0.5056 & 0.2156 & 13.3 & 0.4862 & 1.0 & 0.9507 & 5.5 \\
			7 & 454,544 & 0.4684 & 0.2156 & 13.2 & 0.3887 & 1.0 & 0.9506 & 5.5 \\
			8 & 454,762 & 0.5033 & 0.2155 & 13.2 & 0.5011 & 1.0 & 0.9510 & 5.5 
		\end{tabular}
	\end{minipage}
	\caption{\emph{Wrench} with eight different pairs of starting vertices and evaluation of results achieved.}
\end{figure}

\subsection{\emph{Lock}} \textcolor{sowaswieweiss}{y}

\begin{figure}[h!]
	\begin{minipage}[h!]{.24\linewidth}
		\centering
		\includegraphics[width=0.85\linewidth]{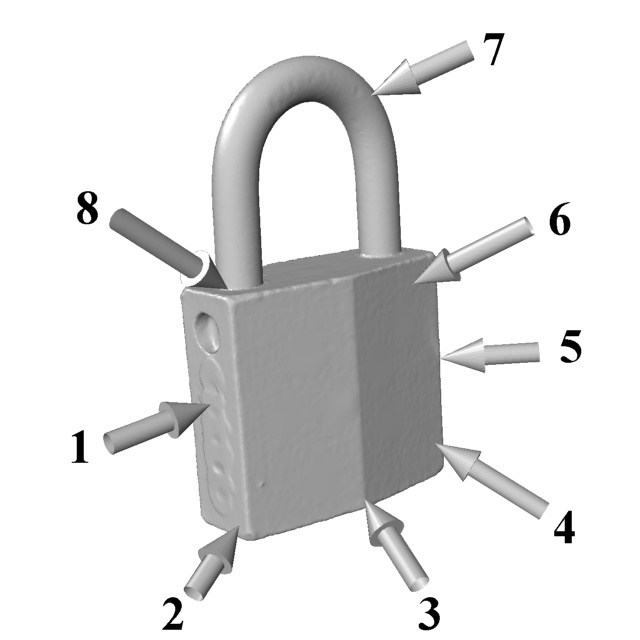}
	\end{minipage}\hfill
	\begin{minipage}[t]{.75\linewidth}
		\centering
		\begin{tabular}{l|rrrrrrrr}
			Pair & $| \mathcal{T} |$ & $E_{\text{max}}$ & $E_{\text{avg}}$ & $E_{\text{RMS}}$ & $Q_{\text{min}}$ & $Q_{\text{max}}$ & $Q_{\text{avg}}$ & $Q_{\text{RMS}}$ \\
			\hline
			1 & 192,308 & 0.4502 & 0.2136 & 11.5 & 0.5373 & 1.0 & 0.9561 & 4.6 \\
			2 & 192,280 & 0.3997 & 0.2136 & 11.5 & 0.5811 & 1.0 & 0.9558 & 4.6 \\
			3 & 192,558 & 0.4296 & 0.2134 & 11.6 & 0.5936 & 1.0 & 0.9566 & 4.7 \\
			4 & 192,708 & 0.4168 & 0.2132 & 11.4 & 0.5858 & 1.0 & 0.9575 & 4.5 \\
			5 & 192,430 & 0.4172 & 0.2134 & 11.5 & 0.5642 & 1.0 & 0.9566 & 4.6 \\
			6 & 192,016 & 0.4162 & 0.2138 & 11.7 & 0.5568 & 1.0 & 0.9549 & 4.6 \\
			7 & 192,256 & 0.4237 & 0.2136 & 11.5 & 0.5630 & 1.0 & 0.9562 & 4.6 \\
			8 & 192,130 & 0.4380 & 0.2137 & 11.6 & 0.4729 & 1.0 & 0.9554 & 4.6 
		\end{tabular}
	\end{minipage}
	\caption{\emph{Lock} with eight different pairs of starting vertices and evaluation of results achieved.}
\end{figure}

\newpage 

\subsection{\emph{Bottle Shampoo}} \textcolor{sowaswieweiss}{y}

\begin{figure}[h!]
	\begin{minipage}[h!]{.24\linewidth}
		\centering
		\includegraphics[width=0.85\linewidth]{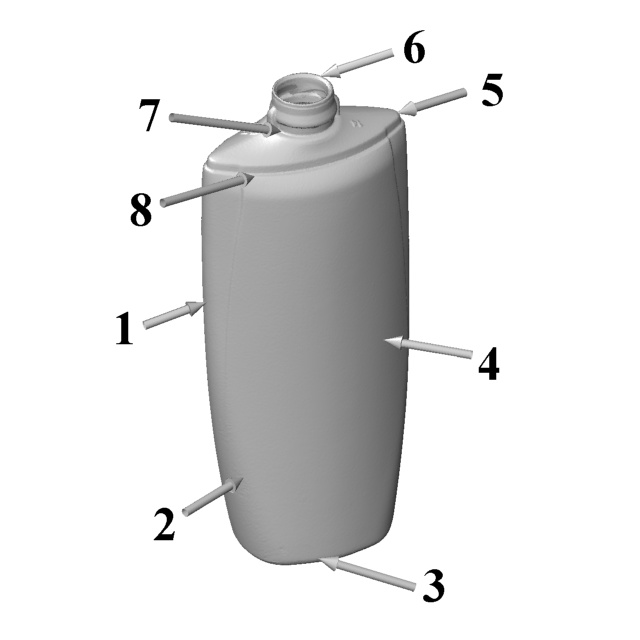}
	\end{minipage}\hfill
	\begin{minipage}[t]{.75\linewidth}
		\centering
		\begin{tabular}{l|rrrrrrrr}
			Pair & $| \mathcal{T} |$ & $E_{\text{max}}$ & $E_{\text{avg}}$ & $E_{\text{RMS}}$ & $Q_{\text{min}}$ & $Q_{\text{max}}$ & $Q_{\text{avg}}$ & $Q_{\text{RMS}}$ \\
			\hline
			1 & 818,160 & 0.5702 & 0.2165 & 14.2 & 0.4270 & 1.0 & 0.9484 & 6.0 \\
			2 & 818,167 & 0.5499 & 0.2164 & 14.2 & 0.4349 & 1.0 & 0.9486 & 5.9 \\
			3 & 817,997 & 0.5734 & 0.2165 & 14.3 & 0.3969 & 1.0 & 0.9485 & 6.0 \\
			4 & 818,122 & 0.5664 & 0.2165 & 14.2 & 0.3876 & 1.0 & 0.9485 & 6.0 \\
			5 & 818,025 & 0.5918 & 0.2165 & 14.2 & 0.3937 & 1.0 & 0.9483 & 6.0 \\
			6 & 818,261 & 0.5841 & 0.2164 & 14.2 & 0.2944 & 1.0 & 0.9486 & 5.9 \\
			7 & 817,836 & 0.5619 & 0.2165 & 14.2 & 0.3317 & 1.0 & 0.9483 & 6.0 \\
			8 & 818,137 & 0.5893 & 0.2165 & 14.2 & 0.3811 & 1.0 & 0.9485 & 6.0 
		\end{tabular}
	\end{minipage}
	\caption{\emph{Bottle Shampoo} with eight different pairs of starting vertices and evaluation of results achieved.}
\end{figure}

\newpage

\section{Collection of Point Clouds for Feature Detection}
\label{sec:CollectionOfPointCloudsForFeatureDetection}
\subsection{\emph{Lock}} \textcolor{sowaswieweiss}{y}

\begin{figure}[h!]
	\begin{subfigure}[t]{.32\textwidth}
		\includegraphics[width=\textwidth]{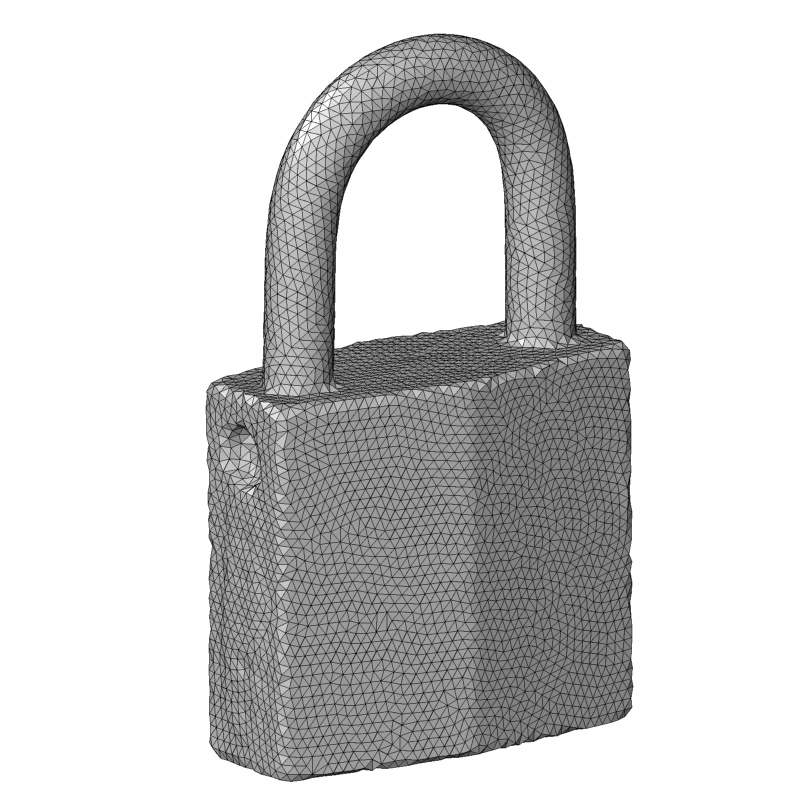}
	\end{subfigure}
	~
	\begin{subfigure}[t]{.32\textwidth}
		\includegraphics[width=\textwidth]{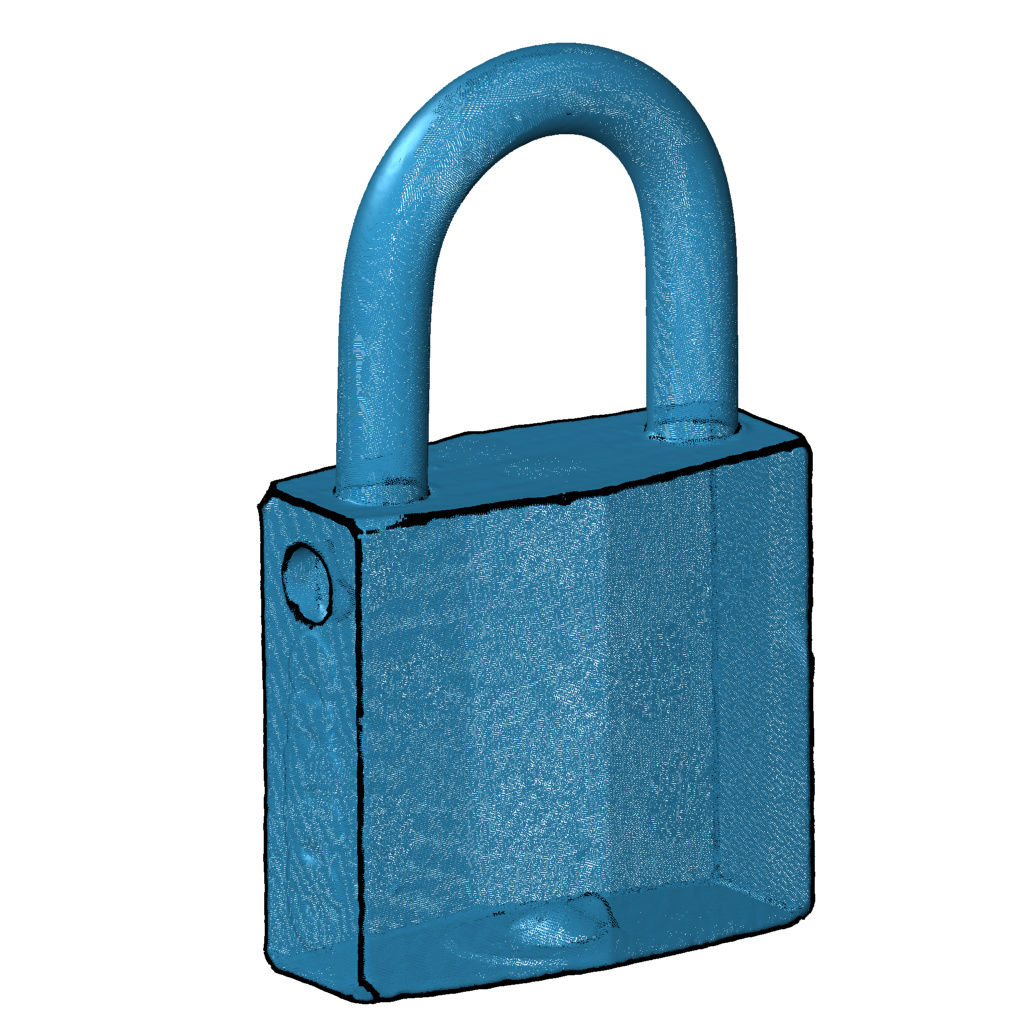}
	\end{subfigure}
	~
	\begin{subfigure}[t]{.32\textwidth}
		\includegraphics[width=\textwidth]{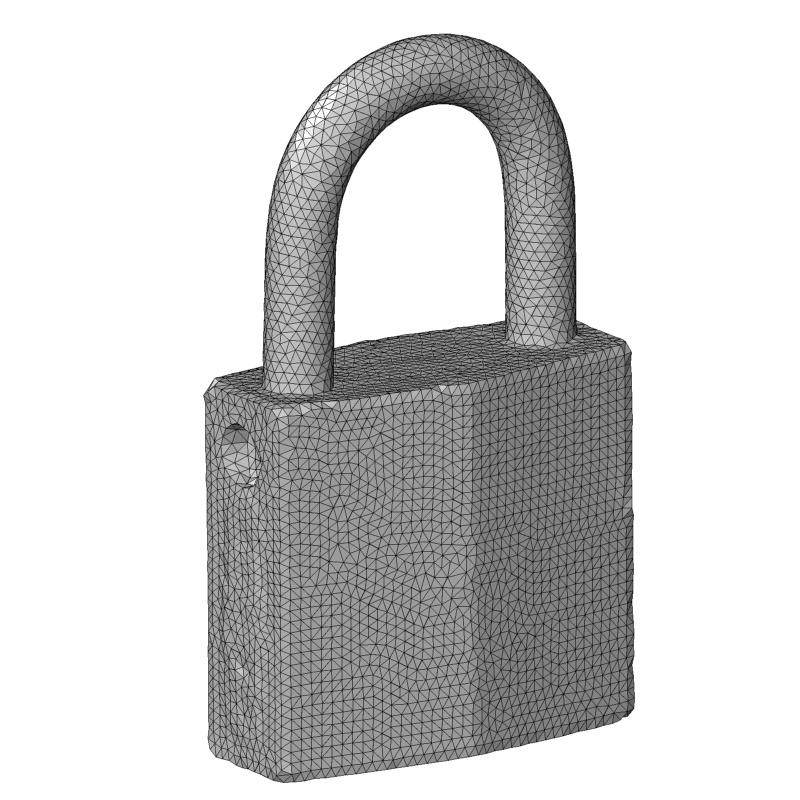}
	\end{subfigure}
	\caption{\emph{Lock}.
		Left: Meshes point cloud without feature detection.
		Middle: Point cloud with features detected for~$\vartheta = 60^{\circ}$ and target edge length~$d = 0.8$.
		Right: Meshed point cloud with features detected before.}
\end{figure}

\vspace{-4mm}

\begin{figure}[h!]
	\centering
	\begin{minipage}{0.45\textwidth}
		\begin{subfigure}[t]{\Histogram}
			\centering
			\includegraphics[width=\textwidth]{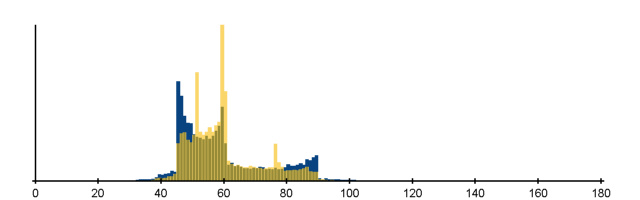}
			\caption{Angle distribution, target=$60^\circ$.}
		\end{subfigure}
		\begin{subfigure}[t]{\Histogram}
			\centering
			\includegraphics[width=\textwidth]{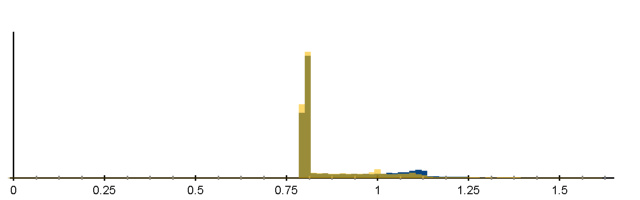}
			\caption{Edge lengths distribution, target=0.8.}
		\end{subfigure}
		\begin{subfigure}[t]{\Histogram}
			\centering
			\includegraphics[width=\textwidth]{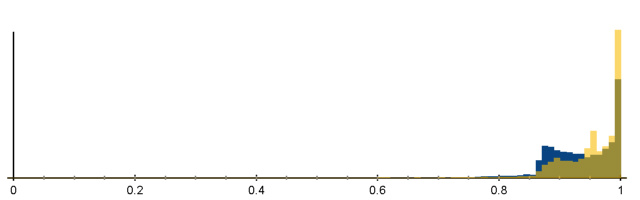}
			\caption{Distribution of quality $Q_t$, target=1.0.}
		\end{subfigure}
	\end{minipage}
	\caption{Histograms for remeshed \emph{Lock}, with (blue) and without (yellow) feature detection,~$\vartheta = 60^{\circ}$.}
\end{figure}

\begin{table}[b!]
	\centering
	\begin{tabular}{lrrrlrrrr}
		$d$ & $\vartheta$ & $| \mathcal{V} |$ & $| \mathcal{T} |$ & $d_{\text{avg}}$ & $d_{\max}$ & $d_{\text{RMS}}$ & $|\text{feature segments}|$ & $|\text{feature vertices}|$\\
		\hline
		0.8 & --- & 5,951 & 11,902 & 0.016 & 0.4609 & 193.6 & --- & --- \\
		0.8 & $60^{\circ}$ & 5,837 & 11,674 & 0.011 & 0.2963 & 142.6 & 473,115 & 423\\
	\end{tabular}
	\caption{Comparison of results achieved for \emph{Lock}, with and without feature detection.}
\end{table}

\newpage

\subsection{\emph{Remote}} \textcolor{sowaswieweiss}{y}

\begin{figure}[h!]
	\begin{subfigure}[t]{.32\textwidth}
		\includegraphics[width=\textwidth]{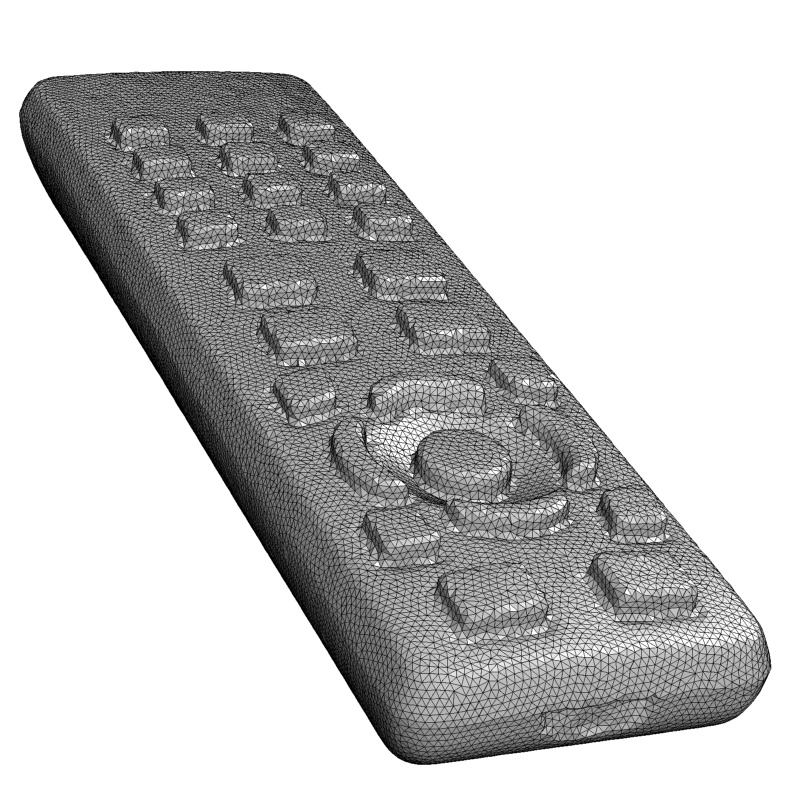}
	\end{subfigure}
	~
	\begin{subfigure}[t]{.32\textwidth}
		\includegraphics[width=\textwidth]{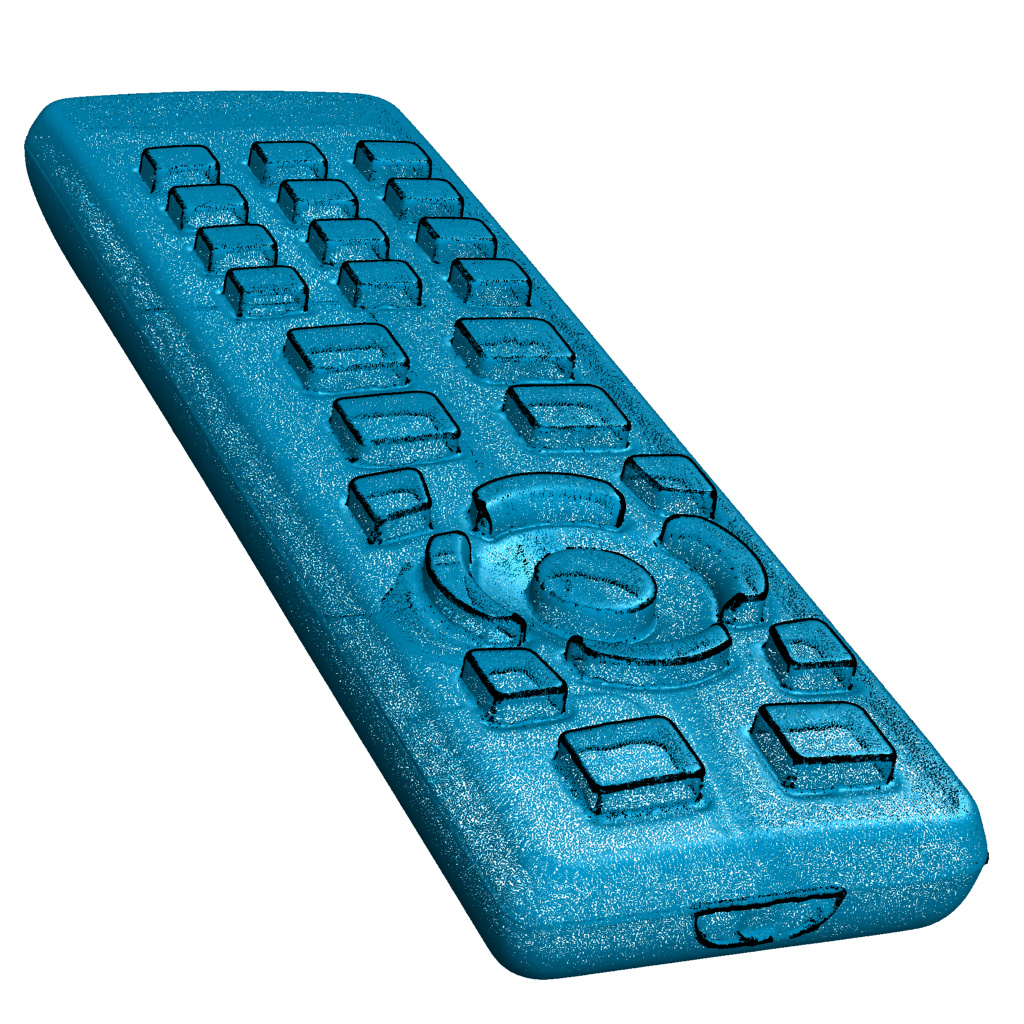}
	\end{subfigure}
	~
	\begin{subfigure}[t]{.32\textwidth}
		\includegraphics[width=\textwidth]{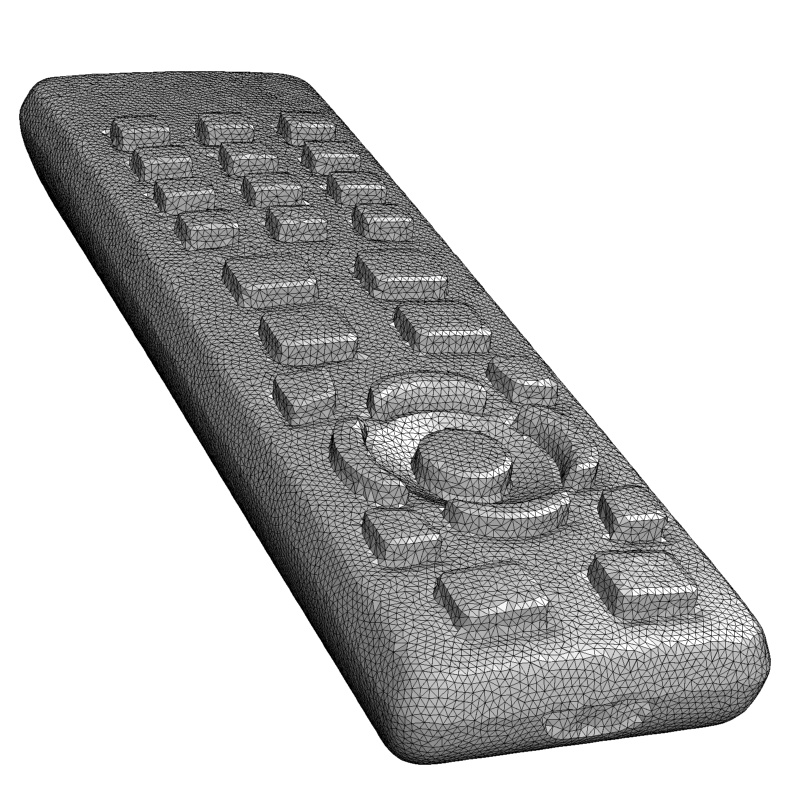}
	\end{subfigure}
	\caption{\emph{Remote}.
		Left: Meshes point cloud without feature detection.
		Middle: Point cloud with features detected for~$\vartheta = 50^{\circ}$ and target edge length~$d = 0.8$.
		Right: Meshed point cloud with features detected before.}
\end{figure}

\vspace{-4mm}

\begin{figure}[h!]
	\centering
	\begin{minipage}{0.45\textwidth}
		\begin{subfigure}[t]{\Histogram}
			\centering
			\includegraphics[width=\textwidth]{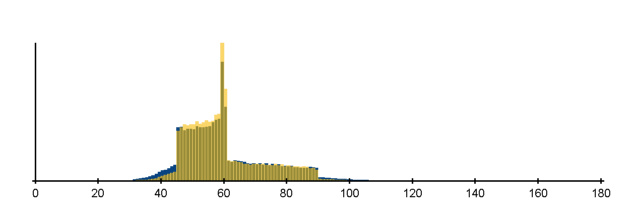}
			\caption{Angle distribution, target=$60^\circ$.}
		\end{subfigure}
		\begin{subfigure}[t]{\Histogram}
			\centering
			\includegraphics[width=\textwidth]{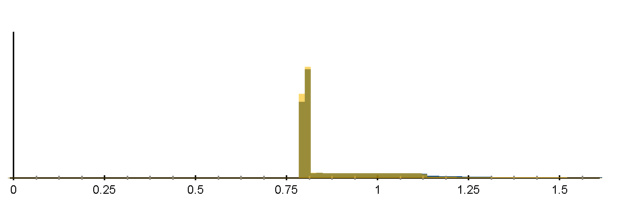}
			\caption{Edge lengths distribution, target=0.8.}
		\end{subfigure}
		\begin{subfigure}[t]{\Histogram}
			\centering
			\includegraphics[width=\textwidth]{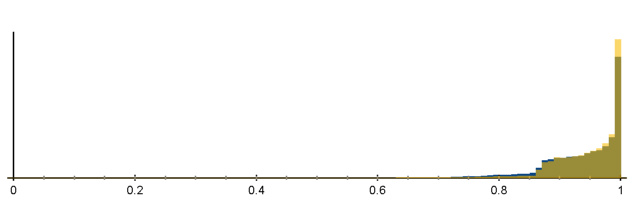}
			\caption{Distribution of quality $Q_t$, target=1.0.}
		\end{subfigure}
	\end{minipage}
	\caption{Histograms for remeshed \emph{Remote}, with (blue) and without (yellow) feature detection,~$\vartheta = 50^{\circ}$.}
\end{figure}

\begin{table}[b!]
	\centering
	\begin{tabular}{lrrrlrrrr}
		$d$ & $\vartheta$ & $| \mathcal{V} |$ & $| \mathcal{T} |$ & $d_{\text{avg}}$ & $d_{\max}$ & $d_{\text{RMS}}$ & $|\text{feature segments}|$ & $|\text{feature vertices}|$ \\
		\hline
		0.8 & --- & 22,595 & 45,186 & 0.0296 & 0.2963 & 192.5 & --- & --- \\
		0.8 & $50^{\circ}$ & 22,239 & 44,474 & 0.0178 & 0.4879 & 162.7 & 1,249,250 & 2,031
	\end{tabular}
	\caption{Comparison of results achieved for \emph{Remote}, with and without feature detection.}
\end{table}

\newpage

\subsection{\emph{Screw}} \textcolor{sowaswieweiss}{y}

\begin{figure}[h!]
	\begin{subfigure}[t]{.32\textwidth}
		\includegraphics[width=\textwidth]{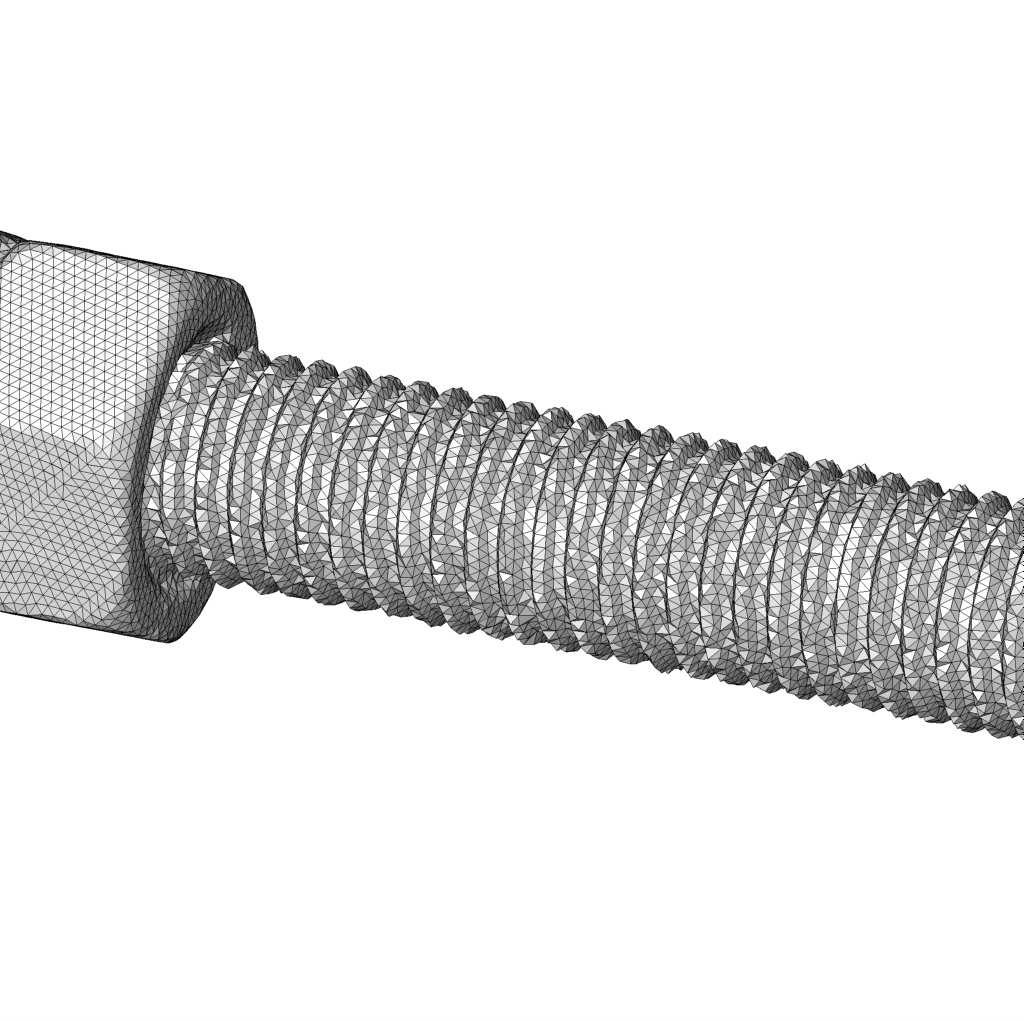}
	\end{subfigure}
	~
	\begin{subfigure}[t]{.32\textwidth}
		\includegraphics[width=\textwidth]{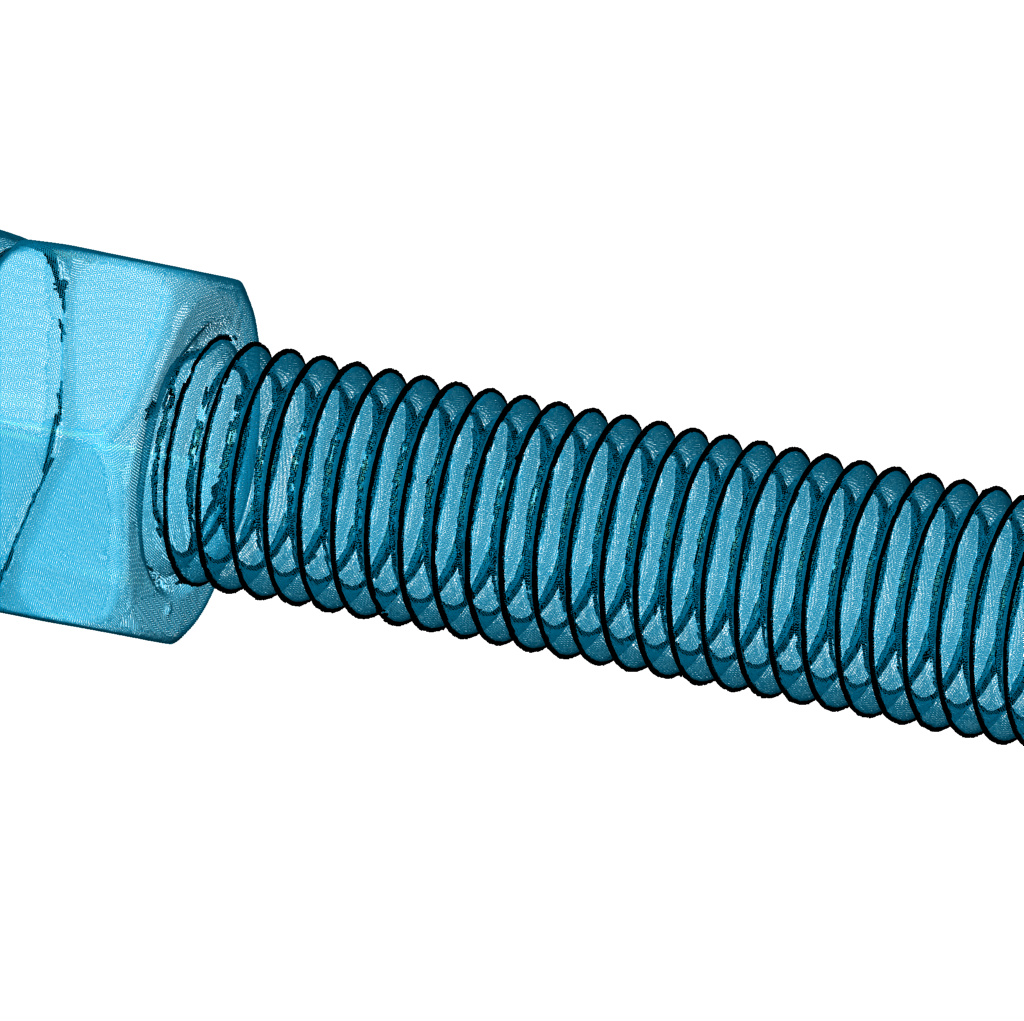}
	\end{subfigure}
	~
	\begin{subfigure}[t]{.32\textwidth}
		\includegraphics[width=\textwidth]{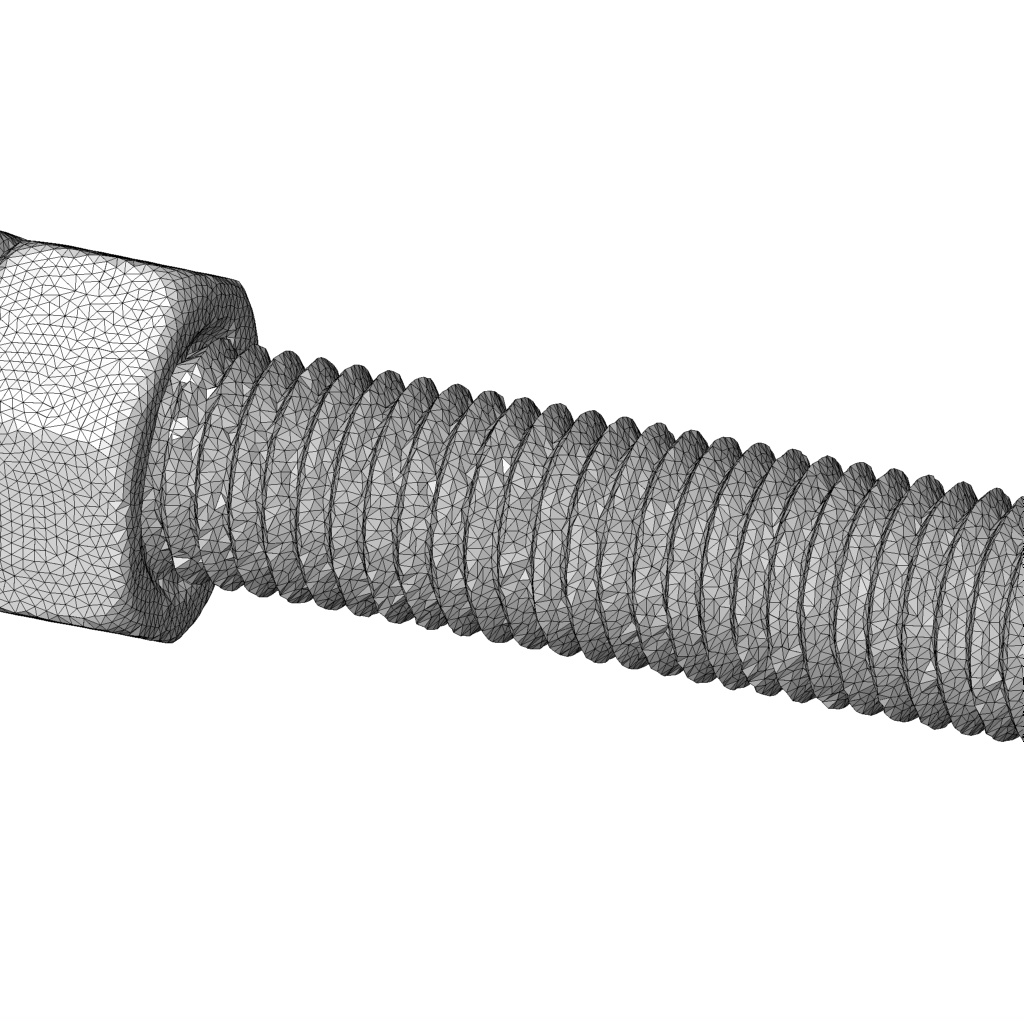}
	\end{subfigure}
	\caption{\emph{Screw}.
		Left: Meshes point cloud without feature detection.
		Middle: Point cloud with features detected for~$\vartheta = 90^{\circ}$ and target edge length~$d = 0.6$.
		Right: Meshed point cloud with features detected before.}
\end{figure}

\vspace{-4mm}

\begin{figure}[h!]
	\centering
	\begin{minipage}{0.45\textwidth}
		\begin{subfigure}[t]{\Histogram}
			\centering
			\includegraphics[width=\textwidth]{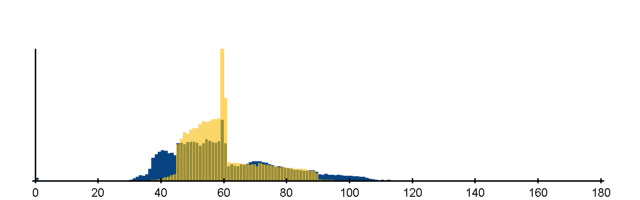}
			\caption{Angle distribution, target=$60^\circ$.}
		\end{subfigure}
		\begin{subfigure}[t]{\Histogram}
			\centering
			\includegraphics[width=\textwidth]{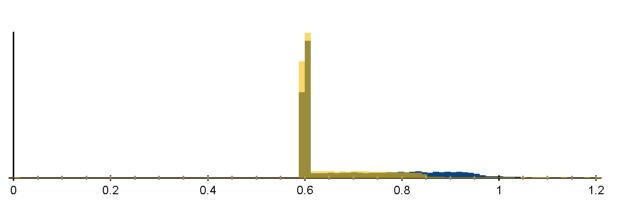}
			\caption{Edge lengths distribution, target=0.6.}
		\end{subfigure}
		\begin{subfigure}[t]{\Histogram}
			\centering
			\includegraphics[width=\textwidth]{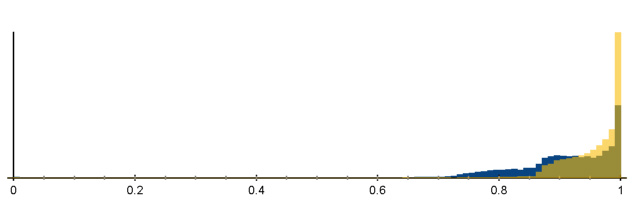}
			\caption{Distribution of quality $Q_t$, target=1.0.}
		\end{subfigure}
	\end{minipage}
	\caption{Histograms for remeshed \emph{Screw}, with (blue) and without (yellow) feature detection,~$\vartheta = 90^{\circ}$.}
\end{figure}

\begin{table}[b!]
	\centering
	\begin{tabular}{lrrrlrrrr}
		$d$ & $\vartheta$ & $| \mathcal{V} |$ & $| \mathcal{T} |$ & $d_{\text{avg}}$ & $d_{\max}$ & $d_{\text{RMS}}$ & $|\text{feature segments}|$ & $|\text{feature vertices}|$\\
		\hline
		0.6 & --- & 13,411 & 26,842 & 0.0438 & 0.4605 & 138.1 & --- & --- \\
		0.6 & $90^{\circ}$ & 13,472 & 26,988 & 0.0190 & 0.3595 & 122.7 & 661,301 & 3,743
	\end{tabular}
	\caption{Comparison of results achieved for \emph{Screw}, with and without feature detection.}
\end{table}

\newpage

\subsection{\emph{Wrench}} \textcolor{sowaswieweiss}{y}

\begin{figure}[h!]
	\centering
	\begin{minipage}{0.45\textwidth}
		\begin{subfigure}{\textwidth}
			\includegraphics[width=1.\textwidth]{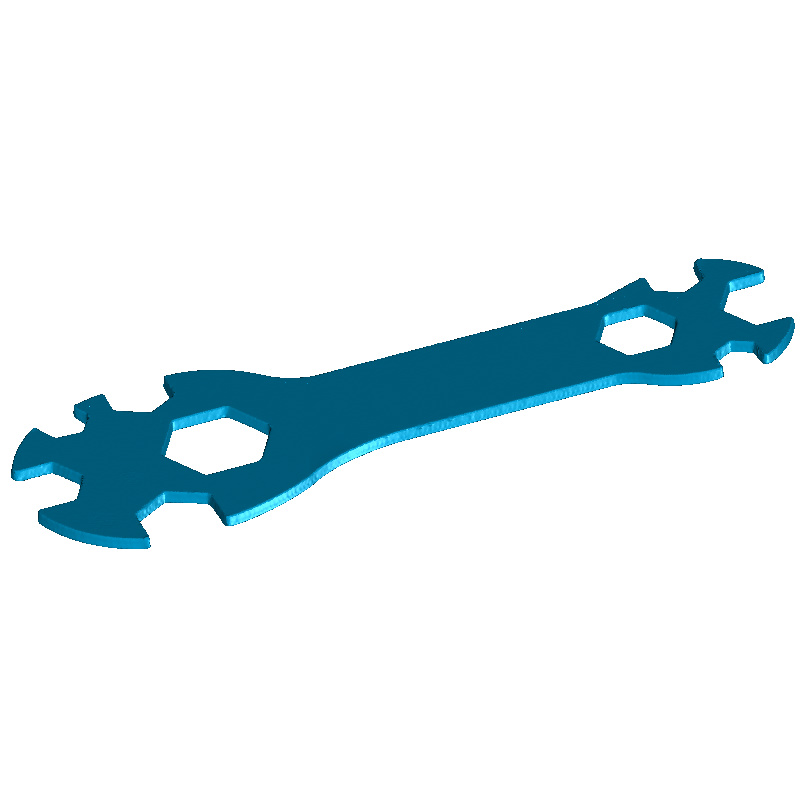}
		\end{subfigure}
	\end{minipage}
	\hfill
	\begin{minipage}{0.5\textwidth}
		\begin{subfigure}[t]{\Histogram}
			\centering
			\includegraphics[width=\textwidth]{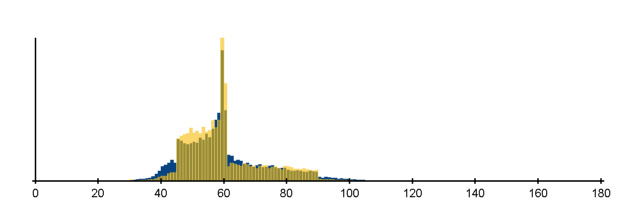}
			\caption{Angle distribution, target=$60^\circ$.}
		\end{subfigure}
		\begin{subfigure}[t]{\Histogram}
			\centering
			\includegraphics[width=\textwidth]{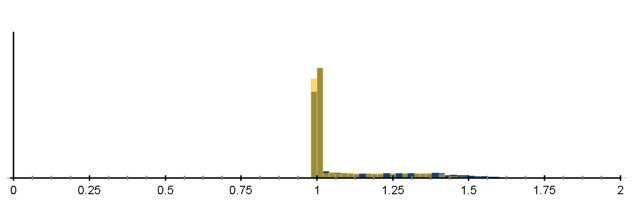}
			\caption{Edge lengths distribution, target=1.0.}
		\end{subfigure}
		\begin{subfigure}[t]{\Histogram}
			\centering
			\includegraphics[width=\textwidth]{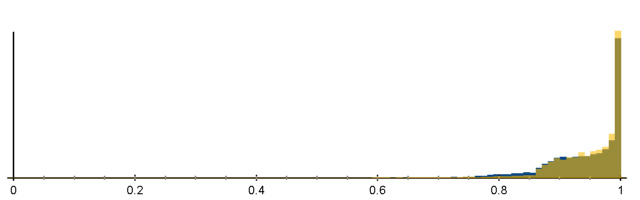}
			\caption{Distribution of quality $Q_t$, target=1.0.}
		\end{subfigure}
	\end{minipage}
	\caption{
		\emph{Wrench}. Point cloud and histograms.
	}
\end{figure}

\begin{table}[b!]
	\centering
	\begin{tabular}{lrrrrlrrr}
		$d$ & $\vartheta$ & $| \mathcal{V} |$ & $| \mathcal{T} |$ & $d_{\text{avg}}$ & $d_{\max}$ & $d_{\text{RMS}}$ & $|\text{feature segments}|$ & $|\text{feature vertices}|$\\
		\hline
		1.0 & --- & 8,946 & 17,896 & 0.0360 & 0.6630 & 207.6 & --- & --- \\
		1.0 & $60^{\circ}$ & 8,871 & 17,746 & 0.0112 & 0.4353 & 189.7 & 583,665 & 1156
	\end{tabular}
	\caption{Comparison of results achieved for \emph{Wrench}, with and without feature detection,~$\vartheta = 60^{\circ}$.}
\end{table}

\newpage

\subsection{\emph{Xiao Jie Jie}} \textcolor{sowaswieweiss}{y}

\begin{figure}[h!]
	\centering
	\begin{minipage}{0.48\textwidth}
		\begin{subfigure}{\textwidth}
			\centering
			\includegraphics[width=.7\textwidth]{xiaojiejie2_SmallPart_d0p4_s0p2_angle70}
		\end{subfigure}
	\end{minipage}
	\hfill
	\begin{minipage}{0.24\textwidth}
		\begin{subfigure}[t]{\Histogram}
			\centering
			\includegraphics[width=\textwidth]{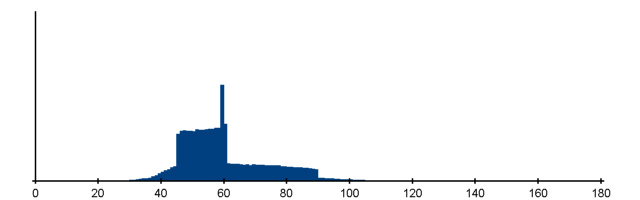}
			\caption{Angle distribution, target=$60^\circ$,~$\vartheta = 40^{\circ}$.}
		\end{subfigure}
		\begin{subfigure}[t]{\Histogram}
			\centering
			\includegraphics[width=\textwidth]{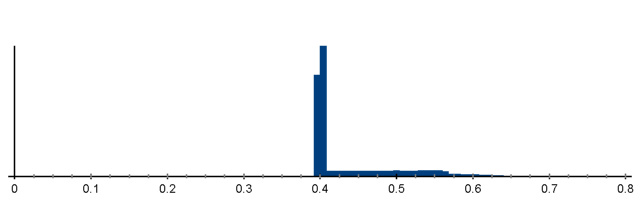}
			\caption{Edge lengths distribution, target=0.4.}
		\end{subfigure}
		\begin{subfigure}[t]{\Histogram}
			\centering
			\includegraphics[width=\textwidth]{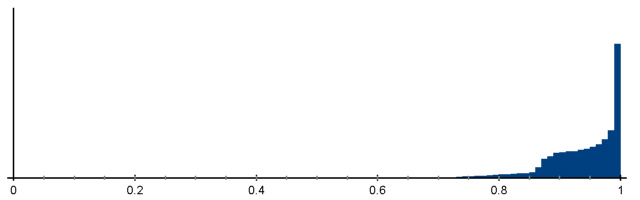}
			\caption{Distribution of quality $Q_t$, target=1.0.}
		\end{subfigure}
	\end{minipage}
	\hfill
	\begin{minipage}{0.24\textwidth}
		\begin{subfigure}[t]{\Histogram}
			\centering
			\includegraphics[width=\textwidth]{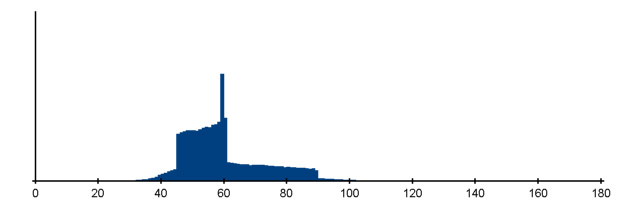}
			\caption{Angle distribution, target=$60^\circ$,~$\vartheta = 50^{\circ}$.}
		\end{subfigure}
		\begin{subfigure}[t]{\Histogram}
			\centering
			\includegraphics[width=\textwidth]{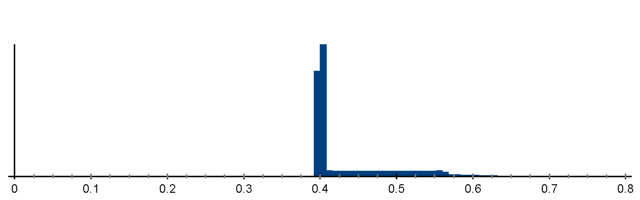}
			\caption{Edge lengths distribution, target=0.4.}
		\end{subfigure}
		\begin{subfigure}[t]{\Histogram}
			\centering
			\includegraphics[width=\textwidth]{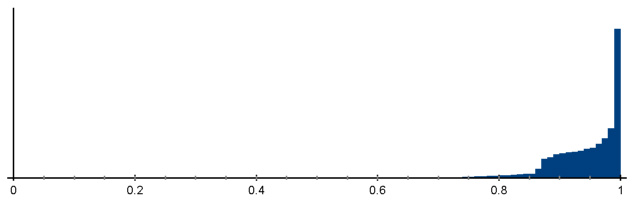}
			\caption{Distribution of quality $Q_t$, target=1.0.}
		\end{subfigure}
	\end{minipage}
	
	\begin{minipage}{0.24\textwidth}
		\begin{subfigure}[t]{\Histogram}
			\centering
			\includegraphics[width=\textwidth]{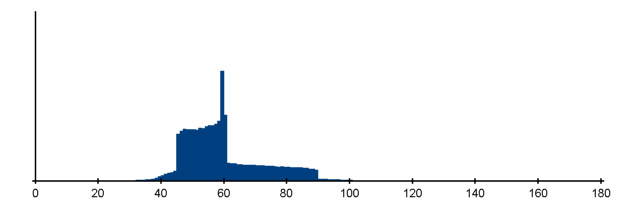}
			\caption{Angle distribution, target=$60^\circ$,~$\vartheta = 60^{\circ}$.}
		\end{subfigure}
		\begin{subfigure}[t]{\Histogram}
			\centering
			\includegraphics[width=\textwidth]{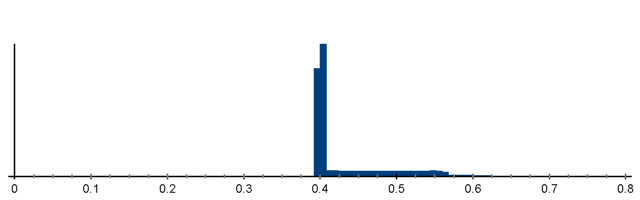}
			\caption{Edge lengths distribution, target=0.4.}
		\end{subfigure}
		\begin{subfigure}[t]{\Histogram}
			\centering
			\includegraphics[width=\textwidth]{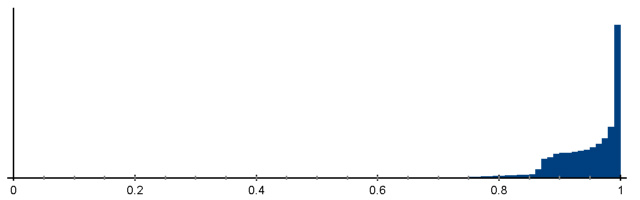}
			\caption{Distribution of quality $Q_t$, target=1.0.}
		\end{subfigure}
	\end{minipage}
	\hfill
	\begin{minipage}{0.24\textwidth}
		\begin{subfigure}[t]{\Histogram}
			\centering
			\includegraphics[width=\textwidth]{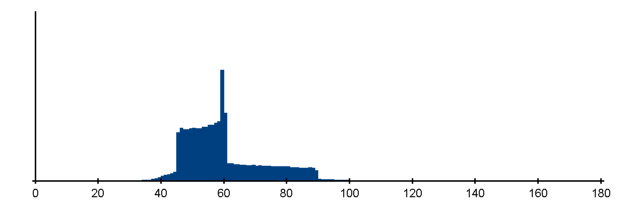}
			\caption{Angle distribution, target=$60^\circ$,~$\vartheta = 70^{\circ}$.}
		\end{subfigure}
		\begin{subfigure}[t]{\Histogram}
			\centering
			\includegraphics[width=\textwidth]{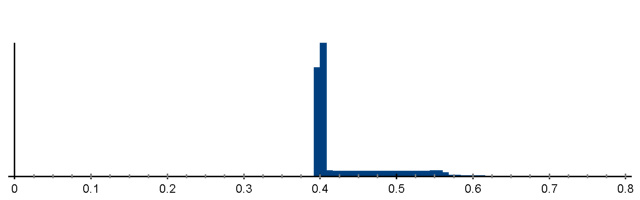}
			\caption{Edge lengths distribution, target=0.4.}
		\end{subfigure}
		\begin{subfigure}[t]{\Histogram}
			\centering
			\includegraphics[width=\textwidth]{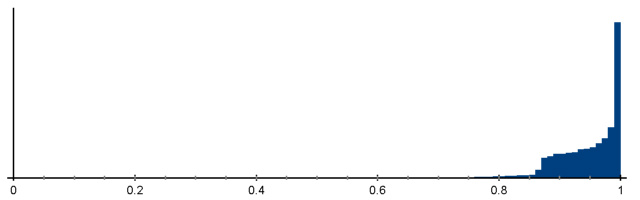}
			\caption{Distribution of quality $Q_t$, target=1.0.}
		\end{subfigure}
	\end{minipage}
	\hfill
	\begin{minipage}{0.24\textwidth}
		\begin{subfigure}[t]{\Histogram}
			\centering
			\includegraphics[width=\textwidth]{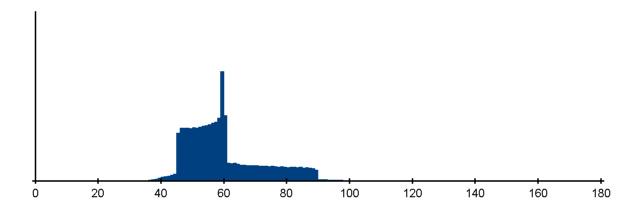}
			\caption{Angle distribution, target=$60^\circ$,~$\vartheta = 80^{\circ}$.}
		\end{subfigure}
		\begin{subfigure}[t]{\Histogram}
			\centering
			\includegraphics[width=\textwidth]{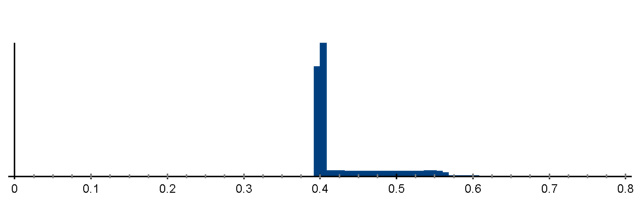}
			\caption{Edge lengths distribution, target=0.4.}
		\end{subfigure}
		\begin{subfigure}[t]{\Histogram}
			\centering
			\includegraphics[width=\textwidth]{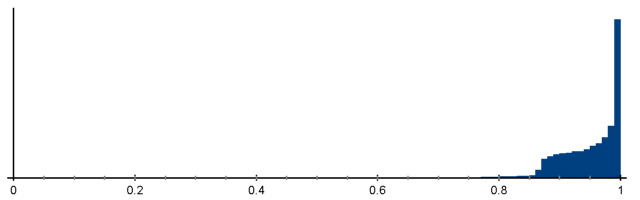}
			\caption{Distribution of quality $Q_t$, target=1.0.}
		\end{subfigure}
	\end{minipage}
	\hfill
	\begin{minipage}{0.24\textwidth}
		\begin{subfigure}[t]{\Histogram}
			\centering
			\includegraphics[width=\textwidth]{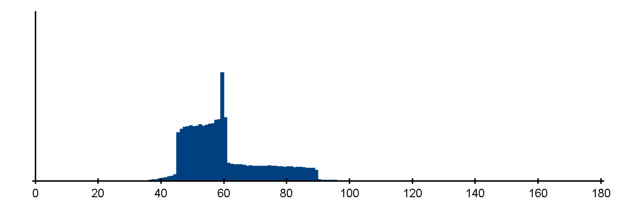}
			\caption{Angle distribution, target=$60^\circ$,~$\vartheta = 90^{\circ}$.}
		\end{subfigure}
		\begin{subfigure}[t]{\Histogram}
			\centering
			\includegraphics[width=\textwidth]{xiaojiejie2_s0p2_d0p4_feature80_histogram_edge}
			\caption{Edge lengths distribution, target=0.4.}
		\end{subfigure}
		\begin{subfigure}[t]{\Histogram}
			\centering
			\includegraphics[width=\textwidth]{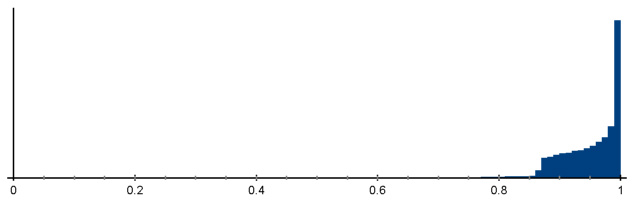}
			\caption{Distribution of quality $Q_t$, target=1.0.}
		\end{subfigure}
	\end{minipage}
	\caption{
		\emph{Xiao Jie Jie}, remesh for $\vartheta = 70^{\circ}$ and histograms for various thresholds.
	}
\end{figure}

\begin{table}[h!]
	\centering
	\begin{tabular}{llrrrrrr}
		$d$ & $\vartheta$ & $| \mathcal{V} |$ & $| \mathcal{T} |$ & $\alpha_{\min}$ & $\alpha_{\max}$ & $\alpha_{\text{avg}}$ & $\alpha_{\text{RMS}}$\\
		\hline
		0.4 & $40^{\circ}$ & 88,822 & 177,573 & $8.2213^{\circ}$ & $156.0215^{\circ}$ & $60^{\circ}$ & 22.3 \\
		0.4 & $50^{\circ}$ & 90,146 & 180,243 & $9.8948^{\circ}$ & $160.2103^{\circ}$ & $60^{\circ}$ & 21.2 \\
		0.4 & $60^{\circ}$ & 90,647 & 181,294 & $8.5513^{\circ}$ & $162.8973^{\circ}$ & $60^{\circ}$ & 20.8 \\
		\hline
		0.4 & $70^{\circ}$ & 90,945 & 181,861 & $22.0826^{\circ}$ & $135.8347^{\circ}$ & $60^{\circ}$ & 20.5 \\
		0.4 & $80^{\circ}$ & 91,239 & 182,454 & $26.9329^{\circ}$ & $124.4023^{\circ}$ & $60^{\circ}$ & 20.2 \\
		0.4 & $90^{\circ}$ & 91,400 & 182,786 & $28.4372^{\circ}$ & $123.1255^{\circ}$ & $60^{\circ}$ & 20.1 \\
		& & & & & & & \\
		$E_{\min}$ & $E_{\max}$ & $E_{\text{avg}}$ & $E_{\text{RMS}}$ & $Q_{\min}$ & $Q_{\max}$ & $Q_{\text{avg}}$ & $Q_{\text{RMS}}$ \\
		\hline
		0.3999 & 2.3119 & 0.4384 & 15.0 & 0.2407 & 1.0000 & 0.9420 & 6.1 \\
		0.3999 & 1.1101 & 0.4341 & 13.7 & 0.1993 & 1.0000 & 0.9474 & 5.6 \\
		0.3999 & 0.8867 & 0.4323 & 13.1 & 0.1723 & 1.0000 & 0.9494 & 5.3 \\
		\hline
		0.3999 & 0.8564 & 0.4311 & 12.8 & 0.4441 & 1.0000 & 0.9508 & 5.1 \\
		0.3999 & 0.8406 & 0.4299 & 12.4 & 0.5571 & 1.0000 & 0.9524 & 4.9 \\
		0.3999 & 0.7903 & 0.4294 & 12.2 & 0.5696 & 1.0000 & 0.9530 & 4.8
	\end{tabular}
	\caption{Comparison of results achieved for \emph{Xiao Jie Jie}, with feature detection.}
\end{table}

\newpage

\section{CAD Models for Feature Detection}
\label{sec:CADModelsForFeatureDetection}

\subsection{\emph{Flange}} \textcolor{sowaswieweiss}{y}

\begin{figure}[h!]
	\centering
	\begin{subfigure}[t]{.3\textwidth}
		\includegraphics[width=\textwidth]{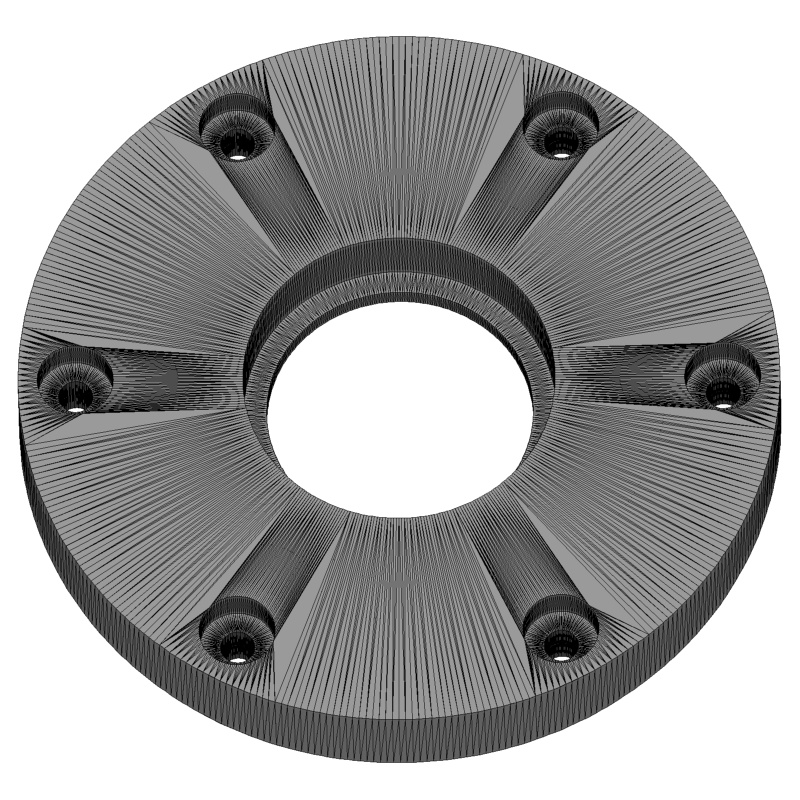}
	\end{subfigure}
	~
	\begin{subfigure}[t]{.3\textwidth}
		\includegraphics[width=\textwidth]{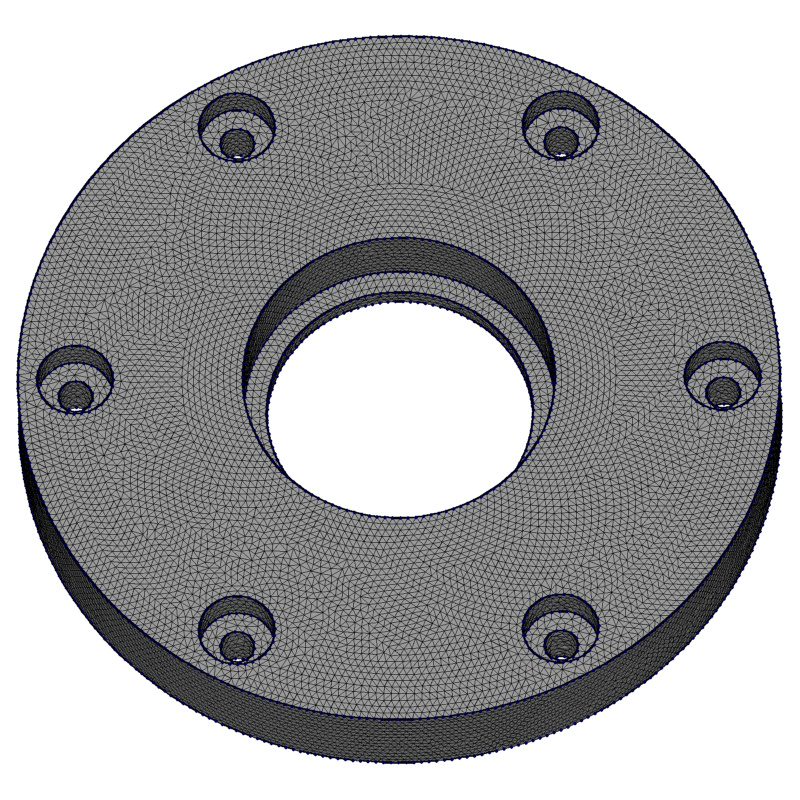}
	\end{subfigure}
	~
	\begin{subfigure}[t]{.3\textwidth}
		\includegraphics[width=\textwidth]{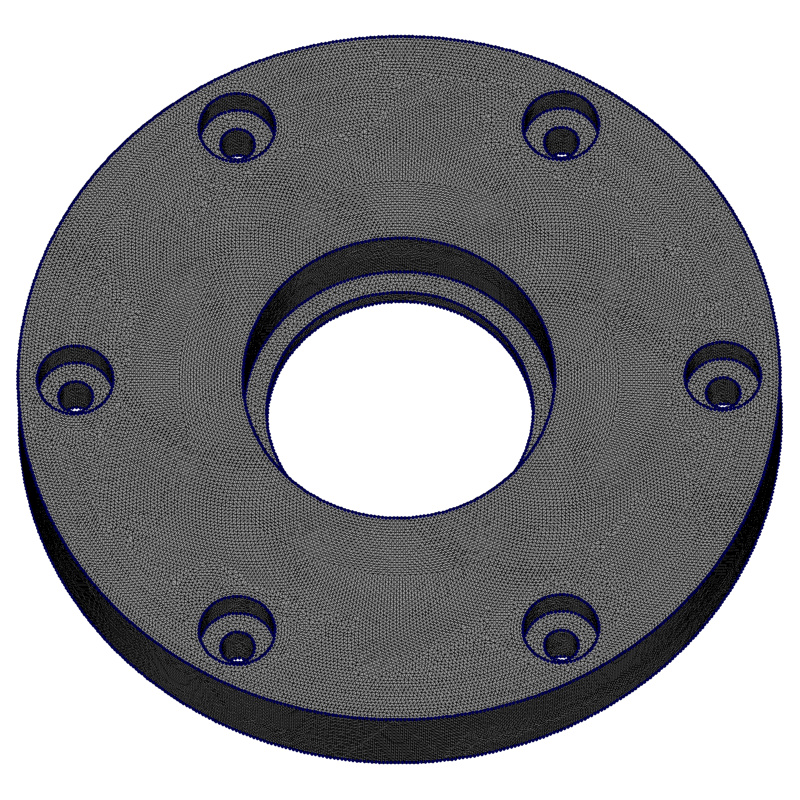}
	\end{subfigure}
	\caption{\emph{Flange}.
		Left: CAD model.
		Middle: Model remeshed with features detected for~$\vartheta = 80^{\circ}$ and target edge length~$d = 2.0$.
		Right: Model remeshed with features detected for~$\vartheta = 80^{\circ}$ and target edge length~$d = 1.0$.}
\end{figure}

\begin{figure}[b!]
	\centering
	\begin{minipage}{0.38\textwidth}
		\begin{subfigure}[t]{\Histogram}
			\centering
			\includegraphics[width=\textwidth]{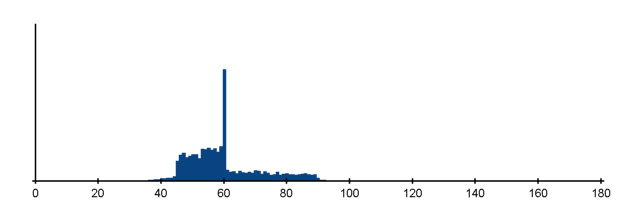}
			\caption{Angle distribution, target=$60^\circ$.}
		\end{subfigure}
		\begin{subfigure}[t]{\Histogram}
			\centering
			\includegraphics[width=\textwidth]{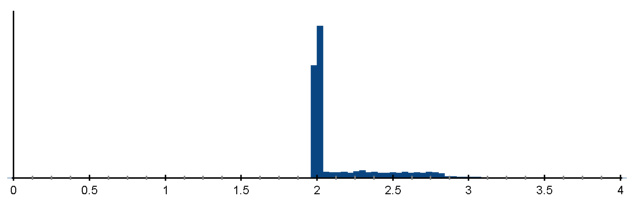}
			\caption{Edge lengths distribution, target=2.0.}
		\end{subfigure}
		\begin{subfigure}[t]{\Histogram}
			\centering
			\includegraphics[width=\textwidth]{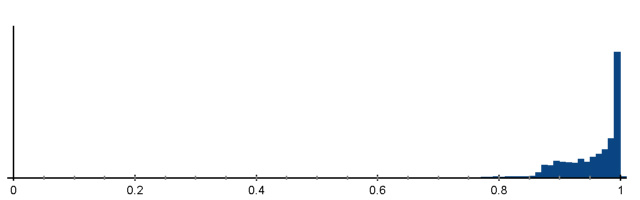}
			\caption{Distribution of quality $Q_t$, target=1.0.}
		\end{subfigure}
	\end{minipage}
	~
	\begin{minipage}{0.38\textwidth}
		\begin{subfigure}[t]{\Histogram}
			\centering
			\includegraphics[width=\textwidth]{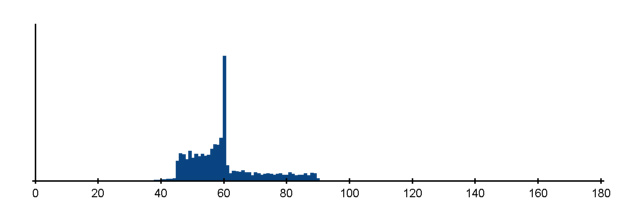}
			\caption{Angle distribution, target=$60^\circ$.}
		\end{subfigure}
		\begin{subfigure}[t]{\Histogram}
			\centering
			\includegraphics[width=\textwidth]{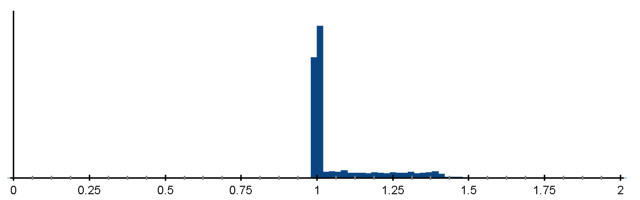}
			\caption{Edge lengths distribution, target=1.0.}
		\end{subfigure}
		\begin{subfigure}[t]{\Histogram}
			\centering
			\includegraphics[width=\textwidth]{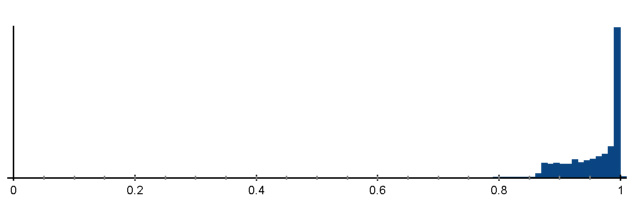}
			\caption{Distribution of quality $Q_t$, target=1.0.}
		\end{subfigure}
	\end{minipage}
	\caption{Histograms for remeshed \emph{Flange},~$\vartheta = 80^{\circ}$.}
\end{figure}

\newpage

\begin{table}[h!]
	\centering
	\begin{tabular}{lllrrrrrr}
		Algorithm & $d$ & $\vartheta$ & $| \mathcal{V} |$ & $| \mathcal{T} |$ & $\alpha_{\min}$ & $\alpha_{\max}$& $\alpha_{\text{avg}}$ & $\alpha_{\text{RMS}}$\\
		\hline
		MeshLab & 2.0 & 80$^{\circ}$ & 23,627 & 47,378 & 14.8313$^{\circ}$ & 122.7730$^{\circ}$ & 60$^{\circ}$ & 18.9 \\
		PMP & 2.0 & 80$^{\circ}$ & 28,432 & 56,888 & 15.9656$^{\circ}$ & 137.0463$^{\circ}$ & 60$^{\circ}$ & 16.3 \\
		ours & 2.0 & 80$^{\circ}$ & 21,633 & 43,290 & 25.8987$^{\circ}$ & 128.2024$^{\circ}$ & 60$^{\circ}$ & 19.7\\
		\hline
		MeshLab & 1.0 & 80$^{\circ}$ & 94,041 & 188,106 & 16.9999$^{\circ}$ & 128.7068$^{\circ}$ & 60$^{\circ}$ & 17.8 \\
		PMP & 1.0 & 80$^{\circ}$ & 115,420 & 230,864 & 25.9868$^{\circ}$ & 123.7041$^{\circ}$ & 60$^{\circ}$ & 17.2 \\
		ours & 1.0 & 80$^{\circ}$ & 87,643 & 175,310 & 25.5484$^{\circ}$ & 128.9030$^{\circ}$ & 60$^{\circ}$ & 18.9\\
		\hline
		MeshLab & 1.0 & -- & 109,371 & 218,766 & 0.3647$^{\circ}$ & 178.7277$^{\circ}$ & 60$^{\circ}$ & 30.6 \\
		PMP & 1.0 & -- & 113,411 & 226,906 & 32.0818$^{\circ}$ & 104.7587$^{\circ}$ & 60$^{\circ}$ & 16.6 \\
		ours & 1.0 & -- & 86,548 & 173,122 & 29.7087$^{\circ}$ & 109.9805$^{\circ}$ & 60$^{\circ}$ & 19.2\\
		& & & & & & & & \\
		$E_{\min}$ & $E_{\text{max}}$ & $E_{\text{avg}}$ & $E_{\text{RMS}}$ & $Q_{\text{min}}$ & $Q_{\text{max}}$ & $Q_{\text{avg}}$ & $Q_{\text{RMS}}$ & \\
		\hline
		0.6472 & 3.7574 & 2.0475 & 14.2 & 0.4284 & 0.9999 & 0.9494 & 6.4 & \\
		1.1490 & 3.3748 & 1.8591 & 12.1 & 0.3950 & 0.9999 & 0.9641 & 3.8 & \\
		2.0 & 4.2169 & 2.1454 & 12.5 & 0.5198 & 1.0000 & 0.9546 & 5.2 & \\
		\hline
		0.3532 & 1.9557 & 1.0241 & 13.2 & 0.4545 & 0.9999 & 0.9562 & 5.1 & \\
		0.5283 & 1.6284 & 0.9240 & 12.6 & 0.5626 & 0.9999 & 0.9601 & 3.8 & \\
		1.0 & 2.0680 & 1.0645 & 11.8 & 0.5128 & 1.0000 & 0.9585 & 4.9 & \\
		\hline
		0.0136 & 1.8675 & 0.9361 & 27.7 & 0.0099 & 1.0 & 0.8823 & 21.4 &  \\
		0.5800 & 1.5010 & 0.9269 & 12.2 & 0.7352 & 1.0 & 0.9626 & 3.5 &  \\
		1.0 & 1.9997 & 1.0666 & 11.4 & 0.6002 & 1.0 & 0.9572 & 4.5 &  
	\end{tabular}
	\caption{Experimental results for \emph{Flange}.}
\end{table}

\begin{table}[h!]
	\centering
	\begin{tabular}{lrrrr}
		Algorithm & $d$ & $\vartheta$ & $d_{\max}$ & $\frac{d_{\max}}{d}$ \\
		\hline
		MeshLab & 1.0 & 80$^{\circ}$ & 0.0590 & 0.0590\\
		PMP     & 1.0 & 80$^{\circ}$ & 0.0289 & 0.0289 \\
		ours	& 1.0 & 80$^{\circ}$ & 0.0608 & 0.0608 \\
		\hline
		MeshLab & 2.0 & 80$^{\circ}$ & 0.1781 & 0.0890\\
		PMP     & 2.0 & 80$^{\circ}$ & 0.1025 & 0.0512\\
		ours	& 2.0 & 80$^{\circ}$ & 0.3505 & 0.1752
	\end{tabular}
	\caption{One-sided Hausdorff distance evaluated on the \emph{Flange}.}
\end{table}

\newpage

\subsection{\emph{Body of Constant Width}} \textcolor{sowaswieweiss}{y}
\label{sec:BodyOfConstantWidth}

\begin{figure}[h!]
	\centering
	\begin{subfigure}[t]{.32\textwidth}
		\includegraphics[width=\textwidth]{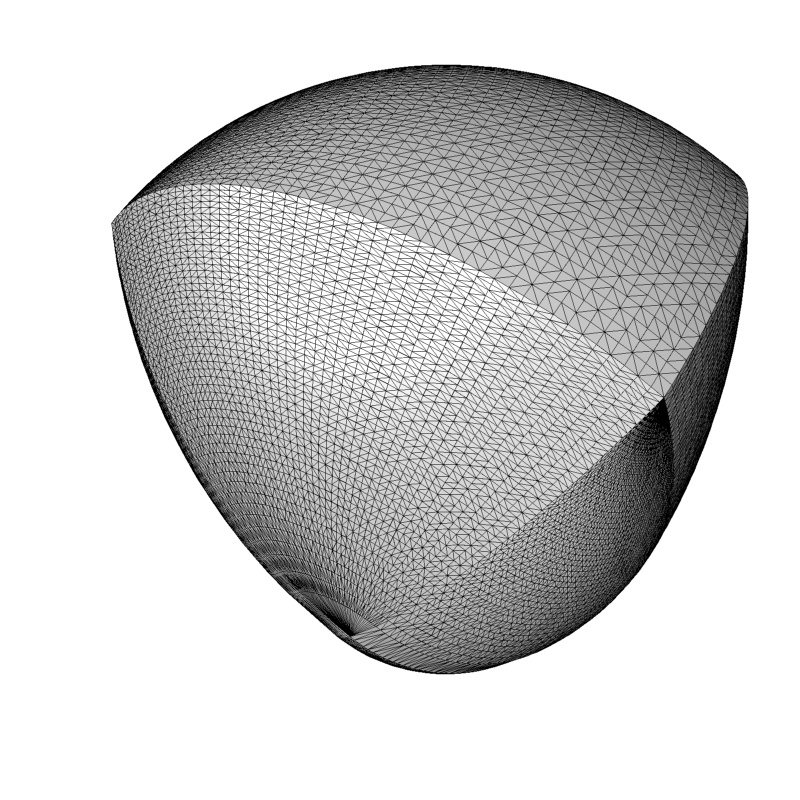}
	\end{subfigure}
	~
	\begin{subfigure}[t]{.32\textwidth}
		\includegraphics[width=\textwidth]{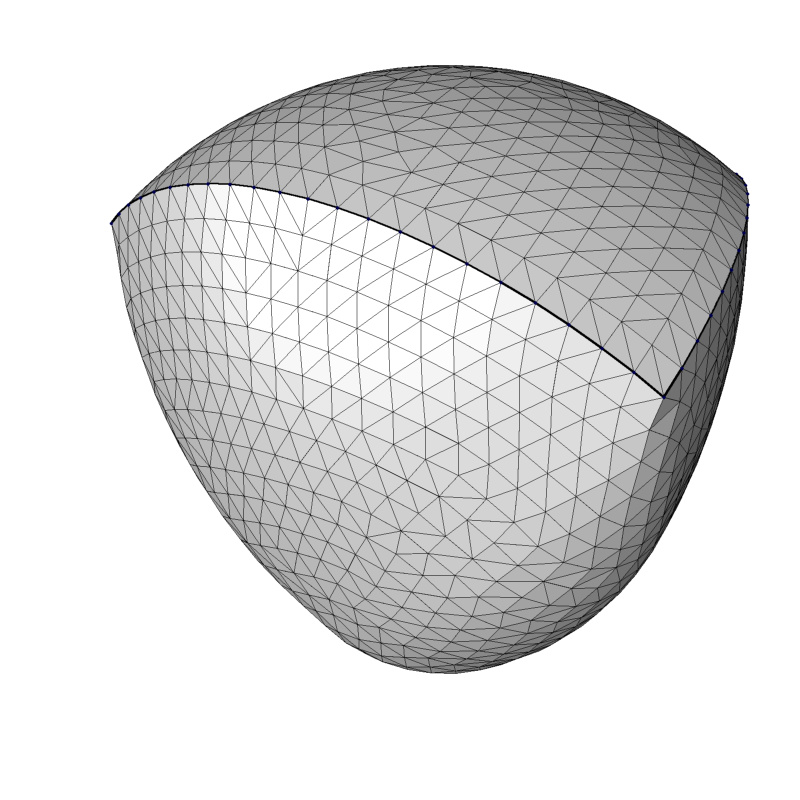}
	\end{subfigure}
	~
	\begin{subfigure}[t]{.32\textwidth}
		\includegraphics[width=\textwidth]{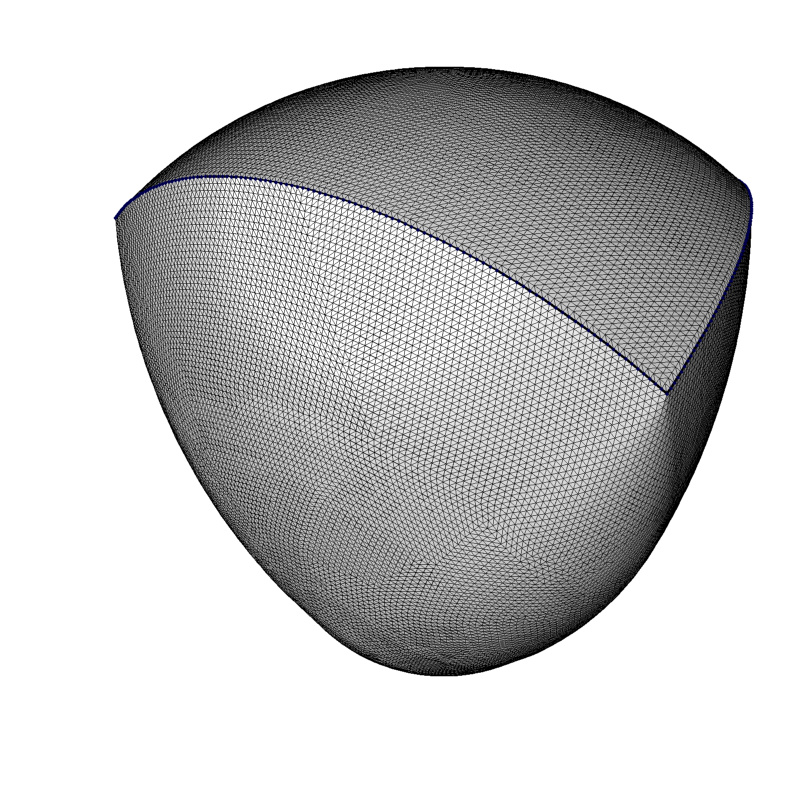}
	\end{subfigure}
	\caption{\emph{Body of Constant Width}.
		Left: CAD model.
		Middle: Model remeshed with features detected for~$\vartheta = 40^{\circ}$ and target edge length~$d = 0.1$.
		Right: Model remeshed with features detected for~$\vartheta = 40^{\circ}$ and target edge length~$d = 0.02$.}
\end{figure}

\begin{figure}[b!]
	\centering
	\begin{minipage}{0.4\textwidth}
		\begin{subfigure}[t]{\Histogram}
			\centering
			\includegraphics[width=\textwidth]{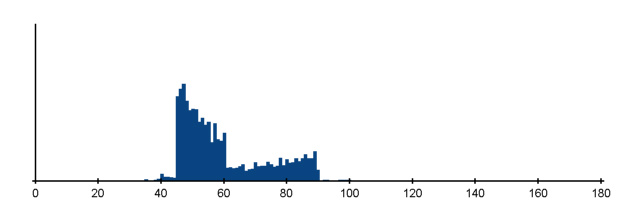}
			\caption{Angle distribution, target=$60^\circ$.}
		\end{subfigure}
		\begin{subfigure}[t]{\Histogram}
			\centering
			\includegraphics[width=\textwidth]{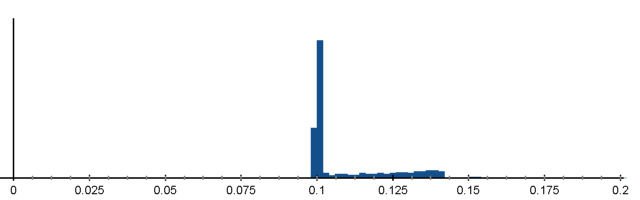}
			\caption{Edge lengths distribution, target=0.1.}
		\end{subfigure}
		\begin{subfigure}[t]{\Histogram}
			\centering
			\includegraphics[width=\textwidth]{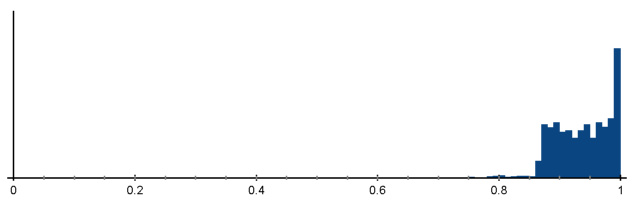}
			\caption{Distribution of quality $Q_t$, target=1.0.}
		\end{subfigure}
	\end{minipage}
	~
	\begin{minipage}{0.4\textwidth}
		\begin{subfigure}[t]{\Histogram}
			\centering
			\includegraphics[width=\textwidth]{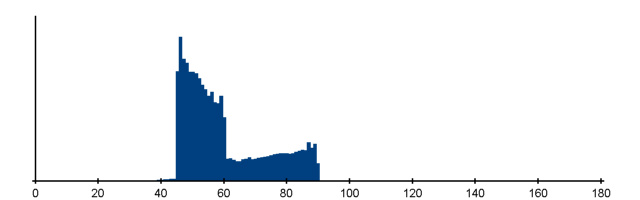}
			\caption{Angle distribution, target=$60^\circ$.}
		\end{subfigure}
		\begin{subfigure}[t]{\Histogram}
			\centering
			\includegraphics[width=\textwidth]{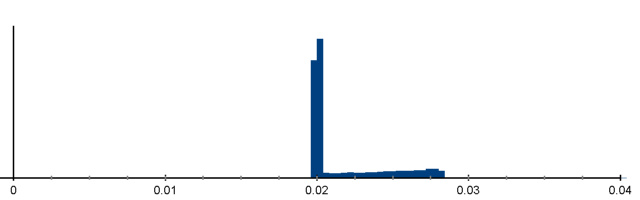}
			\caption{Edge lengths distribution, target=0.02.}
		\end{subfigure}
		\begin{subfigure}[t]{\Histogram}
			\centering
			\includegraphics[width=\textwidth]{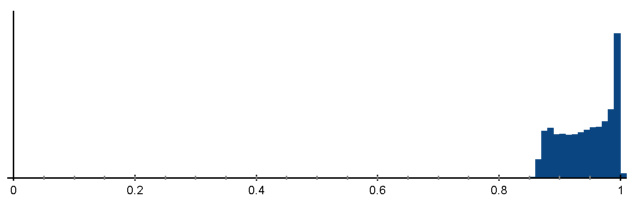}
			\caption{Distribution of quality $Q_t$, target=1.0.}
		\end{subfigure}
	\end{minipage}
	\caption{Histograms for remeshed \emph{Body of Constant Width},~$\vartheta = 40^{\circ}$.}
\end{figure}

\newpage

\begin{table}[h!]
	\centering
	\begin{tabular}{llllrrrrrr}
		Algorithm & $| \text{Iterations}|$ & $d$ & $\vartheta$ & $| \mathcal{V} |$ & $| \mathcal{T} |$ & $\alpha_{\min}$ & $\alpha_{\max}$& $\alpha_{\text{avg}}$ & $\alpha_{\text{RMS}}$\\
		\hline
		MeshLab & 10 & 0.1 & 40$^{\circ}$ & 1,374 & 2,744 & 26.8817$^{\circ}$ & 109.9683$^{\circ}$ & 60$^{\circ}$ & 19.2 \\
		MeshLab & 100 & 0.1 & 40$^{\circ}$ & 1,303 & 2,602 & 27.1838$^{\circ}$ & 106.8252$^{\circ}$ & 60$^{\circ}$ & 11.8 \\
		MeshLab & 1,000 & 0.1 & 40$^{\circ}$ & 1,281 & 2,558 & 32.7818$^{\circ}$ & 106.4673$^{\circ}$ & 60$^{\circ}$ & 10.4 \\
		PMP & 10 & 0.1 & 40$^{\circ}$ & 1,600 & 3,196 & 29.4472$^{\circ}$ & 114.4853$^{\circ}$ & 60$^{\circ}$ & 14.8 \\
		ours & 1 & 0.1 & 40$^{\circ}$ & 1,235 & 2,466 & 29.5060$^{\circ}$ & 120.9878$^{\circ}$ & 60$^{\circ}$ & 23.2\\
		\hline
		MeshLab & 10 & 0.02 & 40$^{\circ}$ & 32,737 & 65,470 & 21.4255$^{\circ}$ & 117.8555$^{\circ}$ & 60$^{\circ}$ & 15.9 \\
		MeshLab & 100 & 0.02 & 40$^{\circ}$ & 31,833 & 63,662 & 22.2400$^{\circ}$ & 114.7592$^{\circ}$ & 60$^{\circ}$ & 10.6 \\
		MeshLab & 1,000 & 0.02 & 40$^{\circ}$ & 31,487 & 62,970 & 23.6324$^{\circ}$ & 110.2686$^{\circ}$ & 60$^{\circ}$ & 8.2 \\
		PMP & 10 & 0.02 & 40$^{\circ}$ & 39,052 & 78,100 & 26.6413$^{\circ}$ & 115.9396$^{\circ}$ & 60$^{\circ}$ & 15.4 \\
		ours & 1 & 0.02 & 40$^{\circ}$ & 31,277 & 62,550 & 30.1203$^{\circ}$ & 115.0667$^{\circ}$ & 60$^{\circ}$ & 22.1\\
		& & & & & & & & & \\
		$E_{\min}$ & $E_{\text{max}}$ & $E_{\text{avg}}$ & $E_{\text{RMS}}$ & $Q_{\text{min}}$ & $Q_{\text{max}}$ & $Q_{\text{avg}}$ & $Q_{\text{RMS}}$ & & \\
		\hline
		0.0561 & 0.1749 & 0.1021 & 13.4 & 0.6525 & 0.9999 & 0.9516 & 4.9 &  &  \\
		0.0512 & 0.1371 & 0.1036 & 10.5 & 0.6835 & 0.9999 & 0.9812 & 2.7 &  &  \\
		0.0566 & 0.1333 & 0.1043 & 10.1 & 0.7250 & 0.9999 & 0.9850 & 2.2 &  &  \\
		0.0603 & 0.1400 & 0.0940 & 11.5 & 0.6441 & 0.9999 & 0.9709 & 3.1 &  &  \\
		0.1 & 0.1884 & 0.1088 & 13.2 & 0.5904 & 1.0000 & 0.9380 & 13.2 &  & \\
		\hline
		0.0094 & 0.0366 & 0.0208 & 11.5 & 0.5774 & 0.9999 & 0.9645 & 4.4 &  &  \\
		0.0096 & 0.0335 & 0.0209 & 9.2 & 0.6041 & 0.9999 & 0.9840 & 2.5 &  &  \\
		0.0102 & 0.0304 & 0.0210 & 8.7 & 0.6223 & 0.9999 & 0.9903 & 1.9 &  &  \\
		0.0123 & 0.0328 & 0.0190 & 11.9 & 0.6155 & 0.9999 & 0.9678 & 3.5 &  &  \\
		0.02 & 0.0394 & 0.0215 & 12.3 & 0.6473 & 1.0000 & 0.9437 & 4.6 & &
	\end{tabular}
	\caption{Experimental results for \emph{Body of Constant Width}.}
\end{table}

\begin{table}[h!]
	\centering
	\begin{tabular}{lrrrrr}
		Algorithm & $|$Iterations$|$ & $d$ & $\vartheta$ & $d_{\max}$ & $\frac{d_{\max}}{d}$ \\
		\hline
		MeshLab & 10 & 0.02 & 40$^{\circ}$ & 0.0046 & 0.2314 \\
		MeshLab & 100 & 0.02 & 40$^{\circ}$ & 0.0046 & 0.2315 \\
		PMP & 10 & 0.02 & 40$^{\circ}$ & 0.0038 & 0.1934 \\
		ours  & 1 & 0.02 & 40$^{\circ}$ & 0.0017 & 0.0868 \\
		\hline
		MeshLab & 10 & 0.1 & 40$^{\circ}$ & 0.0090 & 0.0904 \\
		MeshLab & 100 & 0.1 & 40$^{\circ}$ & 0.0108 & 0.1087 \\
		MeshLab & 1000 & 0.1 & 40$^{\circ}$ & 0.0135 & 0.1351 \\
		PMP & 10 & 0.1 & 40$^{\circ}$ & 0.0248 & 0.2480 \\
		ours & 1 & 0.1 & 40$^{\circ}$ & 0.0140 & 0.1404
	\end{tabular}
	\caption{One-sided Hausdorff distance evaluated on the \emph{Body of Constant Width}.}
\end{table}

\newpage

\subsection{\emph{Joint}} \textcolor{sowaswieweiss}{y}

\begin{figure}[h!]
	\centering
	\begin{subfigure}[t]{.32\textwidth}
		\includegraphics[width=\textwidth]{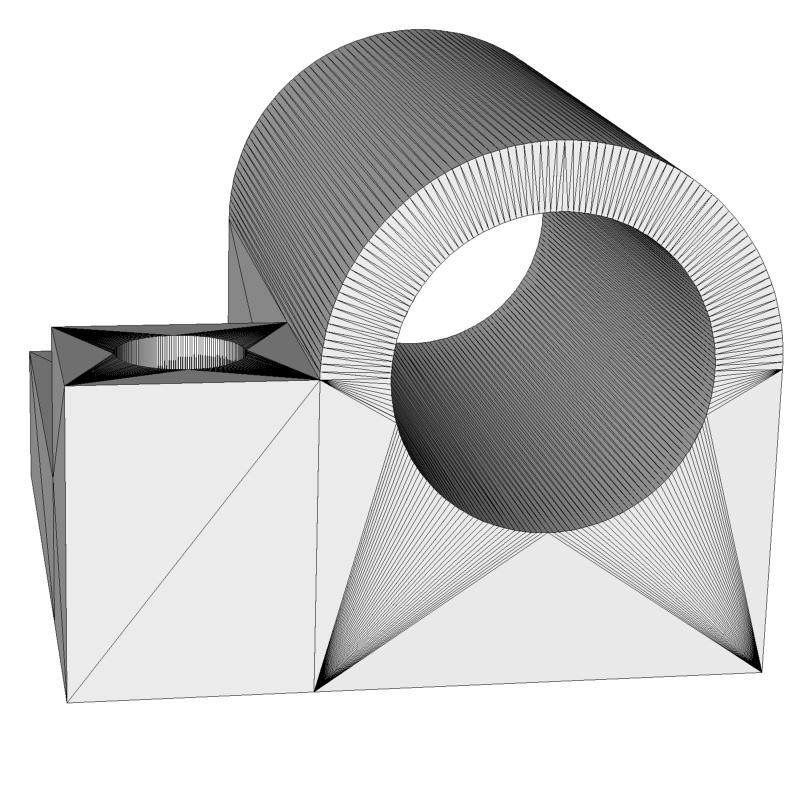}
	\end{subfigure}
	~
	\begin{subfigure}[t]{.32\textwidth}
		\includegraphics[width=\textwidth]{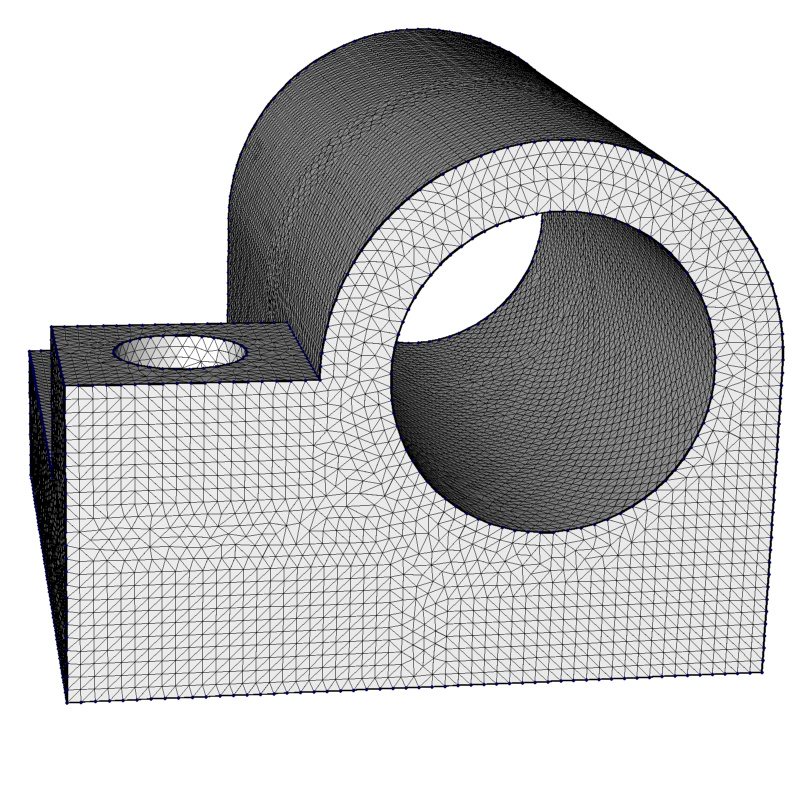}
	\end{subfigure}
	~
	\begin{subfigure}[t]{.32\textwidth}
		\includegraphics[width=\textwidth]{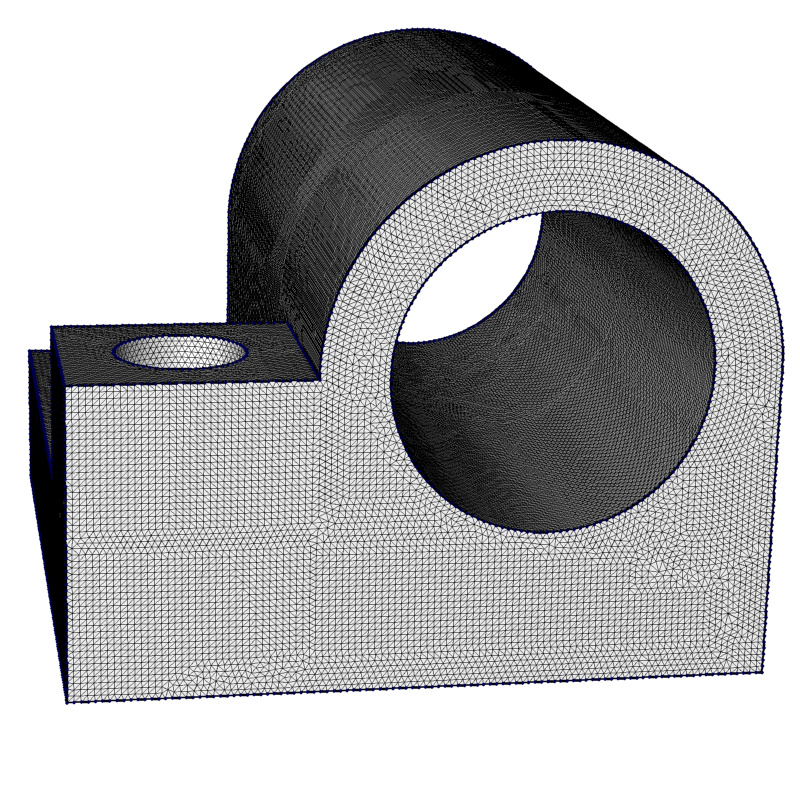}
	\end{subfigure}
	\caption{\emph{Joint}.
		Left: CAD model.
		Middle: Model remeshed with features detected for~$\vartheta = 80^{\circ}$ and target edge length~$d = 0.02$.
		Right: Model remeshed with features detected for~$\vartheta = 80^{\circ}$ and target edge length~$d = 0.01$.
	}
\end{figure}

\begin{figure}[b!]
	\centering
	\begin{minipage}{0.4\textwidth}
		\begin{subfigure}[t]{\Histogram}
			\centering
			\includegraphics[width=\textwidth]{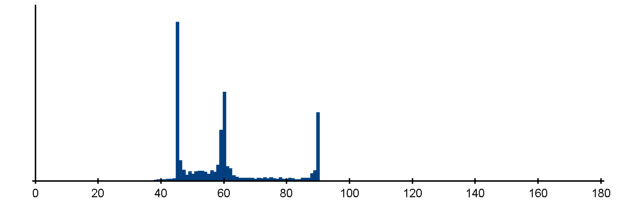}
			\caption{Angle distribution, target=$60^\circ$.}
		\end{subfigure}
		\begin{subfigure}[t]{\Histogram}
			\centering
			\includegraphics[width=\textwidth]{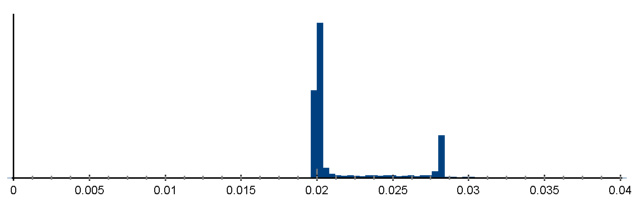}
			\caption{Edge lengths distribution, target=0.02.}
		\end{subfigure}
		\begin{subfigure}[t]{\Histogram}
			\centering
			\includegraphics[width=\textwidth]{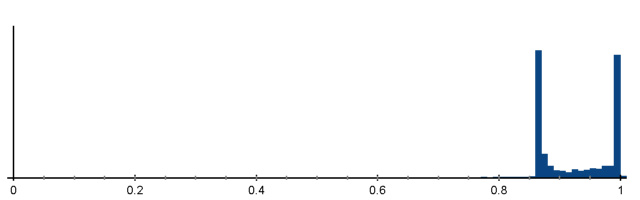}
			\caption{Distribution of quality $Q_t$, target=1.0.}
		\end{subfigure}
	\end{minipage}
	~
	\begin{minipage}{0.4\textwidth}
		\begin{subfigure}[t]{\Histogram}
			\centering
			\includegraphics[width=\textwidth]{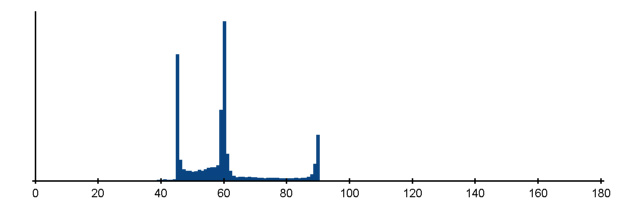}
			\caption{Angle distribution, target=$60^\circ$.}
		\end{subfigure}
		\begin{subfigure}[t]{\Histogram}
			\centering
			\includegraphics[width=\textwidth]{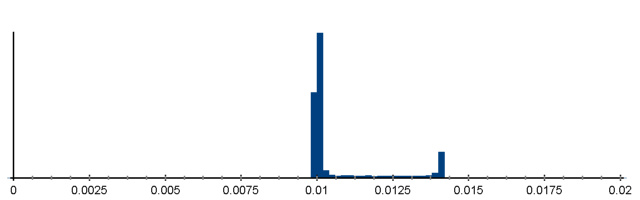}
			\caption{Edge lengths distribution, target=0.01.}
		\end{subfigure}
		\begin{subfigure}[t]{\Histogram}
			\centering
			\includegraphics[width=\textwidth]{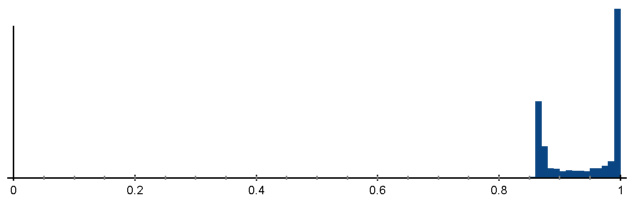}
			\caption{Distribution of quality $Q_t$, target=1.0.}
		\end{subfigure}
	\end{minipage}
	\caption{Histograms for remeshed \emph{Joint},~$\vartheta = 80^{\circ}$.}
\end{figure}

\newpage

\begin{table}[h!]
	\centering
	\begin{tabular}{lllrrrrrr}
		Algorithm & $d$ & $\vartheta$ & $| \mathcal{V} |$ & $| \mathcal{T} |$ & $\alpha_{\min}$ & $\alpha_{\max}$& $\alpha_{\text{avg}}$ & $\alpha_{\text{RMS}}$\\
		\hline
		MeshLab & 0.02 & 80$^{\circ}$ & 19,743 & 39,490 & 17.5321$^{\circ}$ & 120.6853$^{\circ}$ & 60$^{\circ}$ & 19.1 \\
		PMP & 0.02 & 80$^{\circ}$ & 23,046 & 46,096 & 21.4852$^{\circ}$ & 129.5474$^{\circ}$ & 60$^{\circ}$ & 15.1 \\
		ours & 0.02 & 80$^{\circ}$ & 17,966 & 35,936 & 29.2933$^{\circ}$ & 120.4075$^{\circ}$ & 60$^{\circ}$ & 25.7\\
		\hline
		MeshLab & 0.01 & 80$^{\circ}$ & 78,901 & 157,807 & 17.2320$^{\circ}$ & 130.2468$^{\circ}$ & 60$^{\circ}$ & 19.8 \\
		PMP & 0.01 & 80$^{\circ}$ & 90,982 & 181,968 & 24.5734$^{\circ}$ & 125.3222$^{\circ}$ & 60$^{\circ}$ & 15.4 \\
		ours &0.01 & 80$^{\circ}$ & 73,474 & 146,952 & 24.0836$^{\circ}$ & 131.8326$^{\circ}$ & 60$^{\circ}$ & 22.3\\
		& & & & & & & & \\
		$E_{\min}$ & $E_{\text{max}}$ & $E_{\text{avg}}$ & $E_{\text{RMS}}$ & $Q_{\text{min}}$ & $Q_{\text{max}}$ & $Q_{\text{avg}}$ & $Q_{\text{RMS}}$ & \\
		\hline
		0.0085 & 0.0380 & 0.0204 & 14.1 & 0.4913 & 0.9999 & 0.9494 & 5.5 & \\
		0.0115 & 0.0364 & 0.0188 & 11.8 & 0.4918 & 0.9999 & 0.9691 & 3.4 &  \\
		0.02 & 0.0416 & 0.0217 & 14.8 & 0.5960 & 1.0000 & 0.9272 & 6.6 & \\
		\hline
		0.0040 & 0.0193 & 0.0102 & 14.5 & 0.4847 & 0.9999 & 0.9458 & 5.8 & \\
		0.0054 & 0.0165 & 0.0094 & 12.1 & 0.5476 & 0.9999 & 0.9681 & 3.2 & \\
		0.01 & 0.0210 & 0.0106 & 13.3 & 0.4839 & 1.0000 & 0.9448 & 6.3 & 
	\end{tabular}
	\caption{Experimental results for \emph{Joint}.}
\end{table}

\begin{table}[h!]
	\centering
	\begin{tabular}{lrrrr}
		Algorithm & $d$ & $\vartheta$ & $d_{\max}$ & $\frac{d_{\max}}{d}$ \\
		\hline
		MeshLab & 0.01 & 80$^{\circ}$ & 0.0003 & 0.0306\\
		PMP & 0.01 & 80$^{\circ}$ & 0.0002 & 0.0206\\
		ours & 0.01 & 80$^{\circ}$ &0.0005 & 0.0543\\
		\hline
		MeshLab & 0.02 & 80$^{\circ}$ & 0.0008 & 0.0449\\
		PMP & 0.02 & 80$^{\circ}$ & 0.0006 & 0.0308\\
		ours & 0.02 & 80$^{\circ}$ & 0.0015 & 0.0780
	\end{tabular}
	\caption{One-sided Hausdorff distance evaluated on the \emph{Joint}.}
\end{table}

\newpage

\subsection{\emph{Oloid}} \textcolor{sowaswieweiss}{y}
\label{sec:Oloid}

\begin{figure}[h!]
	\centering
	\begin{subfigure}[t]{.32\textwidth}
		\includegraphics[width=\textwidth]{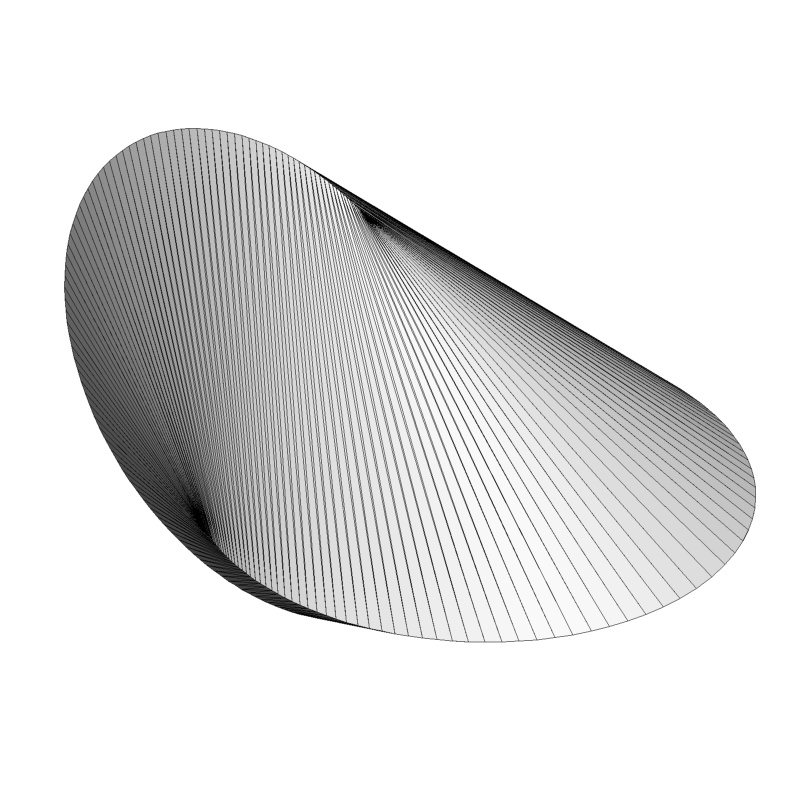}
	\end{subfigure}
	~
	\begin{subfigure}[t]{.32\textwidth}
		\includegraphics[width=\textwidth]{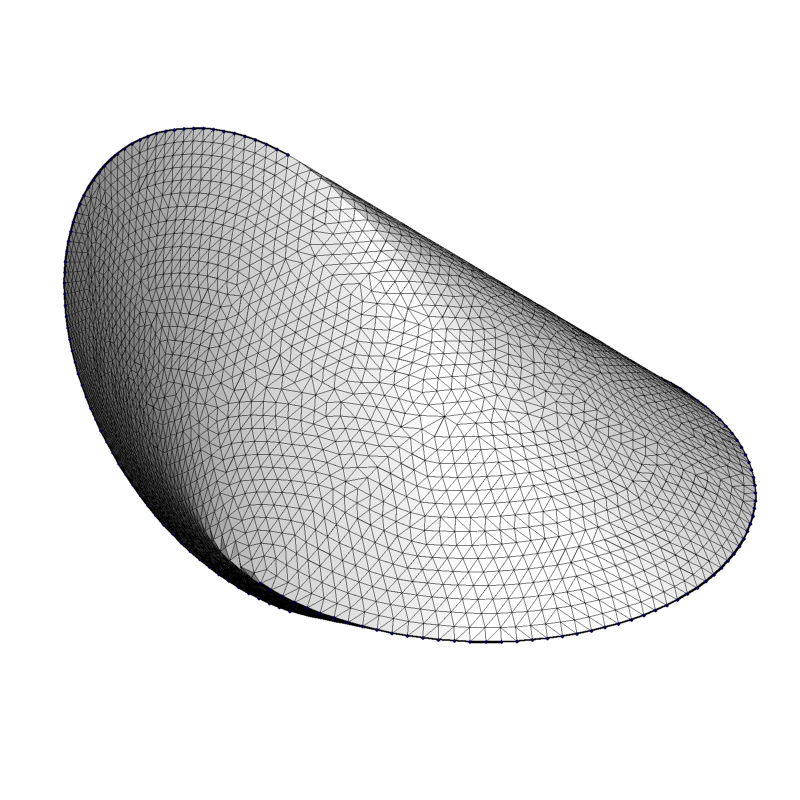}
	\end{subfigure}
	~
	\begin{subfigure}[t]{.32\textwidth}
		\includegraphics[width=\textwidth]{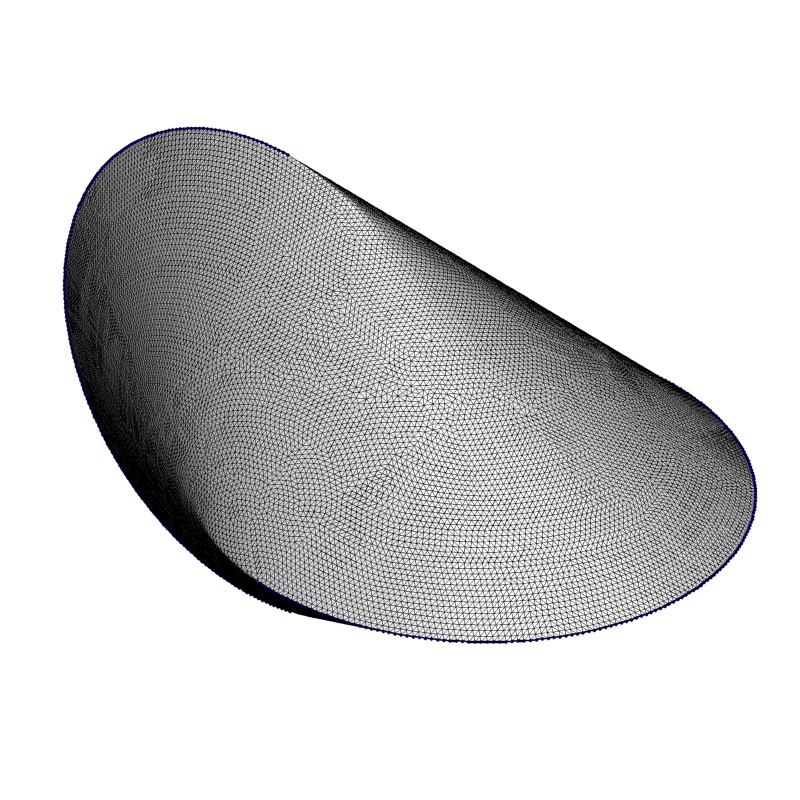}
	\end{subfigure}
	\caption{\emph{Oloid}.
		Left: CAD model.
		Middle: Model remeshed with features detected for~$\vartheta = 80^{\circ}$ and target edge length~$d = 0.05$.
		Right: Model remeshed with features detected for~$\vartheta = 80^{\circ}$ and target edge length~$d = 0.02$.
	}
\end{figure}

\begin{figure}[b!]
	\centering
	\begin{minipage}{0.4\textwidth}
		\begin{subfigure}[t]{\Histogram}
			\centering
			\includegraphics[width=\textwidth]{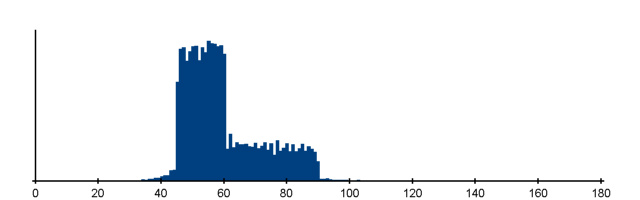}
			\caption{Angle distribution, target=$60^\circ$.}
		\end{subfigure}
		\begin{subfigure}[t]{\Histogram}
			\centering
			\includegraphics[width=\textwidth]{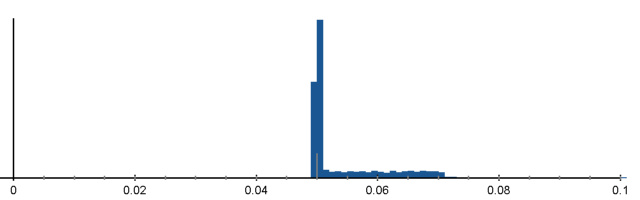}
			\caption{Edge lengths distribution, target=0.05.}
		\end{subfigure}
		\begin{subfigure}[t]{\Histogram}
			\centering
			\includegraphics[width=\textwidth]{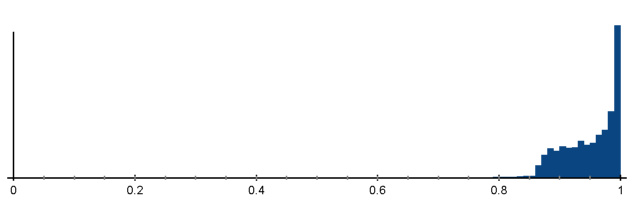}
			\caption{Distribution of quality $Q_t$, target=1.0.}
		\end{subfigure}
	\end{minipage}
	~
	\begin{minipage}{0.4\textwidth}
		\begin{subfigure}[t]{\Histogram}
			\centering
			\includegraphics[width=\textwidth]{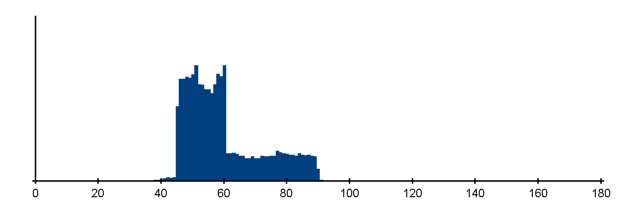}
			\caption{Angle distribution, target=$60^\circ$.}
		\end{subfigure}
		\begin{subfigure}[t]{\Histogram}
			\centering
			\includegraphics[width=\textwidth]{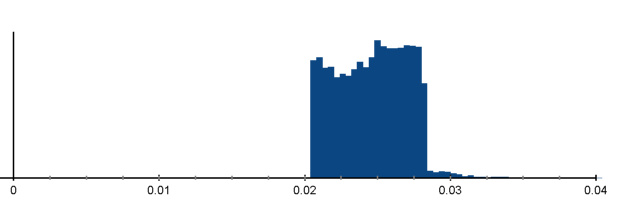}
			\caption{Edge lengths distribution, target=0.02.}
		\end{subfigure}
		\begin{subfigure}[t]{\Histogram}
			\centering
			\includegraphics[width=\textwidth]{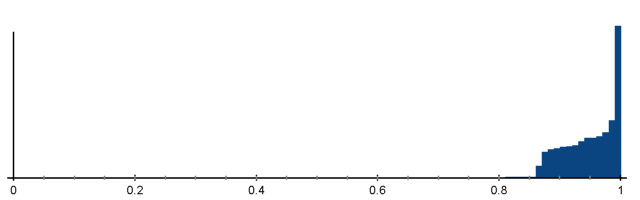}
			\caption{Distribution of quality $Q_t$, target=1.0.}
		\end{subfigure}
	\end{minipage}
	\caption{Histograms for remeshed \emph{Oloid},~$\vartheta = 80^{\circ}$.}
\end{figure}

\newpage

\begin{table}[h!]
	\centering
	\begin{tabular}{llllrrrrrr}
		Algorithm & $| \text{Iterations} |$ & $d$ & $\vartheta$ & $| \mathcal{V} |$ & $| \mathcal{T} |$ & $\alpha_{\min}$ & $\alpha_{\max}$& $\alpha_{\text{avg}}$ & $\alpha_{\text{RMS}}$\\
		\hline
		MeshLab & 10 & 0.05 & 80$^{\circ}$ & 5,980 & 11,956 & 0.0106$^{\circ}$ & 179.9759$^{\circ}$ & 60$^{\circ}$ & 28.8 \\
		MeshLab & 100 & 0.05 & 80$^{\circ}$ & 5,355 & 10,706 & 30.1274$^{\circ}$ & 106.6160$^{\circ}$ & 60$^{\circ}$ & 10.0 \\
		PMP & 10 & 0.05 & 80$^{\circ}$ & 6,709 & 13,414 & 25.3031$^{\circ}$ & 112.5742$^{\circ}$ & 60$^{\circ}$ & 20.5 \\
		ours & 1 & 0.05 & 80$^{\circ}$ & 5,216 & 10,428 & 27.4950$^{\circ}$ & 118.1905$^{\circ}$ & 60$^{\circ}$ & 20.7\\
		\hline
		MeshLab & 10 & 0.02 & 80$^{\circ}$ & 36,218 & 72,437 & 0.0151$^{\circ}$ & 179.8044$^{\circ}$ & 60$^{\circ}$ & 21.4 \\
		MeshLab & 100 & 0.02 & 80$^{\circ}$ & 33,805 & 67,606 & 26.3402$^{\circ}$ & 110.2882$^{\circ}$ & 60$^{\circ}$ & 9.9 \\
		PMP & 10 & 0.02 & 80$^{\circ}$ & 43,933 & 87,862 & 32.6637$^{\circ}$ & 105.2212$^{\circ}$ & 60$^{\circ}$ & 16.8 \\
		ours & 1 & 0.02 & 80$^{\circ}$ & 32,787 & 65,570 & 28.5948$^{\circ}$ & 122.8103$^{\circ}$ & 60$^{\circ}$ & 20.6\\
		\hline
		MeshLab & 10 & 0.02 & -- & 36,607 & 73,210 & 0.0152 & 179.9609$^{\circ}$ & 60$^{\circ}$ & 22.5 \\
		PMP & 10 & 0.02 & -- & 43,933 & 87,862 & 32.6637$^{\circ}$ & 105.2212$^{\circ}$ & 60$^{\circ}$ & 16.8 \\
		ours & 1 & 0.02 & -- & 33,026 & 66,048 & 30.5906$^{\circ}$ & 114.0393$^{\circ}$ & 60$^{\circ}$ & 18.4 \\
		& & & & & & & & & \\
		$E_{\min}$ & $E_{\text{max}}$ & $E_{\text{avg}}$ & $E_{\text{RMS}}$ & $Q_{\text{min}}$ & $Q_{\text{max}}$ & $Q_{\text{avg}}$ & $Q_{\text{RMS}}$ & & \\
		\hline
		4.9551$E^{-4}$ & 0.0849 & 0.0494 & 21.0 & 2.3800$E^{-4}$ & 0.9999 & 0.9225 & 17.6 &  & \\
		0.0264 & 0.0803 & 0.0521 & 9.8 & 0.6996 & 0.9999 & 0.9858 & 2.1 &  &  \\
		0.0296 & 0.0822 & 0.0474 & 13.3 & 0.6331 & 0.9999 & 0.9454 & 3.9 &  & \\
		0.05 & 0.1038 & 0.0538 & 12.2 & 0.6164 & 1.0000 & 0.9501 & 4.7 & & \\
		\hline 
		3.6697$E^{-5}$ & 0.0376 & 0.0202 & 15.8 & 4.5746$E^{-4}$ & 0.9999 & 0.9455 & 9.4 &  &  \\
		0.0099 & 0.0324 & 0.0207 & 9.8 & 0.6440 & 0.9999 & 0.9862 & 1.9 &  &  \\
		0.0112 & 0.0308 & 0.0183 & 12.3 & 0.7291 & 0.9999 & 0.9624 & 3.4 & & \\
		0.02 & 0.0391 & 0.0214 & 11.8 & 0.5727 & 1.0000 & 0.9508 & 4.5 & &\\
		\hline
		8.4272$E^{-5}$ & 0.0404 & 0.0200 & 17.3 & 3.8556$E^{-4}$ & 1.0 & 0.9373 & 11.2 &  &  \\
		0.0113 & 0.0309 & 0.0183 & 12.3 & 0.7292 & 1.0 & 0.9625 & 3.4 &  &  \\
		0.02 & 0.0385 & 0.0212 & 10.9 & 0.6571 & 1.0 & 0.9606 & 4.4 &  &  \\
	\end{tabular}
	\caption{Experimental results for \emph{Oloid}.}
\end{table}

\begin{table}[h!]
	\centering
	\begin{tabular}{lrrrrr}
		Algorithm & $|$Iterations$|$ & $d$ & $\vartheta$ & $d_{\max}$ & $\frac{d_{\max}}{d}$ \\
		\hline
		MeshLab & 10 & 0.02 & 80$^{\circ}$ & 0.0072 & 0.3600\\
		MeshLab & 100 & 0.02 & 80$^{\circ}$ & 0.0085 & 0.4293\\
		PMP & 10 & 0.02 & 80$^{\circ}$ & 0.0154 & 0.7742 \\
		ours & 1 & 0.02 & 80$^{\circ}$ & 0.0071 & 0.35377\\
		\hline
		MeshLab & 10 & 0.05 & 80$^{\circ}$ & 0.0184 & 0.3695\\
		MeshLab & 100 & 0.05 & 80$^{\circ}$ & 0.0205 & 0.4108\\
		PMP & 10 & 0.05 & 80$^{\circ}$ & 0.0129 & 0.2590\\
		ours & 1 & 0.05 & 80$^{\circ}$ & 0.0181 & 0.3628
	\end{tabular}
	\caption{One-sided Hausdorff distance evaluated on the \emph{Oloid}.}
\end{table}

\newpage

\subsection{\emph{Block}} \textcolor{sowaswieweiss}{y}

\begin{figure}[h!]
	\centering
	\begin{subfigure}[t]{.32\textwidth}
		\includegraphics[width=\textwidth]{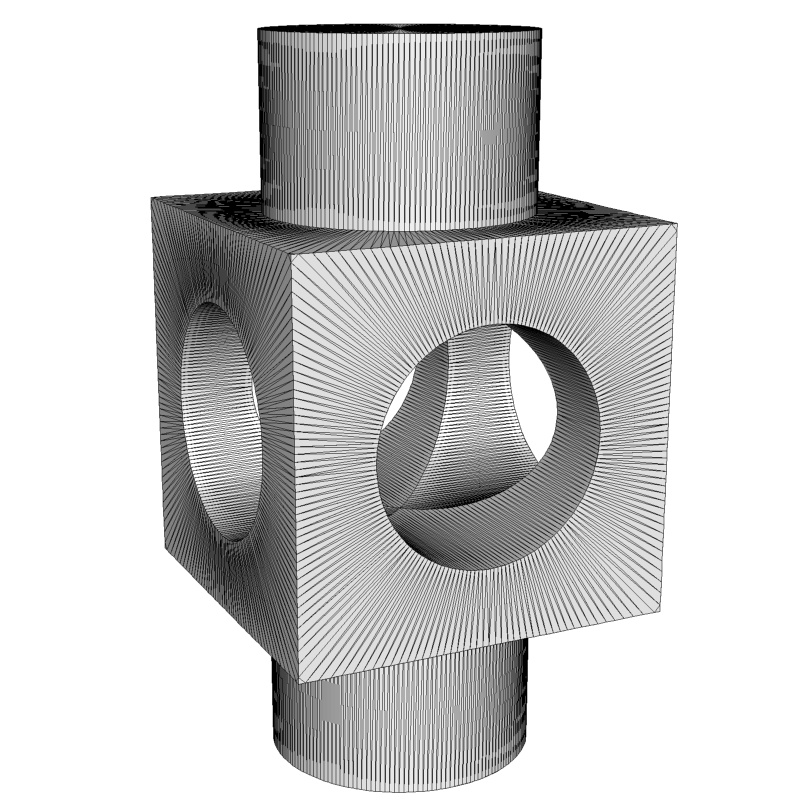}
	\end{subfigure}
	~
	\begin{subfigure}[t]{.32\textwidth}
		\includegraphics[width=\textwidth]{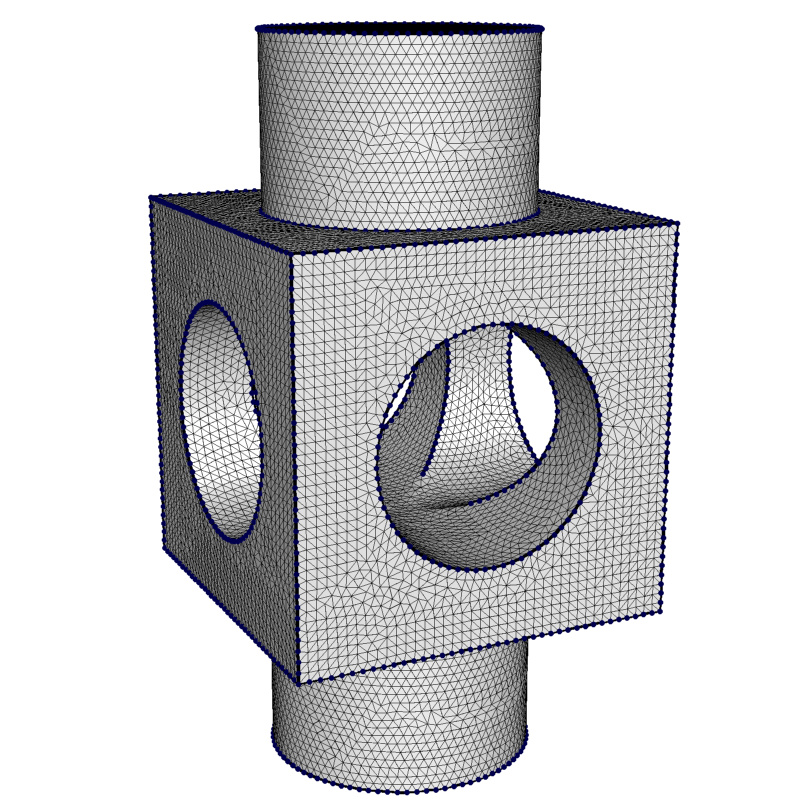}
	\end{subfigure}
	~
	\begin{subfigure}[t]{.32\textwidth}
		\includegraphics[width=\textwidth]{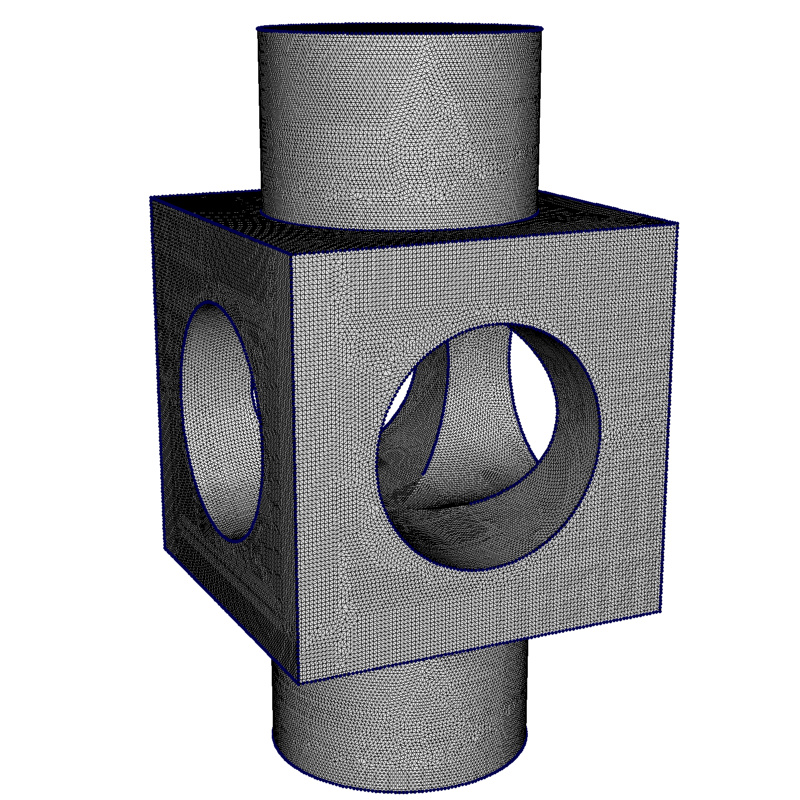}
	\end{subfigure}
	\caption{\emph{Block}.
		Left: CAD model.
		Middle: Model remeshed with features detected for~$\vartheta = 40^{\circ}$ and target edge length~$d = 0.5$.
		Right: Model remeshed with features detected for~$\vartheta = 40^{\circ}$ and target edge length~$d = 0.2$.
	}
\end{figure}

\begin{figure}[b!]
	\centering
	\begin{minipage}{0.38\textwidth}
		\begin{subfigure}[t]{\Histogram}
			\centering
			\includegraphics[width=\textwidth]{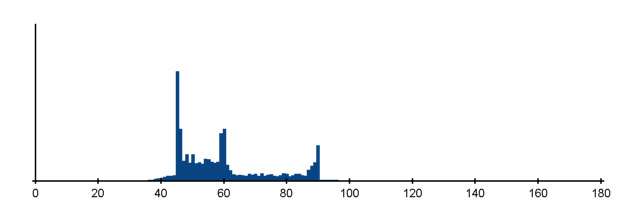}
			\caption{Angle distribution, target=$60^\circ$.}
		\end{subfigure}
		\begin{subfigure}[t]{\Histogram}
			\centering
			\includegraphics[width=\textwidth]{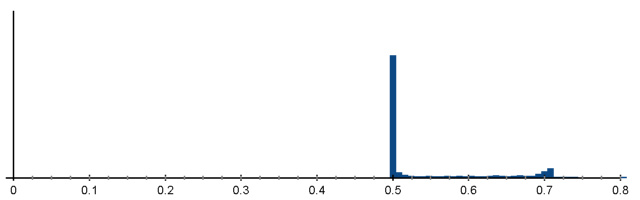}
			\caption{Edge lengths distribution, target=0.5.}
		\end{subfigure}
		\begin{subfigure}[t]{\Histogram}
			\centering
			\includegraphics[width=\textwidth]{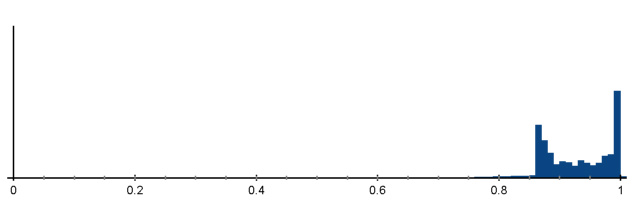}
			\caption{Distribution of quality $Q_t$, target=1.0.}
		\end{subfigure}
	\end{minipage}
	~
	\begin{minipage}{0.38\textwidth}
		\begin{subfigure}[t]{\Histogram}
			\centering
			\includegraphics[width=\textwidth]{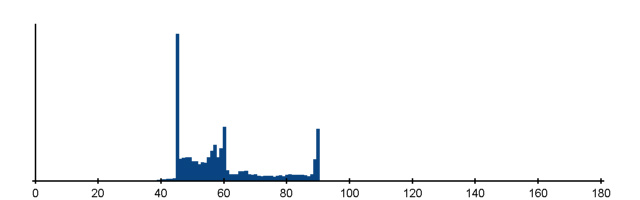}
			\caption{Angle distribution, target=$60^\circ$.}
		\end{subfigure}
		\begin{subfigure}[t]{\Histogram}
			\centering
			\includegraphics[width=\textwidth]{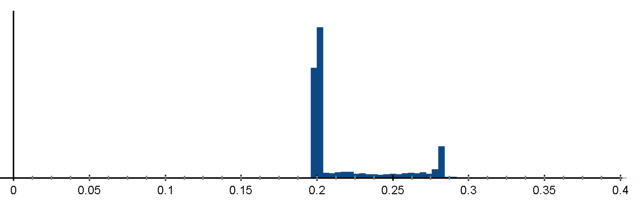}
			\caption{Edge lengths distribution, target=0.2.}
		\end{subfigure}
		\begin{subfigure}[t]{\Histogram}
			\centering
			\includegraphics[width=\textwidth]{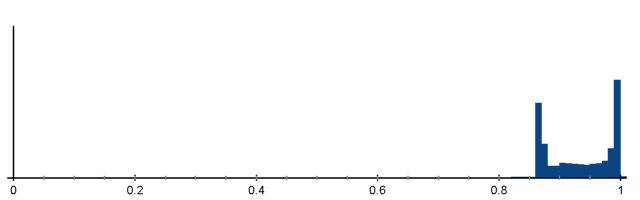}
			\caption{Distribution of quality $Q_t$, target=1.0.}
		\end{subfigure}
	\end{minipage}
	\caption{Histograms for remeshed \emph{Block},~$\vartheta = 40^{\circ}$.}
\end{figure}

\newpage

\begin{table}[h!]
	\centering
	\begin{tabular}{lllrrrrrr}
		Algorithm & $d$ & $\vartheta$ & $| \mathcal{V} |$ & $| \mathcal{T} |$ & $\alpha_{\min}$ & $\alpha_{\max}$& $\alpha_{\text{avg}}$ & $\alpha_{\text{RMS}}$\\
		\hline
		MeshLab & 0.5 & 40$^{\circ}$ & 16,225 & 32,458 & 17.4712$^{\circ}$ & 123.5228$^{\circ}$ & 60$^{\circ}$ & 17.6 \\
		PMP & 0.5 & 40$^{\circ}$ & 20,137 & 40,282 & 26.0368$^{\circ}$ & 123.3726$^{\circ}$ & 60$^{\circ}$ & 17.3 \\
		ours & 0.5 & 40$^{\circ}$ & 14,963 & 29,934 & 25.0851$^{\circ}$ & 120.2231$^{\circ}$ & 60$^{\circ}$ & 25.0\\
		\hline
		MeshLab & 0.2 & 40$^{\circ}$ & 101,174 & 202,356 & 0.0709$^{\circ}$ & 179.8427$^{\circ}$ & 60$^{\circ}$ & 18.4 \\
		PMP & 0.2 & 40$^{\circ}$ & 125,027 & 250,062 & 24.9504$^{\circ}$ & 123.8106$^{\circ}$ & 60$^{\circ}$ & 17.8 \\
		ours & 0.2 & 40$^{\circ}$ & 94,931 & 189,870 & 25.0891$^{\circ}$ & 126.2225$^{\circ}$ & 60$^{\circ}$ & 24.5\\
		& & & & & & & & \\
		$E_{\min}$ & $E_{\text{max}}$ & $E_{\text{avg}}$ & $E_{\text{RMS}}$ & $Q_{\text{min}}$ & $Q_{\text{max}}$ & $Q_{\text{avg}}$ & $Q_{\text{RMS}}$ & \\
		\hline
		0.2094 & 0.8994 & 0.5178 & 13.2 & 0.4619 & 0.9999 & 0.9568 & 5.0 & \\
		0.3021 & 0.8054 & 0.4650 & 12.8 & 0.5663 & 0.9999 & 0.9590 & 3.9 &  \\
		0.4999 & 1.1090 & 0.5481 & 14.6 & 0.5978 & 1.0000 & 0.9294 & 6.1 & \\
		\hline
		0.0018 & 0.4033 & 0.2074 & 14.4 & 0.0015 & 0.9999 & 0.9541 & 5.6 & \\
		0.1179 & 0.3243 & 0.1866 & 13.4 & 0.5622 & 0.9999 & 0.9589 & 3.6 &  \\
		0.1999  & 0.4488 & 0.2173 & 13.8 & 0.5393 & 1.0000 & 0.9331 & 5.9 &
	\end{tabular}
	\caption{Experimental results for \emph{Block}.}
\end{table}

\begin{table}[h!]
	\centering
	\begin{tabular}{lrrrr}
		Algorithm & $d$ & $\vartheta$ & $d_{\max}$ & $\frac{d_{\max}}{d}$ \\
		\hline
		MeshLab & 0.2 & 40$^{\circ}$ & 0.0298 & 0.1493\\
		PMP & 0.2 & 40$^{\circ}$ & 0.0258 & 0.1293\\
		ours & 0.2 & 40$^{\circ}$ & 0.0339 & 0.1698\\
		\hline
		MeshLab & 0.5 & 40$^{\circ}$ & 0.0828 & 0.1657\\
		PMP & 0.5 & 40$^{\circ}$ & 0.0243 & 0.0486\\
		ours & 0.5 & 40$^{\circ}$ & 0.0935 & 0.1871 
	\end{tabular}
	\caption{One-sided Hausdorff distance evaluated on the \emph{Block}.}
\end{table}

\newpage

\subsection{\emph{Fandisk}} \textcolor{sowaswieweiss}{y}

\begin{figure}[h!]
	\centering
	\begin{minipage}{0.45\textwidth}
		\begin{subfigure}{\textwidth}
			\includegraphics[width=1.\textwidth]{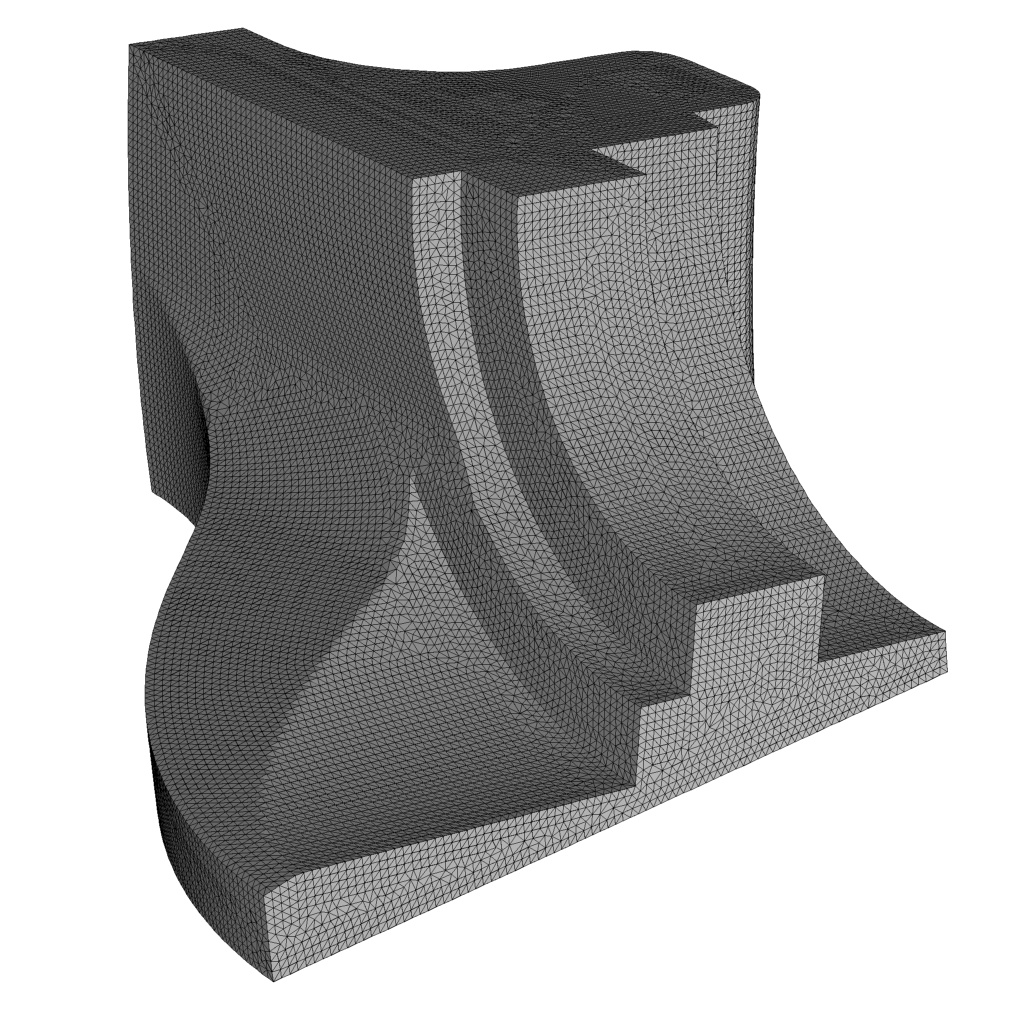}
		\end{subfigure}
	\end{minipage}
	\hfill
	\begin{minipage}{0.5\textwidth}
		\begin{subfigure}[t]{\Histogram}
			\centering
			\includegraphics[width=\textwidth]{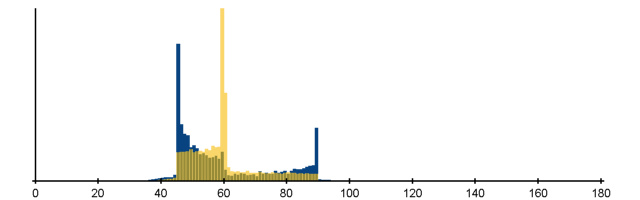}
			\caption{Angle distribution, target=$60^\circ$.}
		\end{subfigure}
		\begin{subfigure}[t]{\Histogram}
			\centering
			\includegraphics[width=\textwidth]{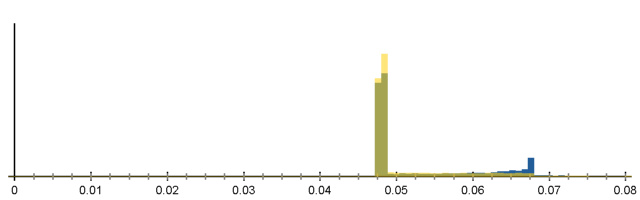}
			\caption{Edge lengths distribution, target=0.048.}
		\end{subfigure}
		\begin{subfigure}[t]{\Histogram}
			\centering
			\includegraphics[width=\textwidth]{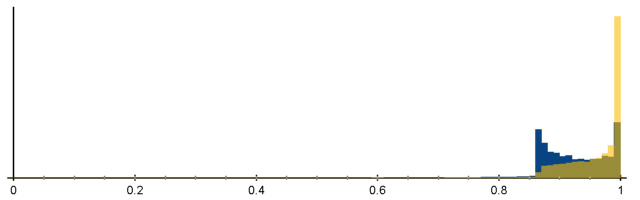}
			\caption{Distribution of quality $Q_t$, target=1.0.}
		\end{subfigure}
	\end{minipage}
	\caption{
		\emph{Fandisk}. Remeshed model and histograms,~$\vartheta = 60^{\circ}$.
	}
\end{figure}

\begin{table}[h!]
	\centering
	\begin{tabular}{lrrrr|lrrrr}
		Algorithm & $d$ & $\vartheta$ & $d_{\max}$ & $\frac{d_{\max}}{d}$ & Algorithm & $d$ & $\vartheta$ & $d_{\max}$ & $\frac{d_{\max}}{d}$ \\
		\hline
		MeshLab & 0.012 & --- & 0.0111 & 0.9302 & MeshLab & 0.012 & 60$^{\circ}$ & 0.0035 & 0.2929\\
		PMP 	& 0.012 & --- & 0.0080 & 0.6721 & PMP 	& 0.012 & 60$^{\circ}$ & 0.0027 & 0.2331\\
		ours 	& 0.012 & --- & 0.0097 & 0.8147 & ours 	& 0.012 & 60$^{\circ}$ & 0.0034 & 0.2835\\
		\hline 
		MeshLab & 0.024 & --- & 0.0216 & 0.9039 & MeshLab & 0.024 & 60$^{\circ}$ & 0.0061 & 0.2561\\
		PMP 	& 0.024 & --- & 0.0180 & 0.7524 & PMP 	& 0.024 & 60$^{\circ}$ & 0.0055 & 0.2302\\
		ours 	& 0.024 & --- & 0.0184 & 0.7679 & ours 	& 0.024 & 60$^{\circ}$ & 0.0072 & 0.2983 \\
		\hline
		MeshLab & 0.036 & --- & 0.0318 & 0.8840 & MeshLab & 0.036 & 60$^{\circ}$ & 0.0060 & 0.1690\\
		PMP 	& 0.036 & --- & 0.0261 & 0.7252 & PMP 	& 0.036 & 60$^{\circ}$ & 0.0088 & 0.2454\\
		ours 	& 0.036 & --- & 0.0302 & 0.8376 & ours	& 0.036 & 60$^{\circ}$ & 0.0085 & 0.2367\\
		\hline
		MeshLab & 0.048 & --- & 0.0461 & 0.9618 & MeshLab & 0.048 & 60$^{\circ}$ & 0.0115 & 0.2405\\
		PMP 	& 0.048 & --- & 0.0334 & 0.6976 & PMP 	& 0.048 & 60$^{\circ}$ & 0.0125 & 0.2612\\
		ours	& 0.048 & --- & 0.0337 & 0.7025 & ours	& 0.048 & 60$^{\circ}$ & 0.0134 & 0.2806
	\end{tabular}
	\caption{One-sided Hausdorff distance evaluated on the \emph{Fandisk}.}
\end{table}

\newpage

\begin{table}[h!]
	\centering
	\begin{tabular}{lllrrrrrr}
		Algorithm & $d$ & $\vartheta$ & $| \mathcal{V} |$ & $| \mathcal{T} |$ & $\alpha_{\min}$ & $\alpha_{\max}$& $\alpha_{\text{avg}}$ & $\alpha_{\text{RMS}}$\\
		\hline
		MeshLab & 0.048 & 60$^{\circ}$ & 28,316 & 56,628 & 18.3621$^{\circ}$ & 123.6706$^{\circ}$ & 60$^{\circ}$ & 16.3 \\
		PMP & 0.048 & 60$^{\circ}$ & 36,135 & 72,266 & 18.2595$^{\circ}$ & 125.1246$^{\circ}$ & 60$^{\circ}$ & 16.3 \\
		ours & 0.048 & $60^{\circ}$ & 26,744 & 53,484 & 19.3842$^{\circ}$ & 122.5733$^{\circ}$ & 60$^{\circ}$ & 25.8\\
		\hline
		MeshLab & 0.048 & --- & 28,124 & 56,244 & 21.0344$^{\circ}$ & 121.2869$^{\circ}$ & 60$^{\circ}$ & 16.4 \\
		PMP & 0.048 & --- & 35,409 & 70,814 & 33.2363$^{\circ}$ & 104.1949$^{\circ}$ & 60$^{\circ}$ & 15.4\\
		ours & 0.048 & --- & 27,295 & 54,586 & 12.8282$^{\circ}$ & 132.3435$^{\circ}$ & 60$^{\circ}$ & 19.0 \\
		\hline
		MeshLab & 0.036 & 60$^{\circ}$ & 52,767 & 105,530 & 19.3825$^{\circ}$ & 124.8048$^{\circ}$ & 60$^{\circ}$ & 16.9 \\
		PMP & 0.036 & 60$^{\circ}$ & 67,435 & 134,866 & 17.3648$^{\circ}$ & 136.9920$^{\circ}$ & 60$^{\circ}$ & 15.8\\
		ours & 0.036 & $60^{\circ}$ & 47,114 & 95,424 & 27.8458$^{\circ}$ & 123.7869$^{\circ}$ & 60$^{\circ}$ & 25.9\\
		\hline
		MeshLab & 0.036 & --- & 51,984 & 103,964 & 23.0220$^{\circ}$ & 117.1838$^{\circ}$ & 60$^{\circ}$ & 16.3 \\
		PMP & 0.036 & --- & 66,275 & 132,546 & 25.1185$^{\circ}$ & 119.1596$^{\circ}$ & 60$^{\circ}$ & 14.6 \\
		ours & 0.036 & --- & 48,788 & 97,572 & 21.9673$^{\circ}$ & 121.1339$^{\circ}$ & 60$^{\circ}$ & 18.8 \\
		\hline
		MeshLab & 0.024 & 60$^{\circ}$ & 115,999 & 231,994 & 18.3622$^{\circ}$ & 128.1835$^{\circ}$ & 60$^{\circ}$ & 18.1 \\
		PMP & 0.024 & 60$^{\circ}$ & 139,962 & 279,920 & 16.9655$^{\circ}$ & 126.8398$^{\circ}$ & 60$^{\circ}$ & 14.8 \\
		ours & 0.024 & 60$^{\circ}$ & 101,514 & 215,024 & 27.8458$^{\circ}$ & 123.7869$^{\circ}$ & 60$^{\circ}$ & 25.9\\
		\hline
		MeshLab & 0.024 & --- & 115,464 & 230,924 & 18.6249$^{\circ}$ & 128.1835$^{\circ}$ & 60$^{\circ}$ & 18.0 \\
		PMP & 0.024 & --- & 138,745 & 277,486 & 33.2062$^{\circ}$ & 102.3007$^{\circ}$ & 60$^{\circ}$ & 14.4 \\
		ours & 0.024 & --- & 110,312 & 220,620 & 22.6942$^{\circ}$ & 125.5307$^{\circ}$ & 60$^{\circ}$ & 18.7 \\
		\hline
		MeshLab & 0.012 & 60$^{\circ}$ & 468,982 & 937,960 & 17.4867$^{\circ}$ & 133.0335$^{\circ}$ & 60$^{\circ}$ & 18.9 \\
		PMP & 0.012 & 60$^{\circ}$ & 561,167 & 1,122,330 & 18.5464$^{\circ}$ & 134.9632$^{\circ}$ & 60$^{\circ}$ & 14.5 \\
		ours & 0.012 & 60$^{\circ}$ & 431,506 & 863,008 & 19.3842$^{\circ}$ & 125.0094$^{\circ}$ & 60$^{\circ}$ & 26.4\\
		\hline
		MeshLab & 0.012 & --- & 466,817 & 933,630 & 16.0072$^{\circ}$ & 133.0326$^{\circ}$ & 60$^{\circ}$ & 18.8 \\
		PMP & 0.012 & --- & 558,622 & 1,117,240 & 33.4158$^{\circ}$ & 101.9205$^{\circ}$ & 60$^{\circ}$ & 14.3 \\
		ours & 0.012 & --- & 443,351 & 886,698 & 19.4246$^{\circ}$ & 134.0945$^{\circ}$ & 60$^{\circ}$ & 18.7 \\
		& & & & & & & & \\
		$E_{\min}$ & $E_{\text{max}}$ & $E_{\text{avg}}$ & $E_{\text{RMS}}$ & $Q_{\text{min}}$ & $Q_{\text{max}}$ & $Q_{\text{avg}}$ & $Q_{\text{RMS}}$ & \\
		\hline
		0.0162 & 0.0864 & 0.0502 & 11.8 & 0.5159 & 0.9999 & 0.9638 & 4.4 &  \\
		0.0162 & 0.0822 & 0.0445 & 12.7 & 0.5128 & 0.9999 & 0.9627 & 3.6 &  \\
		0.0208 & 0.1021 & 0.0528 & 14.4 & 0.5236 & 1.0000 & 0.9249 & 5.5 & \\
		\hline
		0.0243 & 0.0896 & 0.0500 & 12.1 & 0.5503 & 0.9999 & 0.9632 & 4.1 & \\
		0.0278 & 0.0712 & 0.0446 & 12.4 & 0.7333 & 0.9999 & 0.9661 & 3.2 & \\
		0.0447 & 0.1004 & 0.0512 & 11.4 & 0.4788 & 1.0000 & 0.9578 & 4.5 & \\
		\hline
		0.0176 & 0.0759 & 0.0368 & 12.0 & 0.5113 & 0.9999 & 0.9621 & 4.4 &  \\
		0.0207 & 0.0613 & 0.0325 & 12.6 & 0.4147 & 0.9999 & 0.9671 & 3.8 &  \\
		0.0198 & 0.0762 & 0.0395 & 14.3 & 0.5631 & 1.0000 & 0.9247 & 5.4 & \\
		\hline
		0.0180 & 0.0678 & 0.0368 & 16.3 & 0.5968 & 0.9999 & 0.9641 & 4.0 & \\
		0.0206 & 0.0558 & 0.0326 & 12.0 & 0.5840 & 0.9999 & 0.9716 & 2.9 & \\
		0.0359 & 0.0840 & 0.0383 & 11.3 & 0.5477 & 1.0000 & 0.9585 & 4.5 & \\
		\hline
		0.0081 & 0.0448 & 0.0249 & 12.7 & 0.4669 & 0.9999 & 0.9560 & 4.8 &  \\
		0.0081 & 0.0418 & 0.0225 & 12.3 & 0.4776 & 0.9999 & 0.9700 & 3.0 &  \\
		0.0158 & 0.0499 & 0.0263 & 26.3 & 0.5627 & 1.0000 & 0.9227 & 5.3 & \\
		\hline
		0.0116 & 0.0449 & 0.0248 & 12.7 & 0.4669 & 0.9999 & 0.9565 & 4.6 & \\
		0.0140 & 0.0332 & 0.0225 & 12.1 & 0.7629 & 0.9999 & 0.9718 & 2.7 & \\
		0.0239 & 0.0701 & 0.0255 & 11.2 & 0.5311 & 1.0000 & 0.9592 & 4.4 & \\
		\hline
		0.0040 & 0.0254 & 0.0124 & 18.9 & 0.4547 & 0.9999 & 0.9526 & 5.0 &  \\
		0.0040 & 0.0218 & 0.0112 & 12.3 & 0.4462 & 0.9999 & 0.9713 & 2.9 &  \\
		0.0044 & 0.0257 & 0.0131 & 14.2 & 0.5392 & 1.0000 & 0.9221 & 5.3 & \\
		\hline
		0.0038 & 0.0250 & 0.0124 & 12.9 & 0.4164 & 0.9999 & 0.9531 & 4.9 & \\
		0.0069 & 0.0173 & 0.0125 & 12.2 & 0.7625 & 0.9999 & 0.9720 & 2.8 & \\
		0.0119 & 0.0272 & 0.0127 & 11.1 & 0.4470 & 1.0000 & 0.9592 & 4.4 & 
	\end{tabular}
	\caption{Experimental results for \emph{Fandisk}, displaying edges shorter than target edge length $d$.
		MeshLab and PMP were run with~10 iterations each, which corresponds to the standard setting in both.}
\end{table}

\clearpage

\subsection{\emph{Cube}} \textcolor{sowaswieweiss}{y}

\begin{figure}[h!]
	\centering
	\begin{subfigure}[t]{.32\textwidth}
		\includegraphics[width=\textwidth]{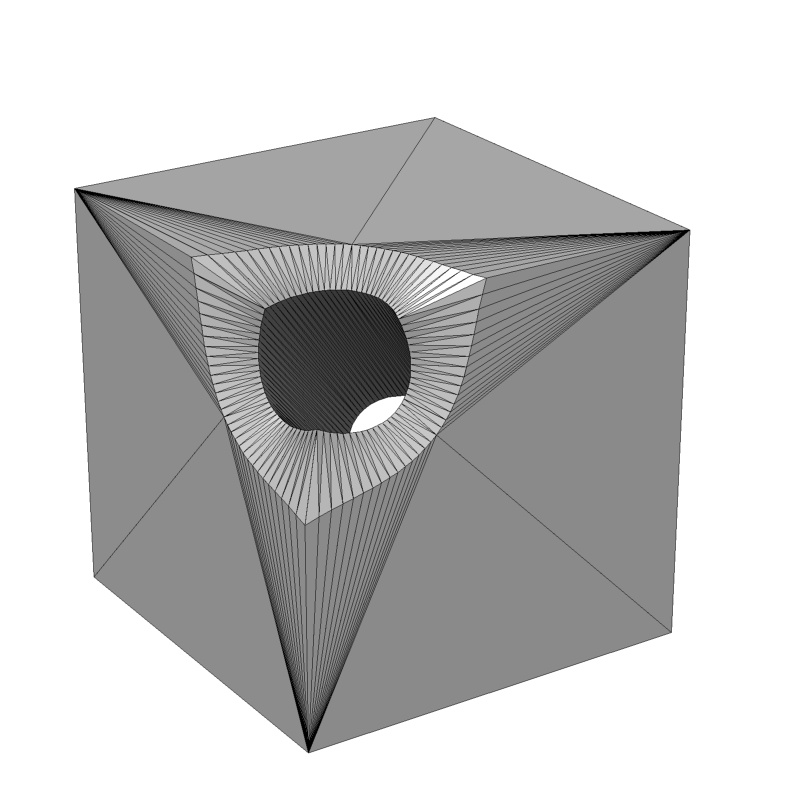}
	\end{subfigure}
	~
	\begin{subfigure}[t]{.32\textwidth}
		\includegraphics[width=\textwidth]{cubeWithHole_0p1_featuresAngle24_remesh}
	\end{subfigure}
	~
	\begin{subfigure}[t]{.32\textwidth}
		\includegraphics[width=\textwidth]{cubeWithHole_0p1_featuresAngle40_remesh}
	\end{subfigure}
	\caption{\emph{Cube}.
		Left: CAD model.
		Middle: Model remeshed with features detected for~$\vartheta = 24^{\circ}$ and target edge length~$d = 0.1$.
		Right: Model remeshed with features detected for~$\vartheta = 40^{\circ}$ and target edge length~$d = 0.1$.
	}
\end{figure}

\begin{table}[h!]
	\centering
	\footnotesize{
		\begin{tabular}{lllrrrrrr}
			Algorithm & $d$ & $\vartheta$ & $| \mathcal{V} |$ & $| \mathcal{T} |$ & $\alpha_{\min}$ & $\alpha_{\max}$& $\alpha_{\text{avg}}$ & $\alpha_{\text{RMS}}$\\
			\hline
			MeshLab & 0.1 & 24$^{\circ}$ & 4,185 & 8,370 & 14.8334$^{\circ}$ & 127.2719$^{\circ}$ & 60$^{\circ}$ & 18.9 \\
			PMP & 0.1 & 24$^{\circ}$ & 5,480 & 10,960 & 20.9229$^{\circ}$ & 135.1919$^{\circ}$ & 60$^{\circ}$ & 18.4 \\
			ours & 0.1 & 24$^{\circ}$ & 3,796 & 7,592 & 28.5518$^{\circ}$ & 115.6221$^{\circ}$ & 60$^{\circ}$ & 26.6 \\
			\hline 
			MeshLab & 0.1 & 40$^{\circ}$ & 3,778 & 7,556 & 28.7024$^{\circ}$ & 120.7229$^{\circ}$ & 60$^{\circ}$ & 27.4 \\
			PMP & 0.1 & 40$^{\circ}$ & 5,470 & 10,940 & 18.4349$^{\circ}$ & 135.0000$^{\circ}$ & 60$^{\circ}$ & 17.9 \\
			ours & 0.1 & 40$^{\circ}$ & 3,778 & 7,556 & 28.7024$^{\circ}$ & 120.7229$^{\circ}$ & 60$^{\circ}$ & 27.4\\
			\hline
			MeshLab & 0.1 & --- & 4,159 & 8,318 & 6.5643$^{\circ}$ & 139.1714$^{\circ}$ & 60$^{\circ}$ & 21.7 \\
			PMP & 0.1 & --- & 5,302 & 10,604 & 35.9060$^{\circ}$ & 98.8441$^{\circ}$ & 60$^{\circ}$ & 14.9\\
			ours & 0.1 & --- & 3,896 & 7,792 & 31.0974$^{\circ}$ & 115.6319$^{\circ}$ & 60$^{\circ}$ & 18.9\\
			\hline
			MeshLab & 0.04 & 24$^{\circ}$ & 26,889 & 53,778 & 2.2081$^{\circ}$ & 122.0278$^{\circ}$ & 60$^{\circ}$ & 17.1 \\
			PMP & 0.04 & 24$^{\circ}$ & 33,817 & 67,634 & 26.3548$^{\circ}$ & 119.3944$^{\circ}$ & 60$^{\circ}$ & 16.6 \\
			ours & 0.04 & 24$^{\circ}$ & 24,331 & 48,662 & 25.2122$^{\circ}$ & 129.5755$^{\circ}$ & 60$^{\circ}$ & 25.1 \\
			\hline
			MeshLab & 0.04 & 40$^{\circ}$ & 25,889 & 53,778 & 23.2081$^{\circ}$ & 122.0278$^{\circ}$ & 60$^{\circ}$ & 17.1 \\
			PMP & 0.04 & 40$^{\circ}$ & 33,781 & 67,562 & 26.2303$^{\circ}$ & 119.5088$^{\circ}$ & 60$^{\circ}$ & 16.6 \\
			ours & 0.04 & 40$^{\circ}$ & 24,667 & 49,334 & 27.2235$^{\circ}$ & 124.2618$^{\circ}$ & 60$^{\circ}$ & 22.1\\
			\hline
			MeshLab & 0.04 & --- & 27,113 & 54,226 & 4.2034$^{\circ}$ & 142.9207$^{\circ}$ & 60$^{\circ}$ & 18.5 \\
			PMP & 0.04 & --- & 33,422 & 66,844 & 32.2670$^{\circ}$ & 107.9337$^{\circ}$ & 60$^{\circ}$ & 16.0 \\
			ours & 0.04 & --- & 24,735 & 49,470 & 30.4114$^{\circ}$ & 116.7102$^{\circ}$ & 60$^{\circ}$ & 18.9\\
			& & & & & & & & \\
			$E_{\min}$ & $E_{\text{max}}$ & $E_{\text{avg}}$ & $E_{\text{RMS}}$ & $Q_{\text{min}}$ & $Q_{\text{max}}$ & $Q_{\text{avg}}$ & $Q_{\text{RMS}}$ & \\
			\hline
			0.0370 & 0.1875 & 0.1037 & 14.5 & 0.4166 & 0.9999 & 0.9511 & 5.9 & \\
			0.0553 & 0.1767 & 0.0907 & 13.1 & 0.4478 & 0.9999 & 0.9566 & 4.9 & \\
			0.1 & 0.2068 & 0.1110 & 15.2 & 0.6420 & 1.0000 & 0.9202 & 5.9 & \\
			\hline
			0.0999 & 0.2046 & 0.1114 & 15.5 & 0.5929 & 0.9999 & 0.9155 & 5.9 &  \\
			0.0546 & 0.1759 & 0.0907 & 12.9 & 0.4330 & 0.9999 & 0.9587 & 4.8 &  \\
			0.1 & 0.2046 & 0.1114 & 15.5 & 0.6419 & 1.0000 & 0.9155 & 11.8 & \\
			\hline
			0.0125 & 0.1698 & 0.1026 & 17.4 & 0.1953 & 0.9999 & 0.9340 & 11.3 & \\
			0.0604 & 0.1314 & 0.0907 & 11.7 & 0.7942 & 0.9999 & 0.9694 & 2.9 &  \\
			0.1 & 0.1927 & 0.1066 & 11.8 & 0.6419 & 1.0000 & 0.9582 & 4.8 & \\
			\hline
			0.0189 & 0.0805 & 0.0408 & 12.8 & 0.5625 & 0.9999 & 0.9600 & 4.5 & \\
			0.0237 & 0.0670 & 0.0364 & 12.2 & 0.5923 & 0.9999 & 0.9636 & 3.5 & \\
			0.04 & 0.0809 & 0.0436 & 14.1 & 0.5062 & 1.0000 & 0.9295 & 5.8 & \\
			\hline
			0.0027 & 0.0736 & 0.0404 & 14.5 & 0.1264 & 0.9999 & 0.9525 & 7.3 & \\
			0.0237 & 0.0670 & 0.0364 & 12.1 & 0.5923 & 0.9999 & 0.9637 & 3.5 & \\
			0.04 & 0.0855 & 0.0431 & 13.1 & 0.5585 & 1.0000 & 0.9445 & 5.6 & \\
			\hline
			0.0027 & 0.0736 & 0.0404 & 14.5 & 0.1264 & 0.9999 & 0.9525 & 7.3 & \\
			0.0238 & 0.0587 & 0.0364 & 11.8 & 0.7125 & 0.9999 & 0.9663 & 3.1 & \\
			0.04 & 0.0789 & 0.0426 & 11.3 & 0.6316 & 1.0000 & 0.9584 & 4.5 & 
		\end{tabular}
	}
	\caption{Experimental results for \emph{Cube}.}
\end{table}

\newpage

\begin{figure}[h!]
	\centering
	\begin{minipage}{0.32\textwidth}
		\begin{subfigure}[t]{\Histogram}
			\centering
			\includegraphics[width=\textwidth]{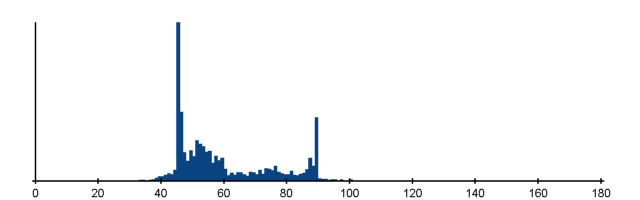}
			\caption{Angle distribution, target=$60^\circ$.}
		\end{subfigure}
		\begin{subfigure}[t]{\Histogram}
			\centering
			\includegraphics[width=\textwidth]{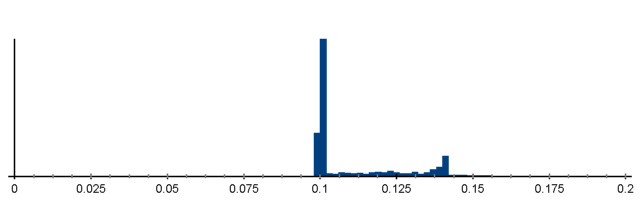}
			\caption{Edge lengths distribution, target=0.1.}
		\end{subfigure}
		\begin{subfigure}[t]{\Histogram}
			\centering
			\includegraphics[width=\textwidth]{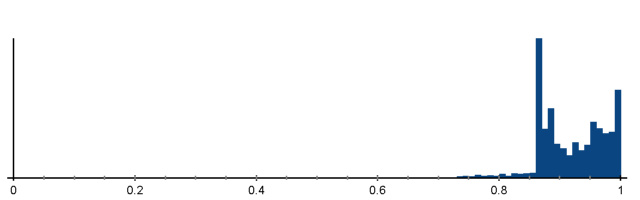}
			\caption{Distribution of quality $Q_t$, target=1.0.}
		\end{subfigure}
	\end{minipage}
	~
	\begin{minipage}{0.32\textwidth}
		\begin{subfigure}[t]{\Histogram}
			\centering
			\includegraphics[width=\textwidth]{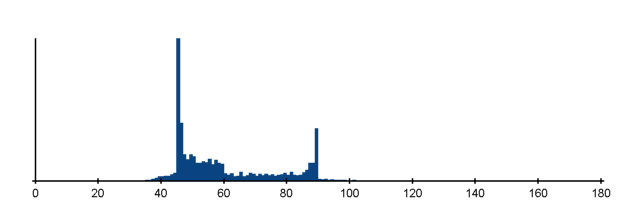}
			\caption{Angle distribution, target=$60^\circ$.}
		\end{subfigure}
		\begin{subfigure}[t]{\Histogram}
			\centering
			\includegraphics[width=\textwidth]{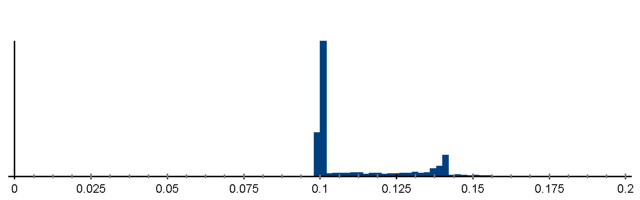}
			\caption{Edge lengths distribution, target=0.1.}
		\end{subfigure}
		\begin{subfigure}[t]{\Histogram}
			\centering
			\includegraphics[width=\textwidth]{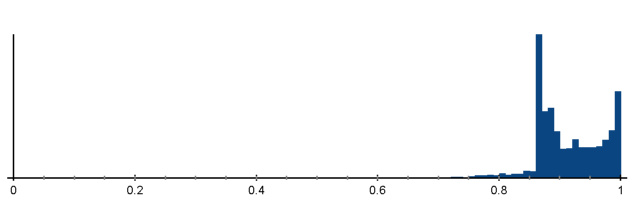}
			\caption{Distribution of quality $Q_t$, target=1.0.}
		\end{subfigure}
	\end{minipage}
	~
	\begin{minipage}{0.32\textwidth}
		\begin{subfigure}[t]{\Histogram}
			\centering
			\includegraphics[width=\textwidth]{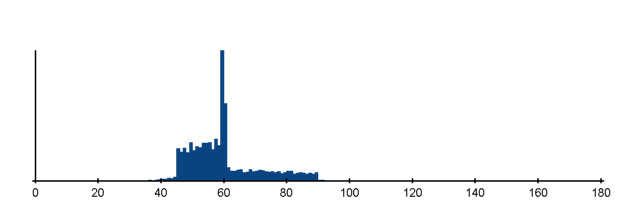}
			\caption{Angle distribution.}
		\end{subfigure}
		\begin{subfigure}[t]{\Histogram}
			\centering
			\includegraphics[width=\textwidth]{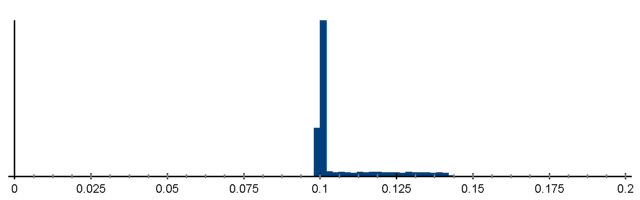}
			\caption{Edge lengths distribution, target=0.1.}
		\end{subfigure}
		\begin{subfigure}[t]{\Histogram}
			\centering
			\includegraphics[width=\textwidth]{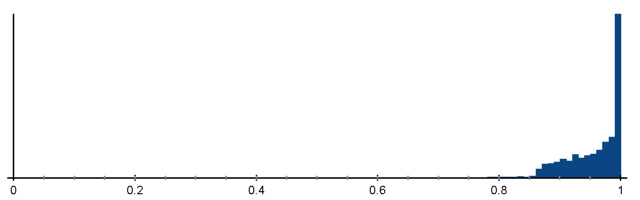}
			\caption{Distribution of quality $Q_t$, target=1.0.}
		\end{subfigure}
	\end{minipage}
	\caption{Histograms for remeshed \emph{Cube} for $d = 0.1$.
		From left to right:~$\vartheta = 24^{\circ}$,~$\vartheta = 40^{\circ}$, no feature detection.}
\end{figure}

\begin{figure}[h!]
	\centering
	\begin{minipage}{0.32\textwidth}
		\begin{subfigure}[t]{\Histogram}
			\centering
			\includegraphics[width=\textwidth]{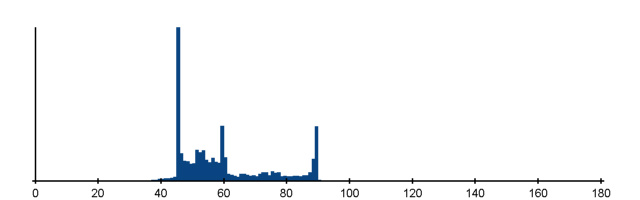}
			\caption{Angle distribution, target=$60^\circ$.}
		\end{subfigure}
		\begin{subfigure}[t]{\Histogram}
			\centering
			\includegraphics[width=\textwidth]{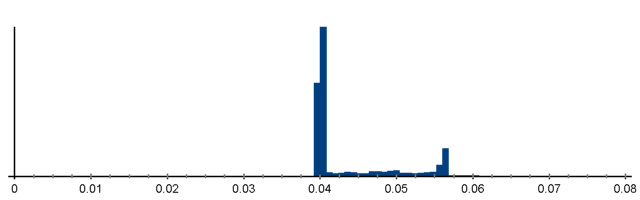}
			\caption{Edge lengths distribution, target=0.04.}
		\end{subfigure}
		\begin{subfigure}[t]{\Histogram}
			\centering
			\includegraphics[width=\textwidth]{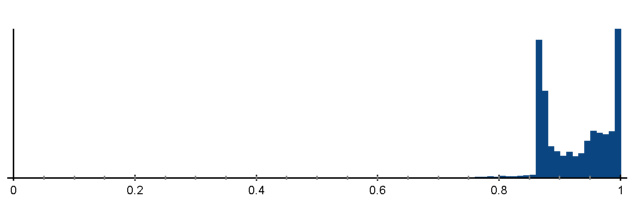}
			\caption{Distribution of quality $Q_t$, target=1.0.}
		\end{subfigure}
	\end{minipage}
	~
	\begin{minipage}{0.32\textwidth}
		\begin{subfigure}[t]{\Histogram}
			\centering
			\includegraphics[width=\textwidth]{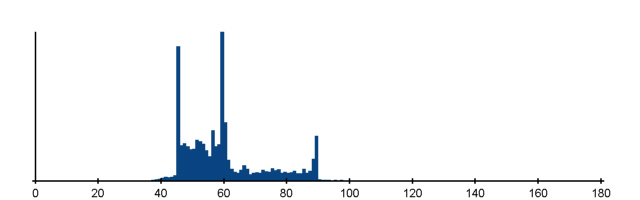}
			\caption{Angle distribution, target=$60^\circ$.}
		\end{subfigure}
		\begin{subfigure}[t]{\Histogram}
			\centering
			\includegraphics[width=\textwidth]{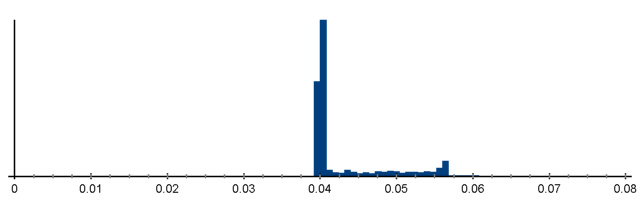}
			\caption{Edge lengths distribution, target=0.04.}
		\end{subfigure}
		\begin{subfigure}[t]{\Histogram}
			\centering
			\includegraphics[width=\textwidth]{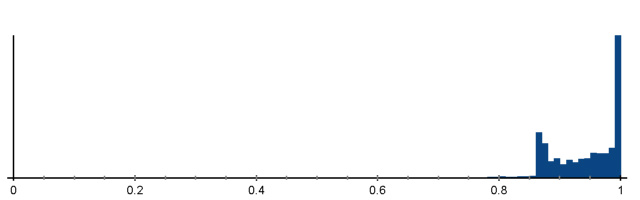}
			\caption{Distribution of quality $Q_t$, target=1.0.}
		\end{subfigure}
	\end{minipage}
	~
	\begin{minipage}{0.32\textwidth}
		\begin{subfigure}[t]{\Histogram}
			\centering
			\includegraphics[width=\textwidth]{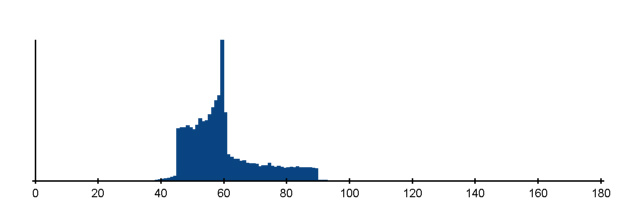}
			\caption{Angle distribution.}
		\end{subfigure}
		\begin{subfigure}[t]{\Histogram}
			\centering
			\includegraphics[width=\textwidth]{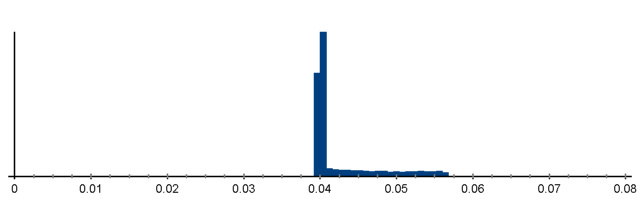}
			\caption{Edge lengths distribution, target=0.04.}
		\end{subfigure}
		\begin{subfigure}[t]{\Histogram}
			\centering
			\includegraphics[width=\textwidth]{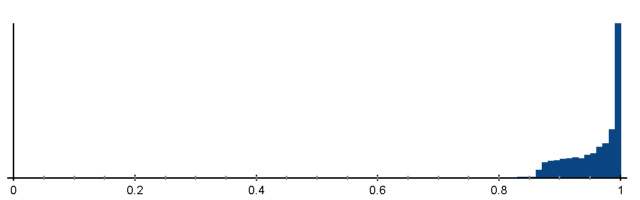}
			\caption{Distribution of quality $Q_t$, target=1.0.}
		\end{subfigure}
	\end{minipage}
	\caption{Histograms for remeshed \emph{Cube} for $d = 0.04$.
		From left to right:~$\vartheta = 24^{\circ}$,~$\vartheta = 40^{\circ}$, no feature detection.}
\end{figure}

\begin{table}[h!]
	\centering
	\begin{tabular}{lrrrr|lrrrr}
		Algorithm & $d$ & $\vartheta$ & $d_{\max}$ & $\frac{d_{\max}}{d}$ & Algorithm & $d$ & $\vartheta$ & $d_{\max}$ & $\frac{d_{\max}}{d}$\\
		\hline
		MeshLab & 0.04 & --- & 0.0347 & 0.8687 & MeshLab & 0.1 & --- & 0.0808 & 0.8082\\
		PMP & 0.04 & --- & 0.0334 & 0.8353 & PMP & 0.1 & --- & 0.2781 & 0.6623\\
		ours & 0.04 & --- & 0.0324 & 0.8122 & ours & 0.1 & --- & 0.0776 & 0.7765\\
		\hline
		MeshLab & 0.04 & 24$^{\circ}$ & 0.0043 & 0.1078 & MeshLab & 0.1 & 24$^{\circ}$ & 0.0062 & 0.0620\\
		PMP & 0.04 & 24$^{\circ}$ & 0.0038 & 0.0951 & PMP &0.1 & 24$^{\circ}$ & 0.0061 & 0.0614\\
		ours & 0.04 & 24$^{\circ}$ & 0.0042 & 0.1059 & ours & 0.1 & 24$^{\circ}$ & 0.0084 & 0.0843\\
		\hline
		MeshLab & 0.04 & 40$^{\circ}$ & 0.0043 & 0.1033 & MeshLab & 0.1 & 40$^{\circ}$ & 0.0190 & 0.1908\\
		PMP & 0.04 & 40$^{\circ}$ & 0.0065 & 0.1637 & PMP & 0.1 & 40$^{\circ}$ & 0.0196 & 0.1961\\
		ours & 0.04 & 40$^{\circ}$ & 0.0097 & 0.2427 & ours & 0.1 & 40$^{\circ}$ & 0.0213 & 0.2134
	\end{tabular}
	\caption{One-sided Hausdorff distance evaluated on the \emph{Cube}.}
\end{table}

\newpage

\subsection{\emph{Boat}} \textcolor{sowaswieweiss}{y}

\begin{figure}[h!]
	\centering
	\begin{subfigure}[t]{.49\textwidth}
		\includegraphics[width=\textwidth]{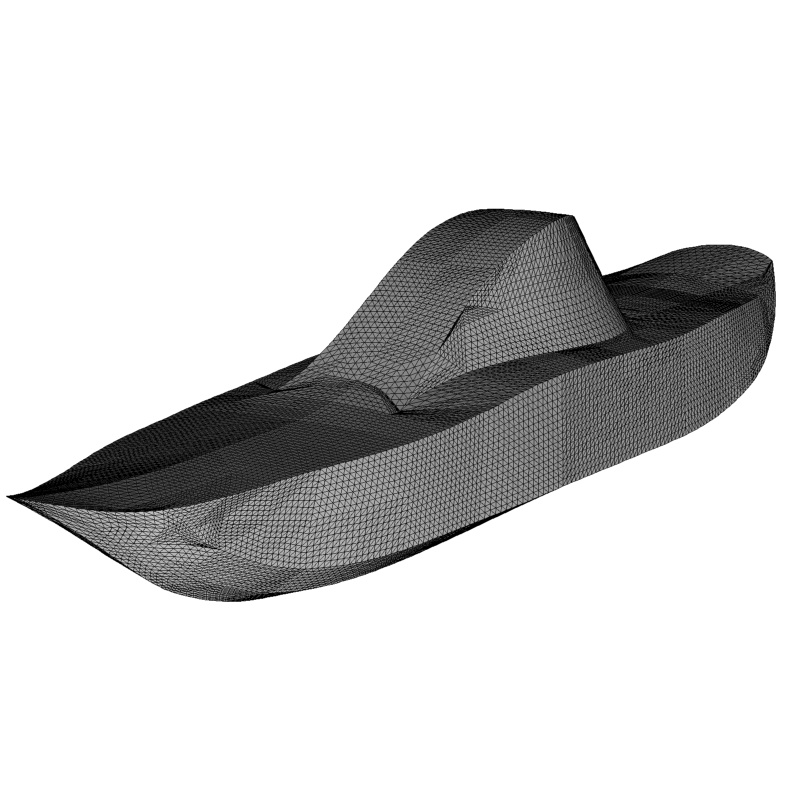}
	\end{subfigure}
	~
	\begin{subfigure}[t]{.49\textwidth}
		\includegraphics[width=\textwidth]{boat_split_remesh_0p1_feature30}
	\end{subfigure}
	\caption{\emph{Boat}.
		Left: CAD model.
		Right: \emph{Boat} model remeshed after splitting along closing features.}
\end{figure}

\begin{table}[h!]
	\centering
	\begin{tabular}{lllrrrrrr}
		Algorithm & $d$ & $\vartheta$ & $| \mathcal{V} |$ & $| \mathcal{T} |$ & $\alpha_{\min}$ & $\alpha_{\max}$& $\alpha_{\text{avg}}$ & $\alpha_{\text{RMS}}$\\
		\hline
		MeshLab & 0.1 & 30$^{\circ}$ & 4,394 & 8,784 & 8.6603$^{\circ}$ & 157.4566$^{\circ}$ & 60$^{\circ}$ & 16.7 \\
		PMP & 0.1 & 30$^{\circ}$ & 5,388 & 10,772 & 11.8547$^{\circ}$ & 130.3723$^{\circ}$ & 60$^{\circ}$ & 15.2\\
		ours & 0.1 & 30$^{\circ}$ & 4,051 & 8,098 & 4.2360$^{\circ}$ & 130.0883$^{\circ}$ & 60$^{\circ}$ & 20.6\\
		& & & & & & & & \\			
		$E_{\min}$ & $E_{\text{max}}$ & $E_{\text{avg}}$ & $E_{\text{RMS}}$  & $Q_{\text{min}}$ & $Q_{\text{max}}$ & $Q_{\text{avg}}$ & $Q_{\text{RMS}}$ & \\
		\hline
		0.0167 & 0.1782 & 0.1027 & 12.9 & 0.2164 & 0.9999 & 0.9617 & 5.4 &  \\
		0.0223 & 0.1672 & 0.092 & 12.1 & 0.3474 & 0.9999 & 0.9694 & 4.1 & \\
		0.0078 & 0.1976 & 0.1079 & 13.3 & 0.1274 & 1.0000 & 0.9500 & 6.2 & 
	\end{tabular}
	\caption{Experimental results for \emph{Boat}.}
\end{table}

\begin{table}[b!]
	\centering
	\begin{tabular}{lrrrr}
		Algorithm & $d$ & $\vartheta$ & $d_{\max}$ & $\frac{d_{\max}}{d}$ \\
		\hline
		MeshLab & 0.1 & 30$^{\circ}$ & 0.0158 & 0.1582\\
		PMP & 0.1 & 30$^{\circ}$ & 0.0143 & 0.1435\\
		ours & 0.1 & 30$^{\circ}$ & 0.0156 & 0.1566
	\end{tabular}
	\caption{One-sided Hausdorff distance evaluated on the \emph{Boat}.}
\end{table}

\newpage

\subsection{\emph{Kitten}} \textcolor{sowaswieweiss}{y}

\begin{figure}[h!]
	\centering
	\begin{subfigure}[t]{0.18\textwidth}
		\includegraphics[width=\textwidth]{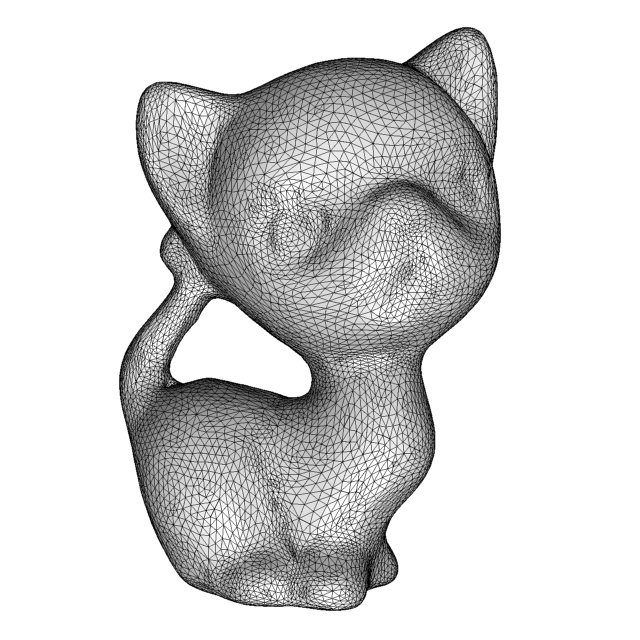}
	\end{subfigure}
	~
	\begin{subfigure}[t]{0.18\textwidth}
		\includegraphics[width=\textwidth]{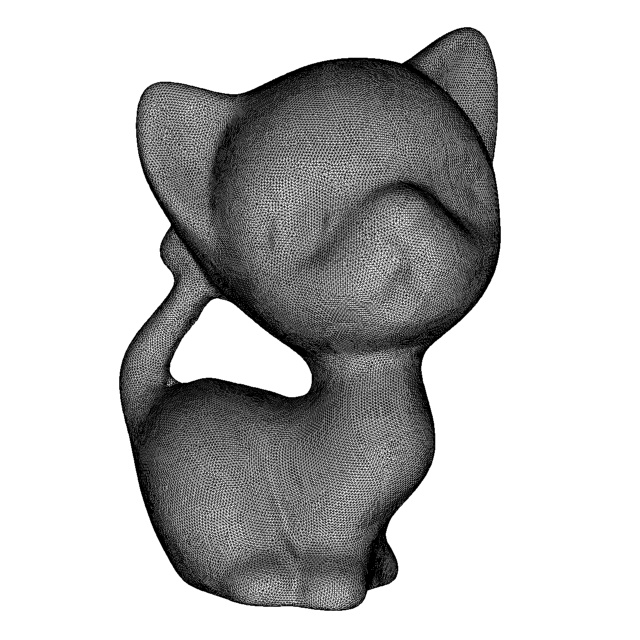}
	\end{subfigure}
	~
	\begin{subfigure}[t]{0.18\textwidth}
		\includegraphics[width=\textwidth]{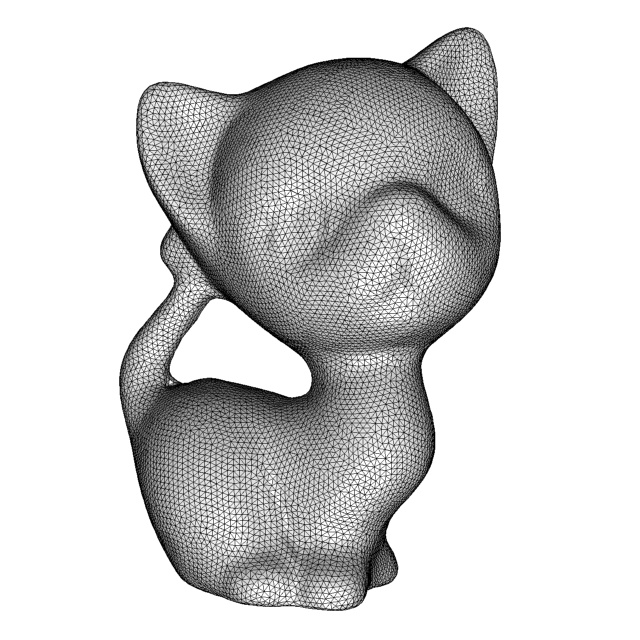}
	\end{subfigure}
	~
	\begin{subfigure}[t]{0.18\textwidth}
		\includegraphics[width=\textwidth]{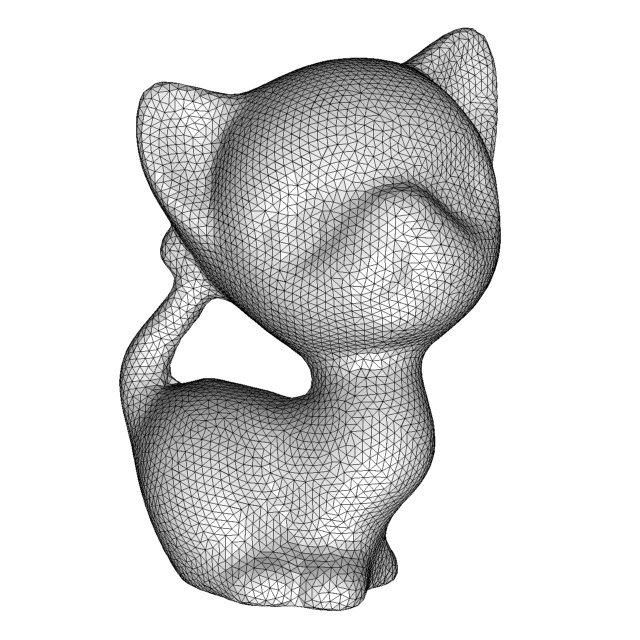}
	\end{subfigure}
	~
	\begin{subfigure}[t]{0.18\textwidth}
		\includegraphics[width=\textwidth]{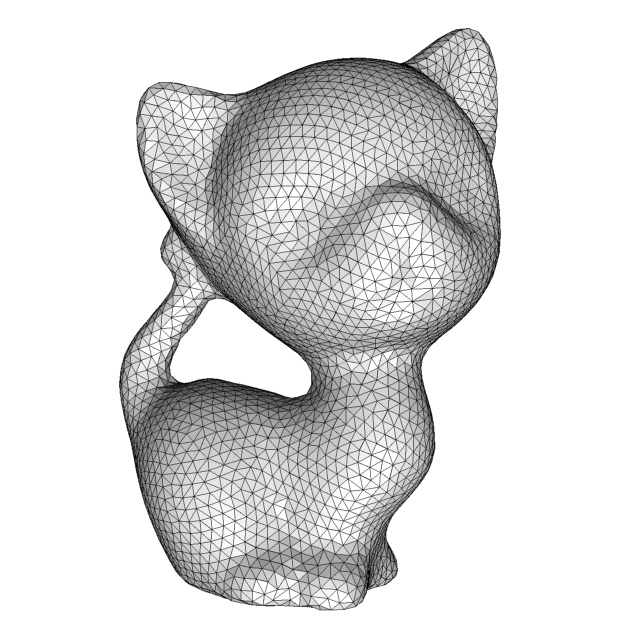}
	\end{subfigure}
	
	\begin{subfigure}[t]{0.18\textwidth}
		\includegraphics[width=\textwidth]{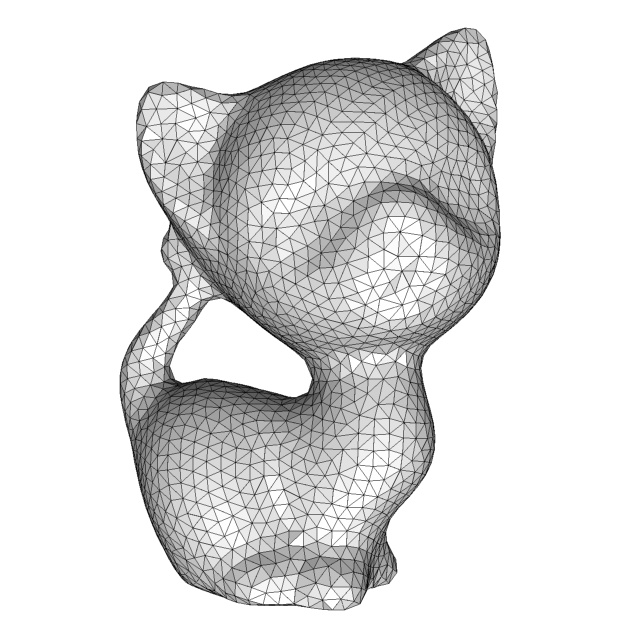}
	\end{subfigure}
	~
	\begin{subfigure}[t]{0.18\textwidth}
		\includegraphics[width=\textwidth]{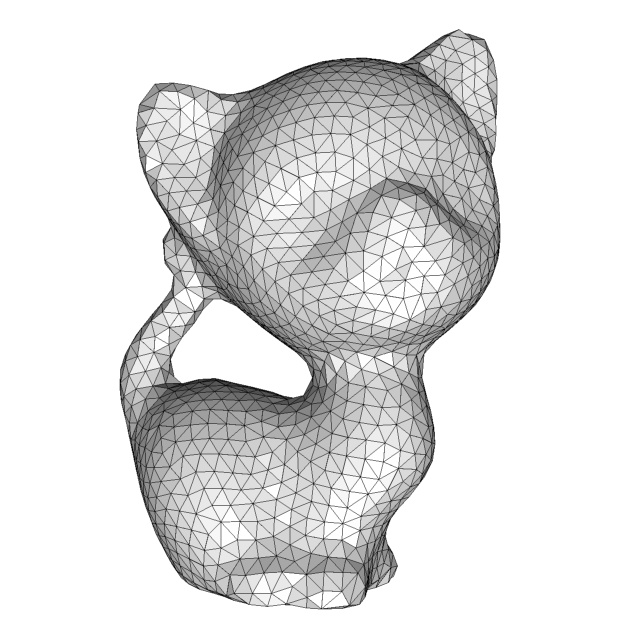}
	\end{subfigure}
	~
	\begin{subfigure}[t]{0.18\textwidth}
		\includegraphics[width=\textwidth]{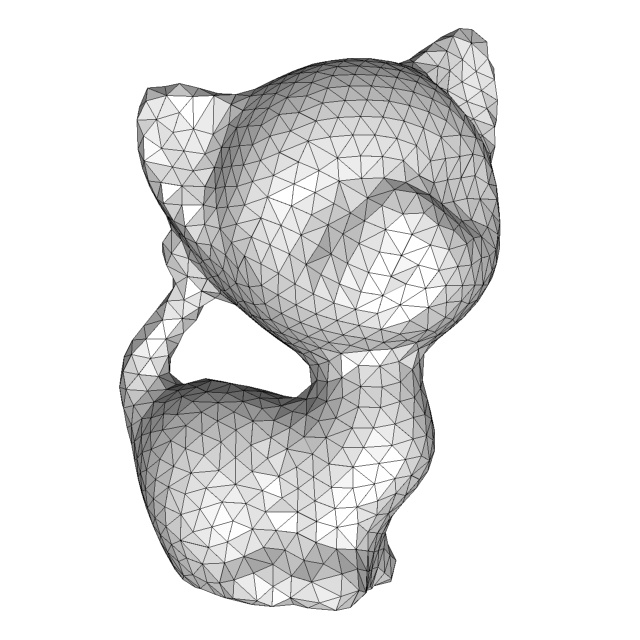}
	\end{subfigure}
	~
	\begin{subfigure}[t]{0.18\textwidth}
		\includegraphics[width=\textwidth]{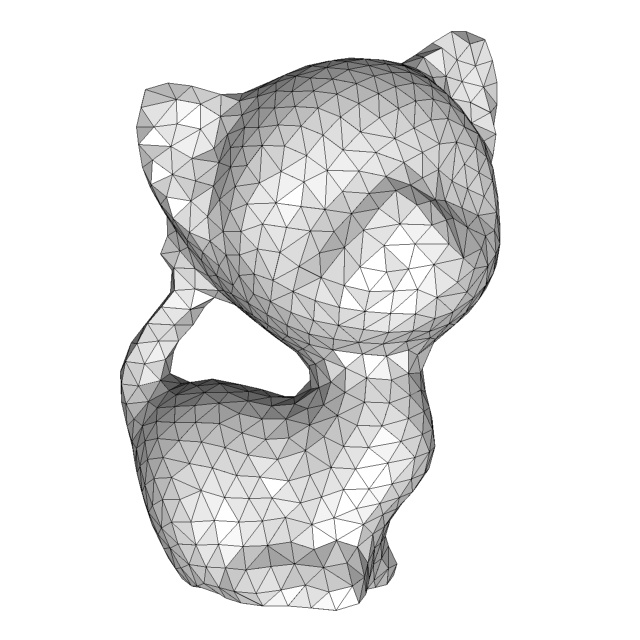}
	\end{subfigure}
	\caption{\emph{Kitten}.
		Upper row, from left to right: input model, followed by results achieved for target edge length $d \in \left\lbrace 0.01, 0.02, 0.03, 0.04 \right\rbrace$.
		Lower row, from left to right: results achieved for target edge length $d \in \left\lbrace 0.05, 0.06, 0.07, 0.08 \right\rbrace$.}
\end{figure}

\begin{figure}[h!]
	\centering
	\begin{minipage}{0.3\textwidth}
		\begin{subfigure}[t]{\Histogram}
			\centering
			\includegraphics[width=\textwidth]{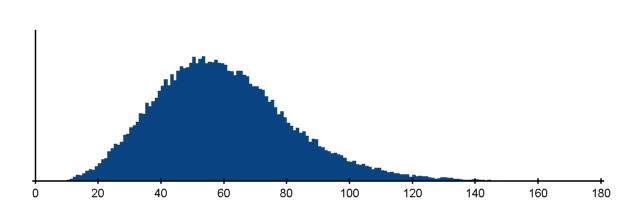}
			\caption*{Angle distribution.}
		\end{subfigure}
		\begin{subfigure}[t]{\Histogram}
			\centering
			\includegraphics[width=\textwidth]{cubeWithHole_0p04_noFeatures_histogram_edge}
			\caption*{Edge lengths distribution.}
		\end{subfigure}
		\begin{subfigure}[t]{\Histogram}
			\centering
			\includegraphics[width=\textwidth]{cubeWithHole_0p04_noFeatures_histogram_quality}
			\caption*{Distribution of quality $Q_t$.}
		\end{subfigure}
		\caption*{Histogram of input.}
	\end{minipage}
	~
	\begin{minipage}{0.3\textwidth}
		\begin{subfigure}[t]{\Histogram}
			\centering
			\includegraphics[width=\textwidth]{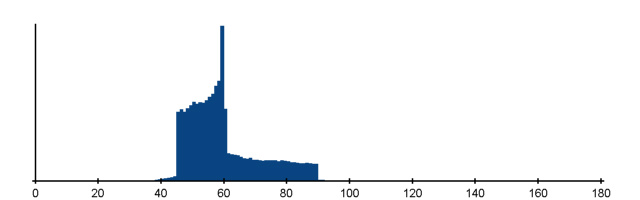}
			\caption*{Angle distribution.}
		\end{subfigure}
		\begin{subfigure}[t]{\Histogram}
			\centering
			\includegraphics[width=\textwidth]{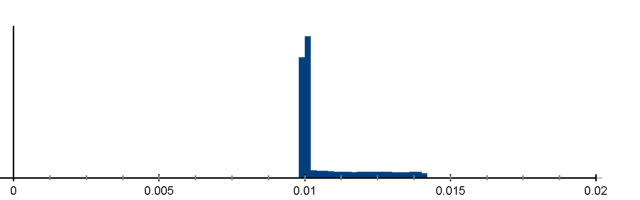}
			\caption*{Edge lengths distribution.}
		\end{subfigure}
		\begin{subfigure}[t]{\Histogram}
			\centering
			\includegraphics[width=\textwidth]{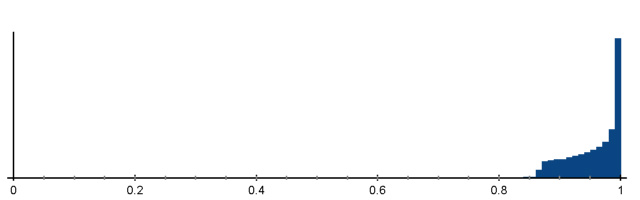}
			\caption*{Distribution of quality $Q_t$.}
		\end{subfigure}
		\caption*{Histograms for $d = 0.01$.}
	\end{minipage}
	~
	\begin{minipage}{0.3\textwidth}
		\begin{subfigure}[t]{\Histogram}
			\centering
			\includegraphics[width=\textwidth]{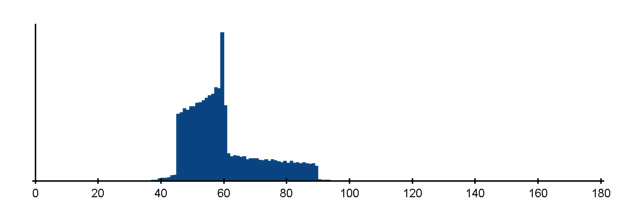}
			\caption*{Angle distribution.}
		\end{subfigure}
		\begin{subfigure}[t]{\Histogram}
			\centering
			\includegraphics[width=\textwidth]{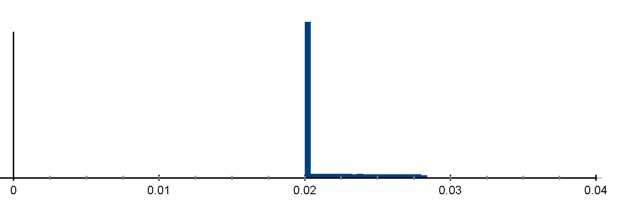}
			\caption*{Edge lengths distribution.}
		\end{subfigure}
		\begin{subfigure}[t]{\Histogram}
			\centering
			\includegraphics[width=\textwidth]{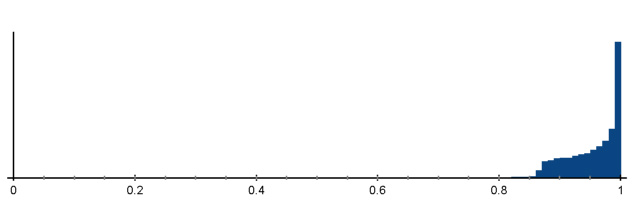}
			\caption*{Distribution of quality $Q_t$.}
		\end{subfigure}
		\caption*{Histograms for $d = 0.02$.}
	\end{minipage}
	
	\begin{minipage}{0.3\textwidth}
		\begin{subfigure}[t]{\Histogram}
			\centering
			\includegraphics[width=\textwidth]{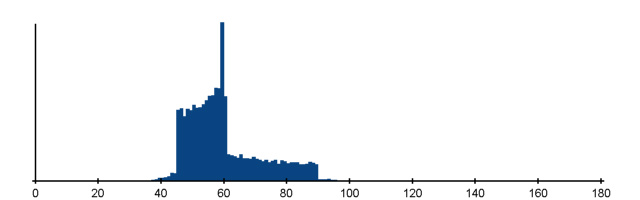}
			\caption*{Angle distribution.}
		\end{subfigure}
		\begin{subfigure}[t]{\Histogram}
			\centering
			\includegraphics[width=\textwidth]{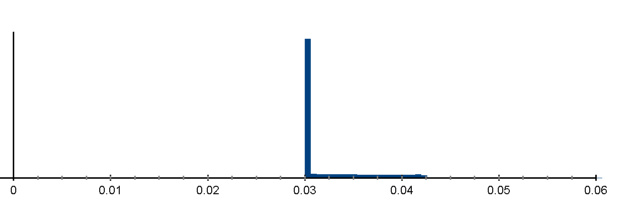}
			\caption*{Edge lengths distribution.}
		\end{subfigure}
		\begin{subfigure}[t]{\Histogram}
			\centering
			\includegraphics[width=\textwidth]{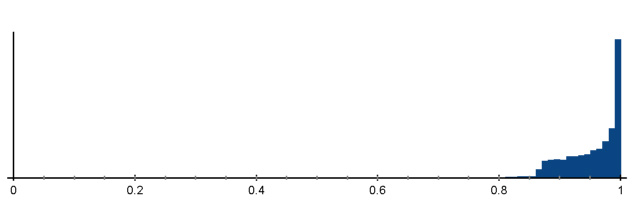}
			\caption*{Distribution of quality $Q_t$.}
		\end{subfigure}
		\caption*{Histograms for $d = 0.03$.}
	\end{minipage}
	~
	\begin{minipage}{0.3\textwidth}
		\begin{subfigure}[t]{\Histogram}
			\centering
			\includegraphics[width=\textwidth]{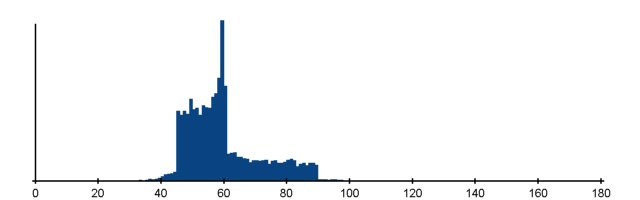}
			\caption*{Angle distribution.}
		\end{subfigure}
		\begin{subfigure}[t]{\Histogram}
			\centering
			\includegraphics[width=\textwidth]{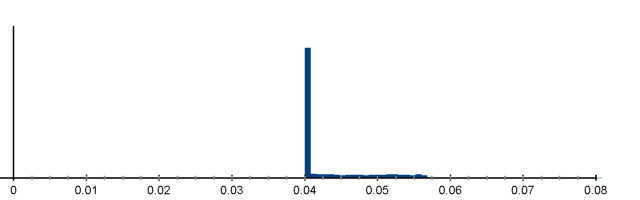}
			\caption*{Edge lengths distribution.}
		\end{subfigure}
		\begin{subfigure}[t]{\Histogram}
			\centering
			\includegraphics[width=\textwidth]{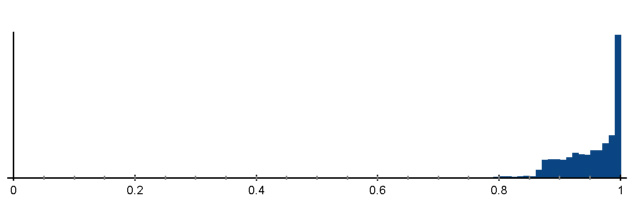}
			\caption*{Distribution of quality $Q_t$.}
		\end{subfigure}
		\caption*{Histograms for $d = 0.04$.}
	\end{minipage}
	~
	\begin{minipage}{0.3\textwidth}
		\begin{subfigure}[t]{\Histogram}
			\centering
			\includegraphics[width=\textwidth]{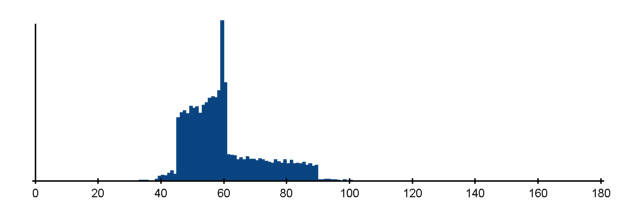}
			\caption*{Angle distribution.}
		\end{subfigure}
		\begin{subfigure}[t]{\Histogram}
			\centering
			\includegraphics[width=\textwidth]{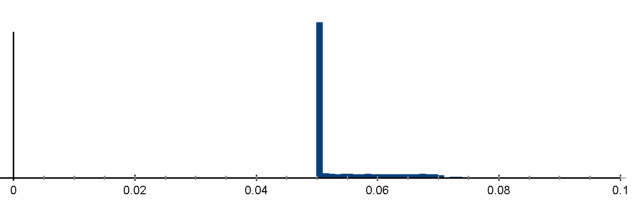}
			\caption*{Edge lengths distribution.}
		\end{subfigure}
		\begin{subfigure}[t]{\Histogram}
			\centering
			\includegraphics[width=\textwidth]{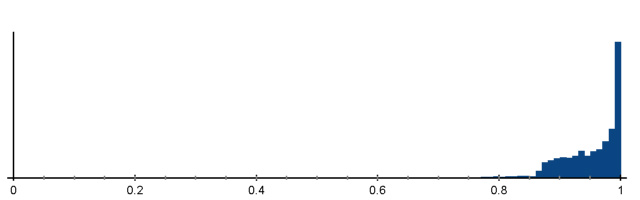}
			\caption*{Distribution of quality $Q_t$.}
		\end{subfigure}
		\caption*{Histograms for $d = 0.05$.}
	\end{minipage}
	
	\begin{minipage}{0.3\textwidth}
		\begin{subfigure}[t]{\Histogram}
			\centering
			\includegraphics[width=\textwidth]{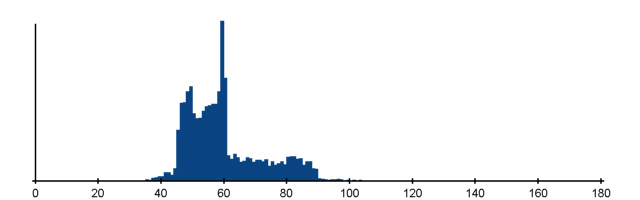}
			\caption*{Angle distribution.}
		\end{subfigure}
		\begin{subfigure}[t]{\Histogram}
			\centering
			\includegraphics[width=\textwidth]{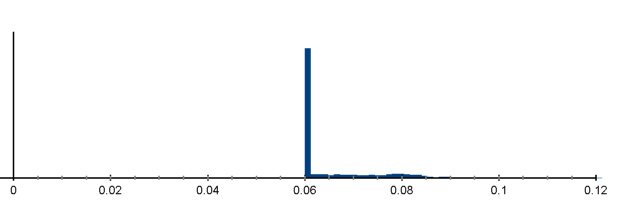}
			\caption*{Edge lengths distribution.}
		\end{subfigure}
		\begin{subfigure}[t]{\Histogram}
			\centering
			\includegraphics[width=\textwidth]{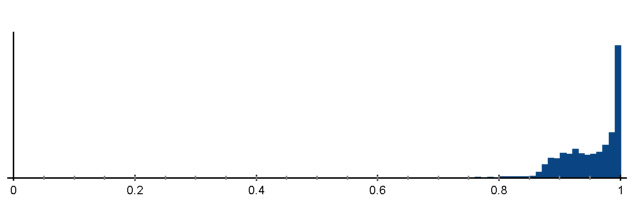}
			\caption*{Distribution of quality $Q_t$.}
		\end{subfigure}
		\caption*{Histograms for $d = 0.06$.}
	\end{minipage}
	~
	\begin{minipage}{0.3\textwidth}
		\begin{subfigure}[t]{\Histogram}
			\centering
			\includegraphics[width=\textwidth]{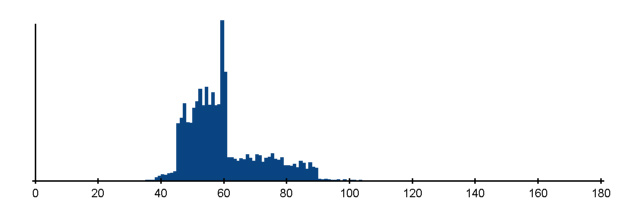}
			\caption*{Angle distribution.}
		\end{subfigure}
		\begin{subfigure}[t]{\Histogram}
			\centering
			\includegraphics[width=\textwidth]{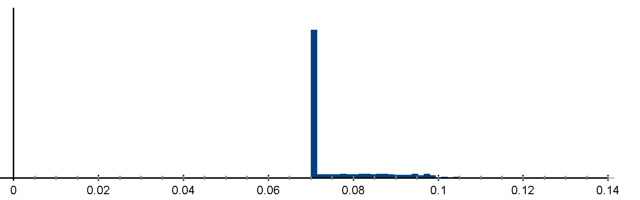}
			\caption*{Edge lengths distribution.}
		\end{subfigure}
		\begin{subfigure}[t]{\Histogram}
			\centering
			\includegraphics[width=\textwidth]{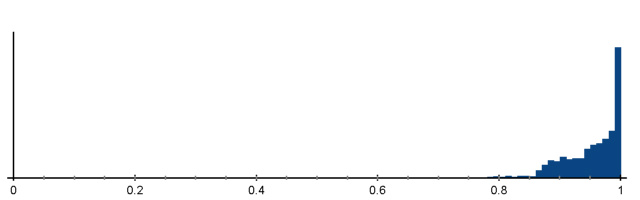}
			\caption*{Distribution of quality $Q_t$.}
		\end{subfigure}
		\caption*{Histograms for $d = 0.07$.}
	\end{minipage}
	~
	\begin{minipage}{0.3\textwidth}
		\begin{subfigure}[t]{\Histogram}
			\centering
			\includegraphics[width=\textwidth]{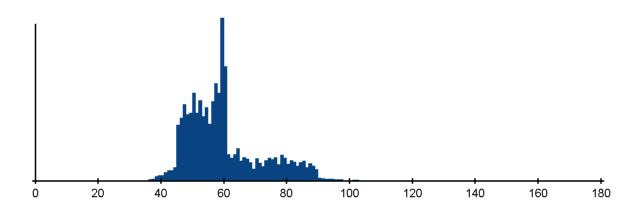}
			\caption*{Angle distribution.}
		\end{subfigure}
		\begin{subfigure}[t]{\Histogram}
			\centering
			\includegraphics[width=\textwidth]{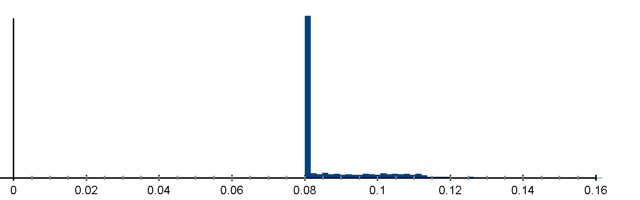}
			\caption*{Edge lengths distribution.}
		\end{subfigure}
		\begin{subfigure}[t]{\Histogram}
			\centering
			\includegraphics[width=\textwidth]{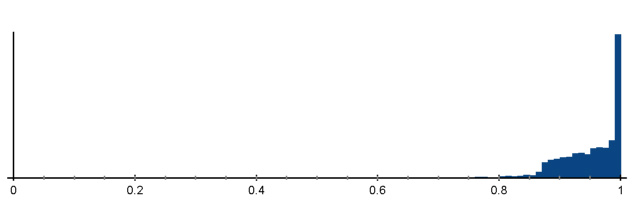}
			\caption*{Distribution of quality $Q_t$.}
		\end{subfigure}
		\caption*{Histograms for $d = 0.08$.}
	\end{minipage}
	\caption{Histograms for \emph{Kitten} model with various target edge lengths and without feature detection.}
\end{figure}

\begin{table}[h!]
	\centering
	\begin{tabular}{llllrrrr}
		& $d$ & $| \mathcal{V} |$ & $| \mathcal{T} |$ & $\alpha_{\min}$ & $\alpha_{\max}$& $\alpha_{\text{avg}}$ & $\alpha_{\text{RMS}}$\\
		\hline
		input & --- &10,000 & 20,000 & 7.6919$^{\circ}$ & 153.3379$^{\circ}$ & 60$^{\circ}$ & 35.9 \\
		remesh & 0.01 & 70,981 & 141,962 & 28.1880$^{\circ}$ & 122.9065$^{\circ}$ & 60$^{\circ}$ & 19.3 \\
		remesh & 0.02 & 17,673 & 35,346 & 30.2108$^{\circ}$ & 112.2197$^{\circ}$ & 60$^{\circ}$ & 19.5 \\
		\hline
		remesh & 0.03 & 7,837 & 15,674 & 28.3165$^{\circ}$ & 115.9218$^{\circ}$ & 60$^{\circ}$ & 19.6 \\
		remesh & 0.04 & 4,387 & 8,774 & 30.1886$^{\circ}$ & 112.2698$^{\circ}$ & 60$^{\circ}$ & 19.8 \\
		remesh & 0.05 & 2,797 & 5,594 & 30.3553$^{\circ}$ & 110.9096$^{\circ}$ & 60$^{\circ}$ & 19.8 \\
		\hline
		remesh & 0.06 & 1,933 & 3,866 & 32.7483$^{\circ}$ & 108.4652$^{\circ}$ & 60$^{\circ}$ & 20.3 \\
		remesh & 0.07 & 1,417 & 2,834 & 31.2655$^{\circ}$ & 112.4897$^{\circ}$ & 60$^{\circ}$ & 19.6 \\
		remesh & 0.08 & 1,080 & 2,160 & 32.0148$^{\circ}$ & 109.0594$^{\circ}$ & 60$^{\circ}$ & 20.0 \\
		& & & & & & & \\
		$E_{\min}$ & $E_{\text{max}}$ & $E_{\text{avg}}$ & $E_{\text{RMS}}$  & $Q_{\text{min}}$ & $Q_{\text{max}}$ & $Q_{\text{avg}}$ & $Q_{\text{RMS}}$ \\
		\hline
		0.0039 & 0.1119 & 0.0287 & 38.2 & 0.2296 & 0.9999 & 0.8438 & 16.4\\
		0.01 & 0.0210 & 0.0106 & 11.3 & 0.5717 & 1.0000 & 0.9566 & 4.4\\
		0.02 & 0.0391 & 0.0213 & 11.5 & 0.6742 & 1.0000 & 0.9558 & 4.5 \\
		\hline
		0.03 & 0.0613 & 0.0320 & 11.8 & 0.6391 & 1.0000 & 0.9552 & 4.7 \\
		0.04 & 0.0782 & 0.0428 & 12.1 & 0.6737 & 1.0000 & 0.9542 & 4.8 \\
		0.05 & 0.9594 & 0.0536 & 12.1 & 0.6864 & 1.0000 & 0.9542 & 4.8 \\
		\hline
		0.06 & 0.1106 & 0.0644 & 12.3 & 0.7091 & 1.0000 & 0.9520 & 4.9 \\
		0.07 & 0.1304 & 0.0751 & 12.1 & 0.6716 & 1.0000 & 0.9549 & 4.7 \\
		0.08 & 0.1485 & 0.0859 & 12.4 & 0.7036 & 1.0000 & 0.9532 & 4.9
	\end{tabular}
	\caption{Experimental results for remeshing \emph{Kitten}.}
\end{table}

\begin{table}[h!]
	\centering
	\begin{tabular}{lrrr|lrrr}
		Algorithm & $d$ & $d_{\max}$ & $\frac{d_{\max}}{d}$ & Algorithm & $d$ & $d_{\max}$ & $\frac{d_{\max}}{d}$ \\
		\hline
		ours & 0.01 & 0.0054 & 0.5406 & ours & 0.06 & 0.0319 & 0.5328\\
		ours & 0.02 & 0.0082 & 0.4145 & ours & 0.08 & 0.0410 & 0.5135\\
		ours & 0.04 & 0.0250 & 0.6266 & & & & 
	\end{tabular}
	\caption{One-sided Hausdorff distance evaluated on the \emph{Kitten}.}
\end{table}

\end{document}

%% file: SplatsOfSizeD.tex
\begin{tikzpicture}
	\coordinate (c1) at (2.5,0);
	\coordinate (c2) at (6.5,.5);
	\coordinate (c3) at (8,4.5);
	\coordinate (c4) at (3,5);
	\draw[black,very thick,fill=grey1,opacity=0.7] (c1) to[out=15, in=170] (c2) to [out=5,in=350] (c3) to[out=165,in=12.5] (c4) to[out=200,in=175] (c1);
	\begin{scope}[shift={(2,3.7)}]
		\begin{scope}[rotate=5]
			\draw[black,very thick,fill=grey2,opacity=.7] (1.59,0) ellipse (\radone cm and \radtwo cm);
			\Ellipse;
			\begin{scope}[scale=.75]
				\draw[fill=black] (2,0) -- (2.25,0) -- (2.25,1.5) -- (2.5,1.5) -- (2.125,2) -- (1.75,1.5) -- (2,1.5) -- (2,0);
			\end{scope}
		\end{scope}
	\end{scope}
	\begin{scope}[shift={(4.5,3.25)}]
		\begin{scope}[rotate=-10]
			\draw[black,very thick,fill=grey2,opacity=.7] (1.59,0) ellipse (\radone cm and \radtwo cm);
			\Ellipse;
			\begin{scope}[scale=.75]
				\draw[fill=black] (2,0) -- (2.25,0) -- (2.25,1.5) -- (2.5,1.5) -- (2.125,2) -- (1.75,1.5) -- (2,1.5) -- (2,0);
			\end{scope}
		\end{scope}
	\end{scope}
	\begin{scope}[shift={(1.25,2.25)}]
		\begin{scope}[rotate=15]
			\draw[black,very thick,fill=grey2,opacity=.7] (1.59,0) ellipse (\radone cm and \radtwo cm);
			\Ellipse;
			\begin{scope}[scale=.75]
				\draw[fill=black] (2,0) -- (2.25,0) -- (2.25,1.5) -- (2.5,1.5) -- (2.125,2) -- (1.75,1.5) -- (2,1.5) -- (2,0);
			\end{scope}
		\end{scope}
	\end{scope}
	\begin{scope}[shift={(4.75,1.85)}]
	\begin{scope}[rotate=-7.5]
		\draw[black,very thick,fill=grey2,opacity=.7] (1.59,0) ellipse (\radone cm and \radtwo cm);
		\Ellipse;
		\begin{scope}[scale=.75]
			\draw[fill=black] (2,0) -- (2.25,0) -- (2.25,1.5) -- (2.5,1.5) -- (2.125,2) -- (1.75,1.5) -- (2,1.5) -- (2,0);
		\end{scope}
	\end{scope}
	\end{scope}
\begin{scope}[shift={(3,2.5)}]
	\begin{scope}[rotate=5]
		\draw[black,very thick,fill=grey2,opacity=.7] (1.59,0) ellipse (\radone cm and \radtwo cm);
		\draw[black,very thick,fill=black] (1.59,0) ellipse (.3cm and .1cm);
		\begin{scope}[scale=.75]
			\draw[fill=black] (2,0) -- (2.25,0) -- (2.25,1.5) -- (2.5,1.5) -- (2.125,2) -- (1.75,1.5) -- (2,1.5) -- (2,0);
		\end{scope}
	\end{scope}
\end{scope}
\begin{scope}[shift={(2.75,1.25)}]
	\begin{scope}[rotate=5]
		\draw[black,very thick,fill=grey2,opacity=.7] (1.59,0) ellipse (\radone cm and \radtwo cm);
		\Ellipse;
		\begin{scope}[scale=.75]
			\draw[fill=black] (2,0) -- (2.25,0) -- (2.25,1.5) -- (2.5,1.5) -- (2.125,2) -- (1.75,1.5) -- (2,1.5) -- (2,0);
		\end{scope}
		\draw[dashed] (1.59,0) -- (0,-.5);
	\end{scope}
\end{scope}
	\node (M) at (8,4){$\mathcal{M}$};
	\node (PI) at (2.2,2.7){$p_i$};
	\node (PJ) at (4.2,4.25){$p_j$};
	\node (PK) at (5,1.4){$p_k$};
	\node (PL) at (7.,1.5){$p_l$};
	\node (PM) at (6.6,3.2){$p_m$};
	\node (P) at (5.1,2.7){$p$};
	\node(S) at (3.5,1.25){$s$};
\end{tikzpicture}

%% file: projectionToTangentPlane.tex
		\tikzset{
		partial ellipse/.style args={#1:#2:#3}{
			insert path={+ (#1:#3) arc (#1:#2:#3)}
		}
	}
\begin{tikzpicture}
	\draw[very thick, fill=grey2, opacity =.7] (0.5,0.5) -- (7,0.5) -- (8,4.5) -- (1.5,4.5) -- (0.5,0.5);
	\begin{scope}[shift={(1.2,0)}]
		\draw[thick] (3,2.5) -- (2,3.7);
		\draw[dashed] (3,2.5) -- (4.5,3.25);
		\draw[dashed] (3,2.5) -- (1.25,2.25);
		\draw[dashed] (3,2.5) -- (4.75,1.85);
		\draw[dashed] (3,2.5) -- (2.75,1.25);
		\draw (2.3,3.5) [partial ellipse=131:185:.55 and .3];
		\draw[fill=black] (2,3.7) circle (.066); 
		\draw[fill=black] (4.5,3.25) circle (.066); 
		\draw[fill=black] (1.25,2.25) circle (.066); 
		\draw[fill=black] (4.75,1.85) circle (.066); 
		\draw[fill=black] (2.75,1.25) circle (.066); 
		\draw[fill=black] (3,2.5) circle (.066); 
		\draw[fill=black] (1.75,3.4) circle (0.066); 
		\draw[dashed] (1.75,3.4) -- (3,2.5);
		\node(pj) at (2,3.95){$\pi(p_j)$};
		\node(projj) at (1.45,3.4){$p_j$};
		\draw[fill=black] (2.75,1.6) circle (0.066);
		\draw[thick] (2.75,1.6) -- (3,2.5);
		\draw (4.7,1.425) [partial ellipse=175:185: 2 and 2];
		\node(pk) at (2.75,1.){$p_k$};
		\node(projk) at (2.2,1.5){$\pi(p_k)$};
		\draw[fill=black] (4.55,3.45) circle (0.066);
		\draw[thick] (4.55,3.45) -- (3,2.5);
		\draw (4.3,3.45) [partial ellipse=320:360: .25 and .25];
		\node(pm) at (4.85,3.2){$p_m$};
		\node(projm) at (4.55,3.725){$\pi(p_m)$};
		\draw[fill=black] (1.2,2.45) circle (0.066);
		\draw[thick] (1.2,2.45) -- (3,2.5);
		\node(pi) at (1.25,1.95){$p_i$};
		\node(proji) at (.75,2.45){$\pi(p_i)$};
		\draw (1.6,2.45) [partial ellipse=180:210: .4 and .4];
		\draw[fill=black] (4.85,2) circle (0.066);
		\draw[thick] (4.85,2) -- (3,2.5);
		\node(pl) at (5,1.7){$p_l$};
		\node(projl) at (5.35,2){$\pi(p_l)$};
		\draw (4.25,2.05) [partial ellipse=355:330: .6 and .5];
		\node(TpM) at (6,4){$T_p\mathcal{M}$};
		\node (d) at (2,2.65){$d_i$};
		\node (d1) at (2,2.15){$d_i$};
		\node (p) at (3.05,2.75){$p$};
	\end{scope}
\end{tikzpicture}

%% file: VoronoiCells.tex
\begin{tikzpicture}
	\draw[very thick, fill=grey2, opacity =.7] (0.5,0.5) -- (7,0.5) -- (8,4.5) -- (1.5,4.5) -- (0.5,0.5);
	\begin{scope}[shift={(1.2,0)}]
		\draw[thin] (3,2.5) -- (2,3.7);
		\draw[fill=black] (2,3.7) circle (.066); 
		\draw[fill=black] (3,2.5) circle (.066); 
		\node(pj) at (2,3.95){$\pi(p_j)$};
		\draw[fill=black] (2.75,1.6) circle (0.066);
		\draw[thin] (2.75,1.6) -- (3,2.5);
		\node(projk) at (2.2,1.5){$\pi(p_k)$};
		\draw[fill=black] (4.55,3.45) circle (0.066);
		\draw[thin] (4.55,3.45) -- (3,2.5);
		\node(projm) at (4.55,3.725){$\pi(p_m)$};
		\draw[fill=black] (1.2,2.45) circle (0.066);
		\draw[thin] (1.2,2.45) -- (3,2.5);
		\node(proji) at (.75,2.45){$\pi(p_i)$};
		\draw[fill=black] (4.85,2) circle (0.066);
		\draw[thin] (4.85,2) -- (3,2.5);
		\draw[fill=black] (4.7,2.5) circle (0.066);
		\draw (4.7,2.5) circle (0.125);
		\node(projl) at (5.35,2){$\pi(p_l)$};
		\node(TpM) at (6,4){$T_p\mathcal{M}$};
		\node (p) at (3.05,2.75){$p$};
		\coordinate (mj) at (3.4,3.5);
		\coordinate (ji) at (2,2.8);
		\coordinate (ik) at (1.9,2);
		\coordinate (kl) at (3.5,1.8);
		\coordinate (lm) at (4.7,2.5);
		\draw[thick] (mj) -- (ji) -- (ik) -- (kl) -- (lm) -- (mj);
		\coordinate (mj1) at (3.7,4.2);
		\coordinate (ji1) at (1,3.2);
		\coordinate (ik1) at (.6,1.1);
		\coordinate (kl1) at (4,1);
		\coordinate (lm1) at (5.4,2.5);
		\draw[dashed] (mj) -- (mj1);
		\draw[dashed] (ji) -- (ji1);
		\draw[dashed] (ik) -- (ik1);
		\draw[dashed] (kl) -- (kl1);
		\draw[dashed] (lm) -- (lm1);
	\end{scope}
\end{tikzpicture}

%% file: IndividualSplatSizes.tex
%
\begin{tikzpicture}
	\coordinate (c1) at (2.5,0);
	\coordinate (c2) at (6.5,.5);
	\coordinate (c3) at (8,4.5);
	\coordinate (c4) at (3,5);
	\draw[black,very thick,fill=grey1,opacity=0.7] (c1) to[out=15, in=170] (c2) to [out=5,in=350] (c3) to[out=165,in=12.5] (c4) to[out=200,in=175] (c1);
	\begin{scope}[shift={(2,3.7)}]
		\begin{scope}[rotate=5]
			\draw[black,very thick,fill=grey2,opacity=.7] (1.59,0) ellipse (1.8cm and .75cm);
			\draw[black,very thick,fill=black] (1.59,0) ellipse (.3cm and .1cm);
			\begin{scope}[scale=.75]
				\draw[fill=black] (2,0) -- (2.25,0) -- (2.25,1.5) -- (2.5,1.5) -- (2.125,2) -- (1.75,1.5) -- (2,1.5) -- (2,0);
			\end{scope}
		\end{scope}
	\end{scope}
	\begin{scope}[shift={(4.5,3.25)}]
		\begin{scope}[rotate=-10]
			\draw[black,very thick,fill=grey2,opacity=.7] (1.59,0) ellipse (1.8cm and .55cm);
			\draw[black,very thick,fill=black] (1.59,0) ellipse (.3cm and .1cm);
			\begin{scope}[scale=.75]
				\draw[fill=black] (2,0) -- (2.25,0) -- (2.25,1.5) -- (2.5,1.5) -- (2.125,2) -- (1.75,1.5) -- (2,1.5) -- (2,0);
			\end{scope}
		\end{scope}
	\end{scope}
	\begin{scope}[shift={(1.25,2.25)}]
		\begin{scope}[rotate=15]
			\draw[black,very thick,fill=grey2,opacity=.7] (1.59,0) ellipse (1.65cm and .5cm);
			\draw[black,very thick,fill=black] (1.59,0) ellipse (.3cm and .1cm);
			\begin{scope}[scale=.75]
				\draw[fill=black] (2,0) -- (2.25,0) -- (2.25,1.5) -- (2.5,1.5) -- (2.125,2) -- (1.75,1.5) -- (2,1.5) -- (2,0);
			\end{scope}
		\end{scope}
	\end{scope}
	\begin{scope}[shift={(4.75,1.85)}]
		\begin{scope}[rotate=-7.5]
			\draw[black,very thick,fill=grey2,opacity=.7] (1.59,0) ellipse (2.cm and .7cm);
			\draw[black,very thick,fill=black] (1.59,0) ellipse (.3cm and .1cm);
			\begin{scope}[scale=.75]
				\draw[fill=black] (2,0) -- (2.25,0) -- (2.25,1.5) -- (2.5,1.5) -- (2.125,2) -- (1.75,1.5) -- (2,1.5) -- (2,0);
			\end{scope}
		\end{scope}
	\end{scope}
	\begin{scope}[shift={(3,2.5)}]
		\begin{scope}[rotate=5]
			\draw[black,very thick,fill=grey2,opacity=.7] (1.59,0) ellipse (2cm and .6cm);
			\draw[black,very thick,fill=black] (1.59,0) ellipse (.3cm and .1cm);
			\begin{scope}[scale=.75]
				\draw[fill=black] (2,0) -- (2.25,0) -- (2.25,1.5) -- (2.5,1.5) -- (2.125,2) -- (1.75,1.5) -- (2,1.5) -- (2,0);
			\end{scope}
		\end{scope}
	\end{scope}
	\begin{scope}[shift={(2.75,1.25)}]
	\begin{scope}[rotate=5]
		\draw[black,very thick,fill=grey2,opacity=.7] (1.59,0) ellipse (1.9cm and .65cm);
		\draw[black,very thick,fill=black] (1.59,0) ellipse (.3cm and .1cm);
		\begin{scope}[scale=.75]
			\draw[fill=black] (2,0) -- (2.25,0) -- (2.25,1.5) -- (2.5,1.5) -- (2.125,2) -- (1.75,1.5) -- (2,1.5) -- (2,0);
		\end{scope}
		\draw[dashed] (1.59,0) -- (0,-.45);
	\end{scope}
\end{scope}
	\node (M) at (8,4){$\mathcal{M}$};
	\node (PI) at (2.2,2.7){$p_i$};
	\node (PI) at (4.2,4.25){$p_j$};
	\node (PI) at (5,1.4){$p_k$};
	\node (PI) at (7,1.5){$p_l$};
	\node (PI) at (6.6,3.2){$p_m$};
	\node (P) at (5.1,2.7){$p$};
	\node(S) at (3.5,1.275){$s_{p_k}$};
\end{tikzpicture}

%% file: SplitBorder_AddInnerVertex.tex
	\begin{tikzpicture}[scale=0.8]
		\node (n1) at (-1.744, 0.798){};
		\node (n2) at (-1.745, -0.201){};
		\node (n3) at (-1.042, -0.912){};
		\node (n4) at (-0.048, -0.804){};
		\node (n5) at (0.782, -0.248){};
		\node (n6) at (0.587, 0.732){};
		\node (n7) at ( -0.038, 1.512){};
		\node (n8) at (-1.038, 1.506){};
		\node (n)  at (-0.243, 0.176){};
		\coordinate (o1) at (-1.87,0.87);
		\coordinate (o2) at (-1.87,-.3);
		\coordinate (o3) at (-1.15,-1.1);
		\coordinate (o4) at (-.0,-1);
		\coordinate (o5) at (1,-.3);
		\coordinate (o6) at (.8,.85);
		\coordinate (o7) at (.1,1.7);
		\coordinate (o8) at (-1.15,1.7);
\draw[fill=lookslikewhite,draw=lookslikewhite] (0,-1.2) circle(.2);
		\fill[grey2] (n1.center) -- (n2.center) -- (n3.center) -- (n4.center) -- (n5.center) -- (n6.center) -- (n7.center) -- (n8.center);
		\draw[very thick, grey2,dashed] (o1) to[out=250,in=125] (o2) to[out=300,in=145] (o3) to[out=350,in=200] (o4) to[out=30,in=230] (o5) to[out=90,in=290] (o6) to[out=125,in=320] (o7) to[out=170,in=15] (o8) to[out=220,in=60] (o1);
		\fill[grey2] (n1.center) -- (n2.center) -- (n3.center) -- (n4.center) -- (n5.center) -- (n6.center) -- (n7.center) -- (n8.center);
		\begin{scope}[shift={(-.55,1.77)}]
			\draw[grey2,very thick] (-.1,-.1) -- (0,0) -- (-.1,.1);
		\end{scope}
		\draw[very thick] (n1.center) -- (n2.center) -- (n3.center) -- (n4.center) -- (n5.center) -- (n6.center) -- (n7.center) -- (n8.center) -- cycle;
		\draw[very thick, dashed, grey1,->] (-.5,1.2) arc (90:450:0.9);
		\draw[dotted, very thick] (n4.center) -- (n.center) -- (n6.center);
		\node[circle,fill=black,inner sep=0pt,minimum size=3.5pt](v)  at (n){};
		\node[circle,fill=black,inner sep=0pt,minimum size=3.5pt](v1) at (n1){};
		\node[circle,fill=black,inner sep=0pt,minimum size=3.5pt](v2) at (n2){};
		\node[circle,fill=black,inner sep=0pt,minimum size=3.5pt](v3) at (n3){};
		\node[circle,fill=black,inner sep=0pt,minimum size=3.5pt](v4) at (n4){};
		\node[circle,fill=black,inner sep=0pt,minimum size=3.5pt](v5) at (n5){};
		\node[circle,fill=black,inner sep=0pt,minimum size=3.5pt](v6) at (n6){};
		\node[circle,fill=black,inner sep=0pt,minimum size=3.5pt](v7) at (n7){};
		\node[circle,fill=black,inner sep=0pt,minimum size=3.5pt](v8) at (n8){};
		\begin{scope}[shift={(-.55,1.77)}]
			\draw[grey2,very thick] (-.1,-.1) -- (0,0) -- (-.1,.1);
		\end{scope}
		\node[left] (v) at (n){$v$};
		\node (rel1) at (1.5,0.2635){$\rightarrow$};
		\begin{scope}[shift={(4,0)}]
			\node (n1) at (-1.744, 0.798){};
			\node (n2) at (-1.745, -0.201){};
			\node (n3) at (-1.042, -0.912){};
			\node (n4) at (-0.048, -0.804){};
			\node (n5) at (0.782, -0.248){};
			\node (n6) at (0.587, 0.732){};
			\node (n7) at ( -0.038, 1.512){};
			\node (n8) at (-1.038, 1.506){};
			\node (n)  at (-0.243, 0.176){};
			\fill[grey2] (n.center) -- (n6.center) -- (n7.center) -- (n8.center) -- (n1.center) -- (n2.center) -- (n3.center) -- (n4.center) -- cycle;
			\fill[grey1] (n.center) -- (n4.center) -- (n5.center) -- (n6.center) -- cycle;
			\draw[very thick, dashed, ->, grey2] (.3,.3) arc (90:450:0.3);
			\draw[very thick] (n1.center) -- (n2.center) -- (n3.center) -- (n4.center) -- (n5.center) -- (n6.center) -- (n7.center) -- (n8.center) -- cycle;
			\draw[very thick] (n4.center) -- (n.center) -- (n6.center);
			\node[circle,fill=black,inner sep=0pt,minimum size=3.5pt](v)  at (n){};
			\node[circle,fill=black,inner sep=0pt,minimum size=3.5pt](v1) at (n1){};
			\node[circle,fill=black,inner sep=0pt,minimum size=3.5pt](v2) at (n2){};
			\node[circle,fill=black,inner sep=0pt,minimum size=3.5pt](v3) at (n3){};
			\node[circle,fill=black,inner sep=0pt,minimum size=3.5pt](v4) at (n4){};
			\node[circle,fill=black,inner sep=0pt,minimum size=3.5pt](v5) at (n5){};
			\node[circle,fill=black,inner sep=0pt,minimum size=3.5pt](v6) at (n6){};
			\node[circle,fill=black,inner sep=0pt,minimum size=3.5pt](v7) at (n7){};
			\node[circle,fill=black,inner sep=0pt,minimum size=3.5pt](v8) at (n8){};
			\draw[grey1, very thick, dashed,->] (0.25,.9) arc (35:90:1);
			\draw[grey1, very thick, dashed] (-.569,1.327) arc (90:285:1);
			\coordinate (avoidv) at (-.6,.1);
			\draw[grey1, very thick, dashed] (0.25,.9) to[out=180,in=90] (avoidv) to[out=90,in=180] (-.17,-.6);
			\coordinate (o1) at (-1.87,0.87);
			\coordinate (o2) at (-1.87,-.3);
			\coordinate (o3) at (-1.15,-1.1);
			\coordinate (o4) at (-.0,-1);
			\coordinate (o5) at (1,-.3);
			\coordinate (o6) at (.8,.85);
			\coordinate (o7) at (.1,1.7);
			\coordinate (o8) at (-1.15,1.7);
			\draw[very thick, grey2,dashed] (o1) to[out=250,in=125] (o2) to[out=300,in=145] (o3) to[out=350,in=200] (o4) to[out=30,in=230] (o5) to[out=90,in=290] (o6) to[out=125,in=320] (o7) to[out=170,in=15] (o8) to[out=220,in=60] (o1);
			\begin{scope}[shift={(-.55,1.77)}]
				\draw[grey2,very thick] (-.1,-.1) -- (0,0) -- (-.1,.1);
			\end{scope}
			\node[left] (v) at (-.343,.176){$v$};
		\end{scope}
	\end{tikzpicture}

%% file: SplitBorder_AddOuterVertex.tex
	\begin{tikzpicture}[scale=0.8]
		\node (n1) at (-1.744, 0.798){};
		\node (n2) at (-1.745, -0.201){};
		\node (n3) at (-1.042, -0.912){};
		\node (n4) at (-0.048, -0.804){};
		\node (n5) at (0.782, -0.248){};
		\node (n6) at (0.587, 0.732){};
		\node (n7) at ( -0.038, 1.512){};
		\node (n8) at (-1.038, 1.506){};
		\node (n)  at (-0.243, 0.176){};
		\coordinate (o1) at (-1.87,0.87);
		\coordinate (o2) at (-1.87,-.3);
		\coordinate (o3) at (-1.15,-1.1);
		\coordinate (o4) at (-.0,-1);
		\coordinate (o5) at (1,-.3);
		\coordinate (o6) at (.8,.85);
		\coordinate (o7) at (.1,1.7);
		\coordinate (o8) at (-1.15,1.7);
			\node (n1) at (-1.744, 0.798){};
			\node (n2) at (-1.745, -0.201){};
			\node (n3) at (-1.042, -0.912){};
			\node (n4) at (-0.048, -0.804){};
			\node (n5) at (0.782, -0.248){};
			\node (n6) at (0.587, 0.732){};
			\node (n7) at ( -0.038, 1.512){};
			\node (n8) at (-1.038, 1.506){};
			\node (n)  at (1.53, 0.41){};
\draw[fill=lookslikewhite,draw=lookslikewhite] (0,-1.2) circle(.2);
			\fill[grey2] (n1.center) -- (n2.center) -- (n3.center) -- (n4.center) -- (n5.center) -- (n6.center) -- (n7.center) -- (n8.center);
			\draw[very thick] (n1.center) -- (n2.center) -- (n3.center) -- (n4.center) -- (n5.center) -- (n6.center) -- (n7.center) -- (n8.center) -- cycle;
			\draw[dotted, very thick] (n5.center) -- (n.center) -- (n6.center);
			\node[circle,fill=black,inner sep=0pt,minimum size=3.5pt](v)  at (n){};
			\node[circle,fill=black,inner sep=0pt,minimum size=3.5pt](v1) at (n1){};
			\node[circle,fill=black,inner sep=0pt,minimum size=3.5pt](v2) at (n2){};
			\node[circle,fill=black,inner sep=0pt,minimum size=3.5pt](v3) at (n3){};
			\node[circle,fill=black,inner sep=0pt,minimum size=3.5pt](v4) at (n4){};
			\node[circle,fill=black,inner sep=0pt,minimum size=3.5pt](v5) at (n5){};
			\node[circle,fill=black,inner sep=0pt,minimum size=3.5pt](v6) at (n6){};
			\node[circle,fill=black,inner sep=0pt,minimum size=3.5pt](v7) at (n7){};
			\node[circle,fill=black,inner sep=0pt,minimum size=3.5pt](v8) at (n8){};
			\node[above] (v) at (n){$v$};
			\draw[very thick, dashed, grey1,->] (-.5,1.2) arc (90:450:0.9);
			\coordinate (o1) at (-1.87,0.87);
			\coordinate (o2) at (-1.87,-.3);
			\coordinate (o3) at (-1.15,-1.1);
			\coordinate (o4) at (-.0,-1);
			\coordinate (o5) at (1,-.3);
			\coordinate (o6) at (.8,.85);
			\coordinate (o7) at (.1,1.7);
			\coordinate (o8) at (-1.15,1.7);
			\draw[very thick, grey2,dashed] (o1) to[out=250,in=125] (o2) to[out=300,in=145] (o3) to[out=350,in=200] (o4) to[out=30,in=230] (o5) to[out=90,in=290] (o6) to[out=125,in=320] (o7) to[out=170,in=15] (o8) to[out=220,in=60] (o1);
			\begin{scope}[shift={(-.55,1.77)}]
				\draw[grey2,very thick] (-.1,-.1) -- (0,0) -- (-.1,.1);
			\end{scope}
			\node (rel1) at (2.,0.2635){$\rightarrow$};
			\begin{scope}[shift={(4.5,0)}]
				\node (n1) at (-1.744, 0.798){};
				\node (n2) at (-1.745, -0.201){};
				\node (n3) at (-1.042, -0.912){};
				\node (n4) at (-0.048, -0.804){};
				\node (n5) at (0.782, -0.248){};
				\node (n6) at (0.587, 0.732){};
				\node (n7) at ( -0.038, 1.512){};
				\node (n8) at (-1.038, 1.506){};
				\node (n)  at (1.53, 0.41){};
				\fill[grey2] (n1.center) -- (n2.center) -- (n3.center) -- (n4.center) -- (n5.center) -- (n6.center) -- (n7.center) -- (n8.center) -- cycle;
				\fill[grey1] (n.center) -- (n5.center) -- (n6.center) -- cycle;
				\draw[very thick] (n1.center) -- (n2.center) -- (n3.center) -- (n4.center) -- (n5.center) -- (n6.center) -- (n7.center) -- (n8.center) -- cycle;
				\draw[very thick] (n5.center) -- (n.center) -- (n6.center);
				\node[circle,fill=black,inner sep=0pt,minimum size=3.5pt](v)  at (n){};
				\node[circle,fill=black,inner sep=0pt,minimum size=3.5pt](v1) at (n1){};
				\node[circle,fill=black,inner sep=0pt,minimum size=3.5pt](v2) at (n2){};
				\node[circle,fill=black,inner sep=0pt,minimum size=3.5pt](v3) at (n3){};
				\node[circle,fill=black,inner sep=0pt,minimum size=3.5pt](v4) at (n4){};
				\node[circle,fill=black,inner sep=0pt,minimum size=3.5pt](v5) at (n5){};
				\node[circle,fill=black,inner sep=0pt,minimum size=3.5pt](v6) at (n6){};
				\node[circle,fill=black,inner sep=0pt,minimum size=3.5pt](v7) at (n7){};
				\node[circle,fill=black,inner sep=0pt,minimum size=3.5pt](v8) at (n8){};
				
				\coordinate (o1) at (-1.87,0.87);
				\coordinate (o2) at (-1.87,-.3);
				\coordinate (o3) at (-1.15,-1.1);
				\coordinate (o4) at (-.0,-1);
				\coordinate (o5) at (1,-.3);
				\coordinate (o6) at (.8,.85);
				\coordinate (o7) at (.1,1.7);
				\coordinate (o8) at (-1.15,1.7);
				\coordinate (umweg) at (1.9,.5);
				\draw[very thick, grey2,dashed] (o1) to[out=250,in=125] (o2) to[out=300,in=145] (o3) to[out=350,in=200] (o4) to[out=30,in=230] (o5) to[out=45,in=250] (umweg) to[out=150,in=320] (o6) to[out=125,in=320] (o7) to[out=170,in=15] (o8) to[out=220,in=60] (o1);
				\begin{scope}[shift={(-.55,1.77)}]
					\draw[grey2,very thick] (-.1,-.1) -- (0,0) -- (-.1,.1);
				\end{scope}
				\draw[very thick, dashed, grey1,->] (-.5,1.2) arc (90:450:0.9);
				\draw[very thick, dashed,grey2,->] (.95,.41) arc (90:450:.18);
				\node[right] (v) at (1.33,.775){$v$};
			\end{scope}
	\end{tikzpicture}

%% file: countryJoin.tex
	\begin{tikzpicture}[scale=1.3]
			\begin{scope}[rotate=90]
	\node (u1) at (0,2.5){};
	\node (u2) at (0,2.3){};
	\node (u3) at (0.26,1.8){};
	\node (u4) at (0.65,1.2){};
	\node (u5) at (1.25,1.2){};
	\node (u6) at (1.7,1.4){};
	\node (u7) at (2,1.8){};
	\node (u8) at (2,2.3){};
	\fill[grey2] (u2.center) -- (u3.center) -- (u4.center) -- (u5.center) -- (u6.center) -- (u7.center) -- (u8.center) to[bend right] (u2.center);
	\node[circle,fill=black,inner sep=0pt,minimum size=3.5pt](v2) at (u2){};
	\node[circle,fill=black,inner sep=0pt,minimum size=3.5pt](v3) at (u3){};
	\node[circle,fill=black,inner sep=0pt,minimum size=3.5pt](v4) at (u4){};
	\node[circle,fill=black,inner sep=0pt,minimum size=3.5pt](v5) at (u5){};
	\node[circle,fill=black,inner sep=0pt,minimum size=3.5pt](v6) at (u6){};
	\node[circle,fill=black,inner sep=0pt,minimum size=3.5pt](v7) at (u7){};
	\node[circle,fill=black,inner sep=0pt,minimum size=3.5pt](v8) at (u8){};
	\draw[very thick] (u2.center) --  (u3.center) -- (u4.center) -- (u5.center) -- (u6.center) -- (u7.center) -- (u8.center);
	\draw[very thick, dashed] (u8.center) to[bend right] (u2.center);
	\node (l1) at (0,0){};
	\node (l2) at (0,0){};
	\node (l3) at (0.45,0.35){};
	\node (l4) at (1,0.5){};
	\node (l5) at (1.5,.6){};
	\node (l6) at (1.9,.4){};
	\node (l7) at (2,0){};
	\fill[grey2] (l2.center) -- (l3.center) -- (l4.center) -- (l5.center) -- (l6.center) -- (l7.center) to[bend left] (l2.center);
	\node[circle,fill=black,inner sep=0pt,minimum size=3.5pt](L2) at (l2){};
	\node[circle,fill=black,inner sep=0pt,minimum size=3.5pt](L3) at (l3){};
	\node[circle,fill=black,inner sep=0pt,minimum size=3.5pt](L4) at (l4){};
	\node[circle,fill=black,inner sep=0pt,minimum size=3.5pt](L5) at (l5){};
	\node[circle,fill=black,inner sep=0pt,minimum size=3.5pt](L6) at (l6){};
	\node[circle,fill=black,inner sep=0pt,minimum size=3.5pt](L7) at (l7){};
	\draw[very thick] (l2.center) -- (l3.center) -- (l4.center) -- (l5.center) -- (l6.center) -- (l7.center);
	\draw[very thick,dashed] (l2.center) to[bend right] (l7.center);
	\node (cl1) at (0.2,0.025){};
	\node (cl2) at (.75,.325){};
	\node (cl3) at (1.3,.45){};
	\node (cl4) at (1.75,.3){};
	\node (cl5) at (1.5,-.05){};
	\node (cl6) at (.75,-.1){};
	\draw[very thick, dashed, grey1] (cl1.center) to[out=40,in=200] (cl2.center) to[out=30,in=180] (cl3.center) to[out=350,in=130] (cl4.center) to[out=300,in=10] (cl5.center) to[out=205,in=350] (cl6.center) to[out=180,in=330] (cl1.center);
	\begin{scope}[shift={(1.2,-.127)}]
		\draw[very thick, grey1] (-.05,-.05) -- (0,0) -- (-.05,.05);
	\end{scope}
	\node (v) at (1.125,.9){};
	\node[circle,fill=black,inner sep=0pt,minimum size=3.5pt] (V) at (v){};
	\node[above] (nameV) at (v){$v$};
	\node (cu1) at (0.35,2.15){};
	\node (cu2) at (0.8,1.4){};
	\node (cu3) at (1.55,1.55){};
	\node (cu4) at (1.8,2.1){};
	\draw[very thick, dashed, grey1] (cu1.center) to[out=290,in=130] (cu2.center) to[out=350,in=210] (cu3.center) to[out=60,in=260] (cu4.center) to[out=150,in=30] (cu1.center);
	\begin{scope}[shift={(1.095,2.33)}]
		\draw[grey1,very thick] (.05,-.05) -- (0,0) -- (.05,.05);
	\end{scope}
	\node (ou1) at (-0.15,2.35){};
	\node (ou2) at (.55,1.1){};
	\node (ou3) at (1.8,1.3){};
	\node (ou4) at (2.1,2.4){};
	\draw[very thick, dashed, grey2] (ou1.center) to[out=290,in=130] (ou2.center) to[out=350,in=210] (ou3.center) to[out=50,in=280] (ou4.center) to[out=150,in=30] (ou1.center);
	\begin{scope}[shift={(1.15,2.705)}]
		\draw[very thick, grey2] (-.05,-.05) -- (0,0) -- (-.05,.05);
	\end{scope}
	\node (ol1) at (-.15,0){};
	\node (ol2) at (1,.65){};
	\node (ol3) at (2.15,0){};
	\draw[dotted, very thick] (l4.center) -- (v.center) -- (u5.center);
	\draw[very thick, dashed, grey2] (ol1.center) to[out=45,in=190] (ol2.center) to[out=20,in=85] (ol3.center) to[out=220,in=320] (ol1.center);
	\begin{scope}[shift={(1.15,-.425)}]
		\draw[very thick,grey2] (.05,.05) -- (0,0) -- (.05,-.05);
	\end{scope}
	\end{scope}
\begin{scope}[shift={(0.5,1)}]
	\draw[thick,->] (0.1,0) -- (.5,0);
\end{scope}
	\begin{scope}[shift={(4.,0)},rotate=90]
		\node (u1) at (0,2.5){};
		\node (u2) at (0,2.3){};
		\node (u3) at (0.26,1.8){};
		\node (u4) at (0.65,1.2){};
		\node (u5) at (1.25,1.2){};
		\node (u6) at (1.7,1.4){};
		\node (u7) at (2,1.8){};
		\node (u8) at (2,2.3){};
		\fill[grey2] (u2.center) -- (u3.center) -- (u4.center) -- (u5.center) -- (u6.center) -- (u7.center) -- (u8.center)	 to[bend right] (u2.center);
		\node[circle,fill=black,inner sep=0pt,minimum size=3.5pt](v1) at (u2){};
		\node[circle,fill=black,inner sep=0pt,minimum size=3.5pt](v1) at (u3){};
		\node[circle,fill=black,inner sep=0pt,minimum size=3.5pt](v1) at (u4){};
		\node[circle,fill=black,inner sep=0pt,minimum size=3.5pt](v1) at (u5){};
		\node[circle,fill=black,inner sep=0pt,minimum size=3.5pt](v1) at (u6){};
		\node[circle,fill=black,inner sep=0pt,minimum size=3.5pt](v1) at (u7){};
		\node[circle,fill=black,inner sep=0pt,minimum size=3.5pt](v8) at (u8){};
		\draw[very thick] (u2.center) --  (u3.center) -- (u4.center) -- (u5.center) -- (u6.center) -- (u7.center) -- (u8.center);
		\draw[very thick, dashed] (u8.center) to[bend right] (u2.center);
		\node (l1) at (0,0){};
		\node (l2) at (0,0){};
		\node (l3) at (0.5,0.4){};
		\node (l4) at (1,0.5){};
		\node (l5) at (1.5,.6){};
		\node (l6) at (1.9,.4){};
		\node (l7) at (2,0){};
		\fill[grey2] (l2.center) -- (l3.center) -- (l4.center) -- (l5.center) -- (l6.center) -- (l7.center) to[bend left] (l2.center);
		\node (cl1) at (0.2,0.025){};
		\node (cl2) at (.75,.325){};
		\node (cl3) at (1.3,.45){};
		\node (cl4) at (1.75,.3){};
		\node (cl5) at (1.5,-.05){};
		\node (cl6) at (.75,-.1){};
		\draw[very thick, dashed, grey1] (cl1.center) to[out=40,in=200] (cl2.center) to[out=30,in=180] (cl3.center) to[out=350,in=130] (cl4.center) to[out=300,in=10] (cl5.center) to[out=205,in=350] (cl6.center) to[out=180,in=330] (cl1.center);
		\begin{scope}[shift={(1.2,-.127)}]
			\draw[very thick, grey1] (-.05,-.05) -- (0,0) -- (-.05,.05);
		\end{scope}
		\node[circle,fill=black,inner sep=0pt,minimum size=3.5pt](L2) at (l2){};
		\node[circle,fill=black,inner sep=0pt,minimum size=3.5pt](L3) at (l3){};
		\node[circle,fill=black,inner sep=0pt,minimum size=3.5pt](L4) at (l4){};
		\node[circle,fill=black,inner sep=0pt,minimum size=3.5pt](L5) at (l5){};
		\node[circle,fill=black,inner sep=0pt,minimum size=3.5pt](L6) at (l6){};
		\node[circle,fill=black,inner sep=0pt,minimum size=3.5pt](L7) at (l7){};
		\draw[very thick] (l2.center) -- (l3.center) -- (l4.center) -- (l5.center) -- (l6.center) -- (l7.center);
		\node (v) at (1.125,.9){};
		\node[circle,fill=black,inner sep=0pt,minimum size=3.5pt] (V) at (v){};
		\node[above] (nameV) at (v){$v$};
		\node (ou1) at (-0.15,2.35){};
		\node (ou2) at (.55,1.1){};
		\node (ou3) at (1.8,1.3){};
		\node (ou4) at (2.1,2.4){};
		\begin{scope}[shift={(1.15,2.705)}]
			\draw[very thick, grey2] (-.05,-.05) -- (0,0) -- (-.05,.05);
		\end{scope}
		\node (ol1) at (-.15,0){};
		\node (ol2) at (1,.65){};
		\node (ol3) at (2.15,0){};
		\node (new1) at (.9,.65){};
		\node (new2) at (1.4,.7){};
		\node (new3) at (1,1){};
		\node (new4) at (1.5,1.1){};
		\draw[dotted, ultra thick] (l4.center) -- (v.center) -- (u5.center);
		\begin{scope}[shift={(1.15,-.425)}]
			\draw[very thick,grey2] (.05,.05) -- (0,0) -- (.05,-.05);
		\end{scope}
		\draw[very thick] (l4.center) -- (v.center) -- (u5.center);
		\draw[very thick,dashed] (l7.center) to[bend left] (l2.center);
		\node (cu1) at (0.35,2.15){};
		\node (cu2) at (0.8,1.4){};
		\node (cu3) at (1.55,1.55){};
		\node (cu4) at (1.8,2.1){};
		\draw[very thick, dashed, grey2] (ou1.center) to[out=290,in=130] (ou2.center) to[out=350,in=170] (new3.center) to[out=250,in=70] (new1.center) to[out=180,in=45] (ol1.center) to[out=320,in=220] (ol3.center) to[out=90,in=350] (new2.center) to[out=70,in=250] (new4.center) to[out=30,in=210] (ou3.center) to[out=50,in=280] (ou4.center) to[out=150,in=30] (ou1.center);
		\draw[very thick, dashed, grey1] (cu1.center) to[out=290,in=130] (cu2.center) to[out=350,in=210] (cu3.center) to[out=60,in=260] (cu4.center) to[out=150,in=30] (cu1.center);
		\begin{scope}[shift={(1.095,2.33)}]
			\draw[grey1,very thick] (.05,-.05) -- (0,0) -- (.05,.05);
		\end{scope}
	\end{scope}
\end{tikzpicture}

%% file: IntersectingSpheres.tex
		
	\tikzset{
		partial ellipse/.style args={#1:#2:#3}{
			insert path={+ (#1:#3) arc (#1:#2:#3)}
		}
	}
	\begin{tikzpicture}[scale=.75]
	\draw[dashed] (2.5,0) [partial ellipse=0:180:1.8 and 0.5];
	\draw[dotted] (2.5,0) -- (4.3,0);
	\draw[fill=grey2, very thick, opacity = .7] (0,0) ellipse (2cm and 1cm);
	\begin{scope}[shift={(1.55,.75)}]
		\begin{scope}[rotate=-25]
			\draw[very thick] (0,0) ellipse (3mm and 5mm);
		\end{scope}
	\end{scope}
	\draw (.4,1.8) -- (1.8,1.2);
	\draw (.4,1.8) -- (1.35,.3);
	\draw (2.5,0) -- (1.8,1.2);
	\draw (2.5,0) -- (1.35,.3);
	\draw[fill=grey2, very thick, opacity = .7] (2.5,0) circle (1.8cm);
	\draw (2.5,0) [partial ellipse=180:360:1.8 and 0.5];
	\draw[very thick] (0,0) [partial ellipse=180:333:2 and 1];
	\draw[grey2,fill=grey2] (2.5,0) circle (.025);
	\node[black!60] (c) at (1.9,1.3) {$c$};
	\node (s) at (-1.25,-1.2){$S$};
	\node (vmw) at (2.5,-.25){$v_{\text{new}}$};
	\node (d) at (3.4,.2){$d$};
	\draw[black,fill=black] (.4,1.8) circle (.025);
	\node (v) at (.4,2){$v$};
	\draw[dotted] (.4,1.8) -- (2.5,0);
	\draw[fill=black] (1.35,.3) circle (.05);
	\node(d1) at (1.1,1.7){$d$};
	\node(d2) at (2.25,.8){$d$};
\end{tikzpicture}

%% file: AddingNewVertex.tex
\begin{tikzpicture}
	\coordinate (c1) at (0.5,0.5);
	\coordinate (c2) at (6.5,.75);
	\coordinate (c3) at (7,5);
	\coordinate (c4) at (1.5,5);
	\draw[black,very thick,fill=grey1,opacity=0.7] (c1) to[out=22.5, in=160] (c2) to [out=95,in=240] (c3) to[out=165,in=20.5] (c4) to[out=220,in=85] (c1);
	\coordinate (v1) at (1.8,3.4);
	\coordinate (v2) at (2.9,3.5);
	\coordinate (v3) at (2.25,4.45);
	\coordinate (v4) at (1.8,2.3);
	\coordinate (v5) at (2.9,1.75);
	\coordinate (v6) at (4,1.6);
	\coordinate (v7) at (4.7,2.2);
	\coordinate (v8) at (3.3,2.6);
	\draw[fill=black] (v1) circle (2pt);
	\draw[fill=black] (v2) circle (2pt);
	\draw[fill=black] (v3) circle (2pt);
	\draw[fill=black] (v4) circle (2pt);
	\draw[fill=black] (v5) circle (2pt);
	\draw[fill=black] (v6) circle (2pt);
	\draw[fill=black] (v7) circle (2pt);
	\draw[fill=black] (v8) circle (3pt);
	\draw[very thick, black] (v2) -- (v1) -- (v3) -- (v2);
	\draw[very thick, black] (v1) -- (v4) -- (v5) -- (v6) -- (v7);
	\draw[very thick, black, dotted] (v2) -- (v8) -- (v5);
	\node(M) at (6.3,4.8){$\mathcal{M}$};
\end{tikzpicture}

%% file: CollectCandidatesTopview.tex
\begin{tikzpicture}
	\coordinate (e1) at (-.5,.3);
	\coordinate (e2) at (7.5,.3);
	\coordinate (e3) at (8,.8);
	\coordinate (e4) at (8,6.2);
	\coordinate (e5) at (7.5,6.7);
	\coordinate (e6) at (-.5,6.7);
	\coordinate (e7) at (-1,6.2);
	\coordinate (e8) at (-1,.8);
	\draw[very thick,fill=grey1] (e1) -- (e2) to[out=0, in=270] (e3) -- (e4) to[out=90, in=0] (e5) -- (e6) to[out=180, in=90] (e7) -- (e8) to[out=270, in=180] (e1);
	\coordinate (i1) at (2.5,2.5);
	\coordinate (i2) at (4.5,2.5);
	\coordinate (i3) at (4.5,4.5);
	\coordinate (i4) at (2.5,4.5);
	\coordinate (a1) at (2.5,.5);
	\coordinate (a2) at (4.5,.5);
	\coordinate (a3) at (6.5,2.5);
	\coordinate (a4) at (6.5,4.5);
	\coordinate (a5) at (4.5,6.5);
	\coordinate (a6) at (2.5,6.5);
	\coordinate (a7) at (.5,4.5);
	\coordinate (a8) at (.5,2.5);
	\begin{scope}[shift={(2.8,3.2)}]
		\draw[fill=grey2,opacity=0.7] (0,0) circle (1);
		\draw[fill=black] (0,0) circle (.1);
	\end{scope}
	\begin{scope}[shift={(-.2,2)}]
		\draw[fill=grey2,opacity=0.7] (0,0) circle (1);
		\draw[fill=black] (0,0) circle (.1);
	\end{scope}
	\begin{scope}[shift={(2.5,1.6)}]
		\draw[fill=grey2,opacity=0.7] (0,0) circle (1);
		\draw[fill=black] (0,0) circle (.1);
	\end{scope}
	\begin{scope}[shift={(7.6,1.75)}]
		\draw[fill=grey2,opacity=0.7] (0,0) circle (1);
		\draw[fill=black] (0,0) circle (.1);
	\end{scope}
	\begin{scope}[shift={(6.5,3.5)}]
		\draw[fill=grey2,opacity=0.7] (0,0) circle (1);
		\draw[fill=black] (0,0) circle (.1);
	\end{scope}
	\node (bInterM) at (5.15,2.5){$\mathcal{M} \cap b$};
	\draw[very thick, dashed] (i1) -- (i2) -- (i3) -- (i4) -- (i1);
	\draw[very thick,dotted] (a1) -- (a2) to[out=0,in=270] (a3) -- (a4) to[out=90,in=0] (a5) -- (a6) to[out=180,in=90] (a7) -- (a8) to[out=270,in=180] (a1);
	\node (pi) at (3.2,3.15){$p_i$};
	\node (pj) at (-.55,2){$p_j$};
	\node (pk) at (2.5,1.3){$p_k$};
	\node (pm) at (7.6,1.45){$p_m$};
	\node (pl) at (6.8,3.5){$p_l$};
	\begin{scope}[shift={(2.5,4.65)}]
		\node(null) at (-.1,0){};
		\node(eins) at (2.1,0){};
		\draw[thick,decorate,decoration=brace] (null) -- (eins) node[midway,yshift=10pt]{$d$};
	\end{scope}
	\begin{scope}[shift={(4.5,4.5)},rotate=45]
		\draw[dotted] (0,0) -- (2,0);
		\begin{scope}[rotate=0]
			\draw[decorate,decoration=brace] (0,0.05) -- (2,0.05) node[midway,yshift=10pt]{$d$};
		\end{scope}
	\end{scope}
	\node(M) at (7.5,6.2){$\mathcal{M}$};
\end{tikzpicture}

%% file: CollectingSplats.tex
\begin{tikzpicture}
	\begin{scope}[scale=.7,shift={(3.45,2.5)}]
		\coordinate (e1) at (1.5,0);
		\coordinate (e2) at (2.5,.5);
		\coordinate (e6) at (0,.5);
		\coordinate (e7) at (1,1);
		\coordinate (c1) at (1.5,.75);
		\coordinate (c2) at (2.5,1.5);
		\coordinate (c3) at (1,2);
		\coordinate (c4) at (0,1.25);
		\draw[thick, black] (e1) -- (e2) -- (e7) -- (e6) -- (e1);
		\draw[thick, black] (e1) -- (c1);
		\draw[thick, black] (e2) -- (c2);
		\draw[thick, black] (e6) -- (c4);
		\draw[thick, black] (e7) -- (c3);
	\end{scope}
	\coordinate (c1) at (0.5,0.5);
\coordinate (c2) at (6.5,.75);
\coordinate (c3) at (7,5);
\coordinate (c4) at (1.5,5);
\draw[black,very thick,fill=grey1,opacity=0.7] (c1) to[out=22.5, in=160] (c2) to [out=95,in=240] (c3) to[out=165,in=20.5] (c4) to[out=220,in=85] (c1);	\coordinate (v1) at (1.8,3.4);
	\coordinate (v2) at (2.9,3.5);
	\coordinate (v3) at (2.25,4.45);
	\coordinate (v4) at (1.8,2.3);
	\coordinate (v5) at (2.9,1.75);
	\coordinate (v6) at (4,1.6);
	\coordinate (v7) at (4.7,2.2);
	\coordinate (v8) at (3.3,2.6);
	\begin{scope}[scale=.7, shift={(3.45,2.5)}]
		\coordinate (e1) at (1.5,0);
		\coordinate (e2) at (2.5,.5);
		\coordinate (e3) at (2.5,2);
		\coordinate (e4) at (1,2.5);
		\coordinate (e5) at (0,2);
		\coordinate (e6) at (0,.5);
		\coordinate (e7) at (1,1);
		\coordinate (e8) at (1.5,1.5);
		\coordinate (c1) at (1.5,.75);
		\coordinate (c2) at (2.5,1.5);
		\coordinate (c3) at (1,2);
		\coordinate (c4) at (0,1.25);
		\draw[draw=grey1,fill=grey2, opacity=.5] (c1) to[out=45, in=200] (c2) to[out=160, in=15] (c3) to[out=210, in=45] (c4) 	to[out=20, in=160] (c1);
		\draw[thick, dotted] (c1) to[out=45, in=200] (c2);
		\draw[thick, dotted] (c2) to[out=160, in=15] (c3);
		\draw[thick, dotted] (c3) to[out=210, in=45] (c4);
		\draw[thick, dotted] (c4) to[out=20, in=160] (c1);
	\end{scope}
	\draw[thick, grey2] (v2) -- (v1) -- (v3) -- (v2);
	\draw[thick, grey2] (v1) -- (v4) -- (v5) -- (v6) -- (v7);
	\draw[thick, grey2] (v2) -- (v8) -- (v5);
	\draw[fill=grey2,grey2] (v1) circle (2pt);
	\draw[fill=grey2,grey2] (v2) circle (2pt);
	\draw[fill=grey2,grey2] (v3) circle (2pt);
	\draw[fill=grey2,grey2] (v4) circle (2pt);
	\draw[fill=grey2,grey2] (v5) circle (2pt);
	\draw[fill=grey2,grey2] (v6) circle (2pt);
	\draw[fill=grey2,black] (v7) circle (2pt);
	\draw[fill=black] (v8) circle (3pt);
	\draw[thick, black] (e3) -- (e4) -- (e5) -- (e8) -- (e3);
	\draw[thick, black] (e8) -- (c1);
	\draw[thick, black] (e3) -- (c2);
	\draw[thick, black] (e4) -- (c3);
	\draw[thick, black] (e5) -- (c4);
	\begin{scope}[scale=.75,shift={(6.8,3.75)}]
		\begin{scope}[rotate=0]
			\draw[grey2,thick,fill=grey2,opacity=.7] (0,0) ellipse (1cm and .4cm);
			\begin{scope}[scale=.5]
				\draw[fill=black] (0,0) -- (.25,0) -- (.25,1.5) -- (.5,1.5) -- (.125,2) -- (-.25,1.5) -- (0,1.5) -- (0,0);
			\end{scope}
		\end{scope}
	\end{scope}
	\begin{scope}[scale=.7,shift={(5.4,3.8)},rotate=-10]
		\coordinate (dist1) at (0,0);
		\coordinate (dist2) at (1.5,.5);
		\draw[black,dotted] (dist1) -- (dist2);
		\node (dist) at (.73,.5){$d$};
	\end{scope}
	\node(M) at (6.3,4.8){$\mathcal{M}$};
\end{tikzpicture}

%% file: NewCandidate.tex
\begin{tikzpicture}
	\coordinate (c1) at (0.5,0.5);
	\coordinate (c2) at (6.5,.75);
	\coordinate (c3) at (7,5);
	\coordinate (c4) at (1.5,5);
	\draw[black,very thick,fill=grey1,opacity=0.7] (c1) to[out=22.5, in=160] (c2) to [out=95,in=240] (c3) to[out=165,in=20.5] (c4) to[out=220,in=85] (c1);
	\coordinate (v1) at (1.8,3.4);
	\coordinate (v2) at (2.9,3.5);
	\coordinate (v3) at (2.25,4.45);
	\coordinate (v4) at (1.8,2.3);
	\coordinate (v5) at (2.9,1.75);
	\coordinate (v6) at (4,1.6);
	\coordinate (v7) at (4.7,2.2);
	\coordinate (v8) at (3.3,2.6);
	\draw[thick, grey2] (v2) -- (v1) -- (v3) -- (v2);
	\draw[thick, grey2] (v1) -- (v4) -- (v5) -- (v6) -- (v7);
	\draw[thick, grey2] (v2) -- (v8) -- (v5);
	\draw[fill=grey2,grey2] (v1) circle (2pt);
	\draw[fill=grey2,grey2] (v2) circle (2pt);
	\draw[fill=grey2,grey2] (v3) circle (2pt);
	\draw[fill=grey2,grey2] (v4) circle (2pt);
	\draw[fill=grey2,grey2] (v5) circle (2pt);
	\draw[fill=grey2,grey2] (v6) circle (2pt);
	\draw[fill=grey2,black] (v7) circle (2pt);
	\draw[fill=black] (v8) circle (3pt);
	\begin{scope}[scale=.75,shift={(6.8,3.75)}]
		\begin{scope}[rotate=0]
			\draw[grey2,thick,fill=grey2,opacity=.7] (0,0) ellipse (1cm and .4cm);
			\begin{scope}[scale=.5]
				\draw[fill=black] (0,0) -- (.25,0) -- (.25,1.5) -- (.5,1.5) -- (.125,2) -- (-.25,1.5) -- (0,1.5) -- (0,0);
			\end{scope}
		\end{scope}
	\end{scope}
	\begin{scope}[scale=.7,shift={(4.8,3.72)},rotate=-10]
		\coordinate (dist1) at (0,0);
		\coordinate (dist2) at (1.5,.5);
		\draw[black, dotted] (dist1) -- (dist2);
		\node (dist) at (.73,.5){$d$};
	\end{scope}
	\begin{scope}[scale=.6,shift={(v7)}]
		\coordinate (begin) at (0,0);
		\coordinate (end) at (-0.4,.95);
		\draw[dotted] (begin) -- (end); 
		\node (d) at (.05,.65){$d$};
	\end{scope}
	\tikzset{
		partial ellipse/.style args={#1:#2:#3}{
			insert path={+ (#1:#3) arc (#1:#2:#3)}
		}
	}
	\begin{scope}[shift={(4,2.4)}]
		\draw[grey2] (0,0) [partial ellipse=90:450:.5cm and 1cm];
		\draw (0,0) [partial ellipse=21:173:.5cm and 1cm];
		\coordinate (secSplat1) at (-.5,.1);
		\coordinate (secSplat2) at (.475,.35);
		\draw[grey2, dashed] (secSplat1) -- (secSplat2);
		\draw[fill=black] (secSplat1) circle (.5pt);
		\draw[fill=black] (secSplat2) circle (.5pt);
		\coordinate (secMfd1) at (-.5,-.1);
		\coordinate (secMfd2) at (.485,.195);
		\draw[grey2, dashed] (secMfd1) to[out=25,in=175] (secMfd2);
		\draw[grey2, fill=grey2] (secMfd1) circle (.5pt);
		\draw[grey2, fill=grey2] (secMfd2) circle (.5pt);
	\end{scope}
		\node(M) at (6.3,4.8){$\mathcal{M}$};
\end{tikzpicture}

%% file: BorderZero.tex
\begin{tikzpicture}[scale=1]
	
	\coordinate (b0) at (-2.5,-2);
	\coordinate (b1) at (-1.5,-1.6);
	\coordinate (b2) at (-0.4,-1.7);
	\coordinate (b3) at (1,-1.3);
	\coordinate (b4) at (1.5,-.55);
	\coordinate (b5) at (2.5,-.1);
	\coordinate (b6) at (2.5,-2);
	\coordinate (b7) at (2.5,1.7);
	\coordinate (b8) at (-2.5,1.7);
	\fill[nearlywhite] (b0) -- (b6) -- (b7) -- (b8) -- cycle;
	\fill[grey2] (b0) -- (b1) -- (b2) -- (b3) -- (b4) -- (b5) -- (b6) -- (b0);
	\draw[very thick, black] (b0) -- (b1) -- (b2) -- (b3) -- (b4) -- (b5);
	\node[fill=black,circle,inner sep=0pt,minimum size=3.5pt] (B1) at (b1){};
	\node[fill=black,circle,inner sep=0pt,minimum size=3.5pt] (B2) at (b2){};
	\node[fill=black,circle,inner sep=0pt,minimum size=3.5pt] (B3) at (b3){};
	\node[fill=black,circle,inner sep=0pt,minimum size=3.5pt] (B4) at (b4){};
	\node[circle,fill=black,inner sep=0pt,minimum size=3.5pt](v1) at (0,0){};
	\node[right] (v) at (0,0){$v$};
	\begin{scope}[shift={(.75,.75)}]
			\begin{scope}[rotate=90]
			\draw[dashed,grey1,very thick,<-] (0,.75) arc (0:360:.75);
		\end{scope}
	\end{scope}

\end{tikzpicture}

%% file: BorderTwo.tex
\begin{tikzpicture}[scale=1]
	\coordinate (r0) at (-2.5,-2);
	\coordinate (r1) at (2.5,-2);
	\coordinate (r2) at (2.5,1.7);
	\coordinate (r3) at (-2.5,1.7);
	\fill[nearlywhite] (r0) -- (r1) -- (r2) -- (r3) -- cycle;
	\coordinate (b0) at (-2.5,-2);
	\coordinate (b1) at (-1.5,-1.6);
	\coordinate (b2) at (-0.4,-1.7);
	\coordinate (b3) at (1,-1.3);
	\coordinate (b4) at (1.5,-.55);
	\coordinate (b5) at (2.5,-.1);
	\coordinate (b6) at (2.5,-2);
	\coordinate (b7) at (2.5,1.7);
	\coordinate (b8) at (-2.5,1.7);
	\fill[nearlywhite] (b0) -- (b6) -- (b7) -- (b8) -- cycle;
	\fill[grey2] (b0) -- (b1) -- (b2) -- (b3) -- (b4) -- (b5) -- (b6) -- (b0);
	\draw[very thick, black] (b0) -- (b1) -- (b2) -- (b3) -- (b4) -- (b5);
	\node[fill=black,circle,inner sep=0pt,minimum size=3.5pt] (B1) at (b1){};
	\node[fill=black,circle,inner sep=0pt,minimum size=3.5pt] (B2) at (b2){};
	\node[fill=black,circle,inner sep=0pt,minimum size=3.5pt] (B3) at (b3){};
	\node[fill=black,circle,inner sep=0pt,minimum size=3.5pt] (B4) at (b4){};
	
	\begin{scope}[rotate=15]
		\node[circle,fill=black,inner sep=0pt,minimum size=3.5pt](v1) at (-.6,0){};
		\node[circle,fill=black,inner sep=0pt,minimum size=3.5pt](v2) at (.6,0){};
		\node[circle,fill=black,inner sep=0pt,minimum size=3.5pt](v3) at (0,.3){};
		\coordinate (l1) at (-.6,-.5);
		\coordinate (l2) at (.6,-.55);
		\coordinate (l3) at (.6,.75);
		\coordinate (l4) at (-.6,.75);
		\coordinate (m1) at (0,-.25);
		\coordinate (m2) at (0,1.25);
		\draw[grey1, very thick, dashed,<-] (m1) -- (l2) to[out=0,in=-30,distance=.8cm] (l3)  -- (m2);
		\draw[grey1, very thick, dashed,<-] (m2) -- (l4) to[out=210,in=180,distance=.8cm] (l1) -- (m1);
		\node[left] (V1) at (v1){$v$};
		\node[right] (V2) at (v2){$v'$};
		\node[above] (V3) at (v3){$v_{\text{new}}$};
		\draw[very thick] (v1) -- (v3) -- (v2);
	\end{scope}
\end{tikzpicture}

%% file: BorderK.tex
\begin{tikzpicture}[scale=1]
	\coordinate (b0) at (-2.5,-2);
	\coordinate (b1) at (-1.5,-1.6);
	\coordinate (b2) at (-0.4,-1.7);
	\coordinate (b3) at (1,-1.3);
	\coordinate (b4) at (1.5,-.55);
	\coordinate (b5) at (2.5,-.1);
	\coordinate (b6) at (2.5,-2);
	\coordinate (b7) at (2.5,1.7);
	\coordinate (b8) at (-2.5,1.7);
	\fill[nearlywhite] (b0) -- (b6) -- (b7) -- (b8) -- cycle;
	\fill[grey2] (b0) -- (b1) -- (b2) -- (b3) -- (b4) -- (b5) -- (b6) -- (b0);
	\draw[very thick, black] (b0) -- (b1) -- (b2) -- (b3) -- (b4) -- (b5);
	\node[fill=black,circle,inner sep=0pt,minimum size=3.5pt] (B1) at (b1){};
	\node[fill=black,circle,inner sep=0pt,minimum size=3.5pt] (B2) at (b2){};
	\node[fill=black,circle,inner sep=0pt,minimum size=3.5pt] (B3) at (b3){};
	\node[fill=black,circle,inner sep=0pt,minimum size=3.5pt] (B4) at (b4){};
	\coordinate (vk) at (-1.5,.5);
	\coordinate (v0) at (-1.3,-.5);
	\coordinate (v1) at (-.5,-1);
	\coordinate (v2) at (.5,-.7);
	\coordinate (v3) at (.9,0);
	\coordinate (v4) at (.5,1);
	\fill[grey1] (vk) -- (v0) -- (v1) -- (v2) -- (v3) -- (v4) to[bend right] (vk);
	\node[circle,fill=black,inner sep=0pt,minimum size=3.5pt](Vk) at (vk){};
	\node[circle,fill=black,inner sep=0pt,minimum size=3.5pt](V0) at (v0){};
	\node[circle,fill=black,inner sep=0pt,minimum size=3.5pt](V1) at (v1){};
	\node[circle,fill=black,inner sep=0pt,minimum size=3.5pt](V2) at (v2){};
	\node[circle,fill=black,inner sep=0pt,minimum size=3.5pt](V3) at (v3){};
	\node[circle,fill=black,inner sep=0pt,minimum size=3.5pt](V4) at (v4){};
	\draw[very thick] (vk) -- (v0) -- (v1) -- (v2) -- (v3) -- (v4);
	\draw[very thick, dashed] (v4) to[bend right] (vk);
	\node[left] (namen)  at (vk){$v_k$};
	\node[left] (name0)  at (v0){$v_0$};
	\node[below] (name1) at (v1){$v_1$};
	\node[below right] (name2) at (v2){$v_2$};
	\node[right] (name3) at (v3){$v_3$};
	\node[right] (name4) at (v4){$v_4$};
	\draw[very thick,dashed,grey2,->] (-.3,.85) arc (90:450:.7);
	\coordinate (nexttov0second) at (-1.5,-.7);
	\coordinate (nexttov1first)  at (-.7,-1.25);
	\draw[dashed,very thick,grey1,<-] (nexttov0second) to[bend right] (nexttov1first);
	\coordinate (nexttov1second) at (-.3,-1.3);
	\coordinate (nexttov2first) at (.5,-1.05);
	\draw[very thick, dashed, grey1,<-] (nexttov1second) to[bend right] (nexttov2first);
	\coordinate (nexttov2second) at (.87,-.7);
	\coordinate (nexttov3first)  at (1.15,-.1);
	\draw[very thick, grey1, dashed,<-] (nexttov2second) to [bend right] (nexttov3first);
	\coordinate (nexttov3second) at (1.2,0.15);
	\coordinate (nexttov4first) at (.9,1.1);
	\draw[very thick, dashed, grey1,<-] (nexttov3second) to[bend right] (nexttov4first);
	\coordinate (nexttov0first) at (-1.7,-.3);
	\coordinate (nexttovnsecond) at (-1.8,.3);
	\draw[very thick, dashed, grey1,->] (nexttov0first) to[bend left] (nexttovnsecond);
	\coordinate (nexttovnfirst) at (-1.66,.8);
	\coordinate (nexttov4second) at (.6,1.37);
	\draw[very thick, dashed, grey1,->] (nexttovnfirst) to[bend left] (nexttov4second);

\end{tikzpicture}

%% file: sm_borderToTriangulate.tex
%
%
	\begin{tikzpicture}
		\draw[sowaswieweiss] (-11,-.5) -- (10.5,-.5) -- (10.5,6) -- (-11,6) -- cycle;
		\coordinate (v1) at (2.5,1);
		\coordinate (v2) at (8,.5);
		\coordinate (v3) at (9.5,3);
		\coordinate (v4) at (9,5);
		\coordinate (v5) at (4,5);
		\coordinate (v6) at (0,4);
		\coordinate (v7) at (6,3);
		\coordinate (v8) at (-3,2);
		\coordinate (v9) at (-6.5,2.5);
		\coordinate (v10) at (-9.5,1.5);
		\coordinate (v11) at (-5,1);
		\coordinate (v12) at (-2,3);
		\draw[thin]  (v6) -- (v8) -- (v9);
		\draw[draw=adarkergrey,fill=adarkergrey] (v1) -- (v2) -- (v3) -- (v4) -- (v5) -- (v6) -- (v7) -- (v1);
		\draw[draw=adarkergrey,fill=adarkergrey] (v2) -- (v1) -- (v11) -- (v10) -- (-10,0.5) -- (v2);
		\draw[draw=adarkergrey,fill=adarkergrey] (-10,0.5) -- (v10) -- (v9) -- (v8) -- (v6) -- (v5) -- (-10,3) -- (-10,.5);
		\draw[very thick] (v9) -- (v8) -- (v6) -- (v7);
		\draw[draw=adarkergrey,fill=adarkergrey,opacity=.8] (v1) -- (v11) -- (v12) -- (v1);
		\draw[very thick] (v1) -- (v7);
		\draw[thin] (v2) -- (v7);
		\draw[thin] (v3) -- (v7);
		\draw[thin] (v4) -- (v7);
		\draw[thin] (v5) -- (v7);
		\draw[thin] (v6) -- (v7);
		\draw[very thick] (v9) -- (v10) -- (v11) -- (v12) -- (v1);
		\draw[thin] (v11) -- (v1);
		\draw[thin] (v1) -- (v2);
		\coordinate (S1) at (-3.4482,2.0344);
		\coordinate (S2) at (-1.7,2.8666);
		\draw[thin] (S2) -- (v6);
		\draw[very thick,->] (v7) -- (6.75,4.5);
		\draw[fill = black] (v7) circle(.05);
		\node (normal) at (7,4.5){$n_b$};
		\node (p) at (5.95,2.75){$p$};
		\node (border) at (4.5,1.75){$\partial R$};
		\draw[thin] (v1) -- (4,.5);
		\draw[thin] (v1) -- (1.5,.5);
		\draw[thin] (v11) -- (-3.5,.5);
		\draw[thin] (v11) -- (-6,.5);
		\draw[thin] (v10) -- (-7.5,.5);
		\draw[thin] (v10) -- (-10,1.25);
		\draw[thin] (v10) -- (-10,1.75);
		\draw[thin] (v9) -- (-10,3);
		\draw[thin] (v9) -- (v6);
		\draw[thin] (v6) -- (-10,3);
		\draw[thin] (v6) -- (v5);
 		\end{tikzpicture}

%% file: sm_cuttingSmallestAngle.tex
%
%
	\begin{tikzpicture}
		\draw[sowaswieweiss] (-1,-1) -- (11,-1) -- (11,6.5) -- (-1,6.5) -- cycle;
		\coordinate (v1) at (2.5,1);
		\coordinate (v2) at (8,.5);
		\coordinate (v3) at (9.5,3);
		\coordinate (v4) at (9,5);
		\coordinate (v5) at (4,5);
		\coordinate (v6) at (0,4);
		\coordinate (v7) at (6,3);
		\draw[draw=adarkergrey,fill=adarkergrey] (v1) -- (v2) -- (v3) -- (v4) -- (v5) -- (v6) -- (v7) -- (v1);
		\draw[thin] (v1) -- (v7);
		\draw[thin] (v2) -- (v7);
		\draw[thin] (v3) -- (v7);
		\draw[thin] (v4) -- (v7);
		\draw[thin] (v5) -- (v7);
		\draw[thin] (v6) -- (v7);
		\coordinate (w1) at (2.5,2);
		\coordinate (w2) at (6.5,1);
		\coordinate (w3) at (8.5,3);
		\coordinate (w4) at (4.5,4);
		\draw[dotted,thick] (4,3.5) -- (4,3.3333);
		\draw[alightergrey,fill=alightergrey,opacity=0.75] (w1) -- (w2) -- (w3) -- (w4) -- (w1);
		\draw[very thick,->] (v7) -- (6.75,4.5);
		\draw[fill = black] (v7) circle(.05);
		\node (normal) at (7,4.5){$n_b$};
		\node (Tp) at (6.425,1.25){$T_p$};
		\node (p) at (6.3,3){$p$};
		\coordinate (b1) at (3,1.875);
		\draw[thin] (v7) -- (b1);
		\draw[thick,dotted] (b1) -- (3,1.2857);
		\coordinate (b2) at (4,3.5);
		\draw[thin] (b2) -- (v7);
		\coordinate (c1) at (5,3.25);
		\coordinate (c2) at (4.5,2.4375);
		\draw[<->] (c1) [out=200,in=100]to (c2);
		\node (alpha) at(5.1,2.9){$\alpha_p$};
	\end{tikzpicture}